\documentclass[11pt,epsfig]{article} 
\usepackage[utf8]{inputenc}
\usepackage[top=1in, left=0.95in, bottom=1.1in, right=0.95in]{geometry}
\usepackage[colorlinks=true,
linkcolor=blue,
urlcolor=red,
citecolor=red]{hyperref}

\usepackage[T1]{fontenc} 

\usepackage[toc]{appendix}
\usepackage{amssymb, amsmath, mathrsfs}
\numberwithin{equation}{section}

\usepackage{graphicx}
\usepackage{tikz}
\usepackage{tikz-3dplot}

\usepackage{caption} 
\usepackage{float}      
\usepackage{placeins}   

\usepackage{cite}

\begin{document}
\begin{center}


{\Large \textbf  {Covariant canonical-spinor amplitudes for partial wave analysis}
}\\[10mm]

Hong Huang $^{a,b,c}$\footnote{huanghong23@mails.ucas.ac.cn}, Yi-Ning Wang $^{ b,c}$\footnote{wangyining@itp.ac.cn},  Jiang-Hao Yu $^{a, b, c, d}$\footnote{jhyu@itp.ac.cn}\\[10mm]

\noindent 
$^a${ \small School of Fundamental Physics and Mathematical Sciences, Hangzhou Institute for Advanced Study, UCAS, Hangzhou 310024, China}  \\
$^b${ \small Institute of Theoretical Physics, Chinese Academy of Sciences,   Beijing 100190, P. R. China}   \\
$^c${ \small School of Physical Sciences, University of Chinese Academy of Sciences,   Beijing 100049, P.R. China} \\
$^d${ \small International Centre for Theoretical Physics Asia-Pacific, Beijing/Hangzhou, China}\\
[10mm]

\end{center}

\begin{abstract}

We propose a covariant orbital-spin decomposed amplitude for the partial wave analysis using the massive spinor-helicity formalism. First we review the traditional-$LS$ method in the little group space and the Zemach tensor method in the double cover of the $\mathrm{SO}(3)$ space. To recover the $\mathrm{SO}(3,1)$ Lorentz covariance, several Lorentz covariant $LS$ tensors have been constructed in several different methods: covariant tensor, covariant projection tensor in pure-spin and general-spin schemes, but performing a intrinsic separation between $LS$ coupling while maintaining covariance is not obvious. We utilize the massive canonical-spinor variables to determine general three-point amplitudes, in which the spin-orbital decomposition is realized in single little group space by projecting little group indices of each particles into one, while the Lorentz covariance is ensured by the spinor form naturally. This covariant spinor method allows direct evaluation in any frame and a streamlined treatment of cascade decays within a single frame without additional alignment rotations in non-covariant treatment. As a benchmark, we implement the method in TF-PWA and analyze $\Lambda_c^+\to\Lambda\pi^+\pi^0$, finding consistent fit results across the helicity, traditional-$LS$, and canonical-spinor amplitudes. This validates the canonical-spinor amplitude as a practical tool for modern partial wave analyses of complex decay chains.

\end{abstract}

\newpage

\tableofcontents  

\section{Introduction}
In high energy physics experiments, multi-body final states are among the most common observables. The same final state often does not originate from a single decay chain, but is produced through several intermediate resonances together with non-resonant processes. Because different resonances can overlap within invariant mass spectra and each exhibits unique angular structures determined by its spin and orbital angular momentum, relying solely on mass peak shapes is generally insufficient to reliably distinguish between them.
Partial Wave Analysis (PWA) provides a systematic framework to address this problem, see Refs.~\cite{Chung:1971ri,Richman:1984gh,Haber:1994pe,Peters:2004qw,Salgado:2013dja} for reviews. Its central idea is to construct the total amplitude built from a set of partial wave amplitudes with definite quantum numbers. 
By exploiting information from invariant mass spectra and angular distributions, PWA can distinguish contributions from different resonances and partial waves, determine their relative magnitudes and phases, and extract information such as fit fraction, interference fractions, and decay asymmetry parameters. 
Therefore, PWA is an essential tool in multi-body hadronic decays for establishing the presence of resonances and determining spin-parity quantum numbers.

In PWA of cascade decays, the total amplitude is typically constructed by connecting a sequence of two-body decay amplitudes via propagators of intermediate resonances. Therefore, the method of building the two-body decay amplitude serves as the foundation of the entire analysis. Early schemes for constructing two-body partial wave amplitudes largely relied on a specific reference frame. Among the most classic is the helicity formalism introduced by Jacob and Wick~\cite{Jacob:1959at,Berman:1965gi,Wick:1962zz}, which defines helicity amplitudes directly in the two-body center-of-momentum (COM) frame. It provides a clear physical picture and has been widely developed and extensively applied in studies of particle polarization~\cite{Richman:1984gh,Haber:1994pe,Boudjema:2009fz,Gao:2010qx,Aguilar-Saavedra:2012bvs,Gratrex:2015hna,Arunprasath:2016tfq,Ajaltouni:2019log,Choi:2019aig}.

The closely related method is the traditional-$LS$ method  in canonical formalism. In the two-body COM frame, it expresses the amplitude as the coupling between the orbital part and the spin part. The angular momentum coupling is carried out in the little group representation fixed by the parent standard momentum, and is implemented explicitly by the Clebsch-Gordan coefficients (CGCs).
This separated structure facilitates the phenomenological inclusion of $L$-dependent barrier factors~\cite{VonHippel:1972fg}. 
It should be emphasized that the helicity basis and the traditional-$LS$ basis are two choices of spin bases for the same two-particle Hilbert space, related by a unitary change of basis, so they yield identical observables when a complete set of partial waves is retained~\cite{Harshman:2004bw}.
However, differences can still arise in practice from different conventions for kinematic and energy dependent factors, and these effects have been explicitly compared~\cite{JPAC:2017vtd,JPAC:2018dfc}.

Besides coupling the orbital and spin parts via CGCs in the little group representation, the Zemach tensor method~\cite{Zemach:1963bc,Zemach:1965ycj} provides an alternative realization of the orbital-spin coupling. It does not use CGCs explicitly in the little group representation. Instead, one first constructs the angular momentum tensors associated with the orbital and spin parts, and then performs the required coupling at the tensor level using the invariant tensors of $\mathrm{SO}(3)$ and $\mathrm{SU}(2)$. These invariant tensors are equivalent in effect to the CGCs coupling in the little group representation. In this way, the Zemach tensor method gives an equivalent tensor realization of the $LS$ partial wave coupling.

However, these non-covariant methods share a fundamental drawback: the definition and calculation of the amplitudes are explicitly dependent on the COM frame. When computing any event, the particle four-momenta must be boosted to this specific reference frame. 
A more subtle yet critical issue arises when the process involves cascade decays with multiple distinct decay chains: the spin states of the same final-state particle in different chains are often defined by different sequences of Lorentz transformations, corresponding to different spin quantization axes and phase conventions. 
Simply adding the amplitudes from different chains coherently may introduce unphysical convention dependence in the interference terms~\cite{Marangotto:2019ucc}. 
To address this issue, the helicity and canonical formalisms differ in their technical treatments~\cite{Habermann:2024sxs}, but they share the same underlying goal: to bring the spin states of the same physical particle appearing in different decay chains into a common choice of quantization axis and phase convention~\cite{Chen:2017gtx,Marangotto:2019ucc,JPAC:2019ufm,Wang:2020giv,Gao:2023jtq,Habermann:2024sxs}. Only after such an alignment of spin states can the amplitudes from different decay chains be coherently summed in a correct and physically meaningful way.

To achieve full covariance for computational simplicity and physical clarity, methods for constructing covariant two-body partial wave amplitudes have been developed. Chung~\cite{Chung:1971ri,Chung:1993da,Chung:1998,Chung:2007nn} pioneered a systematic scheme by constructing covariant orbital and spin tensors, establishing a direct correspondence between helicity coupling amplitudes and the corresponding covariant tensor structures. 
Subsequently, Filippini \textit{et al.}~\cite{Filippini:1995yc} further systematically constructed sets of covariant spin tensor bases suitable for spectroscopic analysis, discussing  their correspondence and equivalence with the covariant helicity approach. 
For practical hadron spectroscopy and charmonium decay analyses, Zou and Bugg~\cite{Zou:2002ar,Zou:2002yy} summarized a widely adopted cookbook style of constructing covariant tensors, referred to as the covariant tensor method. These covariant tensors provide partial wave amplitude expressions for specific decay channels, serving
as one of the main PWA schemes adopted by the BESIII
Collaboration~\cite{BESIII:2023wfi, BESIII:2023sbq, BESIII:2023xac, BESIII:2023htx}. Subsequent research has continuously refined this framework, forming the now mainstream methods of constructing $LS$ amplitudes using covariant tensors~\cite{Filippini:1995yc,Zhu:1999pu,Anisovich:2001ra,Zou:2002ar,Zou:2002yy,Anisovich:2004zz,Dulat:2005in,Anisovich:2006bc,Dulat:2011rn,Matveev:2018gry,Dong:2020mbk,Li:2022qff,Anisovich:2024cmb,Jing:2023rwz,Jing:2024mag}. Besides these methods, another covariant methods that starts from a covariant effective Lagrangian and uses the Feynman rules to systematically generate vertices and propagators, thereby organizing the total amplitude for phenomenological analyses of related reactions and decays~\cite{Benmerrouche:1994uc,Benmerrouche:1996ij,Pascalutsa:1998kcd,Lutz:2003fm,Nakayama:2002mu,Liang:2002tk}.

The methods of constructing $LS$ amplitudes using covariant tensors can essentially be viewed as extending the Zemach tensor method to the Lorentz $\mathrm{SO}(3,1)$ representation. Its key idea is to construct a covariant $LS$ coupling structure and then fully contract it with the external spin wave functions over Lorentz indices, thereby obtaining the corresponding partial wave amplitudes. To obtain a general construction of covariant coupling structures, a systematic approach has been developed based on covariant projection tensors~\cite{Jing:2023rwz,Jing:2024mag}. This approach, referred to as the covariant projection tensor method, provides the required covariant $LS$ coupling structures in a fully general form.

In the covariant projection tensor method, this covariant $LS$ coupling structure can be understood as follows. Taking the standard momentum fixed by the parent momentum as the reference, each angular momentum object is first related between the Lorentz representation and the little group representation by means of the rest frame spin wave functions, so that the angular momentum coupling can be carried out within a common little group representation. After the coupling is completed, the result is mapped back to the Lorentz representation using the rest frame spin wave function associated with the total angular momentum. In this way, one obtains the projection tensor carrying only Lorentz indices, which implements the covariant coupling between the orbital and spin parts.

Depending on the choice of external spin wave functions, all the tensor methods of constructing $LS$ amplitudes, including Zemach tensor, covariant tensor, and covariant projection tensor methods, admit two concrete implementations, referred to as the Pure-spin (PS) scheme and the General-spin (GS) scheme, as follows:
\begin{enumerate}
\item \textbf{Pure-spin scheme (PS-scheme)}~\cite{Chung:1993da,Chung:1998,Chung:2007nn,Li:2022qff,Jing:2023rwz,Jing:2024mag}:
\begin{equation}
  A({k}_3,{p}_1^*,{p}_2^*;L,S) = \underbrace{\Gamma({k}_3,{p}_1^*,{p}_2^*;L,S)}_{\text {$LS$ coupling structure}}\times \underbrace{U({k}_3,{k}_1,{k}_2;s_{3},s_{1},s_{2})}_{\text {pure-spin part}}.
\end{equation}
In this definition, the COM frame amplitude is organized such that the pure-spin part is the pure intrinsic-spin object. As a result, the pure-spin part carries no explicit dependence on the orbital, and the $LS$ separation is explicit in the COM frame, with all angular dependence encoded in the orbital structures.
By applying Lorentz boost, this amplitude can be evaluated in any frame, although the COM momenta are still necessarily introduced. Specifically, the amplitude can be written as
\begin{equation}
\begin{aligned}
  A({p}_3,{p}_1,{p}_2;L,S) 
    =&\Gamma(p_{3},p_{1},p_{2};L,S)\times U({p}_3,{p}_1,{p}_2;s_{3},s_{1},s_{2})  \\
&\times D^{s_1}\Big(L(\mathbf{p}_3)L(\mathbf p_1^\ast)L^{-1}(\mathbf{p}_3)\Big)D^{s_2}\Big(L(\mathbf{p}_3)L(\mathbf p_2^\ast)L^{-1}(\mathbf{p}_3)\Big),
\end{aligned}
\label{eq:psamplitude}
\end{equation}
where  $D^{s_i}\Big(L(\mathbf{p}_3)L(\mathbf p_i^\ast)L^{-1}(\mathbf{p}_3)\Big)$ $(i=1,2)$ are the Lorentz transformation matrices induced by the boost that takes the amplitude from the COM frame to any frame. Although it can be evaluated in any reference frame, the associated Lorentz transformation depends on the particle momenta $\mathbf p_i^\ast$ $(i=1,2)$ in the COM frame. Therefore, the PS-scheme is not fully manifestly covariant at the level of its formal definition.

\item \textbf{General-spin scheme (GS-scheme)}~\cite{Filippini:1995yc,Zhu:1999pu,Anisovich:2001ra,Zou:2002ar,Zou:2002yy,Anisovich:2004zz,Dulat:2005in,Anisovich:2006bc,Dulat:2011rn,Matveev:2018gry,Dong:2020mbk,Jing:2023rwz,Anisovich:2024cmb}:
\begin{equation}
  \mathcal{C}({p}_3,{p}_1,{p}_2;L,S) = \underbrace{\Gamma({p}_3,{p}_1,{p}_2;L,S)}_{\text {covariant $LS$ coupling structure}}\times \underbrace{U({p}_3,{p}_1,{p}_2;s_{3},s_{1},s_{2})}_{\text {general-spin part}}.
\end{equation}
The resulting expressions are manifestly covariant in the sense that they can be evaluated directly in any frame using the corresponding four-momenta.
In this definition, the general-spin part is constructed directly from general-spin structures evaluated with the momenta 
\begin{equation}
    U({p}_3,{p}_1,{p}_2;s_{3},s_{1},s_{2})
    = u(p_3;s_{3})\bar{u}(p_1;s_{1})\bar{u}(p_2;s_{2}) 
    \xrightarrow{\mathrm{COM}} u(k_3;s_{3})\bar{u}(p_1^*;s_{1})\bar{u}(p_2^*;s_{2}). 
\end{equation}
Note that in the COM frame, the general-spin part depends on the relative momentum direction and thus contain residual contributions from the orbital part. 
Consequently, the separation between $L$ and $S$ does not follow the traditional-$LS$ method, since the general-spin part here is not purely intrinsic-spin objects.
Only in the non-relativistic limit, the general-spin part is reduced to $u(k_3;s_{3})\bar{u}(k_1;s_{1})\bar{u}(k_2;s_{2})$, which is the pure spin part $U({k}_3,{k}_1,{k}_2;s_{3},s_{1},s_{2})$ in Eq.~\eqref{eq:psamplitude}, and the covariant amplitude would have clear $LS$ separation.   

\end{enumerate}
Therefore, it can be seen that neither scheme simultaneously provides a manifestly covariant construction that avoids boosts to the two-body COM frame and an $LS$ separation that is identical to the traditional-$LS$ definition. Moreover, when massless gauge bosons are involved, formulations based on polarization vectors, or analogous wave functions, typically require gauge invariance to be enforced explicitly. In practice, one must select a complete set of gauge invariant and linearly independent vertex structures in order to eliminate gauge redundancy and avoid redundant fit parameters, which further increases the implementation complexity.

These limitations motivate the development of a method that simultaneously satisfies:
\begin{itemize}
\item Manifest covariance in the definition, with no reliance on the COM.
\item A complete and clean separation between orbital angular momentum and total spin.
\item A general formula for partial wave amplitudes applicable to particles of any spin, including the case with massless particles.
\end{itemize}
This is precisely the goal of the theoretical framework developed in this work.
Its starting point is an on-shell formulation, where Lorentz covariance is manifest and spin is encoded through $\mathrm{SU}(2)$ little group covariance.
In the on-shell framework, amplitudes are constructed directly in terms of spinor-helicity variables~\cite{Parke:1986gb,Bern:1996je,Dixon:1996wi,Elvang:2013cua,Cheung:2017pzi,Travaglini:2022uwo,Badger:2023eqz}, which removes huge gauge redundancy and trivializes the kinematical on-shell constraints, thereby making explicit calculations much more efficient. For massless particles, three-point amplitudes are entirely fixed by the consistency conditions from little group scaling~\cite{Benincasa:2007xk,Witten:2003nn}, while higher-point amplitudes can be determined from factorization, imposed by locality and unitarity, using gluing techniques or recursive relations~\cite{Britto:2004ap,Britto:2005fq}.

The spinor-helicity formalism has subsequently been extended to processes involving massive particles~\cite{Kleiss:1985yh,Hagiwara:1985yu,Kleiss:1988xr,Dittmaier:1998nn,Schwinn:2005pi,Schwinn:2006ca,Badger:2005jv,Badger:2005zh,Conde:2016vxs,Conde:2016izb,Basile:2024ydc}. In particular, Arkani-Hamed, Huang and Huang~\cite{Arkani-Hamed:2017jhn} developed an on-shell, little group covariant representation of three-point amplitudes for particles of arbitrary mass and spin, formulated in terms of spinor-helicity variables.
In this framework, the four-momentum of a massive particle is written as $p_{\alpha\dot{\alpha}}=\lambda_{\alpha I}\tilde\lambda^{I}_{\dot\alpha}$ with $\mathrm{SU}(2)$ little group index $I$.
With this decomposition, scattering amplitudes can be expressed entirely in terms of spinor variables. 
The resulting amplitudes are manifestly Lorentz invariant and transform covariantly under the little group, and can be calculated in any reference frame without additional boosts.

The present construction is inspired by this on-shell viewpoint and aims at building a covariant $LS$ coupling, referred to as the covariant spinor method in canonical formalism (covariant canonical-spinor scheme). 
In this scheme, the angular momentum coupling is performed in the little group $\mathrm{SU}(2)$ indices carried by the canonical-spinor variables. At the same time, the canonical-spinor variables carry $\mathrm{SL}(2,\mathbb C)$ Lorentz indices, which ensures covariance under general Lorentz transformations. As a result, the spin coupling does not require, as in the covariant tensor method, lifting the structure to the $\mathrm{SO}(3,1)$ Lorentz representation and realizing it through contractions of Lorentz indices.

By contrast, in the traditional-$LS$ method, both the spin coupling and the orbital angular momentum $L$ are defined with respect to the little group associated with the parent standard momentum in the two-body COM frame. Therefore, in practical calculations one needs to boost events back to the two-body COM frame. Here, a little group covariant map $\tau$ is constructed to map the little group representation of each daughter particle into the little group representation of the parent particle. The angular momenta are then coupled by the $\mathrm{SU}(2)$ CGCs, which avoids an explicit boost back to the COM frame.
Accordingly, the amplitude can be written as
\begin{equation}
    \mathcal{A}(p_3,p_1,p_2;L,S)=\underbrace{\mathcal{Y}(p_3,p_1,p_2;L)}_{\text {orbital part}}\cdot\underbrace{\tau_1(p_3,p_1;s_3,s_1)\cdot\tau_2(p_3,p_2;s_3,s_2)}_{\text {spin part}},
\end{equation}
which can be evaluated directly from the four momenta in any frame while preserving the traditional-$LS$ partial wave term by term.
Moreover, the scheme is formulated in a general formula applicable to particles of arbitrary spin, including the case with massless particles.
Finally, this scheme simplifies practical PWA implementations for cascade decays with multiple decay chains. The amplitudes can be evaluated directly from the four-momenta in any frame while keeping an explicit $LS$ separation.

This paper is organized as follows.
In section~\ref{sec:Two different states description---Helicity and Canonical}, we review the definitions of canonical and helicity single-particle states and the corresponding constructions of two-particle states coupled to a total angular momentum $J$. We then construct the helicity and traditional-$LS$ amplitudes, and compare their structures and mutual relations.
In section~\ref{sec:Covariant tensor and Zemach tensor amplitudes}, we systematically revisit the Zemach tensor method, the covariant tensor method, and covariant projection tensor method in both the PS-scheme and GS-scheme, summarize their building blocks and implementation steps, and discuss their respective advantages and limitations in practical amplitude modeling. We also explain how these methods connect to the $LS$ partial wave organization and how their results match the standard $LS$ description in the two-body COM frame.
In section~\ref{sec:Construction schemes of covariant orbital-spin coupling amplitude}, we introduce the covariant spinor scheme in canonical formalism. The corresponding amplitudes are decomposed into $LS$ partial waves and compared with the traditional-$LS$ amplitudes in any frame, demonstrating their equivalence. We then reformulate covariant projection tensor method in the PS-scheme in terms of canonical-spinor variables, present several explicit examples of canonical-spinor amplitude calculations.
In section~\ref{sec:General analysis of cascade decay}, we apply covariant and non-covariant methods to generic cascade decays. 
In section~\ref{sec:The cascade decay of lambda}, we apply the traditional-$LS$ amplitude, the helicity amplitude, and the canonical-spinor amplitude to the specific cascade decay $\Lambda^{+}_{c}\to \Lambda \pi^{+}\pi^{0}$, and demonstrate their numerical consistency.


\section{Helicity and canonical states and amplitudes}
\label{sec:Two different states description---Helicity and Canonical}

\subsection{Description of single-particle and two-particle states}

\subsubsection{Single-particle state}
\label{subsubsec:Description of single-particle}

Relativistic single-particle states can be classified by unitary irreducible representations of the Poincaré group, which is the Wigner particle classification~\cite{Wigner:1939cj}.
To describe the internal degrees of freedom on a given orbit, one first chooses a standard momentum $k^\mu$ and defines the corresponding little group
\begin{equation}
    G_L=\{\Lambda\in \mathrm{SO}^{+}(3,1)\mid \Lambda k = k\}.
\end{equation}
For massive particles with $m>0$, one takes $k^\mu=(m,0,0,0)$. In this case, transformations that leave $k$ invariant can only be spatial rotations, so the little group is $G_L\simeq \mathrm{SO}(3)$. The spin is then labeled by an irreducible representation of $\mathrm{SO}(3)$, namely spin-$j$ with dimension $2j+1$.
For massless particles with $m=0$, one takes $k^\mu=(E,0,0,E)$. The little group that leaves this light-like momentum invariant is $\mathrm{SO}(2)$. The internal degree of freedom is then labeled by an irreducible representation of $\mathrm{SO}(2)$, which is one dimensional and can be characterized by the helicity $\lambda$.

Next, we focus on the massive case, $m>0$, for which the little group is $\mathrm{SO}(3)$. The spin degrees of freedom are therefore described by the usual spin-$j$ irreducible representation of the rotation group. We choose a rest-frame single-particle state $|\mathbf k,j\sigma\rangle$, where $\sigma$ denotes the spin projection along the $z$ axis. The rotation generators $J_i$ satisfy
\begin{equation}
    [J_i,J_j]=i\varepsilon_{ijk}J_k,
\end{equation}
where $i$, $j$, and $k$ run over $x,y,z$, and $\varepsilon_{ijk}$ is the totally antisymmetric tensor. 
The state
$|\mathbf{k}, j\sigma \rangle$ satisfies the following relation with the angular momentum operator and can be written as
\begin{equation}
\begin{aligned}
    J^2|\mathbf{k},j\sigma \rangle=&j(j+1)|\mathbf{k},j\sigma \rangle,\\
    J_z|\mathbf{k},j\sigma \rangle=&\sigma|\mathbf{k},j\sigma \rangle,
\end{aligned}
\end{equation}
where $J^2=J_x^2+J_y^2+J_z^2$.
A finite rotation $R(\alpha,\beta,\gamma)$ on a physical system is associated with a transformation of the state $|\mathbf{k},j\sigma \rangle$. The corresponding unitary operator can be written as
\begin{equation}
	U[R(\alpha,\beta,\gamma)] = e^{-i\alpha J_{z}}e^{-i\beta J_{y}}e^{-i\gamma J_{z}},
\end{equation}
where ($\alpha,\beta,\gamma$) are the standard Euler angles.
The unitary operator preserves the multiplication law
\begin{equation}
    U[R_2R_1]=U[R_2]U[R_1].
\end{equation}
When unitary operator $U[R]$ acts on state $|\mathbf{k},j\sigma \rangle$, it transforms as
\begin{equation}\label{Eq:rotate one partical}
	U[R(\alpha,\beta,\gamma)]|\mathbf{k},j\sigma \rangle = \sum_{\sigma^{\prime}}D^{\sigma^\prime(j)}_{\,\,\sigma}(\alpha,\beta,\gamma)|\mathbf{k}, j\sigma^{\prime} \rangle ,
\end{equation}
where $D^{\sigma^\prime(j)}_{\,\,\sigma}(\alpha,\beta,\gamma)$ can be given by
\begin{equation}
	D^{\sigma^\prime(j)}_{\,\,\sigma}(\alpha,\beta,\gamma) = \langle \mathbf{k},j\sigma^{\prime} | U[R(\alpha,\beta,\gamma)] |\mathbf{k},j\sigma \rangle = e^{-i\sigma^{\prime}\alpha }d^{\sigma^\prime(j)}_{\,\,\sigma}(\beta)e^{-i\sigma\gamma },
\end{equation}
and
\begin{equation}
	d^{\sigma^\prime(j)}_{\,\,\sigma}(\beta) = \langle \mathbf{k},j\sigma^{\prime} | e^{-i\beta J_{y}} |\mathbf{k},j\sigma \rangle.
\end{equation}
Our conventions and useful identities for $D^{\sigma^\prime(j)}_{\,\,\sigma}$ and $d^{\sigma^\prime(j)}_{\,\,\sigma}$ are given in Appendix~\ref{app:D-functions}.

So far we have specified the action of the rotation on rest states, which is governed by the generators $J_i$.
To describe general Lorentz transformations, one needs the full set of infinitesimal generators of the homogeneous Lorentz group.
In addition to spatial rotations, the Lorentz group contains boosts, generated by operators $K_i$.
The six generators $(J_i,K_i)$ satisfy the following commutation relations
\begin{equation}\label{Eq:six L generators}
[J_i,J_j]=i\varepsilon_{ijk}J_k,\qquad
[J_i,K_j]=i\varepsilon_{ijk}K_k,\qquad
[K_i,K_j]=-i\varepsilon_{ijk}J_k.
\end{equation}
A \emph{pure} boost $B(\boldsymbol{\eta})$ is defined as the Lorentz transformation with no accompanying spatial rotation. For a \emph{pure} boost of rapidity $\eta$ along the direction $\hat{\mathbf p}\equiv \mathbf p/|\mathbf p|$, the corresponding unitary operator takes the form
\begin{equation}
U[B(\boldsymbol{\eta})]=e^{-i\eta \mathbf{\hat{p}}\cdot \mathbf{K}},
\end{equation}
where $E+|\mathbf{p}|=me^{\eta}$, $\eta =\text{arctanh}{\frac{|\mathbf{p}|}{E}}$ and $\boldsymbol{\eta}=\hat{\mathbf{p}}\eta$. 
For a general Lorentz transformation $\Lambda$ may contain both rotations and boosts. The unitary operator of the general Lorentz transformation preserves the multiplication
law
\begin{equation}
U[\Lambda_2\Lambda_1]=U[\Lambda_2]\,U[\Lambda_1].
\end{equation}

With the Lorentz generators $(J_i,K_i)$ specified, we can define relativistic single-particle states with momentum $p^\mu=(E,\mathbf p)$ from the rest single-particle states.
We introduce the \textbf{standard boost} $L(\mathbf{p})$ satisfying $p=L(\mathbf{p})k$, and defines
\begin{equation}
|\mathbf{p},j\sigma\rangle\equiv U[L(\mathbf{p})]|\mathbf{k},j\sigma\rangle.
\end{equation}
The choice of $L(\mathbf p)$ is not unique, different choices correspond to different conventions for transporting the spin quantization axis from the rest frame to momentum $\mathbf p$. In the following we focus on two widely used conventions, the canonical-standard boost and the helicity-standard boost.

The \textbf{canonical-standard boost} is taken to be a \emph{pure} boost that sends $k^\mu$ to $p^\mu$ without introducing any rotation. One may write
\begin{equation}\label{Eq:caonical boost}
L_c(\mathbf p)\equiv B(\boldsymbol{\eta})
=R(\phi,\theta,0)L_z(\eta)R^{-1}(\phi,\theta,0),
\end{equation}
where $(\theta,\phi)$ the spherical angles of $\hat{\mathbf p}$, $L_z(\eta)$ is the boost along the $z$-axis, and the subscript $c$ denotes the canonical-standard boost.

\begin{figure}[H]
\centering
\includegraphics[width=\linewidth]{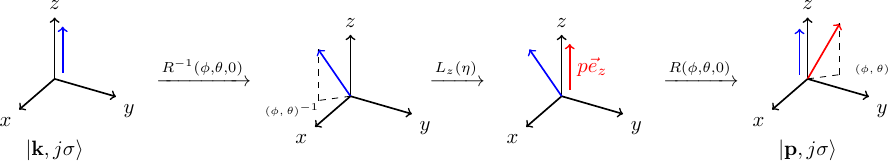}
\caption{Canonical-standard boost for a spin-$j$ particle. The blue arrow indicates the spin polarization direction and the red arrow indicates the momentum direction.}
\label{fig:canonical state}
\end{figure}

\begin{figure}[H]
\centering
\includegraphics[width=\linewidth]{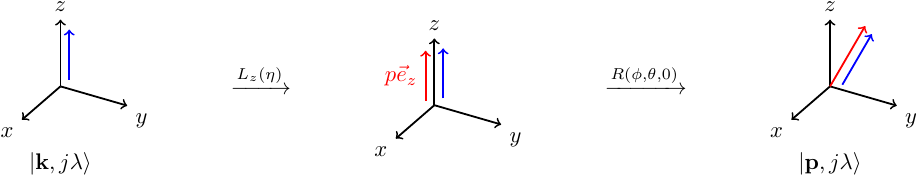}
\caption{Helicity-standard boost for a spin-$j$ particle. The blue arrow indicates the spin polarization direction and the red arrow indicates the momentum direction.}
\label{fig:helicity state}
\end{figure}

The \textbf{helicity-standard boost} is defined by first performing a rotation $R(\phi,\theta,0)$ and then performing a \emph{pure} boost that sends $k^\mu$ to $p^\mu$. 
Using the same rotation $R(\phi,\theta,0)$ that takes $\hat z$ into $\hat{\mathbf p}$, one defines
\begin{equation}\label{Eq:helicity boost}
L_h(\mathbf p)\equiv B(\boldsymbol{\eta})R(\phi,\theta,0)=R(\phi,\theta,0)L_z(\eta),
\end{equation}
where the subscript $h$ denotes the helicity-standard boost. Although these two types of boosts yield the same standard momentum, the quantization axes of the resulting single-particle states are different. For the canonical standard boost, the quantization axis remains the fixed $z$-axis direction, while for the helicity-standard boost, the quantization axis aligns with the direction of the momentum. Their geometric representations are shown in Figs.~\ref{fig:canonical state} and \ref{fig:helicity state}.

The relation between canonical-standard boost and helicity-standard boost can be written as
\begin{equation}\label{Eq:boost relation}
	L_{c}(\mathbf{p}) = L_{h}(\mathbf{p})R^{-1}(\phi,\theta,0).
\end{equation}

Based on the difference in the spin quantization axes, we define two types of single-particle states: the canonical particle state and the helicity particle state.
For the canonical single-particle state, its quantization axis always points along the $z$-axis; therefore, it is obtained by applying the canonical standard boost to the single-particle state at rest.
For the helicity single-particle state, its quantization axis always aligns with the direction of the momentum; hence, it is obtained by applying the helicity standard boost to the single-particle state at rest.
The explicit expressions of these boosted states can be defined through Eqs.~\eqref{Eq:caonical boost} and \eqref{Eq:helicity boost} as follows
\begin{equation}\label{Eq:canonical one}
\begin{aligned}
    |\mathbf{p},j\sigma \rangle \equiv& U[L_{c}(\mathbf{p})] |\mathbf{k},j\sigma \rangle \\
    =& U[R(\phi,\theta,0)]U[L_{z}(\eta)]U^{-1}[R(\phi,\theta,0)]|\mathbf{k},j\sigma \rangle,
\end{aligned}
\end{equation}
and
\begin{equation}\label{Eq:helicity one}
\begin{aligned}
    |\mathbf{p},j\lambda \rangle \equiv& U[L_{h}(\mathbf{p})] |\mathbf{k},j\lambda \rangle \\
=& U[R(\phi,\theta,0)]U[L_{z}(\eta)]|\mathbf{k},j\lambda \rangle .
\end{aligned}
\end{equation}

\begin{figure}[H]
\centering
\includegraphics[scale=1.1]{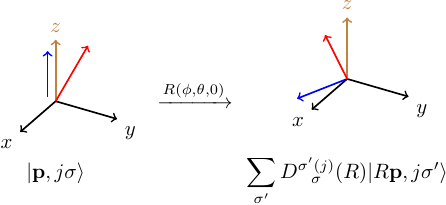}
\caption{Rotation acting on a canonical single-particle state. The brown line indicates the spin quantization axis, the red line represents the momentum direction of the particle, and the blue line denotes the spin polarization direction of the particle. }
\label{fig:rotate canonical state}
\end{figure}

\begin{figure}[H]
\centering
\includegraphics[scale=1.1]{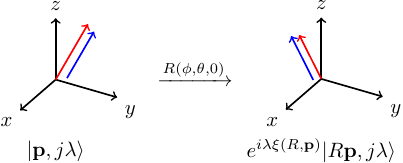}
\caption{Rotation acting on a helicity single-particle state. The red line indicates the momentum direction of the particle, which is also the spin quantization axis, and the blue line denotes the direction of the particle's spin polarization.}
\label{fig:rotate helicity state}
\end{figure}

Due to the different definitions of the quantization axis for the two types of particle states, their transformation properties under rotations also differ. For the canonical single-particle state, because its spin quantization axis is always fixed along the $z$-direction, a rotation will mix its spin projection quantum numbers $\sigma$ via a Wigner $D$-matrix. 
In contrast, for the helicity single-particle state, the quantization axis is always aligned with the momentum direction; after a rotation, the spin quantization axis simply follows the rotated momentum, so the spin projection along that axis remains unchanged.
Their transformation rules under rotations are given by
\begin{equation}\label{Eq:rotate canonical state}
	U[R]|\mathbf{p},j\sigma \rangle = U[RR_{0}]U[L_{z}(\eta)]U^{-1}[RR_{0}]U[R]|\mathbf{k},j\sigma \rangle = \sum_{\sigma^{\prime}}D^{\sigma^\prime(j)}_{\,\,\sigma}(R)|R\mathbf{p},j\sigma^{\prime} \rangle,
\end{equation}
and
\begin{equation}\label{Eq:rotate helicty state}
\begin{aligned}
U[R]|\mathbf{p},j\lambda \rangle 
=& U[RR_{0}]U[L_{z}(\eta)]|\mathbf{k},j\lambda \rangle \\
=& U[R^\prime(\phi^\prime,\theta^\prime,\xi)]U[L_{z}(\eta)]|\mathbf{k},j\lambda \rangle \\
=& U[R^{\prime\prime}(\phi^\prime,\theta^\prime,0)]U[L_{z}(\eta)]U[R_z(\xi(R,\mathbf p))]|\mathbf{k},j\lambda \rangle \\
=&e^{i\lambda\xi(R,\mathbf{p})}|R\mathbf{p},j\lambda \rangle,
\end{aligned}
\end{equation}
where $R^\prime(\phi^\prime,\theta^\prime,\xi)=RR_{0}$, and the azimuthal angle of the momentum $R\mathbf p$ is $(\phi^\prime,\theta^\prime)$. For the rotation $R^\prime(\phi^\prime,\theta^\prime,\xi)$, the first rotation about the $z$ axis is not necessarily zero. Since this initial $z$-axis rotation does not change the final momentum direction, one can use the commutation relation in Eq.~\eqref{Eq:six L generators} to treat it as a rotation about the $z$ axis applied at the state $|\mathbf{k},j\lambda \rangle$, which therefore produces the corresponding phase. 
It can be observed that helicity remains unchanged under rotations.
The geometric difference between the canonical and helicity states is illustrated in Figs.~\ref{fig:rotate canonical state} and \ref{fig:rotate helicity state}.

The relation between the canonical state and the helicity state can be obtained from Eqs.~\eqref{Eq:rotate one partical}, \eqref{Eq:helicity boost}, \eqref{Eq:canonical one}, and \eqref{Eq:helicity one}, and is given by
\begin{equation}\label{Eq:one particle relation}
	\begin{aligned}
		|\mathbf{p},j\lambda \rangle =& U[R(\phi,\theta,0)]U[L_{z}(\eta)]U^{-1}[R(\phi,\theta,0)]U[R(\phi,\theta,0)]|\mathbf{k},j\lambda \rangle \\
		=& U[L_{c}(\mathbf{p})]U[R(\phi,\theta,0)] |\mathbf{k},j\lambda \rangle \\
		=& \sum_{\sigma}D^{\sigma(j)}_{\,\,\lambda}(R)|\mathbf{p},j\sigma \rangle .
	\end{aligned}
\end{equation}

Having established the transformation properties of the relativistic canonical and helicity states under rotations, attention is now turned to their behaviour under Lorentz transformations.
Considering an arbitrary four-momentum $p^\mu=(E,\mathbf{p})$, a proper homogeneous orthochronous Lorentz transformation $\Lambda^{\mu}_{\,\,\nu}$ takes the four-momentum $p^\mu$ into $p^{\prime\mu}$ as follows
\begin{equation}
    p^{\prime\mu}=\Lambda^{\mu}_{\,\,\nu}p^{\nu},
\end{equation}
where the repeated indices summed over.
Under this Lorentz transformation, the state $U[\Lambda] |\mathbf{p}, j\alpha \rangle$ is a linear combination of the states $|\mathbf{ p}^\prime, j\alpha^\prime \rangle$, namely
\begin{equation}
\begin{aligned}
    U[\Lambda] |\mathbf{p}, j\alpha \rangle &= \sum_{\alpha^\prime } C_{\alpha^\prime \alpha} |\mathbf{ p}^\prime, j\alpha^\prime \rangle\\
    &=U[L(\mathbf{ p}^\prime)]\sum_{\alpha^\prime } C_{\alpha^\prime \alpha} |\mathbf{k}, j\alpha^\prime \rangle,
\end{aligned}
\end{equation}
where $\alpha$ represents either the spin projection $\sigma$ (canonical basis) or the helicity $\lambda$ (helicity basis), with the boost $L(\mathbf{p})$ chosen accordingly.
According to Eq.~\eqref{Eq:rotate one partical}, the linear combination over $\alpha^{\prime}$ represents a rotation acting on the single-particle state with standard momentum. 
This rotation is given by 
\begin{equation}\label{Eq:transform state}
\begin{aligned}
    U[\Lambda] |\mathbf{p}, j\alpha \rangle &= U[\Lambda] U[L(\mathbf{p})] |\mathbf{k}, j\alpha \rangle 
= U[L(\mathbf{ p}^\prime)] U\left[ L^{-1}(\mathbf{ p}^\prime)\Lambda L(\mathbf{p}) \right] |\mathbf{k}, j\alpha \rangle,
\end{aligned}
\end{equation}
with 
\begin{equation}\label{Eq:littlegroupro}
    W(\Lambda,p) \equiv L^{-1}(\mathbf{ p}^\prime)\Lambda L(\mathbf{p}),
\end{equation}
where $W(\Lambda,p)$ is a rotation that leaves the standard momentum $k^\mu$ invariant, namely $W^\mu_{\,\,\nu} k^\nu = k^\mu$, and is referred to as the little group rotation. The action of this rotation on the state $|\mathbf{k}, j\alpha \rangle$ is
\begin{equation}
U[W(\Lambda, p)] |\mathbf{k}, j\alpha \rangle = \sum_{\alpha^{\prime}} D^{\alpha^\prime(j)}_{\,\, \alpha}(W(\Lambda, p)) |\mathbf{k}, j\alpha^{\prime} \rangle,
\end{equation}
where $D^{\alpha^\prime(j)}_{\,\,\alpha}(W)$ is the little group representation matrix. Substituting this result into Eq.~\eqref{Eq:transform state}, the Lorentz transformation of the single-particle state becomes
\begin{equation}\label{Eq:Lorentz transformation on a state}
U[\Lambda] |\mathbf{p}, j\alpha \rangle = \sum_{\alpha^{\prime}} D^{\alpha^\prime(j)}_{ \,\,\alpha}(W(\Lambda, p))  |\mathbf{ p}^\prime, j\alpha^{\prime} \rangle.
\end{equation}
This result illustrates how the canonical or helicity quantum numbers transform under Lorentz transformations. The little group rotation describes the additional Wigner rotation induced by the composition of non-collinear Lorentz boosts.
When the Lorentz transformation is a pure boost $B(\boldsymbol{\eta}^\prime)$ whose direction is parallel to $\hat{\mathbf p}$, the transformed momentum remains collinear with $p$,
\begin{equation}
    \bar p=B(\boldsymbol{\eta}')p,\qquad \bar p\parallel p.
\end{equation}
Using Eqs.~\eqref{Eq:caonical boost} and \eqref{Eq:helicity boost}, the canonical little group rotation is
\begin{equation}\label{Eq:little group same direct}
\begin{aligned}
W_c(B(\boldsymbol{\eta}^\prime),p)=&L^{-1}_c(\mathbf{\bar p}) B(\boldsymbol{\eta}^\prime) L_c(\mathbf{p})\\
=&R(\phi,\theta,0)L_z(-\eta^\prime-\eta)L_z(\eta^\prime)L_z(\eta)R^{-1}(\phi,\theta,0)\\
=&\mathbf{I},
\end{aligned}
\end{equation}
and the helicity little group rotation is
\begin{equation}\label{Eq:little group helicity12}
\begin{aligned}
W_h(B(\boldsymbol{\eta}^\prime),p)=&L^{-1}_h(\mathbf{\bar p}) B(\boldsymbol{\eta}^\prime) L_h(\mathbf{p})\\
=&L_z(-\eta^\prime-\eta)L_z(\eta^\prime)L_z(\eta)\\
=&\mathbf{I}.
\end{aligned}
\end{equation}
Therefore, when the Lorentz transformation is a pure boost along the momentum direction $\hat{\mathbf p}$, the canonical and helicity little group rotations are both identity $\textbf{I}$. That is, for the corresponding single-particle states under the Lorentz transformation, the canonical spin projection $\sigma$ and the helicity $\lambda$ remain unchanged, which can be written as
\begin{equation}
    U[B(\boldsymbol{\eta}^\prime)] |\mathbf{p}, j\alpha \rangle = |\mathbf{\bar p}, j\alpha \rangle.
\end{equation}

When the Lorentz transformation is a pure rotation $R$, the canonical little group rotation is 
\begin{equation}
\begin{aligned}
W_c(R,p)=&L^{-1}_c(R\mathbf{p}) R L_c(\mathbf{p})\\
=&L^{-1}_c(R\mathbf{p})  L_c(R\mathbf{p})R\\
=&R,
\end{aligned}
\end{equation}
where for canonical states under a pure rotation, the corresponding little group rotation is $R$, and therefore the spin projection $\sigma$ does not remain unchanged, with the action of the rotation given in Eq.~\eqref{Eq:rotate canonical state}.
In contrast, the helicity little group rotation is
\begin{equation}
\begin{aligned}
W_h(R,p)
&=L_h^{-1}(R\mathbf{p})RL_h(\mathbf{p})\\
&=L_h^{-1}(R\mathbf{p})RR_0L_z(\eta)\\
&=L_h^{-1}(R\mathbf{p})R^\prime(\phi^\prime,\theta^\prime,\xi)L_z(\eta)\\
&=L_h^{-1}(R\mathbf{p})\Big(L_h(R\mathbf{p})\,R_z(\xi(R,\mathbf p))\Big)\\
&=R_z(\xi(R,\mathbf p)),
\end{aligned}
\end{equation}
where $R_z(\xi(R,\mathbf p))$ is a rotation about the helicity direction, which produces a phase and does not change the helicity, with the corresponding result given in Eq.~\eqref{Eq:rotate helicty state}.
Finally, for both the canonical and helicity single particle states, the pure boost along the momentum direction leaves the spin projection $\sigma$ and helicity $\lambda$ unchanged. Under the pure rotation, the helicity $\lambda$ is preserved and the state transforms only by a phase, whereas the canonical spin projection $\sigma$ generally mixes.
\subsubsection{Construction of two-particle states}
\label{subsubsec:Coupling of two-particle states}

Having introduced the canonical and helicity single-particle states, we now consider a two-body system of two massive particles, particle-1 and particle-2, with spins $s_1$ and $s_2$, and construct total angular momentum states with angular momentum $J$ in either the traditional-$LS$ basis or the helicity basis.

In the traditional-$LS$ construction, a common spin quantization axis is fixed, conventionally the $z$-axis, and the relative orbital angular momentum $\boldsymbol{L}$ is defined with respect to the same axis.
The total spin is obtained by adding the individual spins, $\boldsymbol{S}=\boldsymbol{s}_1+\boldsymbol{s}_2,S_z=s_{1z}+s_{2z}$. 
There are $(2s_1+1)\times(2s_2+1)$ subspaces that are described entirely by angular. Here, the angular degrees of freedom and the $(2s_1+1)\times(2s_2+1)$ spin degrees of freedom are independent. Consequently, the states obtained from the $D$-matrix, namely spherical harmonic, decomposition exhibit a $(2s_1+1)\times(2s_2+1)$-fold degeneracy, which must be resolved by further decomposition using CGCs.
Thus, the total angular momentum $J$ is then built through two successive CGCs couplings: first coupling $\boldsymbol{s}_1$ and $\boldsymbol{s}_2$ to form $\boldsymbol{S}$, and then coupling $\boldsymbol{L}$ with $\boldsymbol{S}$ to obtain $\boldsymbol{J}$. 
This construction is based on the complete set of commuting observables $\{\boldsymbol{J}^2,J_z,\boldsymbol{L}^2,\boldsymbol{S}^2,P^2,\boldsymbol{P}\}$. 
In the subsequent discussion, we focus only on the construction of two-particle states in the COM frame; therefore, we omit the $P^2,\boldsymbol{P}$ , and denote the common eigenstates as $|JMLS\rangle$.
By inserting the completeness relation of the product basis formed by the two canonical single-particle spin states, $|JMLS\rangle$ can be written explicitly as
\begin{equation}
\begin{aligned}
    |JMLS \rangle
    &=\sum_{\sigma_{1}\sigma_{2}}|\Omega \sigma_{1}\sigma_{2} \rangle\langle\Omega \sigma_{1}\sigma_{2}|JMLS \rangle\\
    &= \sum_{\sigma_{1}\sigma_{2}\sigma_{L}\sigma_{S}}
       C_{J,M}^{L,\sigma_{L};S,\sigma_{S}}
       C_{S,\sigma_{S}}^{s_1,\sigma_{1};s_2,\sigma_{2}}
       \int d\Omega\, Y^{L}_{\sigma_{L}}(\Omega)\,|\Omega \sigma_{1}\sigma_{2} \rangle.
\end{aligned}
\end{equation}

By contrast, in the helicity construction, the spin quantization axis of each particle is chosen to be along its own momentum direction. When the particle state is rotated, its quantization axis remains aligned with the rotated momentum direction; therefore, helicity possesses rotational invariance.
In this construction, the complete set of commuting observables is $\{ \boldsymbol{J}^2, J_z, \lambda_1, \lambda_2,P^2,\boldsymbol{P}\}$, and the common eigenstates are denoted as $|JM\lambda_1\lambda_2\rangle$. 
When the helicity is fixed, the rotational properties of the entire particle system in the COM frame are completely determined by the momentum direction of particle-1 and its helicity projection along that direction, $\lambda = \lambda_1 - \lambda_2$. In the subspace of fixed $\lambda_i$, only one rotational degree of freedom remains to characterize the state. Therefore, we use the completeness of the $D$-matrices to directly decompose the state into components of different angular momenta,
\begin{equation}
\begin{aligned}
    |JM\lambda_{1}\lambda_{2} \rangle &=\sum_{\lambda_1^\prime\lambda_2^\prime}|\phi\theta\lambda_{1}^\prime\lambda_{2}^\prime \rangle\langle\phi\theta\lambda_{1}^\prime\lambda_{2} ^\prime|JM\lambda_{1}\lambda_{2} \rangle \\
    &=N_{J}\int d\Omega D^{M(J)*}_{\,\,\lambda}(\phi,\theta,0) |\phi\theta\lambda_{1}\lambda_{2} \rangle.
\end{aligned}
\end{equation}

In this way, the helicity construction does not require an explicit $LS$ coupling. 
Next, we will present the explicit construction procedure.

\paragraph{Canonical construction}

The canonical construction is formulated in terms of the canonical two-particle state in the COM frame of the particle-1 and 2. Consider two massive particles, with particle-1 and particle-2 having spins $s_1$ and $s_2$, respectively. The canonical two-particle state is defined by
\begin{equation}\label{Eq:canonical two 1}
	|\phi\theta \sigma_{1}\sigma_{2} \rangle \equiv NU[L_{1,c}(\mathbf{p})]|\mathbf{k}_1,s_{1}\sigma_{1} \rangle U[L_{2,c}(-\mathbf{p})]|\mathbf{k}_2,s_{2}\sigma_{2} \rangle,
\end{equation}
where $N$ is the normalization constant, $U[L_{i,c}(\mathbf{p})]$ $(i=1,2)$ is the canonical-standard boost in Eq.~\eqref{Eq:caonical boost}, and the numerical subscripts are used to distinguish the canonical-standard boosts for different particles. The explicit calculation of the normalization constant $N$ is given in Appendix~\ref{app:Normalization of One- and Two-Particle States}.
The two-particle state $|\phi\theta \sigma_{1}\sigma_{2} \rangle$ satisfies the normalization condition, which can be written as
\begin{equation}\label{Eq:canonical normalization}
	\langle \phi^{\prime}\theta^{\prime}\sigma^{\prime}_{1}\sigma^{\prime}_{2}|\phi\theta \sigma_{1}\sigma_{2} \rangle =\delta(\phi^{\prime}-\phi)\delta(\theta^{\prime}-\theta)\delta^{\sigma^{\prime}_{1}}_{\sigma_{1}}\delta^{\sigma^{\prime}_{2}}_{\sigma_{2}}.
\end{equation}
By coupling the spins $s_{1}$ and $s_{2}$ in Eq.~\eqref{Eq:canonical two 1}, we can define two-particle states with total spin-$S$ as
\begin{equation}\label{Eq:s1s2toS}
	|\phi\theta S\sigma_{S} \rangle = \sum_{\sigma_{1}\sigma_{2}}C_{S,\sigma_{S}}^{s_1,\sigma_{1};s_2,\sigma_{2}}|\phi\theta \sigma_{1}\sigma_{2} \rangle,
\end{equation}
where $C_{S,\sigma_{S}}^{s_1,\sigma_{1};s_2,\sigma_{2}}$ is the CGC. Using the angular momentum selection rules, one obtains
\begin{equation}
    |s_1-s_2|\leq S\leq s_1+s_2.
\end{equation}

Describing the total angular momentum $J$ of the two-particle system requires consideration of both the spins of the particles and their relative orbital motion. In the COM frame, the direction of relative momentum $\mathbf{p}=\mathbf{p}_1-\mathbf{p}_2$ between the two particles determines the angular dependence of the system, which can be expanded in spherical harmonics to describe orbital angular momentum states.

Starting from the state $|\phi \theta S \sigma_{S}\rangle$, where the direction of the relative momentum is specified by the angles $(\theta,\phi)$, the orbital angular momentum eigenstates are defined by projecting $|\phi \theta S \sigma_{S}\rangle$ onto the normalized spherical harmonic $Y^L_{\sigma_{L}}(\theta,\phi)$. Explicitly, this is defined as
\begin{equation}\label{Eq:L spherical harmonic}
	|L\sigma_{L} S\sigma_{S} \rangle \equiv \int d\Omega Y^{L}_{\sigma_{L}}(\Omega) |\Omega S\sigma_{S}\rangle,
\end{equation}
where $\Omega = (\theta,\phi)$ and $d\Omega = d\phi\, d\cos\theta$, and the spherical harmonic $Y^L_{\sigma_{L}}(\theta,\phi)$ is defined in Eq.~\eqref{Eq:define spherical harmonics}.

The state $|L \sigma_{L} S \sigma_{S}\rangle$ now carries both the orbital angular momentum $L$ and the total spin-$S$. Under spatial rotations, it transforms according to the direct product representation of the rotation group: the orbital part transforms under $D^L(R)$, and the spin part under $D^S(R)$. This can be expressed as
\begin{equation}
	U[R]|L\sigma_{L} S\sigma_{S} \rangle = \sum_{\sigma^{\prime}_{L}\sigma^{\prime}_{S}} D^{\sigma^{\prime}_L(L)}_{\,\,\sigma_{L}}(R)D^{\sigma_S^\prime(S)}_{\,\,\sigma_S}(R)|L\sigma^{\prime}_{L} S\sigma_{S}^{\prime} \rangle.
\end{equation}

The orbital angular momentum $L$ and the total spin-$S$ can now be coupled to obtain the total angular momentum $J$, which can be written as
\begin{equation}\label{Eq:cannonical two J}
\begin{aligned}
    |JMLS \rangle
    &= \sum_{\sigma_{L}\sigma_{S}}
       C_{J,M}^{L,\sigma_{L};S,\sigma_{S}}\,|L\sigma_{L} S\sigma_{S} \rangle \\
    &= \sum_{\sigma_{1}\sigma_{2}\sigma_{L}\sigma_{S}}
       C_{J,M}^{L,\sigma_{L};S,\sigma_{S}}
       C_{S,\sigma_{S}}^{s_1,\sigma_{1};s_2,\sigma_{2}}
       \int d\Omega\, Y^{L}_{\sigma_{L}}(\Omega)\,|\Omega \sigma_{1}\sigma_{2} \rangle.
\end{aligned}
\end{equation}
Using the angular momentum selection rules, the following constraints are obtained:
\begin{equation}
    |L-S|\leq J\leq L+S,
    \quad \sigma_{S}=\sigma_{1}+\sigma_{2},
    \quad \sigma_{L}=M-\sigma_{S}.
\end{equation}
From the normalization condition \eqref{Eq:canonical normalization}, the states $|JMLS \rangle$ satisfy 
\begin{equation}
    \langle J^{\prime}M^{\prime}L^{\prime}S^{\prime}|JMLS\rangle
    = \delta_{J}^{J^{\prime}}\delta_{M}^{M^{\prime}}\delta_{L}^{L^{\prime}}\delta_{S}^{S^{\prime}},
\end{equation}
and the completeness relation is given by 
\begin{equation}\label{Eq:canonical completeness}
    \sum_{JMLS}|JMLS\rangle \langle JMLS| = \mathbf{I}.
\end{equation}
The transformation properties of the state $|JMLS \rangle$ under rotations are identical to those of a single-particle state at rest. Its transformation can be expressed as
\begin{equation}
    U[R]|JMLS \rangle
    = \sum_{M^{\prime}} D^{M^\prime(J)}_{\,\,M}(R)\,|JM^{\prime}LS\rangle,
\end{equation}
where $D^{M^\prime(J)}_{\,\,M}(R)$ is the Wigner $D$-matrix for angular momentum $J$. This transformation property shows that $|JMLS \rangle$ forms a complete set of eigenstates of the angular momentum operators $\mathbf{J}^2$ and $J_z$, with the transformation behavior completely determined by the total angular momentum $J$.

\paragraph{Helicity construction}

The helicity construction is formulated in terms of the helicity two-particle state in the COM frame. For two massive particles with helicities $\lambda_1$ and $\lambda_2$, the helicity two-particle state can be defined as 
\begin{equation}\label{Eq:helicity two}
	\begin{aligned}
		|\phi\theta\lambda_{1}\lambda_{2} \rangle &\equiv NU[L_{1,h}(\mathbf{p})]|\mathbf{k}_1,s_{1}\lambda_{1} \rangle U[L_{2,h}(-\mathbf{p})]|\mathbf{k}_2,s_{2}\ {\lambda_{2}} \rangle \\
		&=NU[R]\Big(U[L_{z}(\eta_1)]|\mathbf{k}_1,s_{1}\lambda_{1}\rangle U[L_{z}(-\eta_2)]|\mathbf{k}_2,s_{2}\ {\lambda_{2}} \rangle\Big) \\
        &=NU[R]\Big(|\mathbf{p}_z,s_{1}\lambda_{1}\rangle |-\mathbf{p}_z,s_{2}\ {\lambda_{2}} \rangle\Big)\\
		&=NU[R]|00\lambda_{1}\lambda_{2}\rangle,
	\end{aligned}
\end{equation}
where $N$ is the normalization constant, $U[L_{i,c}(\mathbf{p})]$ $(i=1,2)$ is the helicity-standard boost in Eq.~\eqref{Eq:helicity boost}, and the numerical subscripts are used to distinguish the helicity-standard boosts for different particles. 
Due to the rotational invariance of the helicities $\lambda_{1}$ and $\lambda_{2}$, the action of a rotation $R^{\prime}$ on Eq.~\eqref{Eq:helicity two} gives
\begin{equation}\label{Eq:helicity rotation}
	\begin{aligned}
	U[R^{\prime}]|\Omega\lambda_{1}\lambda_{2} \rangle&=NU[R^{\prime}R]|00\lambda_{1}\lambda_{2}\rangle \\
	&=|R^{\prime}\Omega\lambda_{1}\lambda_{2} \rangle \\
	&=|R^{\prime\prime}\lambda_{1}\lambda_{2} \rangle,
	\end{aligned}
\end{equation}
where $\Omega = (\theta,\phi)$ and $R^{\prime\prime}=R^{\prime}\Omega$.

When the helicities $(\lambda_1,\lambda_2)$ are fixed, the nontrivial rotational dependence of the two-particle state in the COM frame is entirely encoded in the momentum direction of particle-1, parametrized by
$\Omega=(\phi,\theta)$, together with the helicity difference $\lambda=\lambda_1-\lambda_2$.
A rotation about the momentum axis leaves the helicities invariant and only produces a phase
$e^{-i\lambda\alpha}$, hence the third Euler angle can be chosen as $0$.
Therefore, using the completeness of the Wigner $D$-functions for fixed right index $\lambda$, we project the
angle state $|\phi\theta\lambda_1\lambda_2\rangle$ onto definite total angular momentum $J$ and its $z$-component $M$:
\begin{equation}\label{Eq:helicity two J}
	|JM\lambda_{1}\lambda_{2} \rangle
	= N_{J}\int d\Omega\, D^{M(J)*}_{\,\,\lambda}(\phi,\theta,0)\,|\phi\theta\lambda_{1}\lambda_{2} \rangle.
\end{equation}
Using Eqs.~\eqref{Eq:helicity rotation} and \eqref{Eq:helicity two J}, the transformation of the state $|JM\lambda_{1}\lambda_{2} \rangle$ under rotations is given by
\begin{equation}
	\begin{aligned}
	U[R^{\prime}]|JM\lambda_{1}\lambda_{2} \rangle &=N_{J}\int d\Omega D^{M(J)*}_{\,\,\lambda}(\phi,\theta,0) U[R^{\prime}]|\phi\theta\lambda_{1}\lambda_{2} \rangle \\
	&=N_{J}\int d\Omega D^{M(J)*}_{\,\,\lambda}(\Omega) |R^{\prime\prime}\lambda_{1}\lambda_{2} \rangle ,
	\end{aligned}
\end{equation}
where the Wigner $D$-matrix $D^{*M(J)}_{\,\,\,\,\lambda}(\Omega)$ can be written as 
\begin{equation}
	\begin{aligned}
		D^{M(J)*}_{\,\,\lambda}(\Omega) &=D^{M(J)*}_{\,\,\lambda}(R^{\prime-1}R^{\prime\prime}) \\
		&=\sum_{M^{\prime}}D^{M(J)*}_{\,\,M^\prime}(R^{\prime-1})D^{M^\prime(J)*}_{\,\,\lambda}(R^{\prime\prime}) \\
		&=\sum_{M^{\prime}}D^{M^\prime(J)}_{\,\,M}(R^{\prime})D^{M^\prime(J)*}_{\,\,\lambda}(R^{\prime\prime}) .
	\end{aligned}
\end{equation}
Using this relation, one obtains 
\begin{equation}\label{Eq:helicity rotation J}
	U[R^{\prime}]|JM\lambda_{1}\lambda_{2} \rangle = \sum_{M^{\prime}} D^{M^{\prime}(J)}_{\,\,M}(R^{\prime})|JM^{\prime}\lambda_{1}\lambda_{2} \rangle.
\end{equation}
Thus, the state $|JM\lambda_1\lambda_2 \rangle$ with definite total angular momentum $J$ is constructed by coupling the two particles through their helicities $\lambda_1$ and $\lambda_2$. The state $|JM\lambda_1\lambda_2 \rangle$ satisfies the normalization condition
\begin{equation}
    \langle J^{\prime}M^{\prime}\lambda_{1}^{\prime}\lambda_{2}^{\prime}|JM\lambda_{1}\lambda_{2}\rangle
    =\delta_{J}^{J^{\prime}}\delta_{M}^{M^{\prime}}\delta_{\lambda_{1}}^{\lambda_{1}^{\prime}}\delta_{\lambda_{2}}^{\lambda_{2}^{\prime}},
\end{equation}
which fixes the normalization constant to be
\begin{equation}
    N_{J}=\sqrt{\frac{2J+1}{4\pi}}. 
\end{equation}
Finally, the completeness relation takes the form
\begin{equation}\label{nomh}
    \sum_{JM\lambda_{1}\lambda_{2}} | JM\lambda_{1}\lambda_{2}\rangle \langle JM\lambda_{1}\lambda_{2}| =\mathbf{I}.
\end{equation}

\paragraph{Equivalence between two constructions}

Having introduced both the helicity and canonical constructions, we now demonstrate their equivalence. Despite their different starting points, both formulations yield states with total angular momentum $J$. In the following, we derive the explicit transformation between these two equivalent constructions. Starting from the respective two-particle states and using Eqs.~\eqref{Eq:helicity boost}, \eqref{Eq:one particle relation}, and \eqref{Eq:canonical two 1}, we have
\begin{equation}\label{Eq:redefine two-p state}
	\begin{aligned}
		|\phi\theta\lambda_{1}\lambda_{2} \rangle &=NU[L_{1,h}(\mathbf{p})]|\mathbf{k}_1,s_{1}\lambda_{1} \rangle U[L_{2,h}(-\mathbf{p})]|\mathbf{k}_2,s_{2}\ {-\lambda_{2}}\rangle \\ &=NU[L_{1,c}(\mathbf{p})]U[R]|\mathbf{k}_1,s_{1}\lambda_{1} \rangle U[L_{2,c}(-\mathbf{p})]U[R]|\mathbf{k}_2,s_{2}\ {-\lambda_{2}}\rangle \\
		&=\sum_{\sigma_{1}\sigma_{2}}D^{\sigma_{1}(s_{1})}_{\,\,\lambda_{1}}(\phi,\theta,0)D^{\sigma_2(s_{2})}_{\,\,{-\lambda_{2}}}(\phi,\theta,0)U[L_{1,c}(\mathbf{p})]|\mathbf{k}_1,s_{1}\sigma_{1}\rangle U[L_{2,c}(-\mathbf{p})]|\mathbf{k}_2,s_{2}\sigma_{2}\rangle \\
		&=\sum_{\sigma_{1}\sigma_{2}}D^{\sigma_1(s_{1})}_{\,\,\lambda_{1}}(\phi,\theta,0)D^{\sigma_2(s_{2})}_{\,\,{-\lambda_{2}}}(\phi,\theta,0)|\phi\theta \sigma_{1}\sigma_{2} \rangle.
	\end{aligned}
\end{equation}
Substituting this into the definition of the total angular momentum state in Eq.~\eqref{Eq:helicity two J} yields
\begin{equation}\label{Eq:helicity coupling to canonical state}
		\begin{aligned}
			|JM\lambda_{1}\lambda_{2} \rangle & =N_{J}\int d\Omega D^{M(J)*}_{\,\,\lambda}(\phi,\theta,0) |\phi\theta\lambda_{1}\lambda_{2} \rangle \\ &=N_{J}\sum_{\sigma_{1}\sigma_{2}}\int d\Omega D^{M(J)*}_{\,\,\lambda}(\phi,\theta,0)D^{\sigma_1(s_{1})}_{\,\,\lambda_{1}}(\phi,\theta,0)D^{\sigma_2(s_2)}_{\,\,-\lambda_{2}}(\phi,\theta,0)|\phi\theta \sigma_{1}\sigma_{2} \rangle.
		\end{aligned}
\end{equation}
Using Eqs.~\eqref{Eq:D coupling 1}, \eqref{Eq:D coupling 2}, and \eqref{Eq:define spherical harmonics}, coupling the three Wigner $D$-matrices via CGCs yields
\begin{equation}\label{Eq:two wignerD couple}
    D^{\sigma_{1}(s_{1})}_{\,\,\lambda_{1}}D^{\sigma_{2}(s_{2})}_{\,\,-\lambda_{2}} = \sum_{S}C_{S,\sigma_{S}}^{s_1,\sigma_{1};s_2,\sigma_{2}}C^{S,\lambda}_{s_1,\lambda_1;s_2,-\lambda_2}D^{\sigma_{S}(S)}_{\,\,\lambda},
\end{equation}
and
\begin{equation}\label{Eq:two wignerD couple 2}
\begin{aligned}
    D^{M(J)*}_{\,\,\lambda}D^{\sigma_{S}(S)}_{\,\,\lambda}=&\sum_{L}\left(\frac{2L+1}{2J+1}\right)C_{J,M}^{L,\sigma_{L};S,\sigma_{S}}C^{J,\lambda}_{L,0;S,\lambda}D^{\sigma_L(L)*}_{\,\,0}\\
    =&\sum_{L}\sqrt{\frac{4\pi}{2L+1}}\left(\frac{2L+1}{2J+1}\right)C_{J,M}^{L,\sigma_{L};S,\sigma_{S}}C^{J,\lambda}_{L,0;S,\lambda}Y^{L}_{\sigma_{L}},
\end{aligned}
\end{equation}
where
\begin{equation}
    |s_1-s_2|\leq S\leq s_1+s_2,\quad|J-S|\leq L\leq J+S,
\end{equation}
\begin{equation}
    \sigma_{S}=\sigma_{1}+\sigma_2,\quad \lambda=\lambda_1-\lambda_2,\quad M=\sigma_{L}+\sigma_{S},
\end{equation}
and $Y^L_{\sigma_{L}}$ is the normalized spherical harmonic.
Combining Eqs.~\eqref{Eq:helicity coupling to canonical state}, \eqref{Eq:two wignerD couple}, and \eqref{Eq:two wignerD couple 2}, we obtain the transformation between helicity and canonical construction
\begin{equation}\label{Eq:helicity two to canonical two}
	|JM\lambda_{1}\lambda_{2} \rangle = \sum_{LS}\left( \frac{2L+1}{2J+1}\right) ^{\frac{1}{2}}C^{J,\lambda}_{L,0;S,\lambda}C^{S,\lambda}_{s_1,\lambda_1;s_2,-\lambda_2}|JMLS \rangle,
\end{equation}
and the recoupling coefficient
\begin{equation}
    \langle J^\prime M^\prime LS|JM\lambda_1\lambda_2\rangle=\left( \frac{2L+1}{2J+1}\right) ^{\frac{1}{2}}C^{J,\lambda}_{L,0;S,\lambda}C^{S,\lambda}_{s_1,\lambda_1;s_2,-\lambda_2}\delta^{J^\prime}_{J}\delta^{M^\prime}_{M}.
\end{equation}
The inverse relation of Eq.~\eqref{Eq:helicity two to canonical two} is
\begin{equation}\label{Eq:canonical two to helicity two}
	|JMLS \rangle = \sum_{\lambda_{1}\lambda_{2}}\left( \frac{2L+1}{2J+1}\right) ^{\frac{1}{2}}C_{J,\lambda}^{L,0;S,\lambda}C_{S,\lambda}^{s_1,\lambda_1;s_2,-\lambda_2}|JM\lambda_{1}\lambda_{2} \rangle.
\end{equation}

\subsection{Basic amplitudes for two-body decay}
\label{subsec:Two-body decays}

In the following discussion, we consider the two-body decay process $3 \to 1 + 2$, where all particles are massive and particle-1, particle-2, and particle-3 carry spins $s_1$, $s_2$, and $s_3$, respectively. We then compare the canonical, helicity in the COM frame. Finally, we present the two-body decay amplitude in any frame.
\subsubsection{Canonical and helicity decay amplitudes and their equivalence}
\label{subsubsec:Canonical and helicity decay amplitudes and their equivalence}
The canonical decay amplitude in the COM frame, denoted by $A_{\sigma_{3}}^{\sigma_{1}\sigma_{2}}$, can be written as
\begin{equation}
	A_{\sigma_{3}}^{\sigma_{1}\sigma_{2}}(\mathbf{k}_3,\mathbf{p}^*_1,\mathbf{p}^*_2) = \langle \mathbf{p}^*_1,s_1\sigma_{1}; \mathbf{p}^*_2,s_2\sigma_{2}|M|\mathbf{k}_3,s_{3}\sigma_{3}\rangle,\quad\mathbf{p}^*_1=-\mathbf{p}^*_2.
\end{equation}
Within the $LS$ coupling formalism, this amplitude can be expanded in $LS$ partial waves. Using Eq.~\eqref{Eq:cannonical two J}, one has
\begin{equation}\label{Eq:canonical amplitude}
	\begin{aligned}
		A_{\sigma_{3}}^{\sigma_{1}\sigma_{2}} (\mathbf{k}_3,\mathbf{p}^*_1,\mathbf{p}^*_2)&=\sum_{LS} \frac{1}{N}\langle \Omega \sigma_{1}\sigma_{2}|s_{3}\sigma_{3}LS\rangle\langle s_{3}\sigma_{3}LS|M|s_{3}\sigma_{3}\rangle  \\
        &=\sum_{LS} C_{s_3,\sigma_{3}}^{L,\sigma_{L};S,\sigma_{S}}C_{S,\sigma_{S}}^{s_1,\sigma_{1};s_2,\sigma_{2}}  Y^{L}_{\sigma_{L}}(\Omega)\frac{1}{N}\langle s_{3}\sigma_{3}LS|M|s_{3}\sigma_{3}\rangle \\
		&=\sum_{LS} g_{LS}C_{s_3,\sigma_{3}}^{L,\sigma_{L};S,\sigma_{S}}C_{S,\sigma_{S}}^{s_1,\sigma_{1};s_2,\sigma_{2}} Y^{L}_{\sigma_{L}}(\Omega),
	\end{aligned}
\end{equation}
where $g_{LS}$ is called the $LS$ coupling amplitude and is determined from experiment, it can be written as 
\begin{equation}
    g_{LS}=\frac{1}{N}\langle s_{3}\sigma_{3}LS|M|s_{3}\sigma_{3}\rangle.
\end{equation}
From the angular momentum selection rules, the following constraints hold
\begin{equation}
    |s_1-s_2|\leq S\leq s_1+s_2,\quad |s_3-S|\leq L\leq s_3+S.
\end{equation}
The traditional-$LS$ decay amplitude is defined as
\begin{equation}\label{Eq:canonical decay amplitude}
    A_{\sigma_{3}}^{\sigma_{1}\sigma_{2}}(\mathbf{k}_3,\mathbf{p}^*_1,\mathbf{p}^*_2;L,S) \equiv  g_{LS}C_{s_3,\sigma_{3}}^{L,\sigma_{L};S,\sigma_{S}}C_{S,\sigma_{S}}^{s_1,\sigma_{1};s_2,\sigma_{2}}  Y^{L}_{\sigma_{L}}(\Omega).
\end{equation}

Turning to the helicity coupling, the helicity decay amplitude in the COM frame is defined as
\begin{equation}
	A_{\sigma_{3}}^{\lambda_{1}\lambda_{2}}(\mathbf{k}_3,\mathbf{p}^*_1,\mathbf{p}^*_2) = \langle \mathbf{p}^*_1,s_1\lambda_{1}; \mathbf{p}^*_2,s_2\lambda_{2}|M|\mathbf{k}_3,s_{3}\sigma_{3}\rangle.
\end{equation}
In this definition, the quantization axis for the initial state (particle-3) is chosen along its spin direction. Consequently, in its rest frame the helicity $\lambda_3$ is identical to the spin projection $\sigma_{3}$, which is why $\sigma_{3}$ is used in the amplitude.
Using Eqs.~\eqref{Eq:helicity two J} and \eqref{Eq:helicity two to canonical two}, the amplitude can be written as
\begin{equation}\label{Eq:helicity amplitude}
	\begin{aligned}
		A_{\sigma_{3}}^{\lambda_{1}\lambda_{2}} &= \frac{1}{N}\langle \Omega\lambda_{1}\lambda_{2}|s_{3}\sigma_{3}\lambda_{1}\lambda_{2}\rangle\langle s_{3}\sigma_{3}\lambda_{1}\lambda_{2}|M|s_{3}\sigma_{3}\rangle\\
        &= \sqrt{\frac{2s_{3}+1}{4\pi}}D^{\sigma_{3}(s_{3})*}_{\,\,\lambda}(\phi,\theta,0)H^{\lambda_{1}\lambda_{2}},
	\end{aligned}
\end{equation}
where $\theta$ and $\phi$ are the helicity angles of particle-1 in the rest frame of particle-3, and $H^{\lambda_{1}\lambda_{2}}$ is called the helicity coupling amplitude. It can be written as
\begin{equation}
    H^{\lambda_{1}\lambda_{2}} = \frac{1}{N}\langle s_{3}\sigma_{3}\lambda_{1}\lambda_{2}|M|s_{3}\sigma_{3}\rangle.
\end{equation}

Both the canonical decay amplitude and the helicity decay amplitude can be expressed in the $LS$ coupling formalism. The canonical decay amplitude discussed above has already been written in terms of its $LS$ components. For the helicity decay amplitude, using Eqs.~\eqref{Eq:canonical completeness} and \eqref{Eq:helicity two to canonical two}, the helicity coupling amplitude in the $LS$ coupling formalism~\cite{Chung:1998} can be written as
\begin{equation}
	\begin{aligned}
		H^{\lambda_{1}\lambda_{2}} &= \frac{1}{N}\langle s_{3}\sigma_{3}\lambda_{1}\lambda_{2}|M|s_{3}\sigma_{3}\rangle\\
        &= \sum_{LS}\langle s_{3}\sigma_{3}\lambda_{1}\lambda_{2}|s_{3}\sigma_{3}LS\rangle \frac{1}{N}\langle s_{3}\sigma_{3}LS|M|s_{3}\sigma_{3}\rangle \\
        &=\sum_{LS}g_{LS} \left(\frac{2L+1}{2s_{3}+1}\right)^{\frac{1}{2}}C_{J,\lambda}^{L,0;S,\lambda}C_{S,\lambda}^{s_1,\lambda_1;s_2,-\lambda_2}.
	\end{aligned}
\end{equation}
From the normalization relation, it follows that
\begin{equation}
\sum_{\lambda_{1}\lambda_{2}}|H^{\lambda_{1}\lambda_{2}}|^{2}=\sum_{LS}|g_{LS}|^{2}.
\end{equation}
Accordingly, the helicity coupling amplitude for a specific $LS$ partial wave is defined as
\begin{equation}\label{Eq:helicity coupling amplitude}
    H^{\lambda_{1}\lambda_{2}}(L,S) \equiv g_{LS} \left(\frac{2L+1}{2s_{3}+1}\right)^{\frac{1}{2}}C_{J,\lambda}^{L,0;S,\lambda}C_{S,\lambda}^{s_1,\lambda_1;s_2,-\lambda_2}.
\end{equation}

The difference between the helicity decay amplitude and the canonical decay amplitude is that they are expressed in different bases; they are nevertheless physically equivalent. By redefining the canonical state using Eqs.~\eqref{Eq:one particle relation} and \eqref{Eq:redefine two-p state}, the helicity amplitude can be written as
\begin{equation}\label{Eq:redefine helicity canonical}
\begin{aligned}
A_{\sigma_{3}}^{\lambda_{1}\lambda_{2}} =& \langle \mathbf{p}^*_1,s_1\lambda_{1}; \mathbf{p}^*_2,s_2\lambda_{2}|M|\mathbf{k}_3,s_{3}\sigma_{3}\rangle\\
=&\sum_{\sigma_{1}\sigma_{2}}D_{\,\,\lambda_1}^{\sigma_{1}(s_{1})*}(\phi,\theta,0)D_{\,\,-\lambda_2}^{\sigma_{2}(s_{2})*}(\phi,\theta,0)\langle \mathbf{p}^*_1,s_1\sigma_{1}; \mathbf{p}^*_2,s_2\sigma_{2}|M|\mathbf{k}_3,s_{3}\sigma_{3}\rangle\\
=&\sum_{\sigma_{1}\sigma_{2}}D_{\,\,\lambda_1}^{\sigma_{1}(s_{1})*}(\phi,\theta,0)D_{\,\,-\lambda_2}^{\sigma_{2}(s_{2})*}(\phi,\theta,0)A_{\sigma_{3}}^{\sigma_{1}\sigma_{2}}\\
=&\sum_{\sigma_{1}\sigma_{2}}(-1)^{s_2}D_{\,\,\lambda_1}^{\sigma_{1}(s_{1})*}(\phi,\theta,0)D_{\,\,\lambda_2}^{\sigma_{2}(s_{2})*}(\pi+\phi,\pi-\theta,0)A_{\sigma_{3}}^{\sigma_{1}\sigma_{2}}.
\end{aligned}
\end{equation}

\subsubsection{Example of helicity amplitude calculation}
In this subsubsection, we provide some calculation examples of helicity amplitudes. For the example of traditional-$LS$ amplitude, in order to make a comparison with other $LS$ methods, it is presented in subsection~\ref{subsec:Comparison of calculations in different LS couling schemes}.

In a fixed three-point amplitude, the helicity-coupling amplitude depends only on the masses. Therefore, for a given two-body decay process with definite masses and spins, it is merely a constant factor. Hence, we are only concerned with the angular dependence. Finally, we will also illustrate what the helicity-coupling amplitude should be for specific interaction vertices.

The energy and the momentum of the daughter particles are given as
\begin{equation}
    E_{1}=\frac{m^2_3+m_1^2-m_2^2}{2m_3}, \quad E_{2}=\frac{m^2_3+m_2^2-m_1^2}{2m_3}, 
\end{equation}
\begin{equation}
	q=|\mathbf{p}_{1}^*|=|\mathbf{p}_{2}^*| =\frac{\sqrt{((m_{3}+m_{2})^{2}-m_{1}^{2})((m_{3}+m_{1})^{2}-m_{2}^{2})}}{2m_{3}},
\end{equation}
where $m_1$, $m_2$ and $m_3$ are the mass of the particles.

For the two-body decay process $3 \to 1 + 2$ with spins $s_3=1,s_1=1,s_2=0$. The helicity amplitude is
\begin{equation}\label{Eq:ex helicity 110}
\begin{aligned}
A^{\lambda_1\lambda_2}_{\lambda_3}=H^{\lambda_1\lambda_2}D^{\lambda_3(1)*}_{\,\,\lambda}(\phi,\theta,0),
\end{aligned}
\end{equation}
where ${\lambda=\lambda_1-\lambda_2}$. 
When the decay vertex is $C_{VVS}g_{\mu\nu}V_3^\mu V_1^\nu S$, where $C_{VVS}$ is a constant, The above amplitude can be written as
\begin{equation}\label{Eq:ex helicity vertex110}
\begin{aligned}
A^{+0}_+ &= \left[-C_{VVS}\right] D^{+(1)*}_{\,\,+}(\phi,\theta,0),\\
A^{00}_+ &= \left[-C_{VVS}\frac{E_1}{m_1}\right] D^{+(1)*}_{\,\,0}(\phi,\theta,0),\\
A^{-0}_+ &= \left[-C_{VVS}\right] D^{+(1)*}_{\,\,-}(\phi,\theta,0),\\
A^{+0}_0 &= \left[-C_{VVS}\right]D^{0(1)*}_{\,\,+}(\phi,\theta,0),\\
A^{00}_0 &= \left[-C_{VVS}\frac{E_1}{m_1}\right]D^{0(1)*}_{\,\,0}(\phi,\theta,0),\\
A^{-0}_0 &= \left[-C_{VVS}\right]D^{0(1)*}_{\,\,-}(\phi,\theta,0),\\
A^{+0}_- &= \left[-C_{VVS}\right] D^{-(1)*}_{\,\,+}(\phi,\theta,0),\\
A_-^{00} &= \left[-C_{VVS}\frac{E_1}{m_1}\right] D^{-(1)*}_{\,\,0}(\phi,\theta,0),\\
A_-^{-0} &= \left[-C_{VVS}\right] D^{-(1)*}_{\,\,-}(\phi,\theta,0).
\end{aligned}
\end{equation}
By comparing Eqs.~\eqref{Eq:ex helicity 110} and \eqref{Eq:ex helicity vertex110}, the helicity-coupling amplitude can be determined.

For the two-body decay process $3 \to 1 + 2$ with spins $s_3=\frac{1}{2},s_1=\frac{1}{2},s_2=0$. The helicity amplitude is
\begin{equation}\label{Eq:1/21/20helicity}
A^{\lambda_1\lambda_2}_{\lambda_3}=H^{\lambda_1\lambda_2}D^{\lambda_3(\frac{1}{2})*}_{\,\,\lambda}(\phi,\theta,0),
\end{equation}
where $\lambda=\lambda_1-\lambda_2$.
When the decay vertex is $\bar f_1\,\gamma^\mu\left(C_L P_L + C_R P_R\right)f_2\,S$, where $C_{L}$ and $C_R$ are the constant. The left and right-handed projection operators is $P_{L,R}$. The above amplitude can be written as
\begin{equation}\label{1/21/20helicityver}
\begin{aligned}
A^{+\frac{1}{2}0}_{+\frac{1}{2}} &= \left[\frac{C_R\,P_1^- + C_L\,P_1^+}{\sqrt{2}}\right]
D^{+\frac{1}{2}(\frac{1}{2})*}_{\,\,+\frac{1}{2}}(\phi,\theta,0),\\
A^{-\frac{1}{2}0}_{+\frac{1}{2}} &= \left[\frac{C_R\,P_1^+ + C_L\,P_1^-}{\sqrt{2}}\right]
D^{+\frac{1}{2}(\frac{1}{2})*}_{\,\,-\frac{1}{2}}(\phi,\theta,0),\\
A^{+\frac{1}{2}0}_{-\frac{1}{2}} &= \left[\frac{C_R\,P_1^- + C_L\,P_1^+}{\sqrt{2}}\right]
D^{-\frac{1}{2}(\frac{1}{2})*}_{\,\,+\frac{1}{2}}(\phi,\theta,0),\\
A^{-\frac{1}{2}0}_{-\frac{1}{2}} &= \left[\frac{C_R\,P_1^+ + C_L\,P_1^-}{\sqrt{2}}\right]
D^{-\frac{1}{2}(\frac{1}{2})*}_{\,\,-\frac{1}{2}}(\phi,\theta,0),
\end{aligned}
\end{equation}
where
\begin{equation}
P_1^{\pm}=\sqrt{m_3}\frac{E_1+m_1\pm q}{\sqrt{E_1+m_1}},
\qquad
P_2^{\pm}=\frac{1}{\sqrt{2}}\sqrt{\frac{E_2\pm q}{E_2\mp q}}.
\end{equation}
By comparing Eqs.~\eqref{Eq:1/21/20helicity} and \eqref{1/21/20helicityver}, the helicity-coupling amplitude can be determined.

\subsubsection{Two-body decay amplitude in any frame}
\label{subsubsec:Decay amplitude in any frame}

In the above discussion, we have presented the two-body decay amplitudes in the two-body decay COM frame. In practice, however, the momenta of the particles in the decay process are generally not given in the COM frame, but in any frame, such as the lab frame. Therefore, to evaluate the amplitude for a given event in the lab frame, one must boost the event back to the COM frame. Such a boost acts nontrivially on the final-state particles and induces nontrivial Wigner rotation.
Consequently, the amplitude in any frame can be written as the corresponding COM frame amplitude multiplied by the appropriate Wigner $D$-matrices for the final-state particles.

In any frame, within the traditional-$LS$ partial wave expansion, the canonical amplitudes can be organized by the $LS$ coupling amplitudes, the CGCs, the spherical harmonics in the COM frame, and the Wigner $D$-matrices induced by the boost. It can be written as
\begin{equation}
    A_{\sigma_{3}}^{\sigma_{1}\sigma_{2}}(\mathbf{p}_3,\mathbf{p}_1,\mathbf{p}_2) =\sum_{LS\sigma_L\sigma_S\sigma_{1}^{\prime}\sigma_{2}^{\prime}} g_{LS}C_{s_3,\sigma_{3}}^{L,\sigma_{L};S,\sigma_{S}}C_{S,\sigma_{S}}^{s_1,\sigma_{1}^\prime;s_2,\sigma_{2}^\prime} Y^{L}_{\sigma_{L}}(\Omega)D^{\sigma_{1}^{\prime}(s_{1})*}_{\,\,\sigma_{1}}(R_{31})D^{\sigma^{\prime}_{2}(s_{2})*}_{\,\,\sigma_{2}}(R_{32}),
\end{equation}
where $R_{31}$ and $R_{32}$ are the Wigner rotations induced on particle-1 and particle-2 by boosting the amplitude from any frame to the rest frame of particle-3 through the canonical-standard boost of particle-3. This Wigner rotation is exactly the canonical little group rotation in Eq.~\eqref{Eq:littlegroupro}, and it can be written as
\begin{equation}
    R_{3i}=W(L_{3,c}^{-1}(\mathbf{p}_3),p_1).
\end{equation}


By contrast, the helicity amplitudes are organized by the helicity-coupling amplitudes, the helicity Wigner $D$-matrix in the COM frame, and the Wigner $D$-matrices induced by the boost. It can be written as
\begin{equation}
    A_{\lambda_{3}}^{\lambda_{1}\lambda_{2}}(\mathbf{p}_3,\mathbf{p}_1,\mathbf{p}_2) =\sum_{\lambda_{1}^{\prime}\lambda_{2}^{\prime}} \sqrt{\frac{2s_{3}+1}{4\pi}}D^{\lambda_{3}(s_{3})*}_{\,\,\lambda^\prime_1-\lambda^\prime_2}(\phi,\theta,0)H^{\lambda_{1}^\prime\lambda_{2}^\prime}D^{\lambda_{1}^{\prime}(s_{1})*}_{\,\,\lambda_{1}}(R^{\prime}_{31})D^{\lambda^{\prime}_{2}(s_{2})*}_{\,\,\lambda_{2}}(R^{\prime}_{32}),
\end{equation}
where $R{31}^\prime$ and $R_{32}^\prime$ are the Wigner rotations induced on particle-1 and particle-2 by boosting the amplitude from any frame to the rest frame of particle-3 through the helicity standard boost of particle-3. This Wigner rotation is exactly the helicity little group rotation in Eq.~\eqref{Eq:littlegroupro}, and it can be written as
\begin{equation}
    R^\prime_{3i}=W(L_{3,h}^{-1}(\mathbf{p}_3),p_1).
\end{equation}

In what follows, we evaluate this Wigner rotation explicitly for both the canonical and helicity amplitudes, and give the corresponding amplitude expressions in any frame.

\paragraph{Canonical amplitude in any frame}
In any frame, such as the lab frame, we define the canonical amplitude as
\begin{equation}
    A_{\sigma_{3}}^{\sigma_{1}\sigma_{2}}(\mathbf{p}_3,\mathbf{p}_1,\mathbf{p}_2) \equiv \langle \mathbf{p}_1\sigma_{1}; \mathbf{p}_2\sigma_{2}|M|\mathbf{p}_{3}\sigma_{3}\rangle,
\end{equation}
where $\mathbf{p}_3$, $\mathbf{p}_1$ and $\mathbf{p}_2$ are the momenta in the lab frame.
To calculate this amplitude, the system is boosted to the COM frame. 
In the canonical basis, the corresponding boost is $L^{-1}_{3,c}(p_3)$, where the subscript $3$ indicates that this is the standard boost associated with particle-3.

Under the Lorentz transformation, the canonical amplitude becomes
\begin{equation}\label{Eq:boost canonical amplitude}
    \langle \mathbf{p}_1\sigma_{1}; \mathbf{p}_2\sigma_{2}|M|\mathbf{p}_{3}\sigma_{3}\rangle\xrightarrow{L^{-1}_{3,c}(p_3)}  \langle \mathbf{p}^*_1\sigma_{1}^{\prime}; \mathbf{p}^*_2\sigma_{2}^{\prime}|M|\mathbf{k}_{3}\sigma_{3}\rangle,
\end{equation}
where $\mathbf{k}_3$, $\mathbf{p}_1^*$ and $\mathbf{p}_2^*$ are the momenta of particle-3, particle-1 and particle-2 in the two-body decay COM frame.

\begin{figure}[H]
\centering
\includegraphics[scale=0.85]{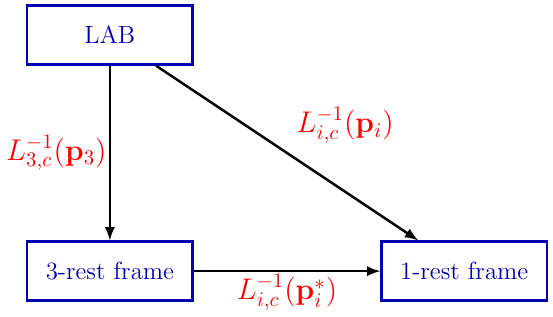}
    \caption{Two boost paths from the lab frame to the rest frame of particle-$i$: a direct boost $L_{i,c}^{-1}(\textbf{p}_i)$ and the two-step boost $L_{i,c}^{-1}(\mathbf{p}_i^\ast)L_{3,c}^{-1}(\mathbf{p}_3)$. Their difference is the Wigner rotation acting on the little group indices of particle-$i$.}
    \label{fig:lab to 3 1frame}
\end{figure}

For particle-3, the boost that brings it from the LAB frame to its rest frame is collinear with its momentum. According to Eq.~\eqref{Eq:little group same direct}, such a collinear boost induces a trivial little group element, so no additional rotation acts on the spin indices. 
In contrast, for particle-1 and particle-2 one obtains nontrivial little group rotations. 
The reason is that the transformation from the LAB frame to the rest frame of particle-$i$ is not unique. 
One option is to reach the particle-$i$ rest frame by a single pure boost. 
Alternatively, one may first boost to the rest frame of particle-3 and then boost to the rest frame of particle-$i$. 
Although both constructions end in the rest frame of particle-$i$ and map $p_i$ to the standard rest momentum $k_i$, they define different spatial axis in that rest frame. 
The two definitions are related by a Wigner rotation. Therefore, when the two-step construction is adopted, the resulting rest-frame state must be supplemented by the corresponding little group rotation.
A simple diagrammatic description for the Wigner rotation is shown in Fig.~\ref{fig:lab to 3 1frame}.

From Eq.~\eqref{Eq:Lorentz transformation on a state}, the transformations of the individual states can be written as
\begin{equation}
\begin{aligned}
U[L^{-1}_{3,c}(\mathbf{p}_3)]|\mathbf{p}_i\sigma_i \rangle
= \sum_{\sigma_{i}^{\prime}} D^{\sigma_{i}^{\prime}(s_{i})}_{\,\,\sigma_{i}}(R_{3i})| \mathbf{p}^*_i\sigma_i^{\prime}\rangle\quad (i=1,2,3),
\end{aligned}
\end{equation}
where $\mathbf{p}^*_3=\mathbf{k}_3$ and $R_{3i}$ is the corresponding Wigner rotation defined by\footnote{The Wigner rotation is also known as the Wigner-Thomas rotation, see e.g. sec~6.7 in Ref.~\cite{Gourgoulhon2013SRGF}}
\begin{equation}\label{Eq:little group rotation1}
    R_{3i} = L^{-1}_{i,c}(\mathbf{p}_{i}^{*}) L^{-1}_{3,c}(\mathbf{p}_{3}) L_{i,c}(\mathbf{p}_{i})\quad (i=1,2,3),
\end{equation}
where the subscript $i$ indicates that this is the standard boost associated with particle-$i$.
In particular, the Wigner rotation for particle-3 is trivial, $R_{33} = \mathbf{I}$. 
Therefore, only two nontrivial Wigner rotation representation matrices act on the amplitude. Explicitly, the transformed amplitude can be written as
\begin{equation}\label{Eq:any frame canonical}
    A_{\sigma_{3}}^{\sigma_{1}\sigma_{2}}(\mathbf{p}_3,\mathbf{p}_1,\mathbf{p}_2) =\sum_{\sigma_{1}^{\prime}\sigma_{2}^{\prime}} A_{\sigma_{3}}^{\sigma^{\prime}_{1}\sigma^{\prime}_{2}}(\mathbf{k}_3,\mathbf{p}^*_1,\mathbf{p}^*_2)D^{\sigma_{1}^{\prime}(s_{1})*}_{\,\,\sigma_{1}}(R_{31})D^{\sigma^{\prime}_{2}(s_{2})*}_{\,\,\sigma_{2}}(R_{32}),
\end{equation}
where $A_{\sigma_{3}}^{\sigma^{\prime}_{1}\sigma^{\prime}_{2}}(\mathbf{k}_3,\mathbf{p}^*_1,\mathbf{p}^*_2)$ are the amplitudes in the COM frame.
With above definitions in place, a detailed derivation of the Wigner rotation $R_{3i}$ is presented below.

For a particle with four-momentum $p$, we introduce the velocity and the Lorentz factor as
\begin{equation}\label{Eq:beta_gamma}
\boldsymbol{\beta} = \frac{\mathbf{p}}{E},
\quad
\beta = |\boldsymbol{\beta}|,
\quad
\gamma = \frac{1}{\sqrt{1-\beta^{2}}}.
\end{equation}
The canonical-standard boost of the particle is
\begin{equation}\label{Eq:canonica boost matrix}
L_c(\mathbf{p}) =
\begin{pmatrix}
\gamma & \gamma\boldsymbol{\beta}^{T} \\
\gamma\boldsymbol{\beta} & \mathbf{I}+\dfrac{\gamma-1}{\beta^{2}}\boldsymbol{\beta}\boldsymbol{\beta}^{T}
\end{pmatrix}.
\end{equation}
Its inverse is
\begin{equation}\label{Eq:inverse canonica boost matrix}
L^{-1}_c(\mathbf{p})=L_c(-\mathbf{p})=
\begin{pmatrix}
\gamma & -\gamma\boldsymbol{\beta}^{T} \\
-\gamma\boldsymbol{\beta} & \mathbf{I}+\dfrac{\gamma-1}{\beta^{2}}\boldsymbol{\beta}\boldsymbol{\beta}^{T}
\end{pmatrix}.
\end{equation}

The composition of the first two Lorentz transformation of Eq.~\eqref{Eq:little group rotation1} is defined as
\begin{equation}\label{Eq:lambda boost}
\Lambda \equiv L^{-1}_{3,c}(\mathbf{p}_{3}) L_{i,c}(\mathbf{p}_{i})=L_{3,c}(-\mathbf{p}_{3}) L_{i,c}(\mathbf{p}_{i}).
\end{equation}
In the rest frame of particle-$i$, the momentum of particle-$i$ is the standard momentum $k_i^{\mu}=(m_i,\mathbf{0})^T$.
The corresponding momentum $p_i^{*}$ in the two-body decay COM frame is obtained by acting with the above Lorentz transformation on $k_i$, namely,
\begin{equation}\label{Eq:lambdakk to p}
p_i^{*}=\Lambda k_i.
\end{equation}
Using Eqs.~\eqref{Eq:beta_gamma}, \eqref{Eq:canonica boost matrix}, \eqref{Eq:inverse canonica boost matrix}, \eqref{Eq:lambda boost} and \eqref{Eq:lambdakk to p}, one can obtain
\begin{equation}
\gamma^*_{i}=\gamma_3\gamma_i\left(1-\boldsymbol{\beta}_3\cdot\boldsymbol{\beta}_i\right),
\quad
\boldsymbol{\beta}^*_{i}=\frac{
-\gamma_3\boldsymbol{\beta}_3
+
\boldsymbol{\beta}_i
+
\dfrac{\gamma_3-1}{\beta_3^{2}}(\boldsymbol{\beta}_3\cdot\boldsymbol{\beta}_i)\boldsymbol{\beta}_3
}{
\gamma_3\left(1-\boldsymbol{\beta}_3\cdot\boldsymbol{\beta}_i\right)
}.
\end{equation}
Therefore, the Wigner rotation induced by the boost can be written as
\begin{equation}
R_{3i}(\boldsymbol{\omega}_i)= L^{-1}_{i,c}(\mathbf{p}_{i}^{*}) L^{-1}_{3,c}(\mathbf{p}_{3}) L_{i,c}(\mathbf{p}_{i}),
\end{equation}
where the $\boldsymbol{\omega}_i$  direction parallel to $-\mathbf p_3 \times \mathbf p_i$.

To obtain the explicit magnitude of the Wigner angle $\omega_i=|\boldsymbol{\omega}_i|$, we first rotate the coordinate system such that the $z$-axis is aligned with the unit vector
\begin{equation}
\hat{\mathbf n}_{3i}
\equiv
\frac{-\mathbf p_3\times \mathbf p_i}{\bigl|\mathbf p_3\times \mathbf p_i\bigr|},
\quad
R_{\hat{\mathbf n}_{3i}}\hat{\mathbf z}=\hat{\mathbf n}_{3i},
\quad\cos\theta_{3i}=-\hat{\mathbf{p}}_3\cdot\hat{\mathbf{p}}_i,
\end{equation}
where $R_{\hat{\mathbf n}_{3i}}$ is a spatial rotation that maps $\hat{\mathbf z}$ to $\hat{\mathbf n}_{3i}$.
In the rotated frame, the two velocities are lie in the $x$-$y$ plane. We may choose coordinates such that
\begin{equation}
\boldsymbol{\beta}_3'=(\beta_3,0,0)
\qquad
\boldsymbol{\beta}_i'=(\beta_i\cos\theta_{3i},\,\beta_i\sin\theta_{3i},\,0),
\end{equation}
where $\theta_{3i}$ is the angle between $\boldsymbol{\beta}_3$ and $\boldsymbol{\beta}_i$ in the lab frame. In the rotated frame, one finds that the boosted velocity $\boldsymbol{\beta}_i^{*\prime}$ also satisfies $(\boldsymbol{\beta}_i^{*\prime})_z=0$.
Therefore, in this rotated frame the Wigner rotation is a pure rotation about the $z$-axis:
\begin{equation}
R_{3i}' = R_z(\omega_{3i})=\begin{pmatrix}
1 & 0 & 0 & 0\\
0 & \cos\omega_{3i} & -\sin\omega_{3i} & 0\\
0 & \sin\omega_{3i} & \cos\omega_{3i} & 0\\
0 & 0 & 0 & 1
\end{pmatrix}.
\end{equation}
A convenient closed form for $\omega_{3i}$ is obtained in terms of rapidities
$\eta_a\equiv \operatorname{arctanh}\beta_a$ ($a=3,i$):
\begin{equation}\label{Eq:wigner angle}
\tan\frac{\omega_{3i}}{2}
=\frac{}{}
\frac{\sin\theta_{3i}\,\sinh\!\left(\frac{\eta_3}{2}\right)\sinh\!\left(\frac{\eta_i}{2}\right)}
{\cosh\!\left(\frac{\eta_3}{2}\right)\cosh\!\left(\frac{\eta_i}{2}\right)
+\cos\theta_{3i}\,\sinh\!\left(\frac{\eta_3}{2}\right)\sinh\!\left(\frac{\eta_i}{2}\right)}.
\end{equation}
This determines the magnitude of the Wigner angle.

Finally, we rotate back to the original orientation. The Wigner rotation in the original frame can be written as
\begin{equation}\label{Eq:canonical wiger rotation}
R_{3i}(\boldsymbol{\omega}_i)
=
R_{\hat{\mathbf{n}}}R_z(\omega_{3i})R_{\hat{\mathbf{n}}}^{-1}.
\end{equation}
With $R_{31}$ and $R_{32}$ determined in this way, the canonical amplitude in any frame becomes
\begin{equation}
    A_{\sigma_{3}}^{\sigma_{1}\sigma_{2}}(\mathbf{p}_3,\mathbf{p}_1,\mathbf{p}_2) =\sum_{LS\sigma_L\sigma_S\sigma_{1}^{\prime}\sigma_{2}^{\prime}} g_{LS}C_{s_3,\sigma_{3}}^{L,\sigma_{L};S,\sigma_{S}}C_{S,\sigma_{S}}^{s_1,\sigma_{1}^\prime;s_2,\sigma_{2}^\prime} Y^{L}_{\sigma_{L}}(\Omega)D^{\sigma_{1}^{\prime}(s_{1})*}_{\,\,\sigma_{1}}(R_{31}(\boldsymbol{\omega}_1))D^{\sigma^{\prime}_{2}(s_{2})*}_{\,\,\sigma_{2}}(R_{32}(\boldsymbol{\omega}_2)).
\end{equation}

\begin{figure}[H]
\centering
\includegraphics[width=\linewidth]{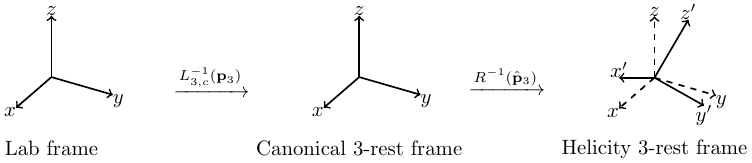}
\caption{The figure illustrates the relationships among the three coordinate systems. The lab frame and the canonical 3-rest frame are connected by a pure boost, which does not change the orientation of the axes; therefore, they share the same set of spatial axes $x, y, z$. For the helicity 3-rest frame, an additional rotation is applied, resulting in a change of the spatial axes. The corresponding new axes are denoted as $x', y', z'$ .}
\label{fig:lab-canonica-helicity}
\end{figure}

\paragraph{Helicity amplitude in any frame}
For the helicity amplitude, the definition in any frame is
\begin{equation}
    A_{\lambda_{3}}^{\lambda_{1}\lambda_{2}}(\mathbf{p}_3,\mathbf{p}_1,\mathbf{p}_2) = \langle \mathbf{p}_1\lambda_{1}; \mathbf{p}_2\lambda_{2}|M|\mathbf{p}_{3}\lambda_{3}\rangle.
\end{equation}
Similarly, the helicity amplitude is boosted to the COM frame. To preserve the helicity of the parent particle, the boost is chosen to be $L^{-1}_h(\mathbf{p}_3)$. 
However, the helicity boost is not a pure boost but a combination of a pure boost and a rotation. This is because, in the rest frame of particle-3, its spin quantization axis must point along the $z$-axis, while originally in the lab frame its spin quantization axis points along the direction of its momentum. Hence, we need to rotate the $z$-axis to align with the momentum direction of particle 3.
This implies that boosting back from the lab frame to the helicity COM frame does not yield the same frame as the canonical COM frame. Consequently, the momenta of particle-1 and particle-2 are different in these two COM frames. We define the momenta in the helicity COM frame as $\mathbf{q}^*_1$, $\mathbf{q}^*_2$, whose relation with the momenta $\mathbf{p}^*_1$, $\mathbf{p}^*_2$ in the canonical COM frame is given by 
\begin{equation}\label{Eq:helicity q canonical p}
\mathbf{q}_i^* = L^{-1}_{3,h}(\mathbf{p}_3) \mathbf{p}_i=R^{-1}(\hat{\mathbf{p}}_3)L^{-1}_{3,c}(\mathbf{p}_3) \mathbf{p}_i=R^{-1}(\hat{\mathbf{p}}_3)\mathbf{p}_i^{*}\quad (i = 1, 2).
\end{equation}
The relations among the three coordinate systems are illustrated in Fig.~\ref{fig:lab-canonica-helicity}.

Under this Lorentz transformation, the helicity amplitude becomes
\begin{equation}\label{Eq:boost helicity amplitude}
    \langle \mathbf{p}_1\lambda_{1}; \mathbf{p}_2\lambda_{2}|M|\mathbf{p}_{3}\lambda_{3}\rangle\xrightarrow{L^{-1}_{3,c}(\mathbf{p}_3)}\langle \mathbf{p}^*_1\tilde\lambda_{1}; \mathbf{p}^*_2\tilde\lambda_{2}|M|\mathbf{k}_{3}\lambda_{3}\rangle\xrightarrow{R^{-1}(\hat{\mathbf{p}}_3)}  \langle \mathbf{q}^*_1\lambda_{1}^{\prime}; \mathbf{q}^*_2\lambda_{2}^{\prime}|M|\mathbf{k}_{3}\lambda_{3}\rangle,
\end{equation}
where $\mathbf{p}_1, \mathbf{p}_2, \mathbf{p}_3$ are the momenta defined in the lab frame with respect to the coordinate axes $\{x, y, z\}$, and the helicities $\lambda_1, \lambda_2, \lambda_3$ correspond to the spin projections along their respective momentum directions. In the COM frame, $\mathbf{k}_3, \mathbf{q}_1, \mathbf{q}_2$ are the momenta defined in the helicity COM frame with respect to the coordinate axes $\{x^\prime, y^\prime, z^\prime\}$, and $\lambda_1^\prime, \lambda_2^\prime, \lambda_3$ again correspond to the spin projections along their respective momentum directions. 
In particular, since the direction of $\mathbf{p}_3$ in the lab frame corresponds to the $z^\prime$-axis direction in the helicity COM frame, the helicity $\lambda_3$ is defined consistently in both frames.

Therefore, due to the helicity mismatch between the two reference frames, the helicity Wigner rotations must exist between the helicity particle states defined in these frames to align their helicities. Applying the boost to the spin states yields
\begin{equation}
\begin{aligned}
U[L^{-1}_{3,h}(\mathbf{p}_3)]|\mathbf{p}_i\lambda_i \rangle
&=  L^{-1}_{3,h}(\mathbf{p}_{3}) L_{i,h}(\mathbf{p}_{i}) |\mathbf{k}_i\lambda_i\rangle\\
&=L_{i,h}(\mathbf{q}_{i}^{*}) L^{-1}_{i,h}(\mathbf{q}_{i}^{*}) L^{-1}_{3,h}(\mathbf{p}_{3}) L_{i,h}(\mathbf{p}_{i})|\mathbf{k}_i\lambda_i\rangle  \\
&=\sum_{\lambda_{i}^{\prime}} D^{\lambda_{i}^{\prime}(s_{i})}_{\,\,\lambda_{i}}(R^{\prime}_{3i})| \mathbf{q}^{*}_i\lambda_i^{\prime}\rangle\quad (i=1,2),
\end{aligned}
\end{equation}
where $R^{\prime}_{3i}$ is the corresponding Wigner rotation defined as
\begin{equation}\label{Eq:little group rotation2}
    R^\prime_{3i} = L^{-1}_{i,h}(\mathbf{q}_{i}^{*}) L^{-1}_{3,h}(\mathbf{p}_{3}) L_{i,h}(\mathbf{p}_{i})\quad (i=1,2).
\end{equation}
The helicity amplitude in the lab frame can be written as
\begin{equation}\label{Eq:helicity in any frame}
    A_{\lambda_{3}}^{\lambda_{1}\lambda_{2}}(\mathbf{p}_3,\mathbf{p}_1,\mathbf{p}_2) =\sum_{\lambda_{1}^{\prime}\lambda_{2}^{\prime}} A_{\lambda_{3}}^{\lambda^{\prime}_{1}\lambda^{\prime}_{2}}(\mathbf{k}_3,\mathbf{q}^{*}_1,\mathbf{q}^{*}_2)D^{\lambda_{1}^{\prime}(s_{1})*}_{\,\,\lambda_{1}}(R^{\prime}_{31})D^{\lambda^{\prime}_{2}(s_{2})*}_{\,\,\lambda_{2}}(R^{\prime}_{32}),
\end{equation}
To factorize $R_{3i}^\prime$, we start from the definition of the helicity boost, 
\begin{align}
L^{-1}_{i,h}(\mathbf{q}_{i}^{*})
&
=R^{-1}(\hat{\mathbf{q}}_{i}^{*})L^{-1}_{i,c}(\mathbf{q}_{i}^{*}),\\
L^{-1}_{3,h}(\mathbf{p}_{3})&=R^{-1}(\mathbf{p}_{3})L_{3,c}^{-1}(\mathbf{p}_{3}),\\
L_{i,h}(\mathbf{p}_{i})&=L_{i,c}(\mathbf{p}_{i})R(\mathbf{p}_{i}).
\end{align}
Substituting these relations into the definition of $R_{3i}^\prime$ gives
\begin{equation}\label{Eq:helicitywignerr}
 R^\prime_{3i}
=R^{-1}(\mathbf{q}_{i}^{*})\Big(L^{-1}_{i,c}(\mathbf{q}_{i}^{*})R^{-1}(\mathbf{p}_{3})L^{-1}_{3,c}(\mathbf{p}_{3})L_{i,c}(\mathbf{p}_{i})\Big)R(\mathbf{p}_{i}).
\end{equation}
Using the property $R\,L^{-1}_c(\mathbf{p})R^{-1}=L_c^{-1}(R\mathbf{p})$ and Eq.~\eqref{Eq:helicity q canonical p}, we have
\begin{equation}
L^{-1}_{i,c}(\mathbf{p}^\prime)\equiv R(\hat{\mathbf{p}}_{3})L^{-1}_{i,c}(\mathbf{q}_{i}^{*})R^{-1}(\hat{\mathbf{p}}_{3}),
\qquad
\mathbf{p}^\prime=R(\hat{\mathbf{p}}_{3})\mathbf{q}_{i}^{*}=\mathbf{p}_i^{*}.
\end{equation}
Therefore, inserting $R^{-1} (\hat{\mathbf{p}}_{3})R(\hat{\mathbf{p}}_{3})$ in Eq.~\eqref{Eq:helicitywignerr}, we obtain
\begin{equation}
\begin{aligned}
    R^\prime_{3i}
=&R^{-1}(\mathbf{q}_{i}^{*})R^{-1} (\hat{\mathbf{p}}_{3})\Big(R(\hat{\mathbf{p}}_{3})L^{-1}_{i,c}(\mathbf{q}_{i}^{*})R^{-1}(\mathbf{p}_{3})L^{-1}_{3,c}(\mathbf{p}_{3})L_{i,c}(\mathbf{p}_{i})\Big)R(\mathbf{p}_{i})\\
=&R^{-1}(\mathbf{q}_{i}^{*})R^{-1} (\hat{\mathbf{p}}_{3}) \Big(L^{-1}_{i,c}(\mathbf{p}^*_i)L^{-1}_{3,c}(\mathbf{p}_{3})L_{i,c}(\mathbf{p}_{i})\Big)R(\mathbf{p}_{i}).
\end{aligned}
\end{equation}
It can be observed that the expression inside the bracket corresponds to the Wigner rotation in the canonical given by Eq.~\eqref{Eq:canonical wiger rotation}. Therefore, the final Wigner rotation for the helicity state is given by
\begin{equation}\label{Eq:helicity wiger rotation}
\begin{aligned}
    R^\prime_{3i}
=&R^{-1}(\mathbf{q}_{i}^{*})R^{-1} (\hat{\mathbf{p}}_{3}) R_{3i}({\boldsymbol{\omega}_i})R(\mathbf{p}_{i}).
\end{aligned}
\end{equation}
Finally, the amplitude in any frame can be written as
\begin{equation}
    A_{\lambda_{3}}^{\lambda_{1}\lambda_{2}}(\mathbf{p}_3,\mathbf{p}_1,\mathbf{p}_2) =\sum_{\lambda_{1}^{\prime}\lambda_{2}^{\prime}} \sqrt{\frac{2s_{3}+1}{4\pi}}D^{\lambda_{3}(s_{3})*}_{\,\,\lambda^\prime_1-\lambda^\prime_2}(\phi,\theta,0)H^{\lambda_{1}^\prime\lambda_{2}^\prime}D^{\lambda_{1}^{\prime}(s_{1})*}_{\,\,\lambda_{1}}(R^{\prime}_{31})D^{\lambda^{\prime}_{2}(s_{2})*}_{\,\,\lambda_{2}}(R^{\prime}_{32}).
\end{equation}

\section{Covariant Zemach and projection tensor amplitudes}
\label{sec:Covariant tensor and Zemach tensor amplitudes}

In this section, tensor constructions of $LS$ amplitudes are discussed. This approach can be traced back to the Zemach tensor method~\cite{Zemach:1965ycj}. In the Zemach tensor method, two-body decays are analyzed in the parent rest frame. Unlike the traditional-$LS$ coupling discussed in section~\ref{sec:Two different states description---Helicity and Canonical}, where angular momenta are coupled in the little group representation using CGCs, the Zemach tensor method first builds angular momentum tensors associated with the orbital and spin parts and then performs the required couplings and projections at the tensor level using the invariant tensors of $\mathrm{SO}(3)$ and $\mathrm{SU}(2)$.

However, the Zemach tensor method is not covariant. To construct covariant partial wave amplitudes, the covariant tensor method is introduced~\cite{Zou:2002ar,Zou:2002yy}. Its key idea is to construct covariant coupling structures and obtain the corresponding partial wave amplitudes by fully contracting them with the external spin wave functions over the Lorentz indices. This idea has been developed into a more systematic framework~\cite{Jing:2023rwz,Jing:2024mag}. Depending on the choice of external spin wave functions, two schemes are commonly used.

The first scheme, referred to as the PS-scheme, returns to the COM frame and defines the final-state spin wave functions at the standard momenta.
In this construction, the amplitude can be written as
\begin{equation}
  A({k}_3,{p}_1^*,{p}_2^*;L,S) = \underbrace{\Gamma({k}_3,{p}_1^*,{p}_2^*;L,S)}_{\text {covariant $LS$ coupling structure}}\times \underbrace{U({k}_3,{k}_1,{k}_2;s_{3},s_{1},s_{2})}_{\text {pure-spin part}}.
\end{equation}
In the COM frame, this scheme allows for a complete separation between orbital and spin. Consequently, it is equivalent to the traditional-$LS$ method. 
By applying a Lorentz boost, this definition can be extended to any frame and evaluated there. However, the boost induces the Wigner rotations, and this rotations depend on the COM frame.

The second scheme is referred to as the GS-scheme, which can be written as
\begin{equation}
  \mathcal{C}({p}_3,{p}_1,{p}_2;L,S) = \underbrace{\Gamma({p}_3,{p}_1,{p}_2;L,S)}_{\text {covariant $LS$ coupling structure}}\times \underbrace{U({p}_3,{p}_1,{p}_2;s_{3},s_{1},s_{2})}_{\text {general-spin part}}.
\end{equation}
In this construction, the amplitude is covariant, and no additional boost back to the COM frame is required. This makes the treatment in any frame more natural. However, from the structure of the amplitude one can directly see that, in this scheme, orbital angular momentum and spin angular momentum cannot be completely separated. Part of the orbital information is mixed into what would otherwise be the general-spin structure. As a result, this method is not fully equivalent to the traditional-$LS$ method. 
Instead, it defines an alternative $LS$ partial wave basis.


In subsection~\ref{subsec:Covariant tensor amplitudes}, we review the Zemach tensor method and then introduce the covariant tensor method, clarifying the correspondence between the two constructions.
In subsection~\ref{subsec:Covariant tensor LS amplitude}, we discuss the covariant projection tensor method both in the PS-scheme and the GS-scheme.  
In subsection~\ref{subsec:Comparison of calculations in different LS couling schemes}, we provide several explicit calculation examples for different $LS$ amplitudes.

\subsection{Covariant tensor amplitudes}
\label{subsec:Covariant tensor amplitudes}

In this subsection, the tensor method for constructing the $LS$ amplitudes is introduced. Its key idea is to formulate angular momentum coupling in tensor representations, so that the coupling is implemented by index contractions rather than by explicit CGCs in the little group.

The discussion starts from the non-relativistic Zemach tensor method~\cite{Zemach:1965ycj}. In this method, each spin is encoded in an appropriate tensor basis, and the angular momentum coupling is implemented by direct contractions of tensor indices.

A covariant extension is then obtained by lifting the construction to Lorentz tensors. In the covariant tensor method~\cite{Zou:2002ar,Zou:2002yy}, the amplitude is built by contracting Lorentz indices in $\mathrm{SO}(3,1)$ representations. This construction is manifestly covariant, while the separation between the orbital and spin parts is not explicit.

In subsubsection~\ref{subsubsec:Zemach angular momentum tensor method}, the construction of the Zemach tensor method is discussed, while in subsubsection~\ref{subsubsec:Covariant tensor method}, the construction of the covariant tensor method is discussed.

\subsubsection{Zemach tensor method}
\label{subsubsec:Zemach angular momentum tensor method}

The non-relativistic Zemach tensor method~\cite{Zemach:1965ycj} formulates the angular momentum coupling in tensor representations. In this formulation, the coupling is implemented by contracting tensor indices with the invariant tensors of $\mathrm{SO}(3)$ and $\mathrm{SU}(2)$.

For contractions between pure $\mathrm{SO}(3)$ tensors, the corresponding invariant contraction tensors are $\varepsilon_{ijk}$ and $\delta_{ij}$. 
For contractions involving both $\mathrm{SU}(2)$ and $\mathrm{SO}(3)$ tensors, the corresponding invariant contraction tensors are $\varepsilon_{\alpha\beta}$, $\delta^{\alpha}_{\beta}$ and $(\sigma_i)^\alpha_{\,\,\beta}$. 
Spatial indices $i,j,k=1,2,3$ label the Cartesian basis of the $\mathrm{SO}(3)$ vector representation, while spinor indices $\alpha,\beta=1,2$ label the fundamental representation of $\mathrm{SU}(2)$. 
With these conventions fixed, contractions with $\varepsilon_{ijk}$, $\delta_{ij}$, $\varepsilon_{\alpha\beta}$, $\delta^{\alpha}_{\beta}$ and $(\sigma_i)^\alpha_{\,\,\beta}$ are equivalent to the CGCs used to couple angular momenta in the little group formulation in Eq.~\eqref{Eq:canonical amplitude}.

Therefore, in the non-relativistic Zemach tensor method, the orbital-spin coupling in a two-body decay amplitude is realized by constructing the orbital tensor and the spin tensors for each external particle, and then contracting them with the $\mathrm{SO}(3)$ or $\mathrm{SU}(2)$ invariant tensors. The following discussion addresses the constructions of the spin tensors and the orbital tensors in turn.

Let us first consider the pure orbital part.
In the two-body decay COM frame, the final-state three-momenta are denoted by $\mathbf{p}^*_1$ and $\mathbf{p}^*_2$, and the relative momentum is defined as
\begin{equation}
    \mathbf{r}=\mathbf{p}^*_1-\mathbf{p}^*_2.
\end{equation}
In section~\ref{sec:Two different states description---Helicity and Canonical}, the construction of the orbital angular momentum $L$ in the $LS$ coupling is given in Eq.~\eqref{Eq:L spherical harmonic}. The pure orbital part is described by the spherical harmonics associated with the direction of the relative momentum, and these spherical harmonics correspond to the pure-$L$ component in the $\mathrm{SO}(3)$ little group representation. 

Therefore, when constructing the tensor structure for the orbital part, the dependence is taken only on the relative momentum, and an appropriate projection is applied to extract the pure-$L$ rank-$L$ tensor structure. This projection is equivalent to taking the symmetric traceless part of the rank-$L$ tensor formed directly from the tensor product of the relative momentum. The resulting symmetric traceless rank-$L$ tensor furnishes the irreducible $\mathrm{SO}(3)$ representation with orbital angular momentum $L$. Consequently, the spherical harmonics admit an equivalent expression in the tensor representation,
\begin{equation}
Y^L_{\sigma_L}(\Omega)\simeq T^{L}_{i_1\cdots i_{L}}(\mathbf{r}\cdots\mathbf{r})\equiv\mathbf{r}_{i_1}\cdots\mathbf{r}_{i_L}-\mathrm{trace\ part},
\end{equation}
where $i_1,\ldots,i_L$ are spatial indices. The relation above holds up to an overall normalization factor, which is omitted for simplicity. 
Here $T^{L}_{i_1\cdots i_{L}}(\mathbf{r}\cdots\mathbf{r})$ denotes the \textbf{symmetric traceless} rank-$L$ tensor constructed from the relative momentum. 
It satisfies the symmetry condition
\begin{equation}
T^{L}_{i_1\cdots i_a\cdots i_b\cdots i_L}
=
T^{L}_{i_1\cdots i_b\cdots i_a\cdots i_L},
\qquad \forall a,b,
\end{equation}
and the tracelessness condition
\begin{equation}
\sum_{i_a i_b}\delta_{i_a i_b}T^{L}_{i_1\cdots i_a\cdots i_b\cdots i_L}=0,
\qquad \forall a\neq b.
\end{equation}
For low values of $L$, the explicit expressions are
\begin{equation}\label{Eq:zemach orbital tensor}
\begin{aligned}
    T^0=&1,\\
    T^1_i(\mathbf{r})=&\mathbf{r}_i,\\
    T^2_{ij}(\mathbf{r}\mathbf{r})=&\mathbf{r}_{i}\mathbf{r}_j-\frac{1}{3}r^2\delta_{ij},\\
    T^3_{ijk}(\mathbf{r}\mathbf{r}\mathbf{r})=&\mathbf{r}_{i}\mathbf{r}_j\mathbf{r}_k-\frac{1}{5}\mathbf{r}^2(\delta_{ij}\mathbf{r}_k+\delta_{jk}\mathbf{r}_i+\delta_{ki}\mathbf{r}_j).\end{aligned}
\end{equation}

Next, the spin part is discussed. For the spin part, it is first necessary to describe particles in the tensor formulation. According to the value of the particle spin, there are two classes, namely integer spin and half-integer spin. Since higher-spin tensors can be obtained by angular momentum coupling of lower-spin objects, only two basic building blocks are required, the spin-1 and spin-$\frac{1}{2}$ spin tensors. In the non-relativistic formulation, they are equivalent to the spin wave functions defined in the particle rest frame, where the spin-1 building block is the three-dimensional polarization vector $(\epsilon_\sigma)_i$ and the spin-$\frac{1}{2}$ building block is the two-component Pauli spinor $(\chi_{\sigma})^{\alpha}$. By taking tensor products of these basic objects and projecting onto the irreducible representation with the desired spin, one obtains the spin tensors required for the coupling of the particles in the pure spin part.

\begin{enumerate}
\item For integer spin, the basic building block is the three-dimensional polarization vector defined in the rest frame. For initial-state particles, the polarization vector is denoted by $(\epsilon_\sigma)_i$ and can be written as
\begin{equation}\label{Eq:three-d polarization}
\begin{aligned}
    \epsilon_{+}=&-\frac{1}{\sqrt{2}}
    \begin{pmatrix}
    1, & i,& 0
    \end{pmatrix},\\
    \epsilon_-=&\quad\frac{1}{\sqrt{2}}
    \begin{pmatrix}
    1, & -i,& 0
    \end{pmatrix},\\
    \epsilon_0=&\quad\quad\quad
    \begin{pmatrix}
    0, & 0,& 1
    \end{pmatrix},
\end{aligned}
\end{equation}
where the subscripts $+$, $-$, and $0$ indicate the allowed values of the polarization label $\sigma$ in $(\epsilon_\sigma)_i$, the symbol $i$ appearing in the vector components denotes the imaginary unit, and the index $i=1,2,3$ labels Cartesian $\mathrm{SO}(3)$ components.
The complex conjugation of these polarization vectors gives the conjugate vector $(\epsilon^{*\sigma})_{i}$, which is used for final-state particles:
\begin{equation}\label{Eq_pol_conjugation}
    (\epsilon_{\sigma })_i^*=(\epsilon^{*\sigma})_i.
\end{equation}
The polarization vectors in Eq.~\eqref{Eq:three-d polarization} are then used as the building blocks for the integer spin-$s$ tensor. 

In the non-relativistic formulation, a spin-$s$ particle is represented by a rank-$s$ symmetric traceless tensor $(T^s_\sigma)_{i_1\cdots i_s}$, where the label $\sigma$ denotes the spin projection of spin-$s$ and takes values $\sigma=-s,\cdots,s$. This label is distinct from the spin-1 polarization label used in $(\epsilon_\sigma)_i$.

The symmetry and tracelessness conditions ensure that $(T^s_\sigma)_{i_1\cdots i_s}$ furnishes the irreducible $\mathrm{SO}(3)$ representation with spin-$s$.  
For an initial-state particle, the spin tensor is taken as $(T^s_\sigma)_{i_1\cdots i_s}$, while for a final-state particle the corresponding object is obtained by complex conjugation, which can be defined as
\begin{equation}
\bigl( T^{s}_{\sigma}\bigr)^*_{i_1\cdots i_s}
\equiv
\bigl(T^{*s,\sigma}\bigr)_{i_1\cdots i_s}.
\end{equation}
To construct the spin-$s$ tensor $(T^s_\sigma)_{i_1\cdots i_s}$ explicitly, one may start from the $s$-fold tensor product of polarization vectors and project onto the symmetric traceless part. It can be defined as
\begin{equation}\label{Eq:spin tensor st}
    (T^s_{\sigma_1\cdots\sigma_s})_{i_1\cdots i_s}\equiv\underbrace{(\epsilon_{\sigma_1})_{i_1}\cdots(\epsilon_{\sigma_s})_{i_s}}_{\text{symmetrized in $i_1,\cdots,i_s$}} - \text{trace part},
\end{equation}
where each label $\sigma_r=+1,0,-1$ $(r=1,\cdots,s)$ specifies the polarization choice of the $r$th spin-1 building block $(\epsilon_{\sigma_r})_{i_r}$. 
For example, some low-spin tensors can be written explicitly as
\begin{equation}
\begin{aligned}
T^0=&1,\\
(T^1_\sigma)_i=&(\epsilon_{\sigma})_i,\\
    (T^2_{\sigma_1\sigma_2})_{ij}= &\frac{1}{2}\Big((\epsilon_{\sigma_1})_{i}(\epsilon_{\sigma_2})_j+(\epsilon_{\sigma_1})_{j}(\epsilon_{\sigma_2})_i\Big)-\frac{1}{3}(\epsilon_{\sigma_1}\cdot\epsilon_{\sigma_2})\delta_{ij},\\
    (T^3_{\sigma_1\sigma_2\sigma_3})_{ijk}=&\frac{1}{3!}\Big((\epsilon_{\sigma_1})_{i}(\epsilon_{\sigma_2})_j(\epsilon_{\sigma_3})_k+\{ijk\}\Big)-\frac{1}{5}\Big(\delta_{ij}V_k+\delta_{jk}V_i+\delta_{ki}V_j\Big),\\
    \text{with}\quad V_k=&\frac{1}{3}\Big((\epsilon_{\sigma_1}\cdot\epsilon_{\sigma_2})(\epsilon_{\sigma_3})_k+(\epsilon_{\sigma_3}\cdot\epsilon_{\sigma_1})(\epsilon_{\sigma_2})_k+(\epsilon_{\sigma_2}\cdot\epsilon_{\sigma_3})(\epsilon_{\sigma_1})_k\Big),\\
\end{aligned}
\end{equation}
where $\{ijk\}$ denotes the set of all $3!$ permutations of the indices $i,j,k$.
After performing the symmetric traceless projection on the vector indices in Eq.~\eqref{Eq:spin tensor st}, and with the chosen normalization convention, the multi-label $\sigma_r$ can be reorganized into a single spin-projection label $\sigma=-s,\cdots,s$. For convenience, the normalized components are chosen according to the following convention. For $0\le\sigma\le s$, define
\begin{equation}\label{Eq:spin s spintensor}
    (T^{s}_\sigma)_{i_1\cdots i_s} \equiv \sqrt{\frac{(2s)!}{2^{s-\sigma}(s+\sigma)!(s-\sigma)!}}(T^{s}_{\underbrace{\scriptstyle 0\cdots 0}_{s-\sigma}\underbrace{\scriptstyle 1\cdots 1}_{\sigma}})_{i_1\cdots i_s},
\end{equation}
where the multi-label subscript on the tensor on the right-hand side specifies the component such that, among the $s$ polarization labels, $\sigma$ of them are equal to $+1$ and the remaining $s-\sigma$ are equal to $0$, and where the symbol $1$ in the subscript denotes the value $+1$ of the spin-1 polarization label. For $0\le\sigma\le s$, define the negative projection components by
\begin{equation}
    (T^{s}_{-\sigma})_{i_1\cdots i_s} \equiv (-1)^\sigma (T^{s}_{\sigma})^*_{i_1\cdots i_s}.
\end{equation}

\item For half-integer spin, the basic building block is the two-component Pauli spinor defined in the rest frame. For initial-state particles, the spinor is denoted by $(\chi_\sigma)^\alpha$ and can be written as
\begin{equation}
    \chi_{+}=\begin{pmatrix}
        1\\
        0
    \end{pmatrix}, 
    \chi_{-}=\begin{pmatrix}
    0 \\
    1
    \end{pmatrix},
\end{equation}
where the subscripts $+$ and $-$ correspond to the spin projection $\sigma = +\frac{1}{2}, -\frac{1}{2}$ of $(\chi_\sigma)^\alpha$. The complex conjugation of these spinors gives the conjugate spinor $(\chi^{*\sigma})_\alpha$, which is used for final-state particles:
\begin{equation}
    (\chi_\sigma)^{\alpha*}=(\chi^{*\sigma})_\alpha.
\end{equation}
In the non-relativistic formulation, a half-integer spin-$(s+\frac{1}{2})$ initial-state particle is described by a mixed tensor $(T^{s+\frac{1}{2}}_{\sigma})_{i_1\cdots i_s}^{\alpha}$, which carries $s$ vector indices $i_1,\dots,i_s$ and one spinor index $\alpha$. To extract the irreducible spin-$(s+\frac{1}{2})$ component from the reducible product space of a rank-$s$ symmetric traceless tensor and a spinor, two essential conditions must be imposed. First, the tensor must be completely symmetric and traceless in all its vector indices. Second, it must satisfy an additional constraint, namely that its contraction with the Pauli matrices vanishes, which can be written as
\begin{equation}\label{Eq:Pauli_trace_condition}
\sum_{i_a} (\sigma_{i_a})^{\alpha}_{\,\,\beta} (T^{s+\frac{1}{2}}_{\sigma})_{i_1\cdots i_a\cdots i_s}^\beta=0, \qquad 1\leq a\leq s,
\end{equation}
where $(\sigma_{i_a})^{\alpha}_{\,\,\beta}$ are the Pauli matrices. The role of this condition is to eliminate the lower spin-$(s-\frac{1}{2})$ components contained in the direct product space $T^s \otimes \chi$, thereby ensuring that the tensor describes only the pure spin-$(s+\frac{1}{2})$ representation. 

An explicit construction that realizes this projection for an initial-state particle is given by the following expression, which can be defined as
\begin{equation}\label{Eq:project to spinor tensor}
\begin{aligned}
(T^{s+\tfrac12}_{\sigma_s\sigma^\prime})_{i_1\cdots i_s}^\alpha
\equiv&(T^s_{\sigma_s})_{i_1\cdots i_s}(\chi_{\sigma^\prime})^{\alpha}
+\frac{i}{s+1}\sum_{a=1}^s\sum_{jk}\varepsilon_{i_a j k}(\sigma_{j})^{\alpha}_{\,\,\beta}
(T^s_{\sigma_s})_{i_1\cdots i_{a-1}k\,i_{a+1}\cdots i_s}(\chi_{\sigma^\prime})^{\beta},
\end{aligned}
\end{equation}
where $\varepsilon_{i_ajk}$ is the totally antisymmetric tensor, the labels $\sigma_s=-s,\cdots,s$ and $\sigma^\prime=-\frac{1}{2},\frac{1}{2}$ in the subscripts of $T^s_{\sigma_s}$ and $\chi_{\sigma^\prime}$ denote spin projection quantum numbers, the symbol $(\sigma_j)^\alpha_{\,\,\beta}$ denotes the Pauli matrices, and the symbol $i$ in the coefficient denotes the imaginary unit.
The two labels $\sigma_s$ and $\sigma^\prime$ can be combined into a single total label $\sigma=-s-\frac{1}{2},\cdots ,s+\frac{1}{2}$ by imposing an appropriate normalization convention, which can be written as
\begin{equation}\label{Eq:CG_decompose_tensor_spinor}
\bigl(T^{s+\frac12}_{\sigma}\bigr)_{i_1\cdots i_s}^\alpha
=
\sqrt{\frac{s+\sigma+\frac12}{2s+1}}
\bigl(T^{s+\frac12}_{\sigma-\frac12,\,+\frac12}\bigr)_{i_1\cdots i_s}^\alpha
+
\sqrt{\frac{s-\sigma+\frac12}{2s+1}}
\bigl(T^{s+\frac12}_{\sigma+\frac12,\,-\frac12}\bigr)_{i_1\cdots i_s}^\alpha\,,
\end{equation}
where the two tensors on the right hand side are the tensors defined in Eq.~\eqref{Eq:project to spinor tensor}, and the subscripts $(\sigma-\frac12,\frac12)$ and $(\sigma+\frac12,-\frac12)$ correspond to $(\sigma_s,\sigma^\prime)$, respectively. 
With this construction, the resulting tensor satisfies both the symmetry and tracelessness conditions in the vector indices and the constraint in Eq.~\eqref{Eq:Pauli_trace_condition}.
The corresponding final-state spin tensor can be obtained by taking the complex conjugate of the spin tensor in Eq.~\eqref{Eq:CG_decompose_tensor_spinor}, which is defined by complex conjugation as
\begin{equation}\label{Eq:final_state_tensor_conjugate}
\bigl( T^{s+\frac12}_{\sigma}\bigr)^{*\alpha}_{i_1\cdots i_s}
\equiv
\bigl(T^{*s+\frac12 ,\sigma}\bigr)_{i_1\cdots i_s\alpha}.
\end{equation}

\end{enumerate}

After constructing the basic tensors above, the amplitude can be obtained by contracting the tensor indices. 
The symbol $\otimes$ is used to denote angular momentum coupling in a unified way. More specifically, if two tensors carry angular momenta $j_1$ and $j_2$, their coupling is written as $T^{j_1}\otimes T^{j_2}$, meaning that the product representation is decomposed into irreducible components with total angular momentum $J$,
\begin{equation}
T^{j_1}\otimes T^{j_2}=\bigoplus_{J=|j_1-j_2|}^{j_1+j_2}T^{J}.
\end{equation}
In the Zemach tensor method, this coupling is implemented by the invariant tensors of $\mathrm{SO}(3)$ and $\mathrm{SU}(2)$. Equivalently, it corresponds to performing the angular momentum coupling and projection in the little group representation using the CGCs.
In what follows, several explicit examples of spin coupling are presented, which can be generalized to more general angular momentum couplings.

For the coupling of two integer spins $s_1$ and $s_2$, the coupling to a total spin-$S$ tensor can be denoted by
\begin{equation}
T^{*s_1}\otimes T^{*s_2}\to T^{S},
\end{equation}
where the allowed values of $S$ satisfy $|s_1-s_2|\le S\le s_1+s_2$.
In particular, for $S=s_1+s_2$, the highest-spin tensor is obtained by taking the direct tensor product of the two spin tensors and then projecting onto the symmetric traceless part. The resulting total spin-$S$ tensor can be written as
\begin{equation}
(T^{S=s_1+s_2,\sigma_1\sigma_2})_{i_1\cdots i_{s_1}j_1\cdots j_{s_2}}
=\underbrace{(T^{*s_1,\sigma_1})_{i_1\cdots i_{s_1}}
(T^{*s_2,\sigma_2})_{j_1\cdots j_{s_2}}}_{\text{symmetrized in all vector indices}}
-\text{trace part}.
\end{equation}
To obtain lower-spin components with $S = s_1 + s_2 - 2k$ ($k = 1,2,\cdots$), $k$ pairs of indices between the two tensors are contracted using $\delta_{ij}$. 
The remaining spin components of the form $S = s_1 + s_2 - 2k - 1$ can be constructed from the rank-$(s_1 + s_2 - 2k)$ tensor by using the antisymmetric tensor $\varepsilon_{ijk}$ to contract one additional index.
As an explicit example, for the coupling of two spin-1 particles, $s_1 = s_2 = 1$, the possible total spins are $S = 0,1,2$, and the corresponding total spin tensors can be written as
\begin{equation}
\begin{aligned}
    T^{S=0,\sigma_1\sigma_2}=&\sum_{ij}(T^{*1,\sigma_1})_i(T^{*1,\sigma_2})_{j}\delta_{ij},\\
    (T^{S=1,\sigma_1\sigma_2})_i=&\sum_{jk}\varepsilon_{ijk}(T^{*1,\sigma_1})_j(T^{*1,\sigma_2})_{k},\\
    (T^{S=2,\sigma_1\sigma_2})_{ij}=&\frac12\Big[(T^{*1,\sigma_1})_i(T^{1,\sigma_2})_{j}+(T^{*1,\sigma_1})_j(T^{*1,\sigma_2})_{i}\Big]-\frac{1}{3}\delta _{ij}T^{S=0,\sigma_1\sigma_2}.
\end{aligned}
\end{equation}
It can be seen that the coupling between different angular momenta does not require explicit evaluation of CGCs, since it is implemented by contracting tensor indices in the corresponding tensor representations.

For the coupling of two half-integer spins $s_1+\frac{1}{2}$ and $s_2+\frac{1}{2}$, the coupling to a total spin-$S$ tensor can be denoted by
\begin{equation}
    T^{*s_1+\frac{1}{2}}\otimes T^{*s_2+\frac{1}{2}}\to T^{S},
\end{equation}
where the allowed values of $S$ satisfy $|s_1 - s_2|\leq S \leq s_1 + s_2 + 1$. 
The highest-spin tensor with $S = s_1 + s_2 + 1$ is obtained by using the tensor $(\sigma_{k})^{\alpha}_{\,\,\beta}$ to contract the two spinor indices into the vector index $k$, which gives a rank-$(s_1+s_2+1)$
tensor. It can be written as
\begin{equation}
    (T^{S=s_1+s_2+1,\sigma_1\sigma_2})_{i_1\cdots i_{s_1}j_1\cdots j_{s_2}k}=\underbrace{(\sigma_{k})^{\alpha}_{\,\,\gamma}\varepsilon^{\gamma\beta}(T^{*s_1+\frac{1}{2},\sigma_1})_{i_1\cdots i_{s_1}\alpha}(T^{*s_2+\frac{1}{2},\sigma_2})_{j_1\cdots j_{s_2}\beta}}_{\text{symmetrized in all vector indices}} -\text{trace part}. 
\end{equation}
The remaining spin part is obtained by contracting with invariant tensors such as $\varepsilon_{\alpha\beta}$, $\delta^{\alpha}_{\beta}$, $\delta_{ij}$, and $\varepsilon_{ijk}$, followed by taking the symmetric traceless part.
As an example, for the coupling of spin-$\frac{1}{2}$ and spin-$\frac{1}{2}$, the possible total spins are $S=0$ and $S=1$. The corresponding tensors can be written as
\begin{equation}
\begin{aligned}
T^{S=0,\sigma_1\sigma_2}=&\varepsilon^{\alpha\beta}(T^{*\frac{1}{2},\sigma_1})_\alpha (T^{*\frac{1}{2},\sigma_2})_\beta,\\
(T^{S=1,\sigma_1\sigma_2})_i
=&(\sigma_i)^{\alpha}_{\,\,\gamma}\varepsilon^{\gamma\beta}(T^{*\frac{1}{2},\sigma_1})_\alpha (T^{*\frac{1}{2},\sigma_2})_\beta.
\end{aligned}
\end{equation}

For the coupling of an integer spin-$s_1$ and a half-integer spin-$s_2+\frac{1}{2}$, the coupling to a total spin-$S$ tensor can be denoted by
\begin{equation}
    T^{*s_1}\otimes T^{*s_2+\frac{1}{2}}\to T^{S},
\end{equation}
where the allowed values of $S$ satisfy $|s_1-s_2-\frac{1}{2}|\leq S\leq s_1 + s_2 + \frac{1}{2}$.
For $s_1 \leq s_2$, one can contract indices in the same way as for two integer spin tensors. 
For $s_1 > s_2$, the lowest total spin component
$S = s_1 - s_2 - \tfrac{1}{2}$
requires the use of tensor $(\sigma_{i})^{\alpha}_{\,\,\beta}$ to contract the spinor index with one of the vector indices. 
As an example, for the coupling of spin-$1$ and spin-$\frac{1}{2}$, the possible total spins are $S = \frac{1}{2}$ and $S = \frac{3}{2}$. The corresponding tensors can be written as
\begin{equation}
\begin{aligned}
    (T^{S=\frac{1}{2},\sigma_1\sigma_2})_\alpha    &= \sum_{i}(\sigma_i)_{\,\,\alpha}^{\beta}(T^{*1,\sigma_1})_i (T^{*\frac{1}{2},\sigma_2})_\beta,\\
    (T^{S=\frac{3}{2},\sigma_1\sigma_2})_{i\alpha } &= (T^{*1,\sigma_1})_i (T^{*\frac{1}{2},\sigma_2})_\alpha-\frac{1}{3}(\sigma_i)^{\beta}_{\,\,\alpha}(T^{S=\frac{1}{2},\sigma_1\sigma_2})_\beta.\\
\end{aligned}
\end{equation}

Having constructed the spin tensors and their coupling relations above, they can be extended to general angular momentum couplings and used to build the corresponding $LS$ partial wave amplitudes. Once the final-state spins have been coupled into a total spin-$S$ tensor, the next step is to couple this tensor with the orbital angular momentum $L$ tensor to obtain a tensor structure with total angular momentum $s_3$, namely
\begin{equation}
T^{S}\otimes T^{L}\to T^{s_3},
\end{equation}
where $T^{s_3}$ denotes the rank-$s_3$ tensor obtained from the $LS$ coupling. Finally, this tensor is contracted with the external spin tensor of the parent particle with spin-$s_3$.
Once all indices have been contracted, the $LS$ amplitude is obtained as
\begin{equation}\label{Eq:zemach tensor amplitude}
    A^{\sigma_1\sigma_2}_{\sigma_3}(\mathbf{k}_3,\mathbf{p}_1^*,\mathbf{p}_2^*;L,S)=\underbrace{T_{\sigma_3}^{s_3}\cdot\Big[(T^{*s_1\sigma_1}\otimes T^{*s_2\sigma_2})^S}_{\text{spin part}}\otimes\underbrace{T^L(\mathbf{p}\cdots\mathbf{p})}_{\text{orbital part}}\Big]^{s_3},
\end{equation}
where the spin part is constructed entirely from spin wave functions defined in the rest frame. It therefore carries no orbital information, and the separation between the orbital part and the spin part is explicit. As a result, the $LS$ partial wave amplitudes of the Zemach tensor method are fully consistent with the traditional-$LS$ amplitudes. Likewise, both approaches are non-covariant. To recover Lorentz covariance, covariant tensor methods have been developed, which will be introduced in the next subsubsection.

\subsubsection{Covariant tensor method}
\label{subsubsec:Covariant tensor method}

In the covariant tensor method~\cite{Zou:2002ar,Zou:2002yy}, the coupling of angular momenta is described by covariant tensors carrying the $\mathrm{SO}(3,1)$ Lorentz indices.
The coupling between different angular momentum tensors is implemented by constructing Lorentz structures and contracting the $\mathrm{SO}(3,1)$ indices of the participating tensors, thereby producing a new tensor with the desired total angular momentum. 
Therefore, to obtain a pure irreducible angular momentum coupling, one typically introduces Lorentz structures transverse to the parent momentum, so that all projections and contractions are performed within the transverse components and the time component does not participate, avoiding admixtures from other angular-momentum components.

Next, the explicit construction of the covariant tensor amplitude is presented. Consider the two-body decay process $3\to1+2$, where the spins of the particles are $s_3$, $s_1$, and $s_2$, respectively, and the four-momenta $p_3^\mu$, $p_1^\mu$, and $p_2^\mu$ are defined in any frame. 
The discussion begins with the orbital part that enters the coupling, for which the relevant definitions are introduced first.

For the orbital part, the Zemach tensor method obtains the pure orbital $L$ irreducible object by taking the symmetric traceless part of the relative three-momentum. To make the construction covariant, introduce the relative four-momentum
\begin{equation}
r^\mu=p_1^\mu-p_2^\mu.
\end{equation}
A natural starting point is the rank-$L$ symmetric-traceless tensor built from $r^\mu$
\begin{equation}
r^L_{\mu_1\cdots\mu_L}\equiv r_{\mu_1}\cdots r_{\mu_L}-\text{trace part}.
\end{equation}
However, symmetric tracelessness only implies that $r^L_{\mu_1\cdots\mu_L}$ furnishes an $\mathrm{SO}(3,1)$ irreducible representation $[\mu_1\cdots\mu_L]$, rather than the pure $\mathrm{SO}(3)$ orbital $L$ component. In the COM frame, the time component transforms as an $\mathrm{SO}(3)$ scalar, tensor components with one or more time indices behave as scalar factors multiplying lower rank spatial tensors, and hence contain contributions from orbital angular momenta smaller than $L$.

To ensure that the constructed tensor indeed lies in the orbital angular momentum $L$ irreducible component, it is required in the COM frame that it has no time components. Equivalently, in any frame, for the symmetric traceless rank-$L$ tensor $r^L_{\mu_1\cdots\mu_L}$ one must subtract a part such that the resulting tensor is orthogonal to the total momentum $p_3$ in each index, namely it satisfies
\begin{equation}\label{Eq:ptothe}
p_3^{\mu_i}\Big[r^L_{\mu_1\cdots \mu_i \cdots\mu_L}-\text{the $p_3$-parallel part}\Big]=0,\qquad i=1,\ldots,L,
\end{equation}
where the expression in the square brackets is the orbital angular momentum $L$ component in terms of the Lorentz indices $\mu_1\cdots\mu_L$.

The explicit form of the covariant orbital tensor can be obtained by constructing a projection tensor and projecting the relative momentum onto the pure-$L$ component. For this purpose, one first introduces the rank-$L$ integer spin wave function associated with the total momentum $p_3$, defined as 
\begin{equation}\label{Eq:rarita wave functions}
    \epsilon^{\mu_1 \mu_2 \cdots \mu_L}_{\sigma_{L}}(p_3;L) \equiv \sum_{\sigma_{L-1},\sigma} C_{L,\sigma_{L}}^{L-1,\sigma_{L-1};1,\sigma}\epsilon^{\mu_1 \mu_2 \cdots \mu_{L-1}}_{\sigma_{L-1}}(p_3;L-1) \epsilon^{\mu_L}_{\sigma}(p_3 ),
\end{equation}
where $C_{L,\sigma_{L}}^{L-1,\sigma_{L-1};1,\sigma}$ denotes the CGCs and $\epsilon^{\mu_L}_{\sigma}(p)$ are the polarization four-vectors. The wave function constructed in this way satisfies the Rarita-Schwinger conditions, namely it is orthogonal to $p_3$ in each index and is symmetric and traceless in the Lorentz indices, which can be written as 
\begin{align}
    p_{3\mu_1}\epsilon^{\mu_1 \mu_2 \cdots \mu_L}_{\sigma_{L}}(p_3;L) =&0,\\
    \epsilon^{\mu_1\cdots \mu_a\cdots\mu_b \cdots \mu_L}_{\sigma_{L}}(p_3;L) =&\epsilon^{\mu_1\cdots \mu_b\cdots\mu_a \cdots \mu_L}_{\sigma_{L}}(p_3;L),\\
    g_{\mu_1\mu_2}\epsilon^{\mu_1 \mu_2 \cdots \mu_L}_{\sigma_{L}}(p_3;L) =&0.
\end{align}
Accordingly, a projection tensor can be constructed and is defined as~\cite{Chung:1993da}
\begin{equation}\label{Eq:projection tensortttt}
    P^{\nu_1\cdots\nu_L}_{\mu_1\cdots\mu_L}(p_3;L)\equiv\sum_{\sigma_L}\epsilon^{\nu_1\cdots\nu_L}_{\sigma_L}(p_3;L)\epsilon_{\mu_1\cdots\mu_L}^{*\sigma_L}(p_3;L).
\end{equation}
This projection tensor also satisfies the Rarita-Schwinger conditions. Acting with this projector tensor on the rank-$L$ tensor formed by the direct product of $L$ relative momenta yields the covariant orbital tensor satisfying the condition in Eq.~\eqref{Eq:ptothe}, which is defined as 
\begin{equation}
    \tilde t^{(L)}_{\mu_1\cdots\mu_L}(p_3,r)\equiv P^{\nu_1\cdots\nu_L}_{\mu_1\cdots\mu_L}(p_3;L)r_{\nu_1}\cdots r_{\nu_L}.
\end{equation}
The tensor $\tilde t^{(L)}_{\mu_1\cdots\mu_L}(p_3,r)$ is a rank-$L$ symmetric traceless Lorentz tensor that represents the pure orbital angular momentum $L$ component in the Lorentz index $\mu_1\cdots\mu_L$. By construction it is also orthogonal to the total momentum $p_3$ in each index,
\begin{equation}
p_3^{\mu_1}\tilde t_{\mu_1\cdots\mu_L}^{(L)}=0.
\end{equation}

Next, several explicit examples are presented. For $L=1$, the projection tensor can be evaluated as
\begin{equation}\label{Eq:transverse_metric}
\begin{aligned}
P^\nu_\mu(p_3;1)=&\sum_{\sigma}\epsilon^\nu_{\sigma}(p_3)\epsilon^{*\sigma}_\mu(p_3)\\=&-g^{\nu}_{\,\,\mu}+\frac{p_3^\nu p_{3\mu}}{p_3^2}\\
=&\tilde g^{\nu}_{\,\,\mu}(p_3).
\end{aligned}
\end{equation}
The result differs from some expressions in the literature by the placement of upper and lower indices. This is because, in the present convention, the projector tensor is defined using two wave functions with distinguished index positions, which leads to a notational difference.

Acting with the projector tensor on the relative momentum then yields the covariant orbital tensor for $L=1$, given by
\begin{equation}\label{eq_transverse_r_def}
\tilde r_\mu(p_3)= \tilde g^{\nu}_{\,\,\mu}(p_3) r_\nu.
\end{equation}
In the rest frame of particle-3, $k_3^\mu=(m_3,\mathbf 0)$, the condition $ k^\mu_3\tilde r_\mu(k_3)=0$ implies $\tilde r^0=0$, hence $\tilde r^\mu=(0,\mathbf r)$. The projection therefore removes the time component, and $\tilde r^\mu$ reduces to the relative three-momentum used in the Zemach tensor method. The transverse metric $\tilde g^{\nu}_{\,\,\mu}(k_3)$ reduces to $\delta_{ij}$ in this frame.

For some low orbital angular momenta, the projector can be worked out explicitly, for example,
\begin{equation}
\begin{aligned}
    P(p_3;0)=&1,\\
    P_{\mu_1}^{\nu_1}(p_3;1)=&\tilde g_{\,\,\mu_1}^{\nu_1},\\
    P_{\mu_1\mu_2}^{\nu_1\nu_2}(p_3;2)=&\frac12\Big(\tilde g^{\nu_1}_{\,\,\mu_1}\tilde g^{\nu_2}_{\,\,\mu_2}
+\tilde g^{\nu_1}_{\,\,\mu_2}\tilde g^{\nu_2}_{\,\,\mu_1}\Big)
-\frac13\,\tilde g^{\nu_1\nu_2}\tilde g_{\mu_1\mu_2},\\
    P_{\mu_1\mu_2\mu_3}^{\nu_1\nu_2\nu_3}(p_3;3)=&\frac{1}{6}\sum_{P\{\mu_1,\mu_2,\mu_3\}}
\tilde g^{\nu_1}_{\,\,\mu_1}\tilde g^{\nu_2}_{\,\,\mu_2}\tilde g^{\nu_3}_{\,\,\mu_3}\\
&-\frac{1}{30}\sum_{P\{\mu_1,\mu_2,\mu_3\}}
\Big(
\tilde g^{\nu_1\nu_2}\tilde g_{\mu_1\mu_2}\tilde g^{\nu_3}_{\,\,\mu_3}
+\tilde g^{\nu_2\nu_3}\tilde g_{\mu_2\mu_3}\tilde g^{\nu_1}_{\,\,\mu_1}
+\tilde g^{\nu_1\nu_3}\tilde g_{\mu_1\mu_3}\tilde g^{\nu_2}_{\,\,\mu_2}
\Big),\\
\end{aligned}
\end{equation}
where $P\{\mu_1,\mu_2,\mu_3\}$ denotes the set of all $3!$ permutations of the indices $\mu_1,\mu_2,\mu_3$, and $\tilde g^{\nu}_{\,\,\mu}$ is defined in Eq.~\eqref{Eq:transverse_metric}. 
The corresponding covariant orbital tensor can be computed explicitly as
\begin{equation}\label{Eq:covtensor L}
\begin{aligned}
\tilde t^{(0)}&=1,\\
\tilde t^{(1)}_{\mu_1}&=\tilde r_{\mu_1},\\
\tilde t^{(2)}_{\mu_1\mu_2}&=\tilde r_{\mu_1}\tilde r_{\mu_2}-\frac13(\tilde r\cdot\tilde r)\tilde g_{\mu_1\mu_2},\\
\tilde t^{(3)}_{\mu_1\mu_2\mu_3}&=\tilde r_{\mu_1}\tilde r_{\mu_2}\tilde r_{\mu_3}
-\frac15(\tilde r\cdot\tilde r)\big(\tilde g_{\mu_1\mu_2}\tilde r_{\mu_3}+\tilde g_{\mu_1\mu_3}\tilde r_{\mu_2}+\tilde g_{\mu_3\mu_2}\tilde r_{\mu_1}\big),\\
\end{aligned}
\end{equation}
with
\begin{equation}
    \tilde g_{\mu\nu}=g_{\mu\mu^\prime}\tilde g^{\mu^\prime}_{\,\,\nu}=-g_{\mu\nu}+\frac{p_{3\mu} p_{3\nu}}{p_3^2},
\end{equation}
where $g_{\mu\nu}$ is the metric tensor, and $\tilde g^{\mu}_{\,\,\nu}$ and $\tilde r_\mu$ are defined in Eqs.~\eqref{Eq:transverse_metric} and \eqref{eq_transverse_r_def}. For brevity, the explicit dependence on $p_3$ is suppressed.

For the spin part, the first step is to specify the external spin wave functions entering the coupling. In the covariant tensor method, the spin wave functions based on the Rarita-Schwinger construction~\cite{Rarita:1941mf}, and canonical spin projection labels are used. For spin-$\frac{1}{2}$, Dirac spinors $u^\alpha_\sigma(p)$ are used. For spin-$1$, polarization four-vectors $\epsilon^\mu_\sigma(p)$ are used. 
Higher spin wave functions are obtained from $\epsilon^\mu_\sigma(p)$ and $u^\alpha_\sigma(p)$ through CGCs couplings as
\begin{align}\label{Eq:rarita schiwinger wave function}
\epsilon^{\mu_1 \mu_2 \cdots \mu_s}_{\sigma_{s}}(p;s) &= \sum_{\sigma_{s-1},\sigma} C_{s,\sigma_{s}}^{s-1,\sigma_{s-1};1,\sigma}\epsilon^{\mu_1 \mu_2 \cdots \mu_{s-1}}_{\sigma_{s-1}}(p;s-1) \epsilon^{\mu_s}_{\sigma}(p ),  \\
u^{\mu_1 \mu_2 \cdots \mu_s\alpha}_{\sigma_{s+\frac{1}{2}}}(p;s + \tfrac{1}{2}) &= \sum_{\sigma_s, \sigma} C_{s+\frac{1}{2},\sigma_{s+\frac{1}{2}}}^{s,\sigma_{s};\frac{1}{2},\sigma}\epsilon^{\mu_1 \mu_2 \cdots \mu_s}_{\sigma_s}(p;s)  u^{\alpha}_{\sigma}(p), 
\end{align}
where $C^{s_1,\sigma_1;s_2,\sigma_2}_{s_3,\sigma_3}$ denotes the CGCs for coupling little group indices.
In the first line, the coupling $(s-1)\otimes1\to s$ combines a rank-$(s-1)$ integer spin wave function with a spin-1 wave function to produce a rank-$s$ integer spin wave function.
Therefore, integer spin wave functions can be generated recursively starting from the spin-1 wave function.
In the second line, the coupling $s\otimes\frac12\to s+\frac12$ combines an integer spin wave function with a Dirac spinor, yielding a half-integer spin wave function.

The basic tensor building blocks for the spin part and the orbital part entering the coupling have now been introduced. In analogy with the Zemach tensor method, the next step is to construct the corresponding coupling structures that implement the angular momentum couplings. The key point is to covariantize the coupling tensors used in the Zemach tensor method and extend them to the $\mathrm{SO}(3,1)$ Lorentz indices. Equivalently, this amounts to projecting the product of two angular momentum tensors onto the tensor with the desired total angular momentum.

First consider the coupling of the two final state spins into a total spin-$S$, namely $s_1+s_2\to S$. When both spins are integer, the Zemach tensor method performs the coupling by contracting the tensor indices of the two final state spin tensors with the $\mathrm{SO}(3)$ invariant tensors $\delta_{ij}$ and $\varepsilon_{ijk}$, yielding a total spin-$S$ spin tensor. The resulting rank-$S$ tensor is required to be symmetric and traceless so that it represents a pure spin-$S$ component.

In the $\mathrm{SO}(3,1)$ covariant extension, as in the discussion of the orbital part, obtaining a total spin-$S$ tensor is not only a matter of enforcing symmetry and tracelessness. It is also necessary to ensure that, when evaluated in the parent rest frame, the time components vanish so that the tensor corresponds to a pure spin-$S$ component. This can be enforced covariantly by requiring the tensor to be orthogonal to the parent momentum $p_3$, which can be written as
\begin{equation}
    p_3^{\mu_1}T_{\mu_1\mu_2\cdots\mu_S}(S)=0,
\end{equation}
where $T_{\mu_1\mu_2\cdots\mu_S}(S)$ denotes the rank-$S$ total spin tensor obtained by coupling the final-state spin wave functions.
By incorporating this transversality condition into the coupling structures, the covariant coupling tensors required for integer spin couplings can be specified. They are given by the transverse metric $\tilde g^{\mu\nu}(p_3)$ and the transverse antisymmetric tensor $\tilde\varepsilon^{\mu\nu\rho}(p_3)$, which are written as 
\begin{align}
    \tilde g^{\mu\nu}(p_3)=& {-}g^{\mu\nu} +\frac{p^\mu_3p^\nu_3}{p_3^2},\\
    \tilde\varepsilon^{\mu\nu\rho}(p_3)=&\varepsilon^{\mu\nu\rho\kappa}p_{3\kappa},
\end{align}
where indices can be raised and lowered with the metric tensor $g_{\mu\nu}$. Since the present convention distinguishes upper and lower indices explicitly in contractions (with repeated indices summed), the resulting notation may differ from that used in some references. One can readily verify that both tensors are transverse to $p_3$, and that, in the parent rest frame, their time components vanish so that they proportional to the invariant tensors $\delta_{ij}$ and $\varepsilon_{ijk}$.

As an explicit example, consider the coupling of two spin-1 tensor wave functions, $s_1=s_2=1$, for which
\begin{equation}
1\otimes1=0\oplus1\oplus2.
\end{equation}
Choose the external spin wave functions in the Lorentz four-vector representation to be $\epsilon_\mu^{*\sigma_1}(p_1)$ and $\epsilon_\mu^{*\sigma_2}(p_2)$ for particle-1 and particle-2. By contracting these two external spin wave functions with the projector tensors in the $1\otimes1\to S$ channels, one obtains three covariant spin tensors corresponding to
the total spins $S=0,1,2$.
The $S=0$ component is
\begin{equation}\label{eq:11_to_S0}
\begin{aligned}
T^{\sigma_1\sigma_2}(p_3,p_1,p_2;0)=&\tilde g^{\mu\nu}(p_3)\epsilon_\mu^{*\sigma_1}(p_1)\epsilon_\nu^{*\sigma_2}(p_2)\\
=&\Big({-}g^{\mu\nu} +\frac{p^\mu_3p^\nu_3}{p_3^2}\Big)\epsilon_\mu^{*\sigma_1}(p_1)\epsilon_\nu^{*\sigma_2}(p_2).
\end{aligned}
\end{equation}
The $S=1$ component is 
\begin{equation}\label{eq:11_to_S1}
\begin{aligned}
    T^{\sigma_1\sigma_2}_{\mu}(p_3,p_1,p_2;1)=&g_{\mu\mu^\prime}\tilde\varepsilon^{\mu^\prime\nu\rho}(p_3)
\epsilon_\nu^{*\sigma_1}(p_1)\epsilon_\rho^{*\sigma_2}(p_2)\\
=&\Big(g_{\mu\mu^\prime}\varepsilon^{\mu^\prime\nu\rho\kappa}p_{3\kappa}\Big)
\epsilon_\nu^{*\sigma_1}(p_1)\epsilon_\rho^{*\sigma_2}(p_2).
\end{aligned}
\end{equation}
The $S=2$ component is 
\begin{equation}\label{eq:11_to_S2}
\begin{aligned}
T^{\sigma_1\sigma_2}_{\mu\nu}(p_3,p_1,p_2;2)
=&\Bigg(\frac12\Big(\tilde g^{\rho}_{\,\,\mu}(p_3)\tilde g^{\kappa}_{\,\,\nu}(p_3)
+\tilde g^{\rho}_{\,\,\nu}(p_3)\tilde g^{\kappa}_{\,\,\mu}(p_3)\Big)
-\frac13\,\tilde g^{\rho\kappa}(p_3)\tilde g_{\mu\nu}(p_3)\Bigg)\epsilon_\rho^{*\sigma_1}(p_1)\epsilon_\kappa^{*\sigma_2}(p_2).
\end{aligned}
\end{equation}

The coupling structures constructed above are used to couple spins $s_1$ and $s_2$ into a total spin-$S$. Conceptually, they implement a projection coupling between two angular momentum tensors, and therefore they apply equally well to the next step, namely coupling the total spin-$S$ with the orbital angular momentum $L$ into the parent spin-$s_3$.

As an example, consider the two-body decay process $3\to1+2$, with $s_3=1$, $s_1=1$, and $s_2=1$. Choose the partial wave $(L,S)=(1,1)$. First, couple the two final-state spins via $1\otimes1\to1$, which yields a total spin-$S$ tensor that can be obtain from Eq.~\eqref{eq:11_to_S1}.
Next, couple the orbital angular momentum $L=1$ with the total spin $S=1$ via $1\otimes1\to s_3=1$, obtaining a rank-$s_3$ tensor structure, which can be written as 
\begin{equation}
\begin{aligned}
    T^{\sigma_1\sigma_2}_{\mu_3}(p_3,p_1,p_2;1,1)=&g_{\mu_3\mu_3^\prime}\tilde\varepsilon^{\mu_3^\prime\mu\nu}(p_3)
T^{\sigma_1\sigma_2}_{\mu}(p_3,p_1,p_2;1)\tilde t^{(1)}_{\nu}(p_3,r)\\
=&\Big(g_{\mu_3\mu_3^\prime}\varepsilon^{\mu_3^\prime\mu\nu\kappa}p_{3\kappa}\Big)\Big(g_{\mu\mu^\prime}\varepsilon^{\mu^\prime\mu_1\mu_2\rho}p_{3\rho}\Big)
\epsilon_{\mu_1}^{*\sigma_1}(p_1)\epsilon_{\mu_2}^{*\sigma_2}(p_2)
\tilde r_{\nu}(p_3)\\
=&-g_{\mu_3\rho}^{\mu_1\mu_2\nu\kappa}\,p_{3\kappa}p_3^{\rho}\,
\epsilon_{\mu_1}^{*\sigma_1}(p_1)\epsilon_{\mu_2}^{*\sigma_2}(p_2)\tilde r_\nu(p_3),
\end{aligned}
\end{equation}
with
\begin{equation}
g_{\mu_3\rho}^{\mu_1\mu_2\nu\kappa}
\equiv
g^{\mu_1}_{\,\,\mu_3}\big(g^{\mu_2\nu}g^{\kappa}_{\,\,\rho}-g^{\kappa\nu}g^{\mu_2}_{\,\,\rho}\big)
-g^{\mu_2}_{\,\,\mu_3}\big(g^{\mu_1\nu}g^{\kappa}_{\,\,\rho}-g^{\kappa\nu}g^{\mu_1}_{\,\,\rho}\big)
+g^{\kappa}_{\,\,\mu_3}\big(g^{\mu_1\nu}g^{\mu_2}_{\,\,\rho}-g^{\mu_2\nu}g^{\mu_1}_{\,\,\rho}\big).
\end{equation}
Finally, contracting this tensor with the external wave function of the initial particle with spin $s_3$ gives the partial wave amplitude for the decay in the $(L,S)=(1,1)$ channel, which can be written as 
\begin{equation}\label{Eq:111covtensor}
\begin{aligned}
    \mathcal{C}^{\sigma_1\sigma_2}_{\sigma_3}(p_3,p_1,p_2;1,1)=&\epsilon^{\mu_3}_{\sigma_3}(p_3)T^{\sigma_1\sigma_2}_{\mu_3}(p_3,p_1,p_2;1,1)\\
    =&-g_{\mu_3\rho}^{\mu_1\mu_2\nu\kappa}p_{3\kappa}p_3^{\rho}\epsilon^{\mu_3}_{\sigma_3}(p_3)
\epsilon_{\mu_1}^{*\sigma_1}(p_1)\epsilon_{\mu_2}^{*\sigma_2}(p_2)\tilde r_\nu(p_3).
\end{aligned}
\end{equation}

It follows that the evaluation of a covariant tensor amplitude can be organized into three parts, namely the external spin wave functions, the orbital part, and the $LS$ coupling part. One can write
\begin{equation}\label{Eq:covtensorampli}
\begin{aligned}
\mathcal{C}(p_3,p_1,p_2;L,S)=\underbrace{P(p_3;s_1,s_2,S)P(p_3;L,S,s_3)}_{\text{$LS$ coupling part}}\,\underbrace{\tilde t^{(L)}(p_3,r)}_{\text{orbital part}}\,
\underbrace{u(p_3;s_3)
\bar u(p_1;s_1)\bar u(p_2;s_2)}_{\text{spin wave function part}} .
\end{aligned}
\end{equation}
After obtaining the covariant tensor amplitude, one encounters an immediate issue. In the COM frame, although the $LS$ coupling part, the orbital part, and the parent spin wave function reduce to the same objects as in the Zemach tensor method, the two final-state spin wave functions become $\bar u(p_i^*;s_i)$ $(i=1,2)$ and carry the final-state COM momenta, rather than being not purely intrinsic-spin objects. The spin wave function part can be written as
\begin{equation}
    u(p_3;s_{3})\bar{u}(p_1;s_{1})\bar{u}(p_2;s_{2}) 
    \xrightarrow{\mathrm{COM}} u(k_3;s_{3})\bar{u}(p_1^*;s_{1})\bar{u}(p_2^*;s_{2}).
\end{equation}
As a result, in the covariant tensor method the spin part can carry orbital dependence, and even for fixed $(L,S)$ the resulting partial waves generally differ from the traditional-$LS$ amplitudes defined in the two body COM frame. Only in the non-relativistic limit, where the final-state spin wave functions reduce to the rest frame spin wave functions, does the covariant method coincide with the Zemach tensor method.

In addition, when half-integer spins are involved, the covariant tensor method typically provides explicit $LS$ coupling structures only for specific processes, rather than a general formula applicable to arbitrary spin. 
To obtain a general computational expression, the covariant projection tensor method has been developed~\cite{Jing:2023rwz}. This method constructs covariant projection tensors for angular momentum, which enables a systematic evaluation of the required coupling structures for arbitrary spins. The next subsection presents the general method for computing the $LS$ coupling part.

\subsection{Covariant projection tensor method}
\label{subsec:Covariant tensor LS amplitude}


In this subsection, the covariant projection tensor method is introduced~\cite{Jing:2023rwz}. In this method, the $\mathrm{SO}(3,1)$ coupling relations are used to construct the covariant projection tensor. 
This covariant projection tensor plays the role for angular momentum coupling in the $\mathrm{SO}(3,1)$ Lorentz representation. 
In the covariant projection tensor method, the coupling in the $\mathrm{SO}(3,1)$ representation is implemented through this projection tensor.
In practice, two schemes are used depending on the COM definition, referred to as the GS-scheme and the PS-scheme. The constructions of these two schemes are discussed next.

\subsubsection{Covariant projection tensor method in the GS-scheme}

Both the GS-scheme and the PS-scheme require the covariant projection tensors and the covariant orbital tensors. Therefore, the discussion begins with the definitions of these two covariant tensors.

In the covariant projection tensor method, angular momentum coupling is required to be implemented directly in the Lorentz representation of $\mathrm{SO}(3,1)$. 
This requires that the little group coupling be transferred to Lorentz indices. The following discussion explains how angular momentum coupling is realized in the Lorentz representation and how it is related to the little group coupling.
A bridge between the little group representation and the Lorentz representation is established by single-particle states, field operators, and spin wave functions.
The single-particle state can be defined using creation operator
\begin{equation}
    |\mathbf{p},s\sigma\rangle\equiv a^{\dagger}(\mathbf{p},s\sigma)|0\rangle,
\end{equation}
where $|0\rangle$ is the vacuum state. 
By Eq.~\eqref{Eq:Lorentz transformation on a state}, the states transform under a Lorentz transformation $\Lambda$ as
\begin{equation}\label{Eq:lorenz on state}
U[\Lambda] |\mathbf{p}, s\sigma \rangle = \sum_{\sigma^{\prime}} D^{\sigma^\prime(s)}_{ \,\,\sigma}(W(\Lambda, p))  | \mathbf{ p}^\prime, s\sigma^{\prime} \rangle,\quad p^\prime=\Lambda p,
\end{equation}
where $W(\Lambda, p)$ is the little group rotation and $D^{\sigma^\prime(s)}_{\,\,\sigma}(W(\Lambda, p))$ is the rotation matrix in the spin-$s$ representation.
For a quantum field operator in the representation $[\ell]$, it can be written as
\begin{equation}
	\Phi^{\ell}(p) = \sum_{\sigma}u^{\ell}_{\sigma}(p;s)a(\mathbf{p},s\sigma),
\end{equation}
where $a(\mathbf{p},s\sigma)$ is the annihilation operator and $u^{\ell}_{\sigma}(p;s)$ is the spin wave function in the corresponding representation $[\ell]$.
Under a Lorentz transformation $\Lambda$, the field operator $\Phi^{\ell}(p)$ transforms as
\begin{equation}\label{Eq:lorentz on filed}
	U[\Lambda]\Phi^{\ell}(p)U^{-1}[\Lambda] = D^{\ell}_{\,\,\ell'}(\Lambda^{-1})\Phi^{\ell'}(\Lambda p),
\end{equation}
where $D^{\ell}_{\,\,\ell'}(\Lambda^{-1})$ is the Lorentz transformation matrix corresponding to the representation $[\ell]$. 
The spin wave function $u^{\ell}_{\sigma}(p;s)$ is  defined as the matrix element of the field operator between the vacuum and the single-particle state. 
\begin{equation}
	u^{\ell}_{\sigma}(p;s) \equiv \langle 0 | \Phi^{\ell}(p) | \mathbf{p},s\sigma \rangle ,
\end{equation}
where the spin wave function is required to satisfy the normalization condition
\begin{equation}
    \bar u^{\sigma_1}_\ell(p;s) u^\ell_{\sigma_2}(p;s)=\delta^{\sigma_1}_{\sigma_2},
\end{equation}
where $\bar u^{\sigma}_\ell(p;s)$ is defined as
\begin{equation}
    \bar u^{\sigma}_\ell(p;s)\equiv\langle\mathbf{p},s\sigma|\bar\Phi_{\ell}(p)|0\rangle,
\end{equation}
with
\begin{equation}
	\bar\Phi_{\ell}(p) \equiv \sum_{\sigma}\bar u^{\sigma}_\ell(p;s)a^\dagger(\mathbf{p},s\sigma).
\end{equation}
Using Eqs.~\eqref{Eq:lorenz on state} and \eqref{Eq:lorentz on filed}, one can show that under a Lorentz transformation $\Lambda$,
\begin{align}\label{Eq:lorenz on wave function}
    \sum_{\sigma^\prime}D^{\sigma^\prime(s)}_{~\sigma}(W(\Lambda, p))u^{\ell}_{\sigma^\prime}( \Lambda p;s)=&D^{\ell}_{\,\ell'}(\Lambda)u^{\ell'}_{\sigma}(p;s),\\
    \sum_{\sigma^\prime}D^{\sigma^\prime(s)*}_{~\sigma}(W(\Lambda, p))\bar u_{\ell}^{\sigma^\prime}( \Lambda p;s)=&D^{\ell^\prime}_{\,\,\ell}(\Lambda^{-1})\bar u_{\ell'}^{\sigma}(p;s)    .
\end{align}
This relation contains both the induced little group rotation and the $\mathrm{SO}(3,1)$ representation matrix, and it provides an explicit correspondence between little group indices and Lorentz indices. With this correspondence, the angular momentum coupling formulated in the little group $\mathrm{SO}(3)$ representation can be rewritten as a tensor coupling in the Lorentz $\mathrm{SO}(3,1)$ representation. This allows the covariant angular momentum projection and contraction to be implemented directly at the level of Lorentz indices.

The two-body decay process $3\to1+2$ is considered, where all particles are massive with spins $s_3$, $s_1$, and $s_2$. In an arbitrary frame, their momenta are $p_3$, $p_1$, and $p_2$. The COM frame is defined as the rest frame of particle-3. The canonical-standard boost $L_{3,c}(\mathbf{p}_3)$ is chosen such that
\begin{equation}
p_3=L_{3,c}(\mathbf{p}_3)k_3,\qquad k_3=(m_3,0,0,0),
\end{equation}
where $m_3$ is the mass of particle-3, and the subscript $3$ on the canonical-standard boost indicates that it is the standard boost associated with particle-3.
By applying the inverse boost, the momenta in the COM frame are obtained as
\begin{equation}\label{Eq:boost momenta ppp}
k_3=L_{3,c}^{-1}(\mathbf{p}_3)p_3,\qquad
p_1^\ast=L_{3,c}^{-1}(\mathbf{p}_3)p_1,\qquad
p_2^\ast=L_{3,c}^{-1}(\mathbf{p}_3)p_2.
\end{equation}
Starting from the coupling of the spin-$s_1$ and spin-$s_2$ to the total spin-$S$. 
In the COM frame, only the intrinsic spin coupling is considered, and the orbital part is therefore omitted. The spin labels are defined with respect to the little group associated with the total momentum in this frame, which is fixed by $k_3$. Accordingly, Eq.~\eqref{Eq:s1s2toS} expresses the spin coupling as
\begin{equation}\label{Eq:p_3 1+2 S}
|S\sigma_S\rangle=\sum_{\sigma_{1}\sigma_{2}}C_{S,\sigma_{S}}^{s_1,\sigma_{1},s_2,\sigma_{2}}|s_1\sigma_{1}\rangle |s_2\sigma_{2} \rangle,
\end{equation}
where the particle states implicitly refer to the standard-momentum states with $\mathbf{k}=0$, denoted by $|\mathbf{k},s\sigma\rangle$, and, for brevity, the corresponding standard-momentum labels are suppressed.
This makes explicit that the two spins are coupled within the little group representation through the CGCs. 

To implement the spin coupling at the level of Lorentz representation, the coupling structure needs to be defined in the COM frame and with respect to the parent standard momentum $k_3$. This is because, at the level of particle states, the spin coupling is carried out in the little group associated with the parent standard momentum $k_3$, and the translation from the corresponding little group coupling to the Lorentz representation coupling should be performed at the same standard momentum. For this purpose, introduce the field operators at the standard momentum $k_3$ corresponding to spin-$s_1$ and spin-$s_2$, transforming in the Lorentz representations $[\ell_1]$ and $[\ell_2]$, written as
\begin{align}
    \Phi^{\ell_1}(k_3)=&\sum_{\sigma_1}u^{\ell_1}_{\sigma_1}(k_3;s_1)a(\mathbf{k},s_1\sigma_1),\\
    \Phi^{\ell_2}(k_3)=&\sum_{\sigma_2}u^{\ell_2}_{\sigma_2}(k_3;s_2)a(\mathbf{k},s_2\sigma_2).
\end{align}
It should be emphasized that the spin wave functions in these expressions are not the external spin wave functions, but the spin wave functions at the standard momentum $k_3$ used to construct the coupling structures. They can be defined as
\begin{equation}\label{eq_one_particle_u_en}
u^{\ell_1}_{\sigma_1}(k_3;s_1)\equiv\langle0|\Phi^{\ell_1}(k_3)|s_1\sigma_1\rangle,
\quad
u^{\ell_2}_{\sigma_2}(k_3;s_2)\equiv\langle0|\Phi^{\ell_2}(k_3)|s_2\sigma_2\rangle.
\end{equation}
For the total spin-$S$ obtained after coupling, the corresponding Lorentz representation is denoted by $[\ell_S]$, and the field operator at the standard momentum $k_3$ is taken as
\begin{equation}
    \bar\Phi^{\ell_S}(k_3)=\sum_{\sigma_S}\bar u^{\ell_S}_{\sigma_S}(k_3;S)a^{\dagger}(\mathbf{k},S\sigma_S).
\end{equation}
The corresponding spin wave functions are defined by
\begin{equation}\label{Eq:S wavefunction}
    \bar{u}_{\ell_S}^{\sigma_S}(k_3;S)\equiv\langle S\sigma_S|\bar\Phi_{\ell_S}(k_3)|0\rangle.
\end{equation}

On this basis, the Lorentz representations $[\ell_1]$ and $[\ell_2]$ carried by the spin wave functions are coupled into the Lorentz representation $[\ell_S]$ associated with the total spin-$S$, which can be written as
\begin{equation}
    [\ell_1] \otimes[\ell_2]\to[\ell_S],\qquad |s_1-s_2|\leq S \leq s_1+s_2.
\end{equation}
Using Eqs.~\eqref{Eq:p_3 1+2 S}, \eqref{eq_one_particle_u_en} and \eqref{Eq:S wavefunction}, the covariant projection tensor is then introduced as 
\begin{equation}\label{Eq:covariant s1s2S}
\begin{aligned}
P_{\ell_S}^{\ell_1\ell_2}(k_3;S,s_1,s_2)=&\langle 0|\bar\Phi_{\ell_S}(k_3)\Phi^{\ell_1}(k_3)\Phi^{\ell_2}(k_3) |0\rangle\\
=&\langle S\sigma_S|\bar{u}_{\ell_S}^{\sigma_S}(k_3;S)
u^{\ell_1}_{\sigma_1}(k_3;s_1)
u^{\ell_2}_{\sigma_2}(k_3;s_2)|s_1\sigma_{1}\rangle |s_2\sigma_{2} \rangle\\
=&\sum_{\sigma_1\sigma_2\sigma_S}
C_{S,\sigma_S}^{s_1,\sigma_1;s_2,\sigma_2}
\bar{u}_{\ell_S}^{\sigma_S}(k_3;S)
u^{\ell_1}_{\sigma_1}(k_3;s_1)
u^{\ell_2}_{\sigma_2}(k_3;s_2).
\end{aligned}
\end{equation}
The meaning of this covariant projection tensor is that it rewrites, at the standard momentum $k_3$ in the COM frame, the spin coupling implemented by CGCs using little group indices into an equivalent covariant object that involves only Lorentz indices. That is, the spin coupling is no longer written explicitly as a contraction over little group indices. Instead, it is encoded into contractions over Lorentz indices, thereby translating the coupling in the little group representation into the coupling in the Lorentz representation.

Its construction can be understood in the following three steps.
\begin{enumerate}
    \item The rest frame spin wave functions at the standard momentum $k_3$, namely $u^{\ell_1}_{\sigma_1}(k_3;s_1)$ and $u^{\ell_2}_{\sigma_2}(k_3;s_2)$, provide the link between Lorentz representations and the corresponding little group representations. In this way, the Lorentz representations associated with the spin-$s_1$ and the spin-$s_2$ are mapped into the same little group representation fixed by $k_3$.
    \item Within this common little group representation, the two spins are coupled by the CGCs, yielding the little group coupling result with total spin-$S$.
    \item The rest frame spin wave function for total spin $S$, namely $\bar{u}_{\ell_S}^{\sigma_S}(k_3;S)$, maps the coupled little group representation back to the Lorentz representation associated with the total spin-$S$.
\end{enumerate}
Therefore, $P_{\ell_S}^{\ell_1\ell_2}$ can be viewed as a covariant projection tensor that projects the tensor product of two Lorentz representations onto the Lorentz representation corresponding to total spin-$S$. When it is completely contracted with the external wave functions over Lorentz indices, it is equivalent to the action of the CGCs in the little group representation.

Under a Lorentz transformation $\Lambda$, the projector tensor in Eq.~\eqref{Eq:covariant s1s2S} satisfies
\begin{equation}\label{eq_P_intertwiner_invariance}
\begin{aligned}
    P^{\ell_1\ell_2}_{\ell_S}(\Lambda k_3;S,s_1,s_2)=&D^{\ell_1}_{\,\ell_1'}(\Lambda)
D^{\ell_2}_{\,\ell_2'}(\Lambda)
P^{\ell_1'\ell_2'}_{\ell_S'}( k_3;S,s_1,s_2)\,
D^{\ell_S'}_{\,\ell_S}(\Lambda^{-1})\\
=&\sum_{\sigma_1\sigma_2\sigma_S}
C_{S,\sigma_S}^{s_1,\sigma_1;s_2,\sigma_2}
\bar{u}_{\ell_S}^{\sigma_S}(\Lambda k_3;S)
u^{\ell_1}_{\sigma_1}(\Lambda k_3;s_1)
u^{\ell_2}_{\sigma_2}(\Lambda k_3;s_2),
\end{aligned}
\end{equation}
which manifests the covariance of the projection tensor. Therefore, when returning to any frame, it retains the same form and can be written as 
\begin{equation}
    P^{\ell_1\ell_2}_{\ell_S}( p_3;S,s_1,s_2)=\sum_{\sigma_1\sigma_2\sigma_S}
C_{S,\sigma_S}^{s_1,\sigma_1;s_2,\sigma_2}
\bar{u}_{\ell_S}^{\sigma_S}(p_3;S)
u^{\ell_1}_{\sigma_1}(p_3;s_1)
u^{\ell_2}_{\sigma_2}(p_3;s_2).
\end{equation}
Once a concrete Lorentz realization is chosen for the representation $\ell$, the corresponding covariant projection tensor can be evaluated explicitly. For example, choosing the Lorentz four-vector representation with indices $\mu,\nu,\cdots$, in the case $s_1=1$, $s_2=1$, and $S=1$, the covariant projection tensor can be written as
\begin{equation}
\begin{aligned}
P^{\mu_1\mu_2}_{\mu_S}(p_{3};1,1,1)=&\sum_{\sigma_1\sigma_2\sigma_S}C^{1,\sigma_1;1,\sigma_2}_{1,\sigma_S}\bar u^{\sigma_S}_{\mu_S}(p_3;1) u_{\sigma_1}^{\mu_1}(p_3;1) u^{\mu_2}_{\sigma_2}(p_3;1)\\
=&\sum_{\sigma_1\sigma_2\sigma_S}C^{1,\sigma_1;1,\sigma_2}_{1,\sigma_S}\epsilon^{*\sigma_S}_{\mu_S}(p_3)\epsilon_{\sigma_1}^{\mu_1}(p_3)\epsilon^{\mu_2}_{\sigma_2}(p_3)\\
=&\frac{i}{\sqrt{2}m_3}g_{\mu_S\mu_S^\prime}\varepsilon^{\mu_S^\prime\mu_1\mu_2\rho}p_{3\rho}.\\
\end{aligned}
\end{equation}
One can see that, after factoring out the overall normalization constant, this covariant projection tensor is precisely the $1\otimes1\to1$ coupling structure used in the covariant tensor method in Eq.~\eqref{eq:11_to_S1}. Therefore, by constructing such projection tensors, the coupling structures required in the covariant tensor method can be obtained systematically for arbitrary spins.

Besides projecting spin-$s_1$ and spin-$s_2$ onto the total spin-$S$, this covariant projection tensor has a more general meaning. It defines an $\mathrm{SO}(3,1)$ covariant tensor that maps the direct product of two angular momentum representations $[\ell_{j_1}]$ and $[\ell_{j_2}]$ into the representation $[\ell_J]$. Therefore, it not only serves as the covariant projection tensor for coupling $s_1$ and $s_2$ to $S$, but can also be used to project the Lorentz representation $[\ell_L]$ of the orbital part $L$ together with the Lorentz representation $[\ell_S]$ of the total spin part $S$ onto the Lorentz representation $[\ell_3]$ associated with the parent spin-$s_3$.
The corresponding covariant projection tensor can be written as 
\begin{equation}\label{Eq:covariant lss3}
  P_{\ell_3}^{\ell_L\ell_S}(p_3;s_3,L,S)=\sum_{\sigma_S\sigma_L\sigma_{3}}
  C_{s_3,\sigma_3}^{L,\sigma_L;S,\sigma_S}
  \bar{u}_{\ell_3}^{\sigma_3}(p_3;s_3)
  u^{\ell_L}_{\sigma_L}(p_3;L)
  u^{\ell_S}_{\sigma_S}(p_3;S).
\end{equation}
Since the representation $[\ell_S]$ of the total spin-$S$ obtained from the projection has been given above, the next step is to discuss the orbital part in the Lorentz representation $[\ell_L]$.

In the two-body COM frame, the orbital part is completely determined by the relative momentum of the final-state particles. Therefore, only the momentum dependence of the particle-1 and particle-2 states is kept, while the spin labels are suppressed. From Eq.~\eqref{Eq:L spherical harmonic}, the orbital angular momentum state, defined in the little group representation associated with the total momentum $k_3$, can be introduced through spherical harmonics as
\begin{equation}\label{Eq:orbital angular momentum3.1}
    |L\sigma_L\rangle=\int d\Omega  Y^L_{\sigma_L}(\Omega) |\mathbf{p}^*_1\mathbf{p}^*_2\rangle,
\end{equation}
where $|\mathbf{p}^*_1\mathbf{p}^*_2\rangle$ denotes the two-particle state without spin labels, namely $|\mathbf{p}^*_1,0\rangle$ and $|\mathbf{p}^*_2,0\rangle$, and the spin labels of particle-1 and particle-2 are suppressed. The orbital angular momentum state is written as $|\mathbf{k},L\sigma_L\rangle$, and for brevity the standard-momentum label of this orbital state is also suppressed.
The spherical harmonic can be viewed as the $\mathrm{SO}(3)$ representation component of the rank-$L$ symmetric traceless tensor built from the relative three-momentum $ \mathbf{p}^*_1-\mathbf{p}^*_2$. Schematically, this structure is represented by
\begin{equation}
Y^L_{\sigma_L}(\Omega)\simeq (\mathbf{p}^*_1-\mathbf{p}^*_2)^{L}_{i_1\cdots i_{L}}\equiv(\mathbf{p}^\ast_1-\mathbf{p}^\ast_2)_{i_1}\cdots(\mathbf{p}^\ast_1-\mathbf{p}^\ast_2)_{i_L}-\mathrm{trace\ part},
\end{equation}
where $i_1,\ldots,i_L$ are spatial indices and $(\mathbf{p}^\ast_1-\mathbf{p}^\ast_2)^{L}_{i_1\cdots i_L}$ denotes the \textbf{symmetric traceless} rank-$L$ tensor constructed from the relative three momentum.
To obtain the covariant formulation, the spherical harmonic in Eq.~\eqref{Eq:L spherical harmonic} is replaced by the rank-$L$ tensor built from the relative four-momentum $(p^*_1-p^*_2)_\mu$. 
For this purpose, in the COM frame with total momentum $k_3$, a Poincar\'e particle state is introduced whose components are carried by Lorentz indices $\mu_1\cdots\mu_L$. It is defined as
\begin{equation}\label{Eq:mu1muL}
    |\mu_1\cdots\mu_L\rangle\equiv\int d\Omega (p^*_1-p^*_2)_{\mu_1\cdots\mu_L}^L|\mathbf{p}^*_1\mathbf{p}^*_2\rangle,
\end{equation}
where $(p^*_1-p^*_2)_{\mu_1\cdots\mu_L}^L$ denotes the \textbf{symmetric traceless} rank-$L$ tensor constructed from the relative four-momentum, and, for brevity, the standard-momentum label of the orbital state $|\mathbf{k},\mu_1\cdots\mu_L\rangle$ is suppressed.

By construction, $(p_1^\ast-p_2^\ast)^L_{\mu_1\cdots\mu_L}$ is a rank-$L$ symmetric traceless tensor, which is an $\mathrm{SO}(3,1)$ irreducible representation $[\mu_1\cdots\mu_L]$. Since the time component transforms as an $\mathrm{SO}(3)$ scalar in the COM frame, tensor components with one or more time indices behave as scalar factors multiplying lower rank spatial tensors, and hence contain contributions from orbital angular momenta smaller than $L$.
Consequently, the restriction of the representation  $[\mu_1\cdots\mu_L]$ to the $\mathrm{SO}(3)$ little group of the total momentum $k_3$ is reducible and decomposes into a direct sum of the $\mathrm{SO}(3)$ irreducible components $[\ell_i]$,
\begin{equation}\label{Eq:mu1muLto elli}
    [\mu_1\cdots\mu_L]=\bigoplus_i[\ell_i],
\end{equation}
where $i$ labels all possible orbital components. The corresponding pure $L$ covariant orbital angular momentum state is denoted by $|\ell_L\rangle$, which represents the orbital angular momentum $L$ irreducible component of the $\mathrm{SO}(3)$ little group. 
It is $\mathrm{SO}(3)$-equivalent to the orbital angular momentum state $|L\sigma_L\rangle$ in Eq.~\eqref{Eq:orbital angular momentum3.1} by
\begin{equation}
|\ell_L\rangle \simeq |L\sigma_L\rangle,
\end{equation}
where both objects describe the orbital angular momentum $L$ irreducible component, but in different realizations. The state $|\ell_L\rangle$ is written in the covariant Lorentz index realization, whereas $|L\sigma_L\rangle$ is written in the little group index realization.

Next, the corresponding projection tensor is introduced, which projects the Lorentz representation $[\mu_1\cdots\mu_L]$ onto the Lorentz representation $[\ell_L]$ that corresponds to the pure orbital angular momentum $L$ irreducible component.
From Eqs.~\eqref{Eq:mu1muL} and \eqref{Eq:mu1muLto elli}, the state $|\mu_1\cdots\mu_L\rangle$ is known to contain the orbital angular momentum $L$ component. It can be defined as
\begin{equation}
|\mu_1\cdots\mu_L\rangle\equiv\bar\Phi_{\mu_1\cdots\mu_L}(k_3)|0\rangle,
\end{equation}
where $\bar\Phi_{\mu_1\cdots\mu_L}(k_3)$ is the field operator that creates the state in Eq.~\eqref{Eq:mu1muL} from the vacuum, carrying Lorentz indices $\mu_1\cdots\mu_L$ and total momentum $k_3$.
Similarly, the pure orbital angular momentum $L$ state $|\ell_L\rangle$ can be defined as
\begin{equation}
\begin{aligned}
    |\ell_L\rangle\equiv&\bar\Phi_{\ell_L} (k_3)|0\rangle,
\end{aligned}
\end{equation}
where $\bar\Phi_{\ell_L}(k_3)$ is the field operator that creates the state in the Lorentz representation $[\ell_L]$ from the vacuum, carrying Lorentz indices $\ell_L$ and total momentum $k_3$.
Therefore, a projection operator exists that projects the Lorentz representation $[\mu_1\cdots\mu
_L]$ onto the pure orbital angular momentum $L$ Lorentz representation $[\ell_L]$. The corresponding orbital projection operator is defined as
\begin{equation}
\begin{aligned}
    P^{\mu_1\cdots\mu_L}_{\ell_L}(k_3;L)\equiv&\langle \mu_1\cdots\mu_L|\ell_L\rangle\\
    =&\langle 0|\bar \Phi_{\ell_L}(k_3)\Phi^{\mu_1\cdots\mu_L}(k_3)|0\rangle.
\end{aligned}
\end{equation}
Using the covariant projection tensor in Eq.~\eqref{Eq:covariant s1s2S}, this projection can be implemented recursively in a fixed coupling order. 
For $L=0$, there is no Lorentz index. In the Lorentz representation associated with orbital angular momentum $L=0$, the corresponding projection tensor is a scalar.

For $L=1$, the corresponding projection tensor is equivalent to coupling the spin-$1$ Lorentz representation $[\mu_1]$ with that of spin-$0$ to obtain the Lorentz representation $[\ell_1]$. 
From Eq.~\eqref{Eq:covariant s1s2S}, the covariant projection tensor can be written as 
\begin{equation}\label{Eq:L=1ellmu1}
\begin{aligned}
        P_{\ell_1}^{\mu_1}(k_3;1,1,0)=&\sum_{\sigma_1\sigma}
  C_{1,\sigma}^{1,\sigma_1;0}
  \bar{u}_{\ell_1}^{\sigma}(k_3;1)
  u^{\mu_1}_{\sigma_1}(k_3;1)\\
  =&\sum_{\sigma_1}
  \bar{u}_{\ell_1}^{\sigma_1}(k_3;1)
  u^{\mu_1}_{\sigma_1}(k_3;1).
\end{aligned}
\end{equation}

For an arbitrary $L>1$, the last $L-1$ indices $\mu_2\cdots\mu_L$ are first projected onto an intermediate orbital angular momentum $L-1$ Lorentz representation $[\ell_{L-1}]$, and the remaining index $\mu_1$ is then treated as a spin-$1$ Lorentz representation $[\mu_1]$ and coupled with $[\ell_{L-1}]$ through the projection tensor realizing $1\otimes(L-1)\to L$. This yields
\begin{equation}\label{Eq:Lprojecttensor}
\begin{aligned}
P^{\mu_1\cdots\mu_L}_{\ell_L}(k_3;L)=&P_{\ell_L}^{\mu_1\ell_{L-1}}(k_3;L,1,L-1)P_{\ell_{L-1}}^{\mu_2\cdots\mu_L}(k_3;L-1),
\end{aligned}
\end{equation}
where $P_{\ell_L}^{\mu_1\ell_{L-1}}(k_3;L,1,L-1)$ is the covariant projection tensor implementing $1\otimes(L-1)\to L$. By Eq.~\eqref{Eq:covariant s1s2S}, it can be written as
\begin{equation}
    P_{\ell_L}^{\mu_1\ell_{L-1}}(k_3;L,1,L-1)=\sum_{\sigma_1\sigma_{L-1}\sigma_{L}}
  C_{L,\sigma_L}^{1,\sigma_1;L-1,\sigma_{L-1}}
  \bar{u}_{\ell_L}^{\sigma_L}(k_3;L)
  u^{\mu_1}_{\sigma_1}(k_3;1)
  u^{\ell_{L-1}}_{\sigma_{L-1}}(k_3;L-1).
\end{equation}
Therefore, $P^{\mu_1\cdots\mu_L}_{\ell_L}(k_3;L)$ is obtained by successive couplings implemented by the projection tensor.
Then, this projection tensor can be used to project the rank-$L$ tensor built from the relative four-momentum, with indices $\mu_1\cdots\mu_L$, onto the desired orbital-$L$ Lorentz representation $[\ell_L]$. The covariant orbital tensor is therefore defined as
\begin{equation}\label{Eq:tensor t}
    t_{\ell_L}^{L}(k_3,p^*_1-p^*_2)\equiv 
    P^{\mu_1\cdots\mu_L}_{\ell_L}(k_3;L)(p^*_1-p^*_2)_{\mu_1}\cdots(p^*_1-p^*_2)_{\mu_{L}},
\end{equation}
where $t_{\ell_L}^{\,L}(k_3,p^*_1-p^*_2)$ is manifestly covariant. In any frame it keeps the same form and can be written as
\begin{equation}
    t_{\ell_L}^{L}(p_3,p_1-p_2)\equiv 
    P^{\mu_1\cdots\mu_L}_{\ell_L}(p_3;L)(p_1-p_2)_{\mu_1}\cdots(p_1-p_2)_{\mu_{L}}.
\end{equation}
After fixing a specific Lorentz realization $[\ell_L]$ of the pure orbital angular momentum $L$, the corresponding covariant orbital tensors can be evaluated explicitly. As a representative example, take $[\ell_L]$ to be realized by Lorentz four-vector indices, namely by Lorentz tensor indices $\mu_1\cdots\mu_L$ that carry the pure $L$ component. In this case, after simplification, the associated projection tensor can be written as 
\begin{equation}
    P^{\nu_1\cdots\nu_L}_{\mu_1\cdots\mu_L}(p_3;L)=\sum_{\sigma_L}\epsilon^{\nu_1\cdots\nu_L}_{\sigma_L}(p_3;L)\epsilon_{\mu_1\cdots\mu_L}^{*\sigma_L}(p_3;L),
\end{equation}
where $\epsilon^{\nu_1\cdots\nu_L}_{\sigma_L}(p_3;L)$ is the spin wave function satisfying the Rarita–Schwinger conditions, given in Eq.~\eqref{Eq:rarita wave functions}.
One can see that this projection tensor is exactly the orbital projector used for the orbital part in Eq.~\eqref{Eq:projection tensortttt} of the covariant tensor method. It projects the rank-$L$ tensor built from the relative momentum onto the subspace that is symmetric and traceless and transverse to the parent momentum in each index, thereby extracting the irreducible component corresponding to the pure orbital angular momentum $L$.

With this choice, the covariant orbital tensors for several low-$L$ cases can be written as 
\begin{equation}
\begin{aligned}
t^{0}=&1,\\
t^{1}_{\mu_1}=&\Big({-}g^{\nu_1}_{\,\,\mu_1}+\frac{p_3^{\nu_1}p_{3\mu_1}}{p^2_3}\Big)(p_1-p_2)_{\nu_1},\\
t^{2}_{\mu_1\mu_2}=&\Big({-}g^{\nu_1}_{\,\,\mu_1}+\frac{p_3^{\nu_1}p_{3\mu_1}}{p^2_3}\Big)\Big({-}g^{\nu_2}_{\,\,\mu_2}+\frac{p_3^{\nu_2}p_{3\mu_2}}{p^2_3}\Big)(p_1-p_2)_{\nu_1}(p_1-p_2)_{\nu_2}\\
&-\frac13\Bigg[\Big({-}g^{\nu_1\nu_2}+\frac{p_3^{\nu_1}p_{3}^{\nu_2}}{p^2_3}\Big)(p_1-p_2)_{\nu_1}(p_1-p_2)_{\nu_2}\Bigg]\Big({-}g_{\mu_1\mu_2}+\frac{p_{3\mu_1}p_{3\mu_2}}{p^2_3}\Big).\\
\end{aligned}
\end{equation}
It follows that the covariant orbital tensors obtained in this realization coincide exactly with the orbital tensors in Eq.~\eqref{Eq:covtensor L} of the covariant tensor method. Equivalently, this projection tensor projects the rank-$L$ tensor constructed from the relative momentum onto the Lorentz representation $[\ell_L]$ associated with the corresponding pure $L$ irreducible component.

After defining the covariant tensors in Eqs.~\eqref{eq_P_intertwiner_invariance}, \eqref{Eq:covariant lss3}, and \eqref{Eq:tensor t}, the construction of the corresponding partial wave amplitudes can be formulated. First, recall the partial wave amplitude in the covariant tensor method in Eq.~\eqref{Eq:covtensorampli}. It can be written as a product of three pieces: the $LS$ coupling part, the orbital part, and the external spin wave function part. The $LS$ coupling part and the orbital part are precisely given by the total spin-$S$ projection tensor, the orbital-spin coupling projection tensor, and the covariant orbital tensor defined above. Therefore, combining these building blocks, the covariant $LS$ coupling structure is defined as
\begin{equation}\label{Eq:Coupling structure}
	\Gamma_{\ell_{3}}^{\ell_{1}\ell_{2}}(k_3,{p}^*_{1},{p}^*_{2};L,S)\equiv P_{\ell_{3}}^{\ell_{L}\ell_{S}}(k_3;s_{3},L,S)P_{\ell_{S}}^{\ell_{1}\ell_{2}}(k_3;S,s_{1},s_{2})t_{\ell_{L}}^{L}(k_3;p^*_{1}-p^*_{2}).
\end{equation}

By contracting this covariant $LS$ coupling structure with the external spin wave functions over the Lorentz indices, the $LS$ partial wave amplitude in the GS-scheme is obtained. It can be written as
\begin{equation}\label{Eq:cscheme ampli}
\begin{aligned}
      \mathcal{C}^{\sigma_1\sigma_2}_{\sigma_3}(k_3,p_1^*,p_2^*;L,S) \equiv& \Gamma_{\ell_{3}}^{\ell_{1}\ell_{2}}(k_3,p_1^*,p_2^*;L,S)\times u_{\sigma_{3}}^{\ell_{3}}(k_3;s_{3})\bar{u}_{\ell_{1}}^{\sigma_{1}}(p^*_1;s_{1})\bar{u}_{\ell_{2}}^{\sigma_{2}}(p^*_2;s_{2})\\
      =&P_{\ell_{S}}^{\ell_{1}\ell_{2}}(k_3;S,s_{1},s_{2})\bar{u}_{\ell_{1}}^{\sigma_{1}}(p^*_1;s_{1})\bar{u}_{\ell_{2}}^{\sigma_{2}}(p^*_2;s_{2})\\
      &\times P_{\ell_{3}}^{\ell_{L}\ell_{S}}(k_3;s_{3},L,S)t_{\ell_{L}}^{L}(k_3;p^*_{1}-p^*_{2})\\
      &\times u_{\sigma_{3}}^{\ell_{3}}(k_3;s_{3}).
\end{aligned}
\end{equation}
The physical content of this construction can be organized into three steps.
\begin{enumerate}
    \item $P_{\ell_S}^{\ell_1\ell_2}(k_3;S,s_1,s_2)\bar u_{\ell_1}^{\sigma_1}(p_1^\ast;s_1)\bar u_{\ell_2}^{\sigma_2}(p_2^\ast;s_2)$ represent projecting the representation $[\ell_1]\otimes[\ell_2]$ carried by the two final state external spin wave functions onto the total spin-$S$ representation $[\ell_S]$, thereby implementing the spin coupling $s_1\otimes s_2\to S$.
    \item $P_{\ell_3}^{\ell_L\ell_S}(k_3;s_3,L,S)t_{\ell_L}^{L}(k_3;p_1^\ast-p_2^\ast)$ represent further projecting $[\ell_L]\otimes[\ell_S]$ onto the representation $[\ell_3]$ of the parent spin $s_3$, implementing the coupling $L\otimes S\to s_3$.
    \item Project the coupling result obtained above onto the parent external spin component $\sigma_3$, and finally obtain a scalar partial wave amplitude with external spin labels $(\sigma_1,\sigma_2,\sigma_3)$.
\end{enumerate}
Since the whole expression implements the coupling and projection explicitly through Lorentz index contractions, it is covariant in form. In any frame it keeps the same form and can be written as
\begin{equation}\label{Eq:covariant LS in COM}
	\mathcal{C}_{\sigma_{3}}^{\sigma_{1}\sigma_{2}}(p_{3},p_{1},p_{2};L,S)=\underbrace{\Gamma_{\ell_{3}}^{\ell_{1}\ell_{2}}(p_{3},p_{1},p_{2};L,S)}_{\text{covariant $LS$ coupling structure}}\times \underbrace{u_{\sigma_{3}}^{\ell_{3}}(p_3;s_{3})\bar{u}_{\ell_{1}}^{\sigma_{1}}(p_1;s_{1})\bar{u}_{\ell_{2}}^{\sigma_{2}}(p_2;s_{2})}_{\text{general-spin part}}.
\end{equation}
The GS-scheme is, in essence, fully equivalent to the covariant tensor method. When the Lorentz representations $[\ell]$ are chosen to be the corresponding four-vector realizations, this equivalence becomes explicit. 

For example, consider the two-body decay with $s_3=1$, $s_1=1$, and $s_2=1$, and select the partial wave $(L,S)=(1,1)$. In this case, the total spin-$S$ covariant projection tensor can be written as
\begin{equation}
\begin{aligned}
P^{\mu_1\mu_2}_{\mu_S}(p_{3};1,1,1)
=&\sum_{\sigma_1\sigma_2\sigma_S}C^{1,\sigma_1;1,\sigma_2}_{1,\sigma_S}\epsilon^{*\sigma_S}_{\mu_S}(p_3)\epsilon_{\sigma_1}^{\mu_1}(p_3)\epsilon^{\mu_2}_{\sigma_2}(p_3)\\
=&\frac{i}{\sqrt{2}m_3}g_{\mu_S\mu_S^\prime}\varepsilon^{\mu_S^\prime\mu_1\mu_2\rho}p_{3\rho},\\
\end{aligned}
\end{equation}
the orbital-spin covariant projection tensor can be written as 
\begin{equation}
\begin{aligned}
P^{\mu_L\mu_S}_{\mu_{3}}(p_{3};1,1,1)
=&\sum_{\sigma_L\sigma_S\sigma_3}C^{1,\sigma_L;1,\sigma_S}_{1,\sigma_3}\epsilon^{*\sigma_3}_{\mu_3}(p_3)\epsilon_{\sigma_L}^{\mu_L}(p_3)\epsilon^{\mu_S}_{\sigma_S}(p_3)\\
=&\frac{i}{\sqrt{2}m_3}g_{\mu_3\mu_3^\prime}\varepsilon^{\mu_3^\prime\mu_L\mu_S\rho}p_{3\rho},\\
\end{aligned}
\end{equation}
and the covariant orbital tensor can be written as 
\begin{equation}
t^{1}_{\mu_L}=\Big({-}g^{\nu_L}_{\,\,\mu_L}+\frac{p_3^{\nu_L}p_{3\mu_L}}{p^2_3}\Big)(p_1-p_2)_{\nu_L}.
\end{equation}
After contracting the covariant $LS$ coupling structure with the external spin wave functions part, the final partial wave amplitude can be written as
\begin{equation}
\begin{aligned}
\mathcal{C}^{\sigma_1\sigma_2}_{\sigma_3}(p_3,p_1,p_2;1,1)=&\Gamma_{\mu_{3}}^{\mu_{1}\mu_{2}}(p_{3},p_{1},p_{2};1,1)\times\epsilon^{\mu_3}_{\sigma_3}(p_3)\epsilon_{\mu_1}^{*\sigma_1}(p_1)\epsilon_{\mu_2}^{*\sigma_2}(p_2)\\
\propto&g_{\mu_3\rho}^{\mu_1\mu_2\mu_L\kappa}p_{3\kappa}p_3^{\rho}\Big({-}g^{\nu_L}_{\,\,\mu_L}+\frac{p_3^{\nu_L}p_{3\mu_L}}{p^2_3}\Big)(p_1-p_2)_{\nu_L}\\
&\times\epsilon^{\mu_3}_{\sigma_3}(p_3)
\epsilon_{\mu_1}^{*\sigma_1}(p_1)\epsilon_{\mu_2}^{*\sigma_2}(p_2),
\end{aligned}
\end{equation}
with
\begin{equation}
g_{\mu_3\rho}^{\mu_1\mu_2\mu_L\kappa}
\equiv
g^{\mu_1}_{\,\,\mu_3}\big(g^{\mu_2\mu_L}g^{\kappa}_{\,\,\rho}-g^{\kappa \mu_L}g^{\mu_2}_{\,\,\rho}\big)
-g^{\mu_2}_{\,\,\mu_3}\big(g^{\mu_1\mu_L}g^{\kappa}_{\,\,\rho}-g^{\kappa\mu_L}g^{\mu_1}_{\,\,\rho}\big)
+g^{\kappa}_{\,\,\mu_3}\big(g^{\mu_1\mu_L}g^{\mu_2}_{\,\,\rho}-g^{\mu_2\mu_L}g^{\mu_1}_{\,\,\rho}\big),
\end{equation}
which coincides exactly with Eq.~\eqref{Eq:111covtensor}.
Consequently, the covariant projection tensor formulation in the covariant projection tensor method provides a systematic completion of the Lorentz structures entering the orbital-spin coupling in the covariant tensor method, and it allows the corresponding coupling structures to be computed for arbitrary spin.

However, the covariant projection tensor amplitude in the GS-scheme has the same issue as the covariant tensor method, namely that the orbital and spin decomposition is not fully explicit. 
The reason is that, in the COM frame, the covariant projection tensor amplitude in the GS-scheme in Eq.~\eqref{Eq:cscheme ampli} evaluates the final-state external spin wave functions at the COM momenta $p_1^\ast$ and $p_2^\ast$, rather than at the rest frame. It can be written as
\begin{equation}
    u(p_3;s_{3})\bar{u}(p_1;s_{1})\bar{u}(p_2;s_{2}) 
    \xrightarrow{\mathrm{COM}} u(k_3;s_{3})\bar{u}(p_1^*;s_{1})\bar{u}(p_2^*;s_{2}),
\end{equation}
where the general-spin part in the COM frame depends on the relative momentum direction and thus contain residual contributions from the orbital part. 
Consequently, the separation between $L$ and $S$ does not follow the traditional-$LS$ method, since the general-spin part here is not purely intrinsic-spin objects.
In this sense, the covariant projection tensor amplitude in the GS-scheme differs from the traditional-$LS$ amplitude in Eq.~\eqref{Eq:canonical amplitude}.
Only in the non-relativistic limit, the general-spin part is reduced to $u(k_3;s_{3})\bar{u}(k_1;s_{1})\bar{u}(k_2;s_{2})$, which is the purely intrinsic-spin objects, and the covariant projection tensor amplitude in the GS-scheme would have clear $LS$ separation.  

Therefore, in order to construct a scheme with a complete orbital-spin separation while using the covariant $LS$ coupling structures, the external spin wave functions must be further processed. Before introducing this treatment, the full amplitude in the GS-scheme is defined by summing the $LS$ partial wave amplitudes over all allowed $(L,S)$ components in the COM definition,
\begin{equation}
    \mathcal{C}^{\sigma_1\sigma_2}_{\sigma_3}(k_3,p^*_1,p^*_2) =\sum_{LS}N_{LS}\mathcal{C}^{\sigma_1\sigma_2}_{\sigma_3}(k_3,p^*_1,p^*_2;L,S),
\end{equation}
where $N_{LS}$ denotes the partial wave coefficient associated with the $(L,S)$ component.
The full covariant projection tensor amplitude in the GS-scheme can also be written as a complete contraction between the covariant coupling structure and the general-spin part,
\begin{equation}\label{Eq:C_scheme_amp}
      \mathcal{C}^{\sigma_1\sigma_2}_{\sigma_3}(k_3,p^*_1,p^*_2) =\underbrace{\Gamma^{\ell_1\ell_2}_{\ell_{3}}(k_3,p^*_1,p^*_2)}_{\text {covariant coupling structure}}
      \times \underbrace{u^{\ell_3}_{\sigma_3}(k_3;s_3) \bar{u}^{\sigma_1}_{\ell_1}(p^*_1;s_1) \bar{u}^{\sigma_2}_{\ell_2}(p^*_2;s_2)}_{\text {general-spin part}},
\end{equation}
with
\begin{equation}
    \Gamma^{\ell_1\ell_2}_{\ell_{3}}(k_3,p^*_1,p^*_2)\equiv\sum_{LS}N_{LS}\Gamma_{\ell_{3}}^{\ell_{1}\ell_{2}}(k_{3},p^*_{1},p^*_{2};L,S).
\end{equation}
In the next subsubsection, the final-state spin wave functions are rewritten such that their momentum dependence is extracted and reassigned to the coupling structure. This leads to the PS-scheme partial-wave amplitudes, where the orbital and spin parts are explicitly separated.

\subsubsection{Covariant projection tensor method in the PS-scheme}
\label{subsubsec:Covariant tensor LS amplitude}
In this subsubsection, the construction of the covariant projection tensor amplitude in the PS-scheme is discussed.
In Eq.~\eqref{Eq:C_scheme_amp}, the general-spin part depends on the relative momentum and therefore carries orbital information. 
To enforce a complete separation between the orbital part and the spin part, the orbital dependence in the final-state wave functions is extracted and absorbed into the orbital part. 
Using Eq.~\eqref{Eq:lorenz on wave function}, the final-state spin wave functions can be written as
\begin{equation}\label{Eq:COM wavefunction boost}
    \bar u^{\sigma_1}_{\ell_1}( p^*_1;s_1)=\bar u^{\sigma_1}_{\ell_1^\prime}( k_1;s_1)D^{\ell_1^\prime}_{\,\,\ell_1}(L^{-1}_{1,c}(\mathbf p_1^*)),\quad
    \bar u^{\sigma_2}_{\ell_2}( p^*_2;s_2)=\bar u^{\sigma_2}_{\ell_2^\prime}( k_2;s_2)D^{\ell_2^\prime}_{\,\,\ell_2}(L^{-1}_{2,c}(\mathbf p_2^*)),
\end{equation}
where $k_1$ and $k_2$ are the standard momenta of the particle-1 and the particle-2.
The boosts are then reassigned to the covariant coupling structure, using Eqs.~\eqref{Eq:C_scheme_amp} and \eqref{Eq:COM wavefunction boost}, the PS-scheme amplitude is obtained as
\begin{equation}\label{Eq:H scheme amplitude c}
\begin{aligned}
      A^{\sigma_1\sigma_2}_{\sigma_3}(k_3,p_1^*,p_2^*) =&\underbrace{\Gamma^{\ell_1\ell_2}_{\ell_{3}}(k_3,p_1^*,p_2^*)D^{\ell_1^\prime}_{\,\,\ell_1}(L^{-1}_{1,c}(\mathbf p_1^*))D^{\ell_2^\prime}_{\,\,\ell_2}(L^{-1}_{2,c}(\mathbf p_2^*))}_{\text {covariant coupling structure}}\\
      &\times \underbrace{u^{\ell_3}_{\sigma_3}(k_3;s_3) \bar{u}^{\sigma_1}_{\ell_1^\prime}(k_1;s_1) \bar{u}^{\sigma_2}_{\ell_2^\prime}(k_2;s_2)}_{\text {pure-spin part}},
\end{aligned}
\end{equation}
where the pure-spin part no longer contains any orbital information, thereby achieving an explicit separation between the orbital and spin part.

For the covariant coupling structure in Eq.~\eqref{Eq:H scheme amplitude c}, 
once all allowed $(L,S)$ partial waves are included, it can be expanded in the set of covariant $LS$ coupling structures $\Gamma^{\ell_1\ell_2}_{\ell_{3}}(k_3,p_1^*,p_2^*;L,S)$ defined in Eq.~\eqref{Eq:Coupling structure}, which can be written as
\begin{equation}\label{Eq:coupling structure expand}
    \Gamma^{\ell_1\ell_2}_{\ell_{3}}(k_3,p_1^*,p_2^*)D^{\ell_1^\prime}_{\,\,\ell_1}(L^{-1}_{1,c}(\mathbf{p}_1^*))D^{\ell_2^\prime}_{\,\,\ell_2}(L^{-1}_{2,c}(\mathbf{p}_2^*))=\sum_{L S}N_{L S}\Gamma^{\ell_1^\prime\ell_2^\prime}_{\ell_{3}}(k_3,p_1^*,p_2^*;L,S),
\end{equation}
where $N_{LS}$ is the partial wave coefficient of the $LS$ amplitude basis.
As a set of pure tensor structures, $\Gamma^{\ell_1\ell_2}_{\ell_{3}}(k_3,p_1^*,p_2^*;L,S)$ does not necessarily (and in practice usually does not) form a complete basis. However, after contracting it with the external spin wave functions, one obtains the scalar amplitudes
\begin{equation}
  \Gamma_{\ell_{3}}^{\ell_{1}\ell_{2}}(k_3,p_1^*,p_2^*;L,S)\times u_{\sigma_{3}}^{\ell_{3}}(k_3;s_{3})\bar{u}_{\ell_{1}}^{\sigma_{1}}(k_1;s_{1})\bar{u}_{\ell_{2}}^{\sigma_{2}}(k_2;s_{2}),
\end{equation}
which reproduce the $LS$ decomposition of the traditional-$LS$ amplitude. In this sense, the decomposition is complete at the amplitude level. Therefore, the structures generated by $\Gamma_{\ell_{3}}^{\ell_{1}\ell_{2}}(k_3,p_1^*,p_2^*;L,S)$ provide a complete basis for the amplitudes. Any additional tensor structures can only yield redundant contributions that become equivalent to this basis after contraction with the external spin wave functions, and they do not generate new independent amplitudes. Based on this observation, it is sufficient to consider only the tensor structures generated by these $\Gamma^{\ell_1\ell_2}_{\ell_{3}}(k_3,p_1^*,p_2^*;L,S)$ to span the full set of amplitudes. Consequently, the covariant projection tensor amplitude in the PS-scheme is defined as
\begin{equation}
  A^{\sigma_1\sigma_2}_{\sigma_3}(k_3,p_1^*,p_2^*;L,S) \equiv\underbrace{\Gamma^{\ell_1\ell_2}_{\ell_{3}}(k_3,p_1^*,p_2^*;L,S)}_{\text{covariant $LS$ coupling structure}}\times \underbrace{u^{\ell_3}_{\sigma_3}(k_3;s_3) \bar{u}^{\sigma_1}_{\ell_1}(k_1;s_1) \bar{u}^{\sigma_2}_{\ell_2}(k_2;s_2)}_{\text{pure-spin part}},
\end{equation}
where the $LS$ coupling structure $\Gamma^{\ell_1\ell_2}_{\ell_{3}}(k_3,p_1^*,p_2^*;L,S)$ is the same as those in Eq.~\eqref{Eq:Coupling structure}, while the external spin wave functions are defined at the standard momenta of the corresponding particles.

In the following, the equivalence between the PS-scheme amplitude and the traditional-$LS$ amplitude is established.
Using Eqs.~\eqref{Eq:Coupling structure}, \eqref{Eq:covariant s1s2S}, \eqref{Eq:covariant lss3} and \eqref{Eq:tensor t}, the amplitude in the COM frame can be expanded as
\begin{equation}
    \begin{aligned}
        A_{\sigma_{3}}^{\sigma_{1}\sigma_{2}}(k_{3},p^{*}_{1},p^{*}_{2};L,S)=&\Gamma_{\ell_{3}}^{\ell_{1}\ell_{2}}(k_{3},p^{*}_{1},p^{*}_{2};L,S)u_{\sigma_{3}}^{\ell_{3}}(k_3;s_{3})\bar{u}_{\ell_{1}}^{\sigma_{1}}(k_1;s_{1})\bar{u}_{\ell_{2}}^{\sigma_{2}}(k_2;s_{2})\\
        =&\sum_{\sigma_1^\prime\sigma_2^\prime\sigma_S^\prime\sigma_L\sigma_S\sigma_3^\prime}C_{s_3,\sigma_{3}^\prime}^{L,\sigma_{L};S,\sigma_{S}}\bar{u}_{\ell_3}^{\sigma_3^\prime}(k_3;s_3)u_{\sigma_L}^{\ell_L}(k_3;L)u_{\sigma_S}^{\ell_S}(k_3;S)\\ 
        &\times C_{S,\sigma_{S}^\prime}^{s_1,\sigma_{1}^\prime;s_2,\sigma_{2}^\prime}\bar{u}_{\ell_S}^{\sigma_S^\prime}(k_3;S)u_{\sigma_1^\prime}^{\ell_1}(k_3;s_1)u_{\sigma_2^\prime}^{\ell_2}(k_3;s_2)\\
        &\times P^{\mu_1\cdots\mu_L}_{\ell_L}(k_3;L)(p^*_1-p^*_2)_{\mu_1}\cdots(p^*_1-p^*_2)_{\mu_{L}}\\
        &\times u^{\ell_3}_{\sigma_3}(k_3;s_3)\bar{u}_{\ell_1}^{\sigma_1}(k_1;s_1)\bar{u}_{\ell_2}^{\sigma_2}(k_2;s_2).
\end{aligned}
\end{equation}
Using the orthonormality relation in Eq.~\eqref{Eq:orthonormal relation of wave function} for the spin wave functions with spins $s_1$, $s_2$, $s_3$, and $S$, 
\begin{equation}
\begin{aligned}
    \bar{u}_{\ell_1}^{\sigma_1}(k_1;s_1)u^{\ell_1}_{\sigma_1^\prime}(k_3;s_1)\propto&\delta_{\sigma_1^\prime}^{\sigma_1},\\
    \bar{u}_{\ell_2}^{\sigma_2}(k_2;s_2)u^{\ell_2}_{\sigma_2^\prime}(k_3;s_2)\propto&\delta_{\sigma_2^\prime}^{\sigma_2},\\
    \bar{u}_{\ell_3}^{\sigma_3^\prime}(k_3;s_3)u^{\ell_3}_{\sigma_3}(k_3;s_3)=&\delta_{\sigma_3^\prime}^{\sigma_3},\\
    \bar{u}_{\ell_S}^{\sigma_S^\prime}(k_3;S)u^{\ell_S}_{\sigma_S}(k_3;S)=&\delta_{\sigma_S^\prime}^{\sigma_S},
\end{aligned}
\end{equation}
where the spin-$s_1$ and spin-$s_2$ wave functions are defined at different standard momenta, the difference is only an overall mass dependent constant. This constant can be absorbed into the $LS$ partial wave coefficients. Therefore, the amplitude can be written as
\begin{equation}\label{Eq:canonical to tensor}
\begin{aligned}
    A_{\sigma_{3}}^{\sigma_{1}\sigma_{2}}(k_{3},p^{*}_{1},p^{*}_{2};L,S)\propto&\sum_{\sigma_L\sigma_S}C_{s_3,\sigma_{3}}^{L,\sigma_{L};S,\sigma_{S}}C_{S,\sigma_{S}}^{s_1,\sigma_{1};s_2,\sigma_{2}}\\
    &\times u_{\sigma_L}^{\ell_L}(k_3;L)P^{\mu_1\cdots\mu_L}_{\ell_L}(k_3;L)(p^*_1-p^*_2)_{\mu_1}\cdots(p^*_1-p^*_2)_{\mu_{L}}.
\end{aligned}
\end{equation}
For the expression in the second line, after simplification it becomes the spherical harmonic with orbital angular momentum $L$, and can be written as
\begin{equation}\label{Eq:relation of tensor t to Y}
    u_{\sigma_L}^{\ell_L}(\mathbf{k}_3;L)P^{\mu_1\cdots\mu_L}_{\ell_L}(\mathbf{k}_3;L)(p^*_1-p^*_2)_{\mu_1}\cdots(p^*_1-p^*_2)_{\mu_{L}}=N_{L}Y^L_{\sigma_L}(\Omega),
\end{equation}
where $N_L$ is the normalization constant.
As an example, for $L=1$, 
\begin{equation}\label{Eq:L=1 t to Y}
\begin{aligned}
    u^{\ell_1}_{\sigma_1}(k_3;1)P^{\mu_1}_{\ell_1}(\mathbf{k}_3;1)(p^*_1-p^*_2)_{\mu_1}=&u^{\ell_1}_{\sigma_1}(k_3;1)P^{\mu_1}_{\ell_1}(k_3;1,1,0)(p^*_1-p^*_2)_{\mu_1}\\
    =&\sum_{\sigma_1^\prime\sigma_1^{\prime\prime}}u^{\ell_1}_{\sigma_1}(k_3;1)C^{1,\sigma_1^{\prime\prime};0}_{1,\sigma_1^{\prime}}\bar{u}^{\sigma^{\prime}_1}_{\ell_1}(k_3;1)u^{\mu_1}_{\sigma^{\prime\prime}_1}(k_3;1)(p^*_1-p^*_2)_{\mu_1}\\
    =&u^{\mu_1}_{\sigma_1}(k_3;1)(p^*_1-p^*_2)_{\mu_1}.
\end{aligned}
\end{equation}
In the rest frame, the spin-1 wave function with the Lorentz four-vector index $\mu$ corresponds to the polarization vector. For the standard momentum $k_3^\mu $, it can be written as
\begin{equation}
    u^\mu_{\sigma}(k_3;1)=\epsilon^{\mu}_{\sigma}(k_3)=\begin{pmatrix}
        0 & \frac{-1}{\sqrt{2}} & \frac{-i}{\sqrt{2}} &0\\
        0 & 0 & 0 & 1\\
        0 & \frac{ 1}{\sqrt{2}} & \frac{-i}{\sqrt{2}} & 0
    \end{pmatrix}_{\sigma}^{\,\,\mu},\quad \bar{u}_{\mu}^\sigma(k_3;1)=\epsilon_{\mu}^{*\sigma}(k_3)=\begin{pmatrix}
        0 & 0 &0\\
        \frac{-1}{\sqrt{2}} & 0 & \frac{ 1}{\sqrt{2}}\\
        \frac{i}{\sqrt{2}} & 0 &  \frac{i}{\sqrt{2}}\\
        0 &  1 & 0
    \end{pmatrix}_{\mu}^{\,\,\sigma}.
\end{equation}
By calculating each component of Eq.~\eqref{Eq:L=1 t to Y}, one finds that the expression is proportional to the spherical harmonics and can be written as
\begin{equation}
\begin{aligned}
u^{\mu_1}_{\sigma_1}(k_3;1)(p^*_1-p^*_2)_{\mu_1}\propto&\begin{pmatrix}
    -\sin{\theta}e^{i\phi}\\
    \cos{\theta}\\
     \sin{\theta}e^{-i\phi}
\end{pmatrix}_{\sigma_1}\\
=&Y^1_{\sigma_1}(\Omega).
\end{aligned}
\end{equation}
In this way, for orbital angular momentum $L$, the same simplification can be performed as above, and the result reduces to the spherical harmonic in Eq.~\eqref{Eq:relation of tensor t to Y}. The normalization constant $N_L$ can be absorbed into the $LS$ partial wave coefficient.
Thus the amplitude in Eq.~\eqref{Eq:canonical to tensor} can be written as
\begin{equation}
    A_{\sigma_{3}}^{\sigma_{1}\sigma_{2}}(k_{3},p^{*}_{1},p^{*}_{2};L,S)\propto \sum_{\sigma_L\sigma_S}C_{s_3,\sigma_{3}}^{L,\sigma_{L};S,\sigma_{S}}C_{S,\sigma_{S}}^{s_1,\sigma_{1};s_2,\sigma_{2}}Y^L_{\sigma_L}(\Omega).
\end{equation}
This shows that the covariant projection tensor amplitude in the PS-scheme is equivalent to the traditional-$LS$ amplitude in the COM frame.
Consequently, the $LS$ expansion of the coupling structure in Eq.~\eqref{Eq:coupling structure expand} is complete once all allowed $(L,S)$ are included. 
In the PS-scheme, the $LS$ partial wave amplitude therefore realizes an explicit separation between the orbital part and the spin part in the COM definition. 
The same construction is extended below to any frame.

From Eq.~\eqref{Eq:any frame canonical}, the canonical amplitude in any frame can be written as the COM frame canonical amplitude multiplied by the corresponding Wigner rotations of the final-state particles, which can be written as 
\begin{equation}
    A_{\sigma_{3}}^{\sigma_{1}\sigma_{2}}(\mathbf{p}_3,\mathbf{p}_1,\mathbf{p}_2;L,S) =\sum_{\sigma_{1}^{\prime}\sigma_{2}^{\prime}} A_{\sigma_{3}}^{\sigma^{\prime}_{1}\sigma^{\prime}_{2}}(\mathbf{k}_3,\mathbf{p}^*_1,\mathbf{p}^*_2;L,S)D^{\sigma_{1}^{\prime}(s_{1})*}_{\,\,\sigma_{1}}(R_{31})D^{\sigma^{\prime}_{2}(s_{2})*}_{\,\,\sigma_{2}}(R_{32}).
\end{equation}
Therefore, in any frame, the covariant projection amplitude in the PS-scheme transforms with the same induced little group rotations as the traditional-$LS$ amplitude.

Since the little group is a subgroup of the Lorentz group, its action on the wave functions at the standard momentum can be expressed as a Lorentz transformation implementing the little group rotation on the standard-momentum wave functions. Using Eq.~\eqref{Eq:lorenz on wave function}, the spin wave function under the induced little group rotation is written as
\begin{equation}
    \sum_{\sigma^\prime}D^{\sigma^\prime(s)}_{\,\,\sigma}(R)u^{\ell}_{\sigma^\prime}(k;s)
    = D^{\ell}_{\,\,\ell^\prime}(R)
      u^{\ell^\prime}_{\sigma}(k;s),\quad
      \sum_{\sigma^\prime}D^{\sigma(s)}_{\,\,\sigma^\prime}(R^{-1})\bar{u}_{\ell}^{\sigma^\prime}(k;s)
    = D^{\ell^\prime}_{\,\,\ell}(R^{-1})
      \bar{u}_{\ell^\prime}^{\sigma}(k;s),
\end{equation}
where $k$ is the standard momentum of the particle.
Finally, in any frame the covariant projection tensor amplitude in the PS-scheme can be written as
\begin{equation}\label{Eq:any frame covariant LS amplitude}
	\begin{aligned}
		A_{\sigma_{3}}^{\sigma_{1}\sigma_{2}}(p_{3},{p}_{1},p_{2};L,S)=&\Gamma_{\ell_{3}}^{\ell_{1}\ell_{2}}(k_{3},p^{*}_{1},p^{*}_{2};L,S)D^{\ell^\prime_{1}}_{\,\,\ell_{1}}(R_{31}^{-1})D^{\ell^\prime_{2}}_{\,\,\ell_{2}}(R_{32}^{-1})  \\
		&\times u_{\sigma_{3}}^{\ell_{3}}(k_3;s_{3})\bar{u}_{\ell^\prime_{1}}^{\sigma_{1}}(k_1;s_{1})\bar{u}_{\ell^\prime_{2}}^{\sigma_{2}}(k_2;s_{2}),
	\end{aligned}
\end{equation}
where $R_{3i}$ ($i=1,2$) are the Wigner rotations obtained from Eq.~\eqref{Eq:little group rotation1}, and can be written as
\begin{equation}\label{Eq:R3i_def}
R_{3i}=L_{i,c}^{-1}(\mathbf p_i^\ast)L_{3,c}^{-1}(\mathbf p_3)L_{i,c}(\mathbf p_i),
\qquad i=1,2,
\end{equation}
where $L_{i,c}(\mathbf p_i)$ is the canonical standard boost for particle-$i$. The matrix $D^{\ell^\prime}_{\,\,\ell}(R)$ denotes the Lorentz transformation matrix in the Lorentz representation $[\ell]$. Although the covariant projection tensor amplitude in the PS-scheme can be evaluated in any frame, the associated Wigner rotation depends on the particle momenta $\mathbf p_i^\ast$ $(i=1,2)$ in the COM frame. Therefore, the PS-scheme is not fully manifestly covariant at the level of its formal definition.

The covariant projection tensor amplitudes both in the PS-scheme and the GS-scheme are related by the corresponding Lorentz transformations, which can be written as
\begin{equation}
\begin{aligned}
A_{\sigma_{3}}^{\sigma_{1}\sigma_{2}}(p_{3},{p}_{1},p_{2};L,S)=&\Gamma_{\ell_{3}}^{\ell_{1}\ell_{2}}(p_{3},p_{1},p_{2};L,S)u_{\sigma_{3}}^{\ell_{3}}(p_3;s_{3})\bar{u}_{\ell^\prime_{1}}^{\sigma_{1}}(p_1;s_{1})\bar{u}_{\ell^\prime_{2}}^{\sigma_{2}}(p_2;s_{2})  \\
&\times D^{\ell^\prime_{1}}_{\,\,\ell_{1}^{}}\Big(L_{3,c}(\mathbf{p}_3)R_{31}^{-1}L_{3,c}^{-1}(\mathbf{p}_3)\Big)D^{\ell^\prime_{2}}_{\,\,\ell_{2}}\Big(L_{3,c}(\mathbf{p}_3)R_{32}^{-1}L_{3,c}^{-1}(\mathbf{p}_3)\Big)\\
=&\Gamma_{\ell_{3}}^{\ell_{1}\ell_{2}}(p_{3},p_{1},p_{2};L,S)u_{\sigma_{3}}^{\ell_{3}}(p_3;s_{3})\bar{u}_{\ell^\prime_{1}}^{\sigma_{1}}(p_1;s_{1})\bar{u}_{\ell^\prime_{2}}^{\sigma_{2}}(p_2;s_{2})  \\
&\times D^{\ell^\prime_{1}}_{\,\,\ell_{1}^{}}\Big(L_{3,c}(\mathbf{p}_3)L_{1,c}(\mathbf p_1^\ast)L_{3,c}^{-1}(\mathbf{p}_3)\Big)D^{\ell^\prime_{2}}_{\,\,\ell_{2}}\Big(L_{3,c}(\mathbf{p}_3)L_{2,c}(\mathbf p_2^\ast)L_{3,c}^{-1}(\mathbf{p}_3)\Big),
\end{aligned}
\end{equation}
where, in the last equality, the first line coincides with the covariant projection tensor amplitude in the GS-scheme, while the second line collects the Lorentz representation matrices that relate the PS-scheme and GS-scheme definitions. In general these matrices are not the identity matrix, so the two schemes define different $(L,S)$ partial waves and therefore need not coincide.

Although the PS-scheme partial wave amplitude achieves the same explicit separation between the orbital part and the spin part as in the traditional-$LS$ construction in the COM frame, it is not manifestly covariant in the strict sense. From the viewpoint of Lorentz transformations, the "covariance" in the PS-scheme refers to the fact that no explicit little group rotation appears on the little group indices. The little group rotations induced by a Lorentz transformation are instead mapped, through the use of the rest frame wave functions, into rotations acting on the Lorentz indices of the rest frame wave functions. However, this procedure does not remove the need to evaluate the Wigner rotations $R_{3i}$ generated by boosts. As a result, the PS-scheme realizes an explicit orbital-spin separation, but it does not provide a fully manifestly covariant construction.

\subsection{Calculations and comparisons of different $LS$ amplitudes}
\label{subsec:Comparison of calculations in different LS couling schemes}

In this subsubsection, we will calculate the amplitudes for the two-body decay process in the traditional-$LS$ amplitude, the covariant projection tensor amplitude in the PS-scheme, the Zemach tensor amplitude and the covaraint tensor amplitude. In the following calculations, we omit the overall constant factors $g_{LS}$ in front of the corresponding $LS$ partial waves, and focus instead on the values of $\sigma_{3},\sigma_{1}\ \text{and}\ \sigma_{2}$ associated with each amplitude.

\paragraph{Different $LS$ amplitude in COM frame}
In the following discussion of the covariant projection tensor amplitude in the PS-scheme, the Lorentz representations are illustrated using Lorentz four-vector indices denoted by $\mu,\nu,\rho\cdots$. 
In the covariant tensor method, the resulting amplitude is equivalent to the covariant projection tensor amplitude in the GS-scheme. Its difference from the PS-scheme originates from the choice of external spin wave functions in the COM frame. Therefore, in the explicit examples of the covariant tensor method, the coupling structure can be taken directly from the PS-scheme construction.

For the two-body decay process $3 \to 1 + 2$ with spins $s_3=1,s_1=1,s_2=0$. The momenta of the particles are $\{k_3,p^*_1,p^*_2\}$, and we define $r=p^*_1-p^*_2$ .
In this process, there are three possible $LS$ partial waves, namely $(L,S) = (0,1)$, $(1,1)$, and $(2,1)$.
\begin{enumerate}
\item For the $(0,1)$ partial wave, the amplitude can be written in the following three schemes:
\begin{enumerate}
\item \textbf{Traditional-$LS$ amplitude}:
  \begin{equation}
	\begin{aligned}
		A_{\sigma_{3}}^{\sigma_{1}}(\mathbf{k}_3,\mathbf{p}^*_1,\mathbf{p}^*_2;0,1)=C_{1,\sigma_{3}}^{0,\sigma_{L};1,\sigma_{S}}C_{1,\sigma_{S}}^{1,\sigma_{1};0,\sigma_{2}} Y^{0}_{\sigma_{L}}(\Omega)=\delta^{\sigma_1}_{\sigma_3}.
	\end{aligned}
\end{equation}
  \item \textbf{The covariant projection tensor amplitude in the PS-scheme}: 
  \begin{equation}
  \begin{aligned}
    A_{\sigma_{3}}^{\sigma_{1}}(k_{3},p^{*}_{1},p^{*}_{2};0,1)=&\Gamma_{\mu_{3}}^{\mu_{1}}(k_{3},p^{*}_{1},p^{*}_{2};0,1)u_{\sigma_{3}}^{\mu_{3}}(k_3;1)\bar u_{\mu_{1}}^{\sigma_{1}}(k_1;1)\\
    =&P_{\mu_{3}}^{\mu_{S}}(k_{3};1,0,1)P_{\mu_{S}}^{\mu_{1}}(k_{3};1,1,0)u_{\sigma_{3}}^{\mu_{3}}(k_3;1)\bar u_{\mu_{1}}^{\sigma_{1}}(k_1;1),\\
  \end{aligned}
  \end{equation}
  where 
\begin{equation}\label{Eq:example Pmu1s}
  \begin{aligned}
    P_{\mu_{S}}^{\mu_{1}}(k_{3};1,1,0)=&\sum_{\sigma_1\sigma_S}C_{1,\sigma_{S}}^{\sigma_{1};0}\bar u^{\sigma_S}_{\mu_S}(k_3;1)u_{\sigma_1}^{\mu_1}(k_3;1)\\
    =&\sum_{\sigma_1}\epsilon_{\sigma_1}^{\mu_1}(k_3)\epsilon^{*\sigma_1}_{\mu_S}(k_3)\\
    =&-g^{\mu_1}_{\,\,\mu_S}+\frac{k^{\mu_1}_{3}k_{3\mu_S}}{m_3^2}\\
    =&\tilde g^{\mu_1}_{\,\,\mu_S}(k_3),
  \end{aligned}
  \end{equation}
  and
  \begin{equation}
  \begin{aligned}
    P_{\mu_{3}}^{\mu_{S}}(k_{3};1,0,1)=&\sum_{\sigma_S\sigma_3}C_{1,\sigma_{3}}^{0;1,\sigma_{S}}\bar u^{\sigma_3}_{\mu_3}(k_3;1)u_{\sigma_S}^{\mu_S}(k_3;1),\\
    =&\sum_{\sigma_S}\epsilon_{\sigma_S}^{\mu_S}(k_3)\epsilon^{*\sigma_S}_{\mu_3}(k_3)\\
    =&-g^{\mu_S}_{\,\,\mu_3}+\frac{k^{\mu_S}_{3}k_{3\mu_3}}{m_3^2}\\
    =&\tilde g^{\mu_S}_{\,\,\mu_3}(k_3).
  \end{aligned}
  \end{equation}
  The amplitude is 
    \begin{equation}
  \begin{aligned}
    A_{\sigma_{3}}^{\sigma_{1}}(k_{3},p^{*}_{1},p^{*}_{2};0,1)=&P_{\mu_{3}}^{\mu_{S}}(k_{3};1,0,1)P_{\mu_{S}}^{\mu_{1}}(k_{3};1,1,0)u_{3,\sigma_{3}}^{\mu_{3}}(k_3;1)\bar u_{1,\mu_{1}}^{\sigma_{1}}(k_1;1),\\
    =&\tilde g^{\mu_1}_{\,\,\mu_3}(k_3)\epsilon_{\sigma_{3}}^{\mu_{3}}(k_3)\epsilon_{\mu_{1}}^{*\sigma_{1}}(k_1)\\
    =&\delta^{\sigma_1}_{\sigma_3}.
  \end{aligned}
  \end{equation}

  \item \textbf{The Zemach tensor amplitude}:
  \begin{equation}
  \begin{aligned}
    A^{\sigma_1}_{\sigma_3}(\mathbf{k}_3,\mathbf{p}^*_1,\mathbf{p}^*_2;0,1)
    =&\sum_{i}(\epsilon_{\sigma_3})_{i}(\epsilon^{*\sigma_1})_i\\
    =&\delta^{\sigma_1}_{\sigma_3}.
  \end{aligned}
\end{equation}
  \item \textbf{The covariant tensor amplitude}:
  \begin{equation}
  \begin{aligned}
    \mathcal{C}^{\sigma_1}_{\sigma_3}(k_3,p^*_1,p^*_2;0,1)
    =&\tilde g^{\mu_1}_{\,\,\mu_3}(k_3)\epsilon_{\sigma_{3}}^{\mu_{3}}(k_3)\epsilon_{\mu_{1}}^{*\sigma_{1}}(p^*_1)\\
    \neq&\delta^{\sigma_1}_{\sigma_3}.
  \end{aligned}
\end{equation}

\end{enumerate}

\item For the $(1,1)$ partial wave, the amplitude can be written in the following four schemes:
\begin{enumerate}
\item \textbf{Traditional-$LS$ amplitude}: 
\begin{equation}
\begin{aligned}
    A_{\sigma_{3}}^{\sigma_{1}}(\mathbf{k}_3,\mathbf{p}^*_1,\mathbf{p}^*_2;1,1)=&C_{1,\sigma_{3}}^{1,\sigma_{L};1,\sigma_{S}}C_{1,\sigma_{S}}^{1,\sigma_{1};0,\sigma_{2}} Y^{1}_{\sigma_{L}}(\Omega)\\
    =&\frac{1}{2}\sqrt{\frac{3}{2\pi}}
    \begin{pmatrix}
    -\cos{\theta} &-\frac{1}{\sqrt{2}}\sin{\theta}e^{i\phi} &0\\
    -\frac{1}{\sqrt{2}}\sin{\theta}e^{-i\phi} & 0 &-\frac{1}{\sqrt{2}}\sin{\theta}e^{i\phi}\\
    0&-\frac{1}{\sqrt{2}}\sin{\theta}e^{-i\phi} &\cos{\theta}
    \end{pmatrix}_{\sigma_3}^{\,\,\sigma_1}.
\end{aligned}
\end{equation}
  \item \textbf{The covariant projection tensor amplitude in the PS-scheme}: 
  \begin{equation}
  \begin{aligned}
    A_{\sigma_{3}}^{\sigma_{1}}(k_{3},p^{*}_{1},p^{*}_{2};1,1)=&\Gamma_{\mu_{3}}^{\mu_{1}}(k_{3},p^{*}_{1},p^{*}_{2};1,1)
    u_{\sigma_{3}}^{\mu_{3}}(k_3;1)\bar{u}_{\mu_{1}}^{\sigma_{1}}(k_1;1)\\
    =&P_{\mu_{3}}^{\mu_{L}\mu_{S}}(k_{3};1,1,1)P_{\mu_{S}}^{\mu_{1}}(k_{3};1,1,0)t_{\mu_{L}}^{1}(k_{3},r)u_{\sigma_{3}}^{\mu_{3}}({k}_3;1)\bar u_{\mu_{1}}^{\sigma_{1}}(k_1;1),
  \end{aligned}
  \end{equation}
  where $P_{\mu_{S}}^{\mu_{1}}(\mathbf{k}_{3};1,1,0)$ can be obtained from Eq.~\eqref{Eq:example Pmu1s},
  \begin{equation}
  \begin{aligned}
    P_{\mu_{3}}^{\mu_{L}\mu_{S}}(k_{3};1,1,1)=&\sum_{\sigma_L\sigma_S\sigma_3}C^{1,\sigma_L;1,\sigma_S}_{1,\sigma_3}\bar u^{\sigma_3}_{\mu_3}(k_3;1) u_{\sigma_L}^{\mu_L}(k_3;1) u^{\mu_S}_{\sigma_S}(k_3;1)\\
    =&\sum_{\sigma_L\sigma_S\sigma_3}C^{1,\sigma_L;1,\sigma_S}_{1,\sigma_3}\epsilon^{*\sigma_3}_{\mu_3}(k_3)\epsilon_{\sigma_L}^{\mu_L}(k_3)\epsilon^{\mu_S}_{\sigma_S}(k_3)\\
    =&\frac{i}{\sqrt{2}m_3}g_{\mu_3\mu_3^\prime}\varepsilon^{\mu_3^\prime\mu_L\mu_S\rho}k_{3\rho}\\
    =&\frac{i}{\sqrt{2}m_3}g_{\mu_3\mu_3^\prime}\tilde\varepsilon^{\mu_3^\prime\mu_L\mu_S}(k_{3}),
  \end{aligned}
  \end{equation}
  and
  \begin{equation}\label{Eq:t1mul}
  \begin{aligned}
    t_{\mu_{L}}^{1}(k_{3},r)=&P^\mu_{\mu_L}(k_3;1,1,0) r_\mu\\
    =&\tilde g^{\mu}_{\,\,\mu_L}(k_3) r_\mu\\
    =&\tilde r_{\mu_L}(k_3).
  \end{aligned}
  \end{equation}
  The amplitude is
\begin{equation}
  \begin{aligned}
    A_{\sigma_{3}}^{\sigma_{1}}(k_{3},p^{*}_{1},p^{*}_{2};1,1)=&P_{\mu_{3}}^{\mu_{L}\mu_{S}}(k_{3};1,1,1)P_{\mu_{S}}^{\mu_{1}}(k_{3};1,1,0)t_{\mu_{L}}^{1}({k}_{3},{r})u_{\sigma_{3}}^{\mu_{3}}({k}_3;1)\bar u_{\mu_{1}}^{\sigma_{1}}({k}_1;1),\\
    =&\frac{i}{\sqrt{2}m_3}g_{\mu_3\mu_3^\prime}\tilde\varepsilon^{\mu_3^\prime\mu_L\mu_1}(k_{3})\tilde r_{\mu_L}(k_3)\epsilon_{\sigma_{3}}^{\mu_{3}}(k_3)\epsilon_{\mu_{1}}^{*\sigma_{1}}(k_1)\\
    \propto&
    \begin{pmatrix}
    -\cos{\theta} &-\frac{1}{\sqrt{2}}\sin{\theta}e^{i\phi} &0\\
    -\frac{1}{\sqrt{2}}\sin{\theta}e^{-i\phi} & 0 &-\frac{1}{\sqrt{2}}\sin{\theta}e^{i\phi}\\
    0&-\frac{1}{\sqrt{2}}\sin{\theta}e^{-i\phi} &\cos{\theta}
    \end{pmatrix}_{\sigma_3}^{\,\,\sigma_1}.
  \end{aligned}
  \end{equation}

  \item \textbf{The Zemach tensor amplitude}:   
  \begin{equation}
  \begin{aligned}
          A^{\sigma_1}_{\sigma_3}(\mathbf{k}_3,\mathbf{p}^*_1,\mathbf{p}^*_2;1,1)
          =&\sum_{ijk}(\epsilon_{\sigma_3})_{k}\varepsilon_{ijk}(\epsilon^{*\sigma_1})_i \mathbf{r}_j\\
          \propto&
    \begin{pmatrix}
    -\cos{\theta} &-\frac{1}{\sqrt{2}}\sin{\theta}e^{i\phi} &0\\
    -\frac{1}{\sqrt{2}}\sin{\theta}e^{-i\phi} & 0 &-\frac{1}{\sqrt{2}}\sin{\theta}e^{i\phi}\\
    0&-\frac{1}{\sqrt{2}}\sin{\theta}e^{-i\phi} &\cos{\theta}
    \end{pmatrix}_{\sigma_3}^{\,\,\sigma_1}.
  \end{aligned}
\end{equation}

  \item \textbf{The covariant tensor amplitude}:
  \begin{equation}
  \begin{aligned}
\mathcal{C}^{\sigma_1}_{\sigma_3}(k_3,p^*_1,p^*_2;1,1)
  =&\frac{i}{\sqrt{2}m_3}g_{\mu_3\mu_3^\prime}\tilde\varepsilon^{\mu_3^\prime\mu_L\mu_1}(k_{3})\tilde r_{\mu_L}(k_3)\epsilon_{\sigma_{3}}^{\mu_{3}}(k_3)\epsilon_{\mu_{1}}^{*\sigma_{1}}(p^*_1)\\
  \not\propto&    
  \begin{pmatrix}
    -\cos{\theta} &-\frac{1}{\sqrt{2}}\sin{\theta}e^{i\phi} &0\\
    -\frac{1}{\sqrt{2}}\sin{\theta}e^{-i\phi} & 0 &-\frac{1}{\sqrt{2}}\sin{\theta}e^{i\phi}\\
    0&-\frac{1}{\sqrt{2}}\sin{\theta}e^{-i\phi} &\cos{\theta}
    \end{pmatrix}_{\sigma_3}^{\,\,\sigma_1}.
  \end{aligned}
\end{equation}

\end{enumerate}

\item For the $(2,1)$ partial wave, the amplitude can be written in the following four schemes:
\begin{enumerate}
\item \textbf{Traditional-$LS$ amplitude}: 
  \begin{equation}
	\begin{aligned}
		A_{\sigma_{3}}^{\sigma_{1}}(\mathbf k_{3},\mathbf p^{*}_{1},\mathbf p^{*}_{2};2,1)=&C_{1,\sigma_{3}}^{2,\sigma_{L};1,\sigma_{S}}C_{1,\sigma_{S}}^{1,\sigma_{1};0,\sigma_{2}} Y^{2}_{\sigma_{L}}(\Omega)\\
        =&    
    \begin{pmatrix}
        \frac{1}{2}(\cos^2{\theta}-\frac{1}{3})&\frac{1}{\sqrt{2}}\cos{\theta}\sin{\theta}e^{i\phi}&\frac{1}{2}\sin^2\theta e^{2i\phi}\\
        -\frac{1}{\sqrt{2}}\cos{\theta}\sin{\theta}e^{-i\phi} & -(\cos^2{\theta}-\frac{1}{3}) & -\frac{1}{\sqrt{2}}\cos{\theta}\sin{\theta}e^{i\phi}\\
        \frac{1}{2}\sin^2\theta e^{-2i\phi} & -\frac{1}{\sqrt{2}}\cos{\theta}\sin{\theta}e^{-i\phi} &\frac{1}{2}(\cos^2{\theta}-\frac{1}{3})
    \end{pmatrix}_{\sigma_3}^{\,\,\sigma_1}.
	\end{aligned}
\end{equation}
  \item \textbf{The covariant projection tensor amplitude in the PS-scheme}: 
  \begin{equation}
  \begin{aligned}
    A_{\sigma_{3}}^{\sigma_{1}}(k_{3},{p}^{*}_{1},{p}^{*}_{2};2,1)=&\Gamma_{\mu_{3}}^{\mu_{1}}({k}_{3},{p}^{*}_{1},{p}^{*}_{2};2,1)
    u_{\sigma_{3}}^{\mu_{3}}({k}_3;1)\bar{u}_{\mu_{1}}^{\sigma_{1}}({k}_1;1)\\
    =&P_{\mu_{3}}^{\mu_{L}\nu_L,\mu_{S}}({k}_{3};1,2,1)P_{\mu_{S}}^{\mu_{1}}({k}_{3};1,1,0)t_{\mu_{L}\nu_L}^{2}({k}_{3},{r})u_{\sigma_{3}}^{\mu_{3}}({k}_3;1)\bar{u}_{\mu_{1}}^{\sigma_{1}}({k}_1;1),
  \end{aligned}
  \end{equation}
  where $P_{\mu_{S}}^{\mu_{1}}(\mathbf{k}_{3};1,1,0)$ can be obtained from Eq.~\eqref{Eq:example Pmu1s},
  \begin{equation}
  \begin{aligned}
            P_{\mu_{3}}^{\mu_{L}\nu_L,\mu_{S}}({k}_{3};1,2,1)=&\sum_{\sigma_L\sigma_S\sigma_3}C^{2,\sigma_L;1,\sigma_S}_{1,\sigma_3}\bar u^{\sigma_3}_{\mu_3}({k}_3;1)u^{\mu_L\nu_L}_{\sigma_L}({k}_3;2)u^{\mu_S}_{\sigma_S}({k}_3;1)\\
            =&\sum_{\sigma_L\sigma_S\sigma_3}C^{2,\sigma_L;1,\sigma_S}_{1,\sigma_3} \epsilon^{*\sigma_3}_{\mu_3}(k_3)\epsilon^{\mu_L\nu_L}_{\sigma_L}(k_3)\epsilon^{\mu_S}_{\sigma_S}(k_3)\\
            =&\frac{1}{2}\left( \tilde{g}^{\mu_L}_{\,\,\mu_3}(k_3) \tilde{g}^{\nu_L \mu_S}(k_3) + \tilde{g}^{\nu_L}_{\,\,\mu_3}(k_3) \tilde{g}^{\mu_L \mu_S}(k_3) \right) - \frac{1}{3} \tilde{g}^{\mu_L \nu_L}(k_3) \tilde{g}^{\mu_S}_{\,\,\mu_3}(k_3),
  \end{aligned}
  \end{equation}
  and
  \begin{equation}
  \begin{aligned}
    t_{\mu_L\nu_L}^{2}({k}_{3},{r})=&P^{\mu_1\mu_2}_{\mu_L\nu_L}({k}_3)r_{\mu_1}r_{\mu_2}\\
    =&\tilde r_{\mu_L}(k_3)\tilde r_{\nu_L}(k_3)-\frac13(\tilde r\cdot\tilde r)\tilde g_{\mu_L\nu_L}(k_3).
  \end{aligned}
  \end{equation}
  The amplitude is
\begin{equation}
  \begin{aligned}
    A_{\sigma_{3}}^{\sigma_{1}}({k}_{3},{p}^{*}_{1},{p}^{*}_{2};2,1)=&P_{\mu_{3}}^{\mu_{L}\nu_L,\mu_{S}}({k}_{3};1,2,1)P_{\mu_{S}}^{\mu_{1}}({k}_{3};1,1,0)t_{\mu_{L}\nu_L}^{2}({k}_{3},{r})u_{\sigma_{3}}^{\mu_{3}}({k}_3;1)\bar{u}_{\mu_{1}}^{\sigma_{1}}({k}_1;1)\\
    =& \epsilon_{\sigma_{3}}^{\mu_{3}}(k_3)\epsilon_{\mu_{1}}^{*\sigma_{1}}(k_1)\Big(\tilde r^{\mu_1}(k_3)\tilde r_{\mu_3}(k_3)-\frac13(\tilde r\cdot\tilde r)\tilde g_{\,\,\mu_3}^{\mu_1}(k_3)\Big)\\
    \propto&\begin{pmatrix}
        \frac{1}{2}(\cos^2{\theta}-\frac{1}{3})&\frac{1}{\sqrt{2}}\cos{\theta}\sin{\theta}e^{i\phi}&\frac{1}{2}\sin^2\theta e^{2i\phi}\\
        -\frac{1}{\sqrt{2}}\cos{\theta}\sin{\theta}e^{-i\phi} & -(\cos^2{\theta}-\frac{1}{3}) & -\frac{1}{\sqrt{2}}\cos{\theta}\sin{\theta}e^{i\phi}\\
        \frac{1}{2}\sin^2\theta e^{-2i\phi} & -\frac{1}{\sqrt{2}}\cos{\theta}\sin{\theta}e^{-i\phi} &\frac{1}{2}(\cos^2{\theta}-\frac{1}{3})
    \end{pmatrix}_{\sigma_3}^{\,\,\sigma_1}.
  \end{aligned}
  \end{equation}

  \item \textbf{The Zemach tensor amplitude}: 
  \begin{equation}
  \begin{aligned}
    A^{\sigma_1}_{\sigma_3}(\mathbf{k}_3,\mathbf{p}^*_1,\mathbf{p}^*_2;2,1)
    =&\sum_{ij}(\epsilon_{\sigma_3})_{j}(\epsilon^{*\sigma_1})_i(\mathbf{r}_{i}\mathbf{r}_j-\frac{1}{3}\mathbf{r}^2\delta_{ij})\\
    \propto&\begin{pmatrix}
        \frac{1}{2}(\cos^2{\theta}-\frac{1}{3})&\frac{1}{\sqrt{2}}\cos{\theta}\sin{\theta}e^{i\phi}&\frac{1}{2}\sin^2\theta e^{2i\phi}\\
        -\frac{1}{\sqrt{2}}\cos{\theta}\sin{\theta}e^{-i\phi} & -(\cos^2{\theta}-\frac{1}{3}) & -\frac{1}{\sqrt{2}}\cos{\theta}\sin{\theta}e^{i\phi}\\
        \frac{1}{2}\sin^2\theta e^{-2i\phi} & -\frac{1}{\sqrt{2}}\cos{\theta}\sin{\theta}e^{-i\phi} &\frac{1}{2}(\cos^2{\theta}-\frac{1}{3})
    \end{pmatrix}_{\sigma_3}^{\,\,\sigma_1}.
  \end{aligned}
\end{equation}

  \item \textbf{The covariant tensor amplitude}: 
  \begin{equation}
  \begin{aligned}
    \mathcal{C}^{\sigma_1}_{\sigma_3}(k_3,p^*_1,p^*_2;2,1)
    =& \epsilon_{\sigma_{3}}^{\mu_{3}}(k_3)\epsilon_{\mu_{1}}^{*\sigma_{1}}(p^*_1)\Big(\tilde r^{\mu_1}(k_3)\tilde r_{\mu_3}(k_3)-\frac13(\tilde r\cdot\tilde r)\tilde g_{\,\,\mu_3}^{\mu_1}(k_3)\Big)\\
    \not\propto&\begin{pmatrix}
        \frac{1}{2}(\cos^2{\theta}-\frac{1}{3})&\frac{1}{\sqrt{2}}\cos{\theta}\sin{\theta}e^{i\phi}&\frac{1}{2}\sin^2\theta e^{2i\phi}\\
        -\frac{1}{\sqrt{2}}\cos{\theta}\sin{\theta}e^{-i\phi} & -(\cos^2{\theta}-\frac{1}{3}) & -\frac{1}{\sqrt{2}}\cos{\theta}\sin{\theta}e^{i\phi}\\
        \frac{1}{2}\sin^2\theta e^{-2i\phi} & -\frac{1}{\sqrt{2}}\cos{\theta}\sin{\theta}e^{-i\phi} &\frac{1}{2}(\cos^2{\theta}-\frac{1}{3})
    \end{pmatrix}_{\sigma_3}^{\,\,\sigma_1}.
  \end{aligned}
\end{equation}

\end{enumerate}
\end{enumerate}

For the two-body decay process $3 \to 1 + 2$ with spins $s_3=\frac{1}{2},s_1=\frac{1}{2},s_2=0$. 
In this process, there are two possible $LS$ partial waves, namely $(L,S) = (0,\frac{1}{2})$, $(1,\frac{1}{2})$.
\begin{enumerate}
\item For the $(0,\frac{1}{2})$ partial wave, the amplitude can be written in the following four schemes:
\begin{enumerate}
\item \textbf{Traditional-$LS$ amplitude}: 
  \begin{equation}
	\begin{aligned}
		A_{\sigma_{3}}^{\sigma_{1}}(\mathbf{k}_3,\mathbf{p}^*_1,\mathbf{p}^*_2;0,\frac{1}{2})=C_{\frac{1}{2},\sigma_{3}}^{0,\sigma_{L};\frac{1}{2},\sigma_{S}}C_{\frac{1}{2},\sigma_{S}}^{\frac{1}{2},\sigma_{1};0,\sigma_{2}} Y^{0}_{\sigma_{L}}(\Omega)=\delta^{\sigma_1}_{\sigma_3}.
	\end{aligned}
\end{equation}
  \item \textbf{The covariant projection tensor amplitude in the PS-scheme}: 
  \begin{equation}
  \begin{aligned}
    A_{\sigma_{3}}^{\sigma_{1}}({k}_{3},{p}^{*}_{1},{p}^{*}_{2};0,\frac{1}{2})=&\Gamma_{\alpha_{3}}^{\alpha_{1}}({k}_{3},{p}^{*}_{1},{p}^{*}_{2};0,\frac{1}{2})u_{\sigma_{3}}^{\alpha_{3}}({k}_3;\frac{1}{2})\bar{u}_{\alpha_{1}}^{\sigma_{1}}({k}_1;\frac{1}{2})\\
    =&P_{\alpha_{3}}^{\alpha_{S}}({k}_{3};\frac{1}{2},0,\frac{1}{2})P_{\alpha_{S}}^{\alpha_{1}}({k}_{3};\frac{1}{2},\frac{1}{2},0)u_{\sigma_{3}}^{\alpha_{3}}({k}_3;\frac{1}{2})\bar{u}_{\alpha_{1}}^{\sigma_{1}}({k}_1;\frac{1}{2}),
    \end{aligned}
  \end{equation}
  where
  \begin{equation}
  \begin{aligned}
      P_{\alpha_{3}}^{\alpha_{S}}({k}_{3};\frac{1}{2},0,\frac{1}{2})=&\sum_{\sigma_3\sigma_S}C^{0;\frac{1}{2},\sigma_S}_{\frac{1}{2},\sigma_3}\bar u^{\sigma_3}_{\alpha_3}({k}_3;\frac{1}{2})u^{\alpha_S}_{\sigma_S}({k}_3;\frac{1}{2})\\
      =&\delta^{\alpha_S}_{\alpha_3},
  \end{aligned}
  \end{equation}
  and
  \begin{equation}\label{Eq:1salphasec3}
      \begin{aligned}
          P_{\alpha_{S}}^{\alpha_{1}}({k}_{3};\frac{1}{2},\frac{1}{2},0)=&\sum_{\sigma_S\sigma_1}C^{\frac{1}{2},\sigma_1;0}_{\frac{1}{2},\sigma_S}\bar u^{\sigma_S}_{\alpha_S}({k}_3;\frac{1}{2})u^{\alpha_1}_{\sigma_1}({k}_3;\frac{1}{2})\\
      =&\delta^{\alpha_1}_{\alpha_S}.
      \end{aligned}
  \end{equation}
  The amplitude is
\begin{equation}
  \begin{aligned}
    A_{\sigma_{3}}^{\sigma_{1}}({k}_{3},{p}^{*}_{1},{p}^{*}_{2};0,\frac{1}{2})=&P_{\alpha_{3}}^{\alpha_{S}}({k}_{3};\frac{1}{2},0,\frac{1}{2})P_{\alpha_{S}}^{\alpha_{1}}({k}_{3};\frac{1}{2},\frac{1}{2},0)u_{\sigma_{3}}^{\alpha_{3}}({k}_3;\frac{1}{2})\bar{u}_{\alpha_{1}}^{\sigma_{1}}({k}_1;\frac{1}{2})\\
    =&\delta^{\alpha_S}_{\alpha_3}\delta^{\alpha_1}_{\alpha_S}u_{\sigma_{3}}^{\alpha_{3}}({k}_3;\frac{1}{2})\bar{u}_{\alpha_{1}}^{\sigma_{1}}({k}_1;\frac{1}{2})\\
    \propto&\delta^{\sigma_1}_{\sigma_3}.
  \end{aligned}
  \end{equation}

  \item \textbf{The Zemach tensor amplitude}:
  \begin{equation}
  \begin{aligned}
    A^{\sigma_1}_{\sigma_3}(\mathbf{k}_3,\mathbf{p}^*_1,\mathbf{p}^*_2;0,\frac{1}{2})
    =&(\chi_{\sigma_3})^{\alpha}(\chi^{*\sigma_1})_\alpha\\
    =&\delta^{\sigma_1}_{\sigma_3}.
  \end{aligned}
\end{equation}
  \item \textbf{The covariant tensor amplitude}: 
  \begin{equation}
  \begin{aligned}
    \mathcal{C}^{\sigma_1}_{\sigma_3}(k_3,p^*_1,p^*_2;0,\frac{1}{2})
    =&\delta^{\alpha_S}_{\alpha_3}\delta^{\alpha_1}_{\alpha_S}u_{\sigma_{3}}^{\alpha_{3}}({k}_3;\frac{1}{2})\bar{u}_{\alpha_{1}}^{\sigma_{1}}({p}^*_1;\frac{1}{2})\\
    \neq&\delta^{\sigma_1}_{\sigma_3}.
  \end{aligned}
\end{equation}
\end{enumerate}

\item For the $(1,\frac{1}{2})$ partial wave, the amplitude can be written in the following four schemes:
\begin{enumerate}
\item \textbf{Traditional-$LS$ amplitude}: 
\begin{equation}
\begin{aligned}
    A_{\sigma_{3}}^{\sigma_{1}}(\mathbf{k}_3,\mathbf{p}^*_1,\mathbf{p}^*_2;1,\frac{1}{2})=&C_{\frac{1}{2},\sigma_{3}}^{1,\sigma_{L};\frac{1}{2},\sigma_{S}}C_{\frac{1}{2},\sigma_{S}}^{\frac{1}{2},\sigma_{1};0,\sigma_{2}} Y^{1}_{\sigma_{L}}(\Omega)\\
    =&\frac{1}{2}\sqrt{\frac{1}{\pi}}
    \begin{pmatrix}
    -\cos{\theta} & -\sin{\theta}e^{i\phi} \\
    -\sin{\theta}e^{-i\phi} & \cos{\theta}     \end{pmatrix}_{\sigma_3}^{\,\,\sigma_1}.
\end{aligned}
\end{equation}
  \item \textbf{The covariant projection tensor amplitude in the PS-scheme}: 
  \begin{equation}
  \begin{aligned}
    A_{\sigma_{3}}^{\sigma_{1}}({k}_{3},{p}^{*}_{1},{p}^{*}_{2};1,\frac{1}{2})=&\Gamma_{\alpha_{3}}^{\alpha_{1}}({k}_{3},{p}^{*}_{1},{p}^{*}_{2};1,\frac{1}{2})u_{\sigma_{3}}^{\alpha_{3}}({k}_3;\frac{1}{2})\bar{u}_{\alpha_{1}}^{\sigma_{1}}({k}_1;\frac{1}{2})\\
    =&P_{\alpha_{3}}^{\mu_{L}\alpha_{S}}({k}_{3};\frac{1}{2},1,\frac{1}{2})P_{\alpha_{S}}^{\alpha_{1}}({k}_{3};\frac{1}{2},\frac{1}{2},0)t_{\mu_{L}}^{1}({k}_{3},{r})u_{\sigma_{3}}^{\alpha_{3}}({k}_3;\frac{1}{2})\bar{u}_{\alpha_{1}}^{\sigma_{1}}({k}_1;\frac{1}{2}),
  \end{aligned}
  \end{equation}
  where $P_{\alpha_{S}}^{\alpha_{1}}(\mathbf{k}_{3};\frac{1}{2},\frac{1}{2},0)$ and $t_{\mu_{L}}^{1}(\mathbf{k}_{3},\mathbf{r})$ can be obtained from Eqs.~\eqref{Eq:1salphasec3} and \eqref{Eq:t1mul}, 
  \begin{equation}
  \begin{aligned}
      P_{\alpha_{3}}^{\mu_{L}\alpha_{S}}({k}_{3};\frac{1}{2},1,\frac{1}{2})=&\sum_{\sigma_3\sigma_L\sigma_S}C^{1,\sigma_L;\frac{1}{2},\sigma_S}_{\frac{1}{2},\sigma_3}\bar u^{\sigma_3}_{\alpha_3}({k}_3;\frac{1}{2})u^{\mu_L}_{\sigma_L}({k}_3;1)u^{\alpha_S}_{\sigma_S}({k}_3;\frac{1}{2})\\
      =&(\gamma_5\tilde g^{\mu_L}_{\,\,\mu_L^\prime}(k_3)\gamma^{\mu_L^\prime})_{\,\,\alpha_3}^{\alpha_S}.
  \end{aligned}
  \end{equation}
  The amplitude is
  \begin{equation}
  \begin{aligned}
    A_{\sigma_{3}}^{\sigma_{1}}({k}_{3},{p}^{*}_{1},{p}^{*}_{2};1,\frac{1}{2})=&P_{\alpha_{3}}^{\mu_{L}\alpha_{S}}({k}_{3};\frac{1}{2},1,\frac{1}{2})P_{\alpha_{S}}^{\alpha_{1}}({k}_{3};\frac{1}{2},\frac{1}{2},0)t_{\mu_{L}}^{1}({k}_{3},{r})u_{\sigma_{3}}^{\alpha_{3}}({k}_3;\frac{1}{2})\bar{u}_{\alpha_{1}}^{\sigma_{1}}({k}_1;\frac{1}{2})\\
    =&\bar{u}_{\alpha_{1}}^{\sigma_{1}}({k}_1;\frac{1}{2})(\gamma_5\tilde g^{\mu_L}_{\,\,\mu_L^\prime}(k_3)\gamma^{\mu_L^\prime})_{\,\,\alpha_3}^{\alpha_1}u_{\sigma_{3}}^{\alpha_{3}}({k}_3;\frac{1}{2})\tilde r_{\mu_L}(k_3)\\
    \propto&
    \begin{pmatrix}
    -\cos{\theta} & -\sin{\theta}e^{i\phi} \\
    -\sin{\theta}e^{-i\phi} & \cos{\theta}     \end{pmatrix}_{\sigma_3}^{\,\,\sigma_1}.
  \end{aligned}
  \end{equation}

  \item \textbf{The Zemach tensor amplitude}:   
  \begin{equation}
  \begin{aligned}
    A^{\sigma_1}_{\sigma_3}(\mathbf{k}_3,\mathbf{p}^*_1,\mathbf{p}^*_2;1,\frac{1}{2})
    =&\sum_{i} (\chi^{*\sigma_1})_\alpha(\sigma_i)_{\,\,\beta}^{\alpha}(\chi_{\sigma_3})^{\beta}\mathbf{r}_i\\
        \propto&
    \begin{pmatrix}
    -\cos{\theta} & -\sin{\theta}e^{i\phi} \\
    -\sin{\theta}e^{-i\phi} & \cos{\theta}     \end{pmatrix}_{\sigma_3}^{\,\,\sigma_1}.
  \end{aligned}
\end{equation}
  \item \textbf{The covariant tensor amplitude}: 
  \begin{equation}
  \begin{aligned}
    \mathcal{C}^{\sigma_1}_{\sigma_3}(k_3,p^*_1,p^*_2;1,\frac{1}{2})
    =& \bar{u}_{\alpha_{1}}^{\sigma_{1}}({p}^*_1;\frac{1}{2})(\gamma_5\tilde g^{\mu_L}_{\,\,\mu_L^\prime}(k_3)\gamma^{\mu_L^\prime})_{\,\,\alpha_3}^{\alpha_1}u_{\sigma_{3}}^{\alpha_{3}}({k}_3;\frac{1}{2})\tilde r_{\mu_L}(k_3)\\
    \not\propto&
    \begin{pmatrix}
    -\cos{\theta} & -\sin{\theta}e^{i\phi} \\
    -\sin{\theta}e^{-i\phi} & \cos{\theta}     \end{pmatrix}_{\sigma_3}^{\,\,\sigma_1}.
  \end{aligned}
\end{equation}
\end{enumerate}
\end{enumerate}

Through the above comparison, we note that for the traditional-$LS$ amplitudes, the covariant projection tensor amplitude in the PS-scheme, and the Zemach tensor amplitudes, they are completely equivalent in the COM frame and use the same $LS$ component. 

By contrast, the covariant tensor amplitudes are not equivalent to these above amplitudes in the COM frame.
Only in the non-relativistic limit, the covariant tensor amplitudes are equivalent to the other three amplitudes.


\paragraph{Different $LS$ amplitude in any frame}
Since the traditional-$LS$ amplitudes, the covariant projection tensor amplitudes in the PS-scheme, and the Zemach tensor amplitudes have been shown to be equivalent in the COM frame, boosting them to any frame induces the same little group rotations, which can be written as 
\begin{equation}
    R_{31} = L^{-1}_{1,c}(\mathbf{p}_{1}^{*}) L^{-1}_{3,c}(\mathbf{p}_{3}) L_{1,c}(\mathbf{p}_{1}),
\end{equation}
where only the rotation for particle-$1$ is written explicitly, since in the examples below particle-$1$ is the only final-state particle that carries spin.
In the following, the covariant projection tensor amplitudes in the PS-scheme is therefore used as the representative scheme for comparisons with the covariant projection tensor amplitudes in the GS-scheme in any frame.

For the two-body decay process $3 \to 1 + 2$ with spins $s_3=1,s_1=1,s_2=0$. The momenta of the particles are $\{p_3,p_1,p_2\}$.
In this process, there are three possible $LS$ partial waves, namely $(L,S) = (0,1)$, $(1,1)$, and $(2,1)$.
\begin{enumerate}
\item For the $(0,1)$ partial wave, the amplitude can be written in the following three schemes:
\begin{enumerate}
\item \textbf{The covariant projection tensor amplitude in PS-scheme}:
\begin{equation}
\begin{aligned}
A_{\sigma_{3}}^{\sigma_{1}}({p}_3,{p}_1,{p}_2;0,1)=&\tilde g^{\mu_1}_{\,\,\mu_3}(k_3)\epsilon_{\sigma_{3}}^{\mu_{3}}(k_3)\epsilon_{\mu_{1}^\prime}^{*\sigma_{1}}(k_1)D^{\mu_1^\prime}_{\,\,\mu_1}(R^{-1}_{31})\\
=&\tilde g^{\mu_1}_{\,\,\mu_3}(p_3)\epsilon_{\sigma_{3}}^{\mu_{3}}(p_3)\epsilon_{\mu_{1}^\prime}^{*\sigma_{1}}(k_1)D^{\mu_1^\prime}_{\,\,\mu_1}(R^{-1}_{31}L^{-1}_{3,c}(\mathbf{p}_3))\\
=&\tilde g^{\mu_1}_{\,\,\mu_3}(p_3)\epsilon_{\sigma_{3}}^{\mu_{3}}(p_3)\epsilon_{\mu_{1}^\prime}^{*\sigma_{1}}(p_1)D^{\mu_1^\prime}_{\,\,\mu_1}(L_{3,c}(\mathbf{p}_{3})L_{1,c}(\mathbf{p}_{1}^{*})L^{-1}_{3,c}(\mathbf{p}_3)).
\end{aligned}
\end{equation}
  \item \textbf{The covariant projection tensor amplitude in GS-scheme}:
  \begin{equation}
  \begin{aligned}
    \mathcal{C}^{\sigma_1}_{\sigma_3}(p_3,p_2,p_1;0,1)=\tilde g^{\mu_1}_{\,\,\mu_3}(p_3)\epsilon_{\sigma_{3}}^{\mu_{3}}(p_3)\epsilon_{\mu_{1}}^{*\sigma_{1}}(p_1).
  \end{aligned}
\end{equation}
\end{enumerate}

\item For the $(1,1)$ partial wave, the amplitude can be written in the following three schemes:
\begin{enumerate}
\item \textbf{The covariant projection tensor amplitude in PS-scheme}: 
\begin{equation}
\begin{aligned}
A_{\sigma_{3}}^{\sigma_{1}}({p}_3,{p}_1,{p}_2;1,1)=&\frac{i}{\sqrt{2}m_3}g_{\mu_3\mu_3^\prime}\tilde\varepsilon^{\mu_3^\prime\mu_L\mu_1}(p_{3})\tilde r_{\mu_L}(p_3)\epsilon_{\sigma_{3}}^{\mu_{3}}(p_3)\epsilon_{\mu_{1}^\prime}^{*\sigma_{1}}(p_1)\\
&\times D^{\mu_1^\prime}_{\,\,\mu_1}(L_{3,c}(\mathbf{p}_{3})L_{1,c}(\mathbf{p}_{1}^{*})L^{-1}_{3,c}(\mathbf{p}_3)).
\end{aligned}
\end{equation}
  \item \textbf{The covariant projection tensor amplitude in GS-scheme}: 
\begin{equation}
\begin{aligned}
\mathcal{C}^{\sigma_1}_{\sigma_3}(p_3,p_2,p_1;1,1)=
\frac{i}{\sqrt{2}m_3}g_{\mu_3\mu_3^\prime}\tilde\varepsilon^{\mu_3^\prime\mu_L\mu_1}(p_{3})\tilde r_{\mu_L}(p_3)\epsilon_{\sigma_{3}}^{\mu_{3}}(p_3)\epsilon_{\mu_{1}}^{*\sigma_{1}}(p_1).
  \end{aligned}
\end{equation}
\end{enumerate}

\item For the $(2,1)$ partial wave, the amplitude can be written in the following three schemes:
\begin{enumerate}
\item \textbf{The covariant projection tensor amplitude in PS-scheme}: 
  \begin{equation}
	\begin{aligned}
A_{\sigma_{3}}^{\sigma_{1}}({p}_3,{p}_1,{p}_2;2,1)=&\epsilon_{\sigma_{3}}^{\mu_{3}}(p_3)\epsilon_{\mu_{1}^\prime}^{*\sigma_{1}}(p_1)\Big(\tilde r^{\mu_1}(p_3)\tilde r_{\mu_3}(p_3)-\frac13(\tilde r\cdot\tilde r)\tilde g_{\,\,\mu_3}^{\mu_1}(p_3)\Big)\\
&\times D^{\mu_1^\prime}_{\,\,\mu_1}(L_{3,c}(\mathbf{p}_{3})L_{1,c}(\mathbf{p}_{1}^{*})L^{-1}_{3,c}(\mathbf{p}_3)).
	\end{aligned}
\end{equation}
  \item \textbf{The covariant projection tensor amplitude in GS-scheme}: 
  \begin{equation}
  \begin{aligned}
    \mathcal{C}^{\sigma_1}_{\sigma_3}(p_3,p_2,p_1;2,1)=\epsilon_{\sigma_{3}}^{\mu_{3}}(p_3)\epsilon_{\mu_{1}}^{*\sigma_{1}}(p_1)\Big(\tilde r^{\mu_1}(p_3)\tilde r_{\mu_3}(p_3)-\frac13(\tilde r\cdot\tilde r)\tilde g_{\,\,\mu_3}^{\mu_1}(p_3)\Big).
  \end{aligned}
\end{equation}
\end{enumerate}
\end{enumerate}

For the two-body decay process $3 \to 1 + 2$ with spins $s_3=\frac{1}{2},s_1=\frac{1}{2},s_2=0$. In this process, there are two possible $LS$ partial waves, namely $(L,S) = (0,\frac{1}{2})$, $(1,\frac{1}{2})$.
\begin{enumerate}
\item For the $(0,\frac{1}{2})$ partial wave, the amplitude can be written in the following three schemes:
\begin{enumerate}
\item \textbf{The covariant projection tensor amplitude in PS-scheme}: 
  \begin{equation}
	\begin{aligned}
A_{\sigma_{3}}^{\sigma_{1}}({p}_3,{p}_1,{p}_2;0,\frac{1}{2})=&\delta^{\alpha_S}_{\alpha_3}\delta^{\alpha_1}_{\alpha_S}u_{\sigma_{3}}^{\alpha_{3}}({p}_3;\frac{1}{2})\bar{u}_{\alpha_{1}^\prime}^{\sigma_{1}}({p}_1;\frac{1}{2})\\
&\times D^{\alpha_1^\prime}_{\,\,\alpha_1}(L_{3,c}(\mathbf{p}_{3})L_{1,c}(\mathbf{p}_{1}^{*})L^{-1}_{3,c}(\mathbf{p}_3)).
	\end{aligned}
\end{equation}
  \item \textbf{The covariant projection tensor amplitude in GS-scheme}:
  \begin{equation}
  \begin{aligned}
    \mathcal{C}^{\sigma_1}_{\sigma_3}(p_3,p_2,p_1;0,\frac{1}{2})= \delta^{\alpha_S}_{\alpha_3}\delta^{\alpha_1}_{\alpha_S}u_{\sigma_{3}}^{\alpha_{3}}({p}_3;\frac{1}{2})\bar{u}_{\alpha_{1}}^{\sigma_{1}}({p}_1;\frac{1}{2}).
  \end{aligned}
\end{equation}
\end{enumerate}

\item For the $(1,\frac{1}{2})$ partial wave, the amplitude can be written in the following three schemes:
\begin{enumerate}
\item \textbf{The covariant projection tensor amplitude in PS-scheme}: 
\begin{equation}
\begin{aligned}
A_{\sigma_{3}}^{\sigma_{1}}({p}_3,{p}_1,{p}_2;1,\frac{1}{2})=&\bar{u}_{\alpha_{1}^\prime}^{\sigma_{1}}({p}_1;\frac{1}{2})(\gamma_5\tilde g^{\mu_L}_{\,\,\mu_L^\prime}(p_3)\gamma^{\mu_L^\prime})_{\,\,\alpha_3}^{\alpha_1}u_{\sigma_{3}}^{\alpha_{3}}({p}_3;\frac{1}{2})\tilde r_{\mu_L}(p_3)\\
&\times D^{\alpha_1^\prime}_{\,\,\alpha_1}(L_{3,c}(\mathbf{p}_{3})L_{1,c}(\mathbf{p}_{1}^{*})L^{-1}_{3,c}(\mathbf{p}_3)).
\end{aligned}
\end{equation}
  \item \textbf{The covariant projection tensor amplitude in GS-scheme}: 
  \begin{equation}
  \begin{aligned}
    \mathcal{C}^{\sigma_1}_{\sigma_3}(p_3,p_2,p_1;1,\frac{1}{2})=  \bar{u}_{\alpha_{1}}^{\sigma_{1}}({p}_1;\frac{1}{2})(\gamma_5\tilde g^{\mu_L}_{\,\,\mu_L^\prime}(p_3)\gamma^{\mu_L^\prime})_{\,\,\alpha_3}^{\alpha_1}u_{\sigma_{3}}^{\alpha_{3}}({p}_3;\frac{1}{2})\tilde r_{\mu_L}(p_3).
  \end{aligned}
\end{equation}
\end{enumerate}
\end{enumerate}
By comparing the two expressions, one can see that the covariant projection tensor amplitude in PS-scheme contains an additional Lorentz transformation factor relative to the covariant projection tensor amplitude in GS-scheme, namely $L_{3,c}(\mathbf{p}_{3})L_{1,c}(\mathbf{p}_{1}^{*})L^{-1}_{3,c}(\mathbf{p}_3)$. This factor is generally not the identity matrix, so the $(L,S)$ partial wave bases in the two constructions do not match term by term, and the extracted $LS$ partial wave coefficients therefore need not be the same. 



\section{On-shell construction of covariant orbital-spin coupling amplitude}
\label{sec:Construction schemes of covariant orbital-spin coupling amplitude}

In the previous section, methods for constructing $LS$ amplitudes using tensors were presented, and each of them has certain limitations. For example, the covariant projection tensor method in PS-scheme, although the covariant projection tensor amplitudes in PS-scheme can be evaluated in any frame, their expressions still involve momenta defined in the COM frame. As a result, the formulation is not fully covariant in form.
For the covariant tensor method (GS-scheme), although the covariant tensor amplitudes retain a manifestly Lorentz covariant form, the general-spin part carries orbital information when one returns to the COM frame, and therefore the traditional separation between the orbital part and the spin part is not realized. Only in the non-relativistic limit does it agree with the traditional-$LS$ method.

Moreover, when the process involves massless gauge bosons, the use of polarization vectors or similar spin wave functions typically requires gauge invariance constraints to be imposed explicitly. One must then choose a complete set of vertex structures that satisfy these constraints and are linearly independent, in order to avoid gauge redundancy and redundant fit parameters, which makes practical implementations more cumbersome.



To address the issues of the methods discussed above, we introduce our covariant $LS$ amplitudes constructed in the covariant canonical-spinor scheme. They admit a proper separation between orbital and spin parts, do not require boosting to the COM frame, and can be applied directly to processes involving massless particles.

In subsection~\ref{subsec:Spinor-Canonical scheme}, the covariant canonical-spinor scheme is presented.
In subsection~\ref{subsec:Covariant tensor method}, we further compare the covariant canonical-spinor scheme with the covariant projection tensor method in the PS-scheme, and reformulate their expressions using our spinor construction to obtain a genuinely covariant form.
In subsection~\ref{subsec:Examples of spinor-canonical amplitude calculation}, explicit examples covariant canonical-spinor amplitudes calculations are given.

\subsection{Spinorial construction}
\label{subsec:Spinor-Canonical scheme}

In the spinorial construction, spinor variables are taken as the basic building blocks. Before introducing these spinor variables, it is useful to begin from the homogeneous Lorentz group. The six generators of homogeneous Lorentz transformations in Eq.~\eqref{Eq:six L generators} can be combined as
\begin{equation}
A_i=J_i+iK_i,\quad B_i=J_i-iK_i\qquad(i=1,2,3).
\end{equation}
They satisfy the following commutation relations
\begin{equation}\label{Eq: A B commutation}
[A_i,A_j]=i\varepsilon_{ijk}A_k,\quad[B_i,B_j]=i\varepsilon_{ijk}B_k,\quad [A_i,B_j]=0.
\end{equation}
These commutation relations imply that $\mathrm{SO}(3,1)\simeq \mathrm{SU}(2)_{L}\otimes \mathrm{SU}(2)_{R}$. Correspondingly, the momentum in the Lorentz four-vector representation $p^\mu=(E,\mathbf{p})$ can be described as bispinors $p_{\alpha\dot{\alpha}}$. The corresponding conversion matrix is $\sigma_{\mu\alpha\dot{\alpha}}$, which can be written as
\begin{equation}
    \sigma_{0\alpha\dot{\alpha}}=\begin{pmatrix}
        1&0\\
        0&1\\
    \end{pmatrix},\quad \sigma_{1\alpha\dot{\alpha}}=\begin{pmatrix}
        0&1\\
        1&0\\
    \end{pmatrix},\quad \sigma_{2\alpha\dot{\alpha}}=\begin{pmatrix}
        0&-i\\
        i&0\\
    \end{pmatrix},\quad \sigma_{3\alpha\dot{\alpha}}=\begin{pmatrix}
        1&0\\
        0&-1
    \end{pmatrix}.
\end{equation}
Using this matrix on $p^\mu$ gives the $2\times2$ matrix $p_{\alpha\dot{\alpha}}$, which can be written explicitly as
\begin{equation}\label{Eq:bispinor momentun}
p_{\alpha\dot{\alpha}}=p^{\mu}\sigma_{\mu\alpha\dot{\alpha}}=\begin{pmatrix}
        p^0+p^3&p^1-ip^2\\
        p^1+ip^2&p^0-p^3
    \end{pmatrix}.
\end{equation}
Using the on-shell condition, one has $\det{p_{\alpha\dot{\alpha}}}=m^{2}$. Therefore, there are two cases, corresponding to massless and massive particles.

For a massless particle, satisfying $\det{p_{\alpha\dot{\alpha}}}=0$, the matrix $p_{\alpha\dot{\alpha}}$ has rank one. Thus, it can be decomposed as the direct product of two spinors, $\lambda$ and $\tilde{\lambda}$, as
\begin{equation}
    p_{\alpha \dot{\alpha}}=\lambda_{\alpha}\tilde{\lambda}_{\dot{\alpha}}.
\end{equation}

For massive particle, satisfying $\det{p_{\alpha\dot{\alpha}}}=m^{2}$, the matrix $p_{\alpha\dot{\alpha}}$ has rank two instead of one. Thus, it can be decomposed as the sum of two rank-one bispinors as
\begin{equation}
    p_{\alpha \dot{\alpha}}=\lambda_{\alpha I}\tilde{\lambda}^I_{\dot{\alpha}},
\end{equation}
where $\lambda_{\alpha I}$ and $\tilde{\lambda}^I_{\dot{\alpha}}$ are spinors carrying the $\mathrm{SU}(2)$ little group index $I=\frac{1}{2},-\frac{1}{2}$ and the $\mathrm{SL}(2,\mathbb{C})$ Lorentz group index $\alpha/\dot{\alpha}=1,2$. 
In this work, the discussion mainly focuses on spinors of massive particles.

Under a Lorentz transformation, the spinor variables transform under the $\mathrm{SL}(2,\mathbb{C})$ Lorentz group and the $\mathrm{SU}(2)$ little group. The explicit form of the transformation is
\begin{equation}
\begin{aligned}
    \lambda_{\alpha I}(p)\to\lambda_{\alpha I}(\Lambda p)= W_{I}^{J}\Lambda^\beta_\alpha\lambda_{\beta J}( p),\quad
    \tilde{\lambda}^I_{\dot{\alpha}}(p)\to\tilde{\lambda}^I_{\dot{\alpha}}(\Lambda p)= (W^{-1})_{J}^{I}\tilde{\Lambda}^{\dot{\beta}}_{\dot{\alpha}}\tilde{\lambda}^J_{\dot{\beta}}(p),
\end{aligned}
\end{equation}
where $W$ and $W^{-1}$ are $\mathrm{SU}(2)$ matrices, and $\Lambda$ and $\tilde{\Lambda}$ are $\mathrm{SL}(2,\mathbb{C})$ matrices.
The antisymmetric tensor $\varepsilon_{\alpha\beta}=-\varepsilon^{\alpha\beta}=\begin{pmatrix}
    0&-1\\
    1&0\\
\end{pmatrix}$ is used to raise and lower $\mathrm{SL}(2,\mathbb{C})$ and $\mathrm{SU}(2)$ spinor indices:
\begin{equation}
    \psi^{\alpha}=\varepsilon^{\alpha\beta}\psi_{\beta},\quad \psi_{\alpha}=\psi^{\beta}\varepsilon_{\alpha\beta},\quad\varepsilon^{\alpha\beta}\varepsilon_{\beta\gamma}=\delta^{\alpha}_{\gamma}.
\end{equation}

The explicit form of the spinor variables at a general momentum $p$, one can obtain them by boosting the spinors defined at the standard momentum. The standard momentum can be written in bispinor form as
\begin{equation}
    k_{\alpha\dot{\alpha}}=k^{\mu}\sigma_{\mu\alpha\dot{\alpha}}=\lambda_{\alpha I}(k)\tilde{\lambda}^I_{\dot{\alpha}}(k)=\begin{pmatrix}
    m&0\\
    0&m
    \end{pmatrix}.
\end{equation}
In general, there are many equivalent spinor solutions; here a particular one is chosen, given by
\begin{equation}
    \lambda_{\alpha I}(k)=\begin{pmatrix}
        \sqrt{m}&0\\
        0&\sqrt{m}
    \end{pmatrix},\quad
    \tilde{\lambda}^I_{\dot{\alpha}}(k) =\begin{pmatrix}
        \sqrt{m}&0\\
        0&\sqrt{m}
    \end{pmatrix},
\end{equation}
which is consistent with the conventional choice.
By applying a Lorentz transformation, the standard momentum $k$ is mapped to a general momentum $p$, which in two-component form can be written as 
\begin{equation}
    p_{\alpha\dot{\alpha}}=L(\mathbf{p})^{\beta}_{\alpha}\tilde{L}(\mathbf{p})^{\dot{\beta}}_{\dot{\alpha}}k_{\beta\dot{\beta}},
\end{equation}
where $L(\mathbf{p})^{\beta}_{\alpha}$ and $\tilde{L}(\mathbf{p})^{\dot{\beta}}_{\dot{\alpha}}$ are $\mathrm{SL}(2,\mathbb{C})$ Lorentz boosts.
The corresponding transformation of the spinor variables is then given by
\begin{equation}\label{Eq:boost on spinor}
        \lambda_{\alpha I}(p)=L(\mathbf{p})^{\beta}_{\alpha}\lambda_{\beta I}(k),\quad
        \tilde{\lambda}^I_{\dot{\alpha}}(p)=\tilde{L}(\mathbf{p})^{\dot{\beta}}_{\dot{\alpha}}\tilde{\lambda}^I_{\dot{\beta}}(k).
\end{equation}

Since subsubsection~\ref{subsubsec:Description of single-particle} discusses both the helicity and the canonical types of boost, it is natural to introduce the corresponding spinor-helicity variables and canonical-spinor variables generated by these two types of boost, respectively. From Eqs.~\eqref{Eq:caonical boost} and \eqref{Eq:helicity boost} it follows that both types of boost can be factorized into the product of a pure boost along the $z$-axis and rotation. In the left-handed representation, the two-dimensional rotation and the pure boost along the $z$-axis can be written as
\begin{equation}
    R_{z}=\begin{pmatrix}
        e^{-i\frac{\phi}{2}} & 0\\
        0 & e^{i\frac{\phi}{2}}
    \end{pmatrix},\ 
    R_{y}=\begin{pmatrix}
        \cos{\frac{\theta}{2}} & -\sin{\frac{\theta}{2}}\\
        \sin{\frac{\theta}{2}} & \cos{\frac{\theta}{2}}
    \end{pmatrix},\ 
    B_{z}=\begin{pmatrix}
        e^{\frac{\eta}{2}} & 0 \\
        0 & e^{-\frac{\eta}{2}}
    \end{pmatrix},
\end{equation}
where $E+|\mathbf{p}|=me^{\eta}$ and $\eta =\text{arctanh}{\frac{|\mathbf{p}|}{E}}$. 
For the rotation part of the boost, we take 
\begin{equation}
\begin{aligned}
    R(\phi,\theta,-\phi)&=R_{z}(\phi)R_{y}(\theta)R_{z}(-\phi)\\
    &=\begin{pmatrix}
        c & -s^{*}\\
        s & c
    \end{pmatrix},
\end{aligned}
\end{equation} 
where $c\equiv \cos{\frac{\theta}{2}}$ and $s\equiv \sin{\frac{\theta}{2}}e^{i\phi}$. 
Here, it should be noted that the choice of rotation $R(\phi,\theta,-\phi)$ differs from the rotation $R(\phi,\theta,0)$ in Eq.~\eqref{Eq:caonical boost}. Specifically, the first rotation about the $z$-axis is taken to be $-\phi$. For both the helicity boost and the canonical boost, this initial rotation around the $z$-axis does not affect the momentum after the boost. To align with the convention of the helicity variables commonly used in the literature, we therefore adopt the rotation $R(\phi,\theta,-\phi)$.
Analogously to Eqs.~\eqref{Eq:caonical boost} and \eqref{Eq:helicity boost}, the two-dimensional helicity and canonical boosts in the left-handed representation are
\begin{equation}
    \begin{aligned}
        L_{h}(\mathbf{p}) &= R(\phi,\theta,-\phi)B_{z}(\eta)\\
        &=\begin{pmatrix}
            ce^{\frac{\eta}{2}} & (-s^*)e^{-\frac{\eta}{2}}\\
            se^{\frac{\eta}{2}} & ce^{-\frac{\eta}{2}}
        \end{pmatrix},
    \end{aligned}
\end{equation}
and
\begin{equation}
    \begin{aligned}
        L_{c}(\mathbf{p}) &= R(\phi,\theta,-\phi)B_{z}(\eta)R(\phi,\theta,-\phi)^{-1}\\
        &=\begin{pmatrix}
            c^{2}e^{\frac{\eta}{2}}+|s|^{2}e^{-\frac{\eta}{2}} & cs^*e^{\frac{\eta}{2}}-cs^*e^{-\frac{\eta}{2}}\\
            cse^{\frac{\eta}{2}}-cse^{-\frac{\eta}{2}} & |s|^{2}e^{\frac{\eta}{2}}+c^{2}e^{-\frac{\eta}{2}}
        \end{pmatrix}.
    \end{aligned}
\end{equation}
Using the helicity boost and the canonical boost, the spinor-helicity variables and canonical-spinor variables for a general momentum $p$ can be obtained from Eq.~\eqref{Eq:boost on spinor} and written as
\begin{equation}
\begin{aligned}
    \lambda_{\alpha I}(p)_{h}&=L_{h}(\mathbf{p})\lambda_{\alpha I}(k)\\
    &=\sqrt{m}\begin{pmatrix}
        ce^{\frac{\eta}{2}} & (-s^*)e^{-\frac{\eta}{2}}\\
        se^{\frac{\eta}{2}} & ce^{-\frac{\eta}{2}}
    \end{pmatrix},
\end{aligned}
\end{equation}
and
\begin{equation}
    \begin{aligned}
        \lambda_{\alpha I}(p)_{c}&=L_{c}(\mathbf{p})\lambda_{\alpha I}(k)\\
        &=\sqrt{m}\begin{pmatrix}
            c^{2}e^{\frac{\eta}{2}}+|s|^{2}e^{-\frac{\eta}{2}} & cs^*e^{\frac{\eta}{2}}-cs^*e^{-\frac{\eta}{2}}\\
            cse^{\frac{\eta}{2}}-cse^{-\frac{\eta}{2}} & |s|^{2}e^{\frac{\eta}{2}}+c^{2}e^{-\frac{\eta}{2}}
        \end{pmatrix}.
    \end{aligned}
\end{equation}
Unless otherwise stated, the canonical-spinor variables will be used and denoted simply as $\lambda_{\alpha I},\tilde{\lambda}^I_{\dot{\alpha}}$. 
Conventionally, angle brackets $\langle\cdots\rangle$ and square brackets $[\cdots]$ are used to denote spinor contractions. 
The angle and square brackets corresponding to left- and right-handed spinors are defined by
\begin{equation}\label{Eq:spinor canonical variables}
\begin{aligned}
         |p_{I}\rangle &= \lambda_{\alpha I}=\sqrt{m}\begin{pmatrix}
            c^{2}e^{\frac{\eta}{2}}+|s|^{2}e^{-\frac{\eta}{2}} & cs^*e^{\frac{\eta}{2}}-cs^*e^{-\frac{\eta}{2}}\\
            cse^{\frac{\eta}{2}}-cse^{-\frac{\eta}{2}} & |s|^{2}e^{\frac{\eta}{2}}+c^{2}e^{-\frac{\eta}{2}}
        \end{pmatrix}_{\alpha I},\\[5pt]
        \quad [p^{I}| &= \tilde{\lambda}_{\dot{\alpha}}^{I} = \sqrt{m}\begin{pmatrix}
            c^{2}e^{\frac{\eta}{2}}+|s|^{2}e^{-\frac{\eta}{2}} & cs^*e^{\frac{\eta}{2}}-cs^*e^{-\frac{\eta}{2}}\\
            cse^{\frac{\eta}{2}}-cse^{-\frac{\eta}{2}} & |s|^{2}e^{\frac{\eta}{2}}+c^{2}e^{-\frac{\eta}{2}}
        \end{pmatrix}_{\,\,\dot{\alpha}}^{I},\\[5pt]
        \langle p_{I}|&=\lambda^\alpha_{I}=\sqrt{m}\begin{pmatrix}cse^{\frac{\eta}{2}}-cse^{-\frac{\eta}{2}}
             & -(c^{2}e^{\frac{\eta}{2}}+|s|^{2}e^{-\frac{\eta}{2}})\\
            |s|^{2}e^{\frac{\eta}{2}}+c^{2}e^{-\frac{\eta}{2}} & -(cs^*e^{\frac{\eta}{2}}-cs^*e^{-\frac{\eta}{2}})
        \end{pmatrix}_{I}^{\,\,\alpha},\\[5pt]
        |p^{I}]&=\tilde{\lambda}^{\dot{\alpha} I}=\sqrt{m}\begin{pmatrix}cs^*e^{\frac{\eta}{2}}-cs^*e^{-\frac{\eta}{2}}
             & |s|^{2}e^{\frac{\eta}{2}}+c^{2}e^{-\frac{\eta}{2}}\\
            -(c^{2}e^{\frac{\eta}{2}}+|s|^{2}e^{-\frac{\eta}{2}}) & -(cse^{\frac{\eta}{2}}-cse^{-\frac{\eta}{2}})
        \end{pmatrix}^{\dot{\alpha} I}.
\end{aligned}
\end{equation}
Some basic spinor contractions can be written as
\begin{equation}\label{Eq:spinor contraction basic}
    \langle p_Iq_J\rangle=\lambda^\alpha_I(p)\lambda_{\alpha J}(q)=\varepsilon^{\alpha\beta}\lambda_{\beta I}(p)\lambda_{\alpha J}(q), \quad[p^Iq^J]=\tilde{\lambda}_{\dot{\alpha}}^{I}(p)\tilde{\lambda}^{\dot{\alpha} J}(q)=\varepsilon_{\dot{\alpha}\dot{\beta}}\tilde{\lambda}^{\dot{\beta}I}(p)\tilde{\lambda}^{\dot{\alpha}J}(q),
\end{equation}
\begin{equation}
    \langle q_I|p|k^J]=\lambda^{\alpha}_I(q)p_{\alpha\dot{\alpha}}\tilde{\lambda}^{\dot{\alpha}J}(k),\quad [q^I|p_1p_2|k^J]=\tilde{\lambda}_{\dot{\alpha}}^{I}(q)p_1^{\dot{\alpha}\alpha}p_{2\alpha\dot{\beta}}\tilde{\lambda}^{\dot{\beta} J}(k).
\end{equation}
For the little group indices $I,J$, raising and lowering is performed with $\varepsilon_{IJ}/\varepsilon^{IJ}$, which allows conversion between upper and lower indices. More general contractions with multiple momenta inserted between spinors can be obtained following the above rules. In what follows, for notational simplicity spinors will be labeled by the particle index, so that $\langle p_i|\to\langle i|$ denotes the spinor variables associated with the $i$-th particle.

The above summarizes the basic kinematics of particle scattering amplitudes. 
For an $n$-point scattering process, the amplitude is a function of the external momenta and the little group indices, $(p_a,\sigma_a)$, $a=1,\cdots,n$. 
Under a Lorentz transformation $\Lambda$, it transforms with the momenta and the corresponding little group actions as
\begin{equation}\label{Eq:lorentz on amplitude traf}
    A_{\sigma_a}(p_a)=\prod _{a}D^{\sigma_a^\prime(s_a)}_{\,\,\sigma_a}(W)A_{\sigma_a^\prime}\bigg((\Lambda p)_a\bigg),
\end{equation}
where $D^{\sigma_a^\prime(s_a)}_{\,\,\sigma_a}(W)$ is the spin-$s_a$ representation matrix of the little group rotation associated with particle-$a$. Motivated by this, momenta are decomposed into spinor variables carrying both Lorentz and little group indices. 
By contracting the Lorentz indices of these spinor variables, amplitudes can be constructed that are Lorentz invariant and transform under the little group. The resulting amplitudes depend only on these spinor variables and, by construction, satisfy Eq~\eqref{Eq:lorentz on amplitude traf}.

For example, in the case $n=3$, consider the three-point amplitude $3\to1+2$, where each particle is massive and has spins $s_3=\frac{1}{2},s_1=\frac{1}{2},s_2=1$. The amplitude constructed from spinor variables is
\begin{equation}\label{Eq:example of spinor amplitude 1}
\begin{aligned}
    \mathcal{A}^{\{J_1\},\{I_1I_2\}}_{\{K_1\}}=&\kappa_1\langle1^{J_1}2^{(I_1}\rangle[2^{I_2)}3_{K_1}]+\kappa_2[1^{J_1}2^{(I_1}]\langle2^{I_2)}3_{K_1}\rangle\\
    &+\kappa_3\langle1^{J_1}2^{(I_1}\rangle\langle2^{I_2)}3_{K_1}\rangle+\kappa_4[1^{J_1}2^{(I_1}][2^{I_2)}3_{K_1}],
\end{aligned}
\end{equation}
where $(I_1I_2)$ denotes symmetrization of the indices and $\kappa_{1,2,3,4}$ are coefficients determined by experiment. 
It can be seen that the amplitude constructed in this way satisfies the Lorentz invariance and the required little group transformation properties 
\begin{equation}
    \mathcal{A}^{\{J_1\},\{I_1I_2\}}_{\{K_1\}}(p_3,p_1,p_2)=(W_3)^{K^\prime_1}_{K_1}(W^{-1}_1)^{J_1}_{J^\prime_1}(W^{-1}_2)^{I_1}_{I^\prime_1}(W^{-1}_2)^{I_2}_{I^\prime_2}\mathcal{A}^{\{J_1^\prime\},\{I_1^\prime I_2^\prime\}}_{\{K_1^\prime\}}(\Lambda p_3,\Lambda p_1,\Lambda p_2).
\end{equation}
Moreover, this three-point amplitude has four $LS$ partial waves
$(L,S)=(0,\frac{1}{2}),(1,\frac{1}{2}),(1,\frac{3}{2}),(2,\frac{3}{2})$
in total, so there are four independent structures for the amplitude. 
Consequently, the four independent bases constructed from spinor contractions form a complete set.

These four spinor contractions represent the basic building blocks of the three-point amplitude with spins $s_3=\frac{1}{2},s_1=\frac{1}{2},s_2=1$.

To make contact with the conventional description based on Feynman rules, consider the most general vertex between two fermions and one vector,
$\mathcal{L}_{\mathrm{int}}= \bar\psi_1\Gamma^\mu\psi_3V_\mu$, whose coupling structure $\Gamma^\mu$ can be written as
\begin{equation}\label{Eq:general-fermion-vector-vertex}
    \Gamma_\mu= \gamma_\mu(g_LP_L+g_RP_R)
    + i\sigma_{\mu\nu}p_2^{\nu}
      \bigl( f_L P_L + f_R P_R \bigr),
\end{equation}
where $P_{L,R}=\frac{1}{2}(1\mp \gamma_5)$, $\sigma_{\mu\nu}=\tfrac{i}{2}[\gamma_\mu,\gamma_\nu]$, and $p_2$ denotes the momentum carried by the vector field $V^\mu$.
The gamma matrices in our convention are given by
\begin{equation}
    \gamma_\mu=\begin{pmatrix}
        0 & \sigma_{\mu\alpha\dot{\alpha}}\\
        \bar{\sigma}_\mu^{\dot{\alpha}\alpha} & 0\\
    \end{pmatrix},\ 
    \gamma_5=\gamma^5=\begin{pmatrix}
        -1 & 0\\
        0 & 1
    \end{pmatrix},
\end{equation}
where $\sigma_{\mu\alpha\dot{\alpha}}=(\delta_{\alpha\dot{\alpha}},\sigma_{\alpha\dot{\alpha}})$ and $\bar{\sigma}_\mu^{\dot{\alpha}\alpha}=(\delta^{\dot{\alpha}\alpha},-\sigma^{\dot{\alpha}\alpha})$.
The projection operators $P_{L,R}$ can be written as
\begin{equation}
    P_L=\begin{pmatrix}
        1 & 0\\
        0 & 0\\
    \end{pmatrix},\quad
    P_R=\begin{pmatrix}
        0 & 0\\
        0 & 1\\
    \end{pmatrix}.
\end{equation}
The decay amplitude for the process $\psi_3\to\psi_1 V_2$ can be written as
\begin{equation}
\mathcal{A}=\epsilon^{*\mu}\bar{u}(p_1)\big(\gamma_\mu (g_LP_L+g_RP_R)+i\sigma_{\mu\nu}p^\nu_2 (f_LP_L+f_RP_R)\big)u(p_3),
\end{equation}
where the four-component spinors are
\begin{equation}
    u(p)=\begin{pmatrix}
        |p_K\rangle\\
        |p_K]
    \end{pmatrix},\quad
    \bar{u}(p)=\begin{pmatrix}
        -\langle p^J| & [p^J| 
    \end{pmatrix},
\end{equation}
and the polarization vectors in bispinor form are
\begin{equation}
\epsilon^*_{\alpha\dot{\alpha}}=\epsilon^{*\mu}\sigma_{\mu\alpha\dot{\alpha}}=-\frac{|p^{(I_1}\rangle[p^{I_2)}|}{m},\quad
\epsilon^{*\dot{\alpha}\alpha}=\epsilon^{*\mu}\bar{\sigma}^{\dot{\alpha}\alpha}_{\mu}=-\frac{|p^{(I_1}]\langle p^{I_2)}|}{m},
\end{equation}
where $(I_1I_2)$ denotes symmetrization of the indices. The amplitude written in the spinor variables form is
\begin{equation}
\begin{aligned}
    \mathcal{A}^{\{J_1\},\{I_1I_2\}}_{\{K_1\}}=&-g_R\langle1^{J_1}|\epsilon^*_{\alpha\dot{\alpha}}|3_{K_1}]+g_L[1^{J_1}|\epsilon^{*\dot{\alpha}\alpha}|3_{K_1}\rangle-\frac{f_R}{2}[1^{J_1}|\epsilon^{*\dot{\alpha}\alpha}p_{2\alpha\dot{\beta}}|3_{K_1}]+\frac{f_L}{2}\langle1^{J_1}|\epsilon^*_{\alpha\dot{\alpha}}p^{2\dot{\alpha}\beta}|3_{K_1}\rangle\\
    =&\frac{g_R\langle1^{J_1}2^{(I_1}\rangle[2^{I_2)}3_{K_1}]-g_L[1^{J_1}2^{(I_1}]\langle2^{I_2)}3_{K_1}\rangle}{m_2}\\
    &+\frac{f_R}{2}[1^{J_1}2^{(I_1}][2^{I_2)}3_{K_1}]-\frac{f_L}{2}\langle1^{J_1}2^{(I_1}\rangle\langle2^{I_2)}3_{K_1}\rangle.\\
\end{aligned}
\end{equation}
Taking $\kappa_1=\frac{g_R}{m_2},\kappa_2=-\frac{g_L}{m_2},\kappa_3=-\frac{f_L}{2}$ and $\kappa_4=\frac{f_R}{2}$ in Eq.~\eqref{Eq:example of spinor amplitude 1}, the corresponding expression is obtained.

More generally, three-point amplitudes for massive particles with arbitrary spins are discussed in Ref.~\cite{Arkani-Hamed:2017jhn}. 
Now consider the more general two-body decay amplitude $3\to1 + 2$, where the amplitude is a function of the canonical-spinor variables. The particle spins are $s_{1}$, $s_{2}$, and $s_{3}$, respectively. The amplitude can be written as
\begin{equation}
    \mathcal{A}_{\{I^{(3)}_{1}\cdots I^{(3)}_{2s_{3}}\}}^{\{I^{(1)}_{1}\cdots I^{(1)}_{2s_{1}}\},\{I^{(2)}_{1}\cdots I^{(2)}_{2s_{2}}\}}\left((\lambda_{\alpha I},\tilde{\lambda}^I_{\dot{\alpha}})_{1};(\lambda_{\alpha I},\tilde{\lambda}^I_{\dot{\alpha} })_{2};(\lambda_{\alpha I},\tilde{\lambda}^I_{\dot{\alpha} })_{3}\right).
\end{equation}
Under a Lorentz transformation, the amplitude transforms as
\begin{equation}
    \begin{aligned}
        \mathcal{A}_{\{I^{(3)}_{1}\cdots I^{(3)}_{2s_{3}}\}}^{\{I^{(1)}_{1}\cdots I^{(1)}_{2s_{1}}\},\{I^{(2)}_{1}\cdots I^{(2)}_{2s_{2}}\}}(p_i)&\xrightarrow{\Lambda} \left(W_{I^{(3)}_{1}}^{K^{(3)}_{1}}\cdots W_{I^{(3)}_{2s_{3}}}^{K^{(3)}_{2s_{3}}}\right)\left((W^{-1})^{K^{(1)}_{1}}_{I^{(1)}_{1}}\cdots (W^{-1})_{K^{(1)}_{2s_{1}}}^{I^{(1)}_{2s_{1}}}\right) \times\\
        &\left((W^{-1})_{K^{(2)}_{1}}^{I^{(2)}_{1}}\cdots (W^{-1})_{K^{(2)}_{2s_{2}}}^{I^{(2)}_{2s_{2}}}\right) \ \mathcal{A}_{\{K^{(3)}_{1}\cdots K^{(3)}_{2s_{3}}\}}^{\{K^{(1)}_{1}\cdots K^{(1)}_{2s_{1}}\},\{K^{(2)}_{1}\cdots K^{(2)}_{2s_{2}}\}}(\Lambda p_i).
    \end{aligned}
\end{equation}
As discussed in section~\ref{sec:Two different states description---Helicity and Canonical}, amplitudes can be decomposed into $LS$ partial waves. Accordingly, the amplitude constructed from canonical-spinor variables can be decomposed as
\begin{equation}
	\mathcal{A}_{\{I^{(3)}_{1}\cdots I^{(3)}_{2s_{3}}\}}^{\{I^{(1)}_{1}\cdots I^{(1)}_{2s_{1}}\},\{I^{(2)}_{1}\cdots I^{(2)}_{2s_{2}}\}} = \sum_{LS}N_{LS} \mathcal{A}_{\{I^{(3)}_{1}\cdots I^{(3)}_{2s_{3}}\}}^{\{I^{(1)}_{1}\cdots I^{(1)}_{2s_{1}}\},\{I^{(2)}_{1}\cdots I^{(2)}_{2s_{2}}\}}(L,S),
\end{equation}
where $N_{LS}$ is the coefficient related to energy and mass, which can be measured from experiment. Here $\mathcal{A}(L,S)$ denotes the canonical-spinor amplitude in the $LS$ expansion.

The decay amplitude in any frame can be expressed as the amplitude in the COM frame, transformed by a little group rotation. The COM frame amplitude itself can be decomposed into orbital ($L$) and spin ($S$) parts. Consequently, the amplitude in any frame can be separated into covariant $L$ and $S$ components via a spinor decomposition. The corresponding $LS$ amplitude is defined as
\begin{equation}\label{Eq:sphe1}
	\mathcal{A}_{\{I^{(3)}_{1}\cdots I^{(3)}_{2s_{3}}\}}^{\{I^{(1)}_{1}\cdots I^{(1)}_{2s_{1}}\},\{I^{(2)}_{1}\cdots I^{(2)}_{2s_{2}}\}}(L,S)\equiv\mathcal{S}_{\{K\}}^{\{I^{(1)}_{1}\cdots I^{(1)}_{2s_{1}}\},\{I^{(2)}_{1}\cdots I^{(2)}_{2s_{2}}\}}\mathcal{Y}^L_{\{I\}}C_{s_{3},\{I^{(3)}_{1}\cdots I^{(3)}_{2s_{3}}\}}^{S,\{K\};L,\{I\}}.
\end{equation}
Here
\begin{itemize}
	\item $\mathcal{Y}^L_{\{I\}}$ (defined in Eq.~\eqref{Eq:spinor Y}), which encodes the orbital angular momentum $L$ structure. For $L$ (orbital angular momentum quantum number), the function $\mathcal{Y}^{L}_{\{I\}}$ carries $2L$ symmetric little group $\mathrm{SU}(2)$ indices $\{I\}=(I_1\cdots I_{2L})$.
	\item $\mathcal{S}_{\{K\}}^{\{I^{(1)}_{1}\cdots I^{(1)}_{2s_{1}}\},\{I^{(2)}_{1}\cdots I^{(2)}_{2s_{2}}\}}$ (defined in Eq.~\eqref{Eq:spinor S}), which couples the spins $s_{1}$ and $s_{2}$ into a total spin-$S$, not only carries $2s_1$ and $2s_2$ symmetric little group indices $\{I^{(1)}_{1}\cdots I^{(1)}_{2s_{1}}\},\{I^{(2)}_{1}\cdots I^{(2)}_{2s_{2}}\}$ associated with the individual spins, but also includes $2S$ symmetric little group indices $\{K\}=(K_1\cdots K_{2S})$ corresponding to the total spin-$S$ representation resulting from the coupling.
\end{itemize}
By coupling the indices $\{I\}$ carried by $\mathcal{Y}^{L}_{\{I\}}$ with the indices $\{K\}$ carried by $\mathcal{S}_{\{K\}}^{\{I^{(1)}_{1}\cdots I^{(1)}_{2s_{1}}\},\{I^{(2)}_{1}\cdots I^{(2)}_{2s_{2}}\}}$ via the $\mathrm{SU}(2)$ CGCs, the final amplitude $\mathcal{A}(L,S)$ is obtained in Eq.~\eqref{Eq:sphe1}.

For the canonical-spinor amplitude, we can choose to normalize the amplitude $\mathcal{A}_{\{I^{(3)}\}}^{\{I^{(1)}\}\{I^{(2)}\}}(L,S)$, which allows separate extraction of normalization factors for the orbital part $\mathcal{Y}^L_{\{I\}}$ and the spin part $\mathcal{S}_{\{K\}}^{\{I^{(1)}\},\{I^{(2)}\}}$. The normalized amplitudes satisfy
\begin{equation}\label{Eq:normalized spinor canonical amplitude}
	\int d\Phi \bigg(\mathcal{A}_{\{I^{(3)}\}}^{\{I^{(1)}\}\{I^{(2)}\}}(L,S)\bigg)^*\mathcal{A}_{\{I^{(3)}\}}^{\{I^{(1)}\}\{I^{(2)}\}}(L^\prime,S^\prime)=\delta^{S^{\prime}}_S\delta^{L^{\prime}}_L.
\end{equation}
Before constructing the orbital and spin parts of the amplitude, the $\mathrm{SU}(2)$ CGCs are first defined as 
\begin{equation}\label{Eq:SU(2) CGC}
	C_{s,\{J\}}^{s_{1},\{I\};s_{2},\{K\}}\equiv\sqrt{\dfrac{(a+b)!(b+c)!(a+c+1)!}{b!(a+b+c+1)!a!c!}}\delta_{(J_{1}\cdots J_{a}}^{(I_{1}\cdots I_{a}}\varepsilon^{I_{a+1}\cdots I_{a+b}),(K_{1}\cdots K_{b}}\delta_{J_{a+1}\cdots J_{a+c})}^{K_{b+1}\cdots K_{b+c})},
\end{equation}
with
\begin{equation}
	\begin{aligned}
		a=&s_{1}+s-s_{2},\\
		b=&s_{1}+s_{2}-s,\\
		c=&s_{2}+s-s_{1},\\
        \delta^{I_{1}\cdots I_{a}}_{J_{1}\cdots J_{a}}=&\delta^{I_1}_{J_1}\cdots\delta^{I_a}_{J_a},\\
        \varepsilon^{I_{1}\cdots I_{b},K_{1}\cdots K_{b}}=&\varepsilon^{I_{1}K_{1}}\cdots \varepsilon^{I_{b}K_{b}},\\
	\end{aligned}
\end{equation}
where $(J_{1}\cdots J_{a+c})$,$(I_{1}\cdots I_{a+b})$ and $(K_{1}\cdots K_{b+c})$ denote symmetrization over the corresponding indices. In subsection~\ref{subsec:Examples of CGC calculation}, there are some examples of calculating the $\mathrm{SU}(2)$ CGC.

For the orbital part $\mathcal{Y}^L_{\{I\}}$, it is convenient to begin in the COM frame. 
In this frame, the orbital structure depends explicitly on the momenta of the final particles. 
Momentum conservation ensures that the momenta of the two final particles satisfy $\mathbf{p}^*_1 = -\mathbf{p}^*_2$. 
Therefore, the relative motion is fully characterized by the momentum of a single particle; here particle-1 is chosen.
The most elementary building block of the orbital part can be constructed from the momentum of particle-1, which can be written as 
\begin{equation}
    \langle3_{I_1}|p_{1}|3_{I_2}].
\end{equation}
Owing to the Lorentz covariance of the canonical-spinor amplitude, such a building block, once properly constructed, can be promoted to any frame while still encoding the correct orbital angular momentum information. As a result, the orbital structure $\mathcal{Y}^L_{\{I\}}$ in any frame can be defined as
\begin{equation}\label{Eq:spinor Y}
	\mathcal{Y}^L_{\{I\}}\equiv\sqrt{\frac{(2(2L+1)!)}{\pi|2\mathbf{p}_{1\mathrm{com}}|^{2L+1}m^{2L-1}_{3}L!^{2}}}(\langle3|p_{1}|3]^{L})_{\{I\}},
\end{equation}
with
\begin{equation}
    (\langle3|p_{1}|3]^{L})_{\{I\}}=\langle3_{(I_{1}}|p_{1}|3_{I_{2}}]\cdots\langle3_{I_{2L-1}}|p_{1}|3_{I_{2L})}],
\end{equation}
where $L$ is the orbital angular momentum, $|\mathbf{p}_{1\mathrm{com}}|$ is the magnitude of the relative momentum in the COM frame, $m_3$ is the mass of the parent particle (particle-3), and $(I_{1}\cdots I_{2L})$ denotes symmetrization over the indices. The normalization factor in front of $(\langle3|p_1|3]^L)_{\{I\}}$ arises from the normalization condition in Eq.~\eqref{Eq:normalized spinor canonical amplitude}.

For the spin part $\mathcal{S}_{\{K\}}^{\{I^{(1)}\},\{I^{(2)}\}}$, it is convenient to begin in the COM frame. 
In this frame, appropriate spinor contractions are constructed whose structure matches the required number of little group indices. 
Additionally, these contractions must reduce to a form proportional to the identity matrix in the COM frame, since no further little group transformation is needed there.
To this end, a set of spinor structures, denoted by $\tau$, is identified that satisfies these criteria. 
The explicit form of $\tau$ is
\begin{equation}\label{tau}
\begin{aligned}
    	\tau_{J}^{I}(p_{3},p_{1})&\equiv\langle1^{I}3_{J}\rangle - [1^{I}3_{J}]\\
        &\overset{\mathrm{COM}}{=}\sqrt{2m_{3}(E_{1\mathrm{com}}+m_{1})}\begin{pmatrix}
        1&0\\
        0&1
        \end{pmatrix}.
\end{aligned}
\end{equation}
In the COM frame, this tensor is proportional to the identity matrix.
In any frame, its form remains covariant and it behaves as an effective little group transformation acting on the spin indices.

With this building block, the spin part $\mathcal{S}_{\{K\}}^{\{I^{(1)}\},\{I^{(2)}\}}$ can be defined by coupling the $\tau$ tensors for particle-1 and particle-2 through CGCs. 
The spin part of the amplitude is defined as 
\begin{equation}\label{Eq:spinor S}
\begin{aligned}
    	\mathcal{S}_{\{K\}}^{\{I^{(1)}_{1}\cdots I^{(1)}_{2s_{1}}\},\{I^{(2)}_{1}\cdots I^{(2)}_{2s_{2}}\}}\equiv&\frac{1}{(m_{3}(E_{1\mathrm{com}}+m_{1}))^{s_{1}}(m_{3}(E_{2\mathrm{com}}+m_{2}))^{s_{2}}}(\tau^{2s_{1}}_{1})_{\{J^{(1)}\}}^{\{I^{(1)}\}}(\tau^{2s_{2}}_{2})_{\{J^{(2)}\}}^{\{I^{(2)}\}}\\
        &\times C_{S,\{K\}}^{s_{1},\{J^{(1)}\};s_{2},\{J^{(2)}\}},
\end{aligned}
\end{equation}
with
\begin{equation}\label{Eq:define tau 1}
    \begin{aligned}
    (\tau^{2s_{1}}_{1})_{\{J^{(1)}\}}^{\{I^{(1)}\}} &= \tau_{(J^{(1)}_1}^{(I^{(1)}_{1}}\tau_{J^{(1)}_2}^{I^{(1)}_{2}}\cdots\tau_{J^{(1)}_{2s_{1}})}^{I^{(1)}_{2s_{1}})}(p_3,p_1),\\
    (\tau^{2s_{2}}_{2})_{\{J^{(2)}\}}^{\{I^{(2)}\}} &= \tau_{(J^{(2)}_1}^{(I^{(2)}_{1}}\tau_{J^{(2)}_2}^{I^{(2)}_{2}}\cdots\tau_{J^{(2)}_{2s_{2}})}^{I^{(2)}_{2s_{2}})}(p_3,p_2),
\end{aligned}
\end{equation}
where $(J^{(1)}_{1}\cdots J_{2s_1}^{(1)})$, $(I^{(1)}_{1}\cdots I^{(1)}_{2s_1})$, $(J^{(2)}_{1}\cdots J^{(2)}_{2s_2})$ and $(I^{(2)}_{1}\cdots I^{(2)}_{2s_2})$ denote symmetrization over the corresponding indices. The normalization factor in front of $\tau$ arises from the normalization condition in Eq.~\eqref{Eq:normalized spinor canonical amplitude}. 

After constructing the orbital part $\mathcal{Y}^L_{\{I\}}$ and the spin part $\mathcal{S}_{\{K\}}^{\{I^{(1)}\},\{I^{(2)}\}}$, these can be combined to obtain the canonical-spinor amplitude in the $LS$ expansion, given in Eq.~\eqref{Eq:sphe1}. 

The canonical-spinor amplitude and the canonical amplitude are represented in two different bases. 
The canonical-spinor amplitude is defined in the canonical-spinor basis $|s,\{I\}\rangle$, while the canonical amplitude is defined in the canonical basis $|s,\sigma\rangle$.
Between the two bases there exists a corresponding transformation, which can be written as
\begin{equation}\label{Eq:little group indices I to sigma}
    |s,\sigma\rangle
    = \sqrt{\frac{(2s)!}{(s+\sigma)!(s-\sigma)!}}\,|s,\{I\}\rangle,
\end{equation}
where the indices $\{I\}$ denotes the symmetric combination $(I_{1}\cdots I_{2s})$, with
$I_k = \pm \tfrac{1}{2}$ for $k=1,\cdots,2s$, and represents any choice of $\{I\}$ such that
\begin{equation}
    \sigma=\sum^{2s}_{k=1}I_{k}.
\end{equation}
Therefore, the relation between the canonical amplitude in Eq.~\eqref{Eq:canonical decay amplitude} and the canonical-spinor amplitude in Eq.~\eqref{Eq:sphe1} can be written as 
\begin{equation}\label{Eq:amplitude index conversion}
	\begin{aligned}
		\mathcal{A}_{\sigma_{3}}^{\sigma_{1}\sigma_{2}}(L,S)=&N\sqrt{\frac{(2s_{3})!}{(s_{3}+\sigma_{3})!(s_{3}-\sigma_{3})!}}\sqrt{\frac{(2s_{1})!}{(s_{1}+\sigma_{1})!(s_{1}-\sigma_{1})!}}\sqrt{\frac{(2s_{2})!}{(s_{2}+\sigma_{2})!(s_{2}-\sigma_{2})!}}\\
		&\times \mathcal{A}_{\{I_{3}\}}^{\{I_{1}\},\{I_{2}\}}(p_{3},p_{1},p_{2},L,S),
	\end{aligned}
\end{equation}
where $N$ is the coefficient related to energy and mass. A detailed discussion of how to determine $N$ in the different bases is given in Appendix~\ref{app:Numerically relate different bases}. The factors in front of the spin part $\mathcal{S}_{\{K\}}^{\{I^{(1)}\},\{I^{(2)}\}}$ and the orbital part $\mathcal{Y}^L_{\{I\}}$ in Eq.~\eqref{Eq:sphe1} can be chosen such that $N = 1$. For this new choice, the spin and orbital parts will be denoted by $\mathcal{S}_{\{K\}}^{\prime\{I^{(1)}\},\{I^{(2)}\}}$ and $\mathcal{Y}^{\prime L}_{\{I\}}$, respectively.

The spin part $\mathcal{S}_{\{K\}}^{\prime\{I^{(1)}\},\{I^{(2)}\}}$ is defined as
\begin{equation}\label{Eq:spinor Sprime}
\begin{aligned}
    	\mathcal{S}_{\{K\}}^{\prime\{I^{(1)}\},\{I^{(2)}\}}\equiv&(\tau^{\prime2s_{1}}_{1})_{\{J^{(1)}\}}^{\{I^{(1)}\}}(\tau^{\prime2s_{2}}_{2})_{\{J^{(2)}\}}^{\{I^{(2)}\}}C_{S,\{K\}}^{s_{1},\{J^{(1)}\};s_{2},\{J^{(2)}\}},
\end{aligned}
\end{equation}
where $\tau^\prime$ is defined as 
\begin{equation}\label{Eq:define tau}
	\begin{aligned}
		\tau_{J^{(1)}}^{\prime I^{(1)}}(p_{3},p_{1})&\equiv\frac{1}{\sqrt{2m_{3}(E_{1\mathrm{com}}+m_{1})}}(\langle1^{I^{(1)}}3_{J^{(1)}}\rangle-[1^{I^{(1)}}3_{J^{(1)}}]),\\
		\tau_{J^{(2)}}^{\prime I^{(2)}}(p_{3},p_{2})&\equiv\frac{1}{\sqrt{2m_{3}(E_{2\mathrm{com}}+m_{2})}}(\langle2^{I^{(2)}}3_{J^{(2)}}\rangle-[2^{I^{(2)}}3_{J^{(2)}}]).
	\end{aligned}
\end{equation}

The orbital part $\mathcal{Y}^{\prime L}_{\{I\}}$ is defined as 
\begin{equation}\label{Eq:Yprime}
	\mathcal{Y}^{\prime L}_{\{I\}}\equiv \frac{(-1)^L}{(m_3|\mathbf{p}_{1\mathrm{com}}|)^{L}} \sqrt{\frac{(2L+1)}{4\pi a_{L}}}\langle3|p_{1}|3]^{L}.
\end{equation}
with
\begin{equation}
	a_{L}= \prod_{k=1}^{L} \frac{2k}{2k - 1}, \quad a_{0}=1.
\end{equation}
Using Eq.~\eqref{Eq:little group indices I to sigma} to convert the indices of the orbital part $\mathcal{Y}^{\prime L}_{\{I\}}$, the spherical harmonic is obtained, which can be written as
\begin{equation}
    Y^{L}_{\sigma_L}=\sqrt{\frac{(2L)!}{(L+\sigma_L)!(L-\sigma_L)!}}\mathcal{Y}^{\prime L}_{\{I\}}.
\end{equation}
Under this choice, the factor $N$ reduces to 1. 

This allows a direct comparison with the traditional-$LS$ amplitude in Eq.~\eqref{Eq:canonical decay amplitude}, which can be written as 
\begin{equation}
    \begin{aligned}
        A_{\sigma_{3}}^{\sigma_{1}\sigma_{2}}(\mathbf{p}_{3},\mathbf{p}_{1},\mathbf{p}_{2};L,S) =&\underset{\text{little group}}{\underbrace{D_{\,\,\sigma_{1}}^{\sigma_{1}^{\prime}(s_1)*}(R_{31})\  D_{\,\,\sigma_{2}}^{\sigma_{2}^{\prime}(s_2)*}(R_{32})}}\underset{\text{The CGC of $s_{1}s_{2}\to S$}}{\underbrace{C_{S,\sigma_S}^{s_{1},\sigma_1^\prime;s_{2},\sigma_2^\prime}}}\\
        &\times\underset{\text{The CGC of $LS\to s_3$}}{\underbrace{C_{s_{3},\sigma_3}^{L,\sigma_L;S,\sigma_S}}} \quad\underset{\text{spherical harmonic}}{\underbrace{Y^{L}_{\sigma_L}(\Omega)}},
    \end{aligned}
\end{equation}
\begin{equation}
    \begin{aligned}
        \mathcal{A}_{\{I^{(3)}\}}^{\prime\{I^{(1)}\},\{I^{(2)}\}}(p_{3},p_{1},p_{2};L,S)=&\underset{\text{little group}}{\underbrace{(\tau^{\prime2s_{1}}_{1})_{\{J^{(1)}\}}^{\{I^{(1)}\}}(\tau^{\prime 2s_{2}}_{2})_{\{J^{(2)}\}}^{\{I^{(2)}\}}}}
        \underset{\text{The CGC of $s_{1}s_{2}\to S$}}{\underbrace{C_{S,\{K\}}^{s_{1},\{J^{(1)}\};s_{2},\{J^{(2)}\}}}}\\
        &\times\underset{\text{The CGC of $LS\to s_3$}}{\underbrace{C_{s_{3},\{I^{(3)}\}}^{S,\{K\};L,\{I\}}}}\quad
        \underset{\text{spherical harmonic}}{\underbrace{\mathcal{Y}^{\prime L}_{\{I\}}}}.
    \end{aligned}
\end{equation}
By comparing the two $LS$ partial wave amplitudes in any frame, one finds that the traditional-$LS$ amplitude is not manifestly Lorentz covariant. It requires boosting all momenta to the two body COM frame, evaluating the amplitude once in the COM frame, and then using little group rotations to transform the amplitude to any frame. By contrast, the canonical-spinor amplitude is defined in an arbitrary reference frame. It does not require boosting the event back to the COM frame, and the amplitude can be evaluated directly from the momenta in the chosen frame. 

Moreover, when returning to the COM frame, $\tau$ in the canonical-spinor construction becomes the identity $1$, and the spin part no longer contains momentum information, so it is fully equivalent to the traditional $LS$ partial waves. Meanwhile, in any frame, the covariant structure $\tau^{\prime I}_J$ ensures that the spin coupling is carried out within the same little group representation, and therefore the final result is consistent with the traditional-$LS$ amplitude.


\subsection{Comparison with the covariant projection tensor method}
\label{subsec:Covariant tensor method}

In subsection~\ref{subsec:Covariant tensor LS amplitude}, the covariant projection tensor method is presented in both the GS scheme and the PS scheme. Although both schemes lead to expressions that can be evaluated in any frame, neither scheme simultaneously achieves an explicit $LS$ separation and a manifestly covariant definition.
Therefore, we rewrite the  the covariant projection tensor method in spinor form to make the covariance and the $LS$  separation explicit.

In Ref.~\cite{Jing:2023rwz}, angular momentum coupling in the Lorentz representation is organized using irreducible representations $(s_L,s_R)$ of $\mathrm{SO}(3,1)$, where $s_L$ and $s_R$ are integers or half-integers.
For an IRREP $(s_L,s_R)$, the complex conjugate representation is $(s_R,s_L)$. 
Therefore, for a representation $[\ell]$ the following convention is adopted,
\begin{equation}\label{eq:self_conj_rep_label}
	[\ell]~\equiv~\left\{\begin{array}{cc}
		~(s_L,s_R)~ &\text{for $s_L=s_R$} \\
		\\
		~(s_L,s_R)\oplus(s_R,s_L)~ &\text{for $s_L \neq s_R$}
	\end{array}\right..
\end{equation}
For the spin wave function at the standard momentum $k^{\mu}=(m,\mathbf{0})$, consider a massive particle with spin-$s$. 
Its associated IRREP $[\ell]=(s_L,s_R)=[l]\otimes[r]$ is defined as
\begin{equation}\label{Eq:LR wave function}
\begin{aligned}
    u^{\ell}_{\sigma}({k};s)\equiv u^{\ell}_{\sigma}({k};[\ell],s)&= u^{lr}_{\sigma}({k};(s_L,s_R),s),\\
    \bar{u}_{\ell}^{\sigma}({k};s)\equiv \bar{u}_{\ell}^{\sigma}({k};[\ell],s)&= \bar{u}_{lr}^{\sigma}({k};(s_L,s_R),s),
\end{aligned}
\end{equation}
where $l=-s_L,-s_L+1,\cdots ,s_L$ and $r=-s_R,-s_R+1,\cdots ,s_R$ correspond to the IRREP indices of $\mathrm{SU}(2)_L$ and $\mathrm{SU}(2)_R$, respectively. 
Note that, in our convention, the placement of the indices on the wave functions $u$ and $\bar u$ differs from Ref.~\cite{Jing:2023rwz}.

The spin wave functions $u^{lr}_{\sigma}({k};(s_L,s_R),s)$ and $\bar{u}_{lr}^{\sigma}({k};(s_L,s_R),s)$ are essentially the CGCs\footnote{A more detailed derivation can be found in Section~VII of Chapter~V in Ref.~\cite{Weinberg_1995}.}, which can be written as
\begin{equation}\label{Eq:wavefunctionCGC}
    u^{lr}_{\sigma}({k};(s_L,s_R),s)=C^{s_L,\sigma_l;s_R,\sigma_r}_{s,\sigma},\quad \bar{u}_{lr}^{\sigma}({k};(s_L,s_R),s)=C_{s_L,\sigma_l;s_R,\sigma_r}^{s,\sigma}.
\end{equation}
The spin wave functions satisfy the orthonormality relation
\begin{equation}\label{Eq:orthonormal relation of wave function}
    \bar{u}^{\sigma_1}_{\ell}({k};s)u^{\ell}_{\sigma_2}({k};s)=\delta_{\sigma_1}^{\sigma_2}.
\end{equation}
From Eq.~\eqref{Eq:lorenz on wave function}, the spin wave function in any frame can be written as
\begin{equation}
\begin{aligned}
u^{lr}_{\sigma}({p};(s_L,s_R),s)=&D^{l}_{\,\,l^\prime}(L(\mathbf{p}))D^{r}_{\,\,r^\prime}(L(\mathbf{p}))u^{l^\prime r^\prime}_{\sigma}({k};(s_L,s_R),s),\\
\bar{u}_{lr}^{\sigma}({p};(s_L,s_R),s)=&D^{l^\prime}_{\,\,l}(L^{-1}(\mathbf{p}))D^{r^\prime}_{\,\,r}(L^{-1}(\mathbf{p}))\bar{u}_{l^\prime r^\prime}^{\sigma}({k};(s_L,s_R),s).
\end{aligned}
\end{equation}
From Eqs.~\eqref{Eq:boost on spinor} and \eqref{Eq:spinor canonical variables}, it follows that the spin wave functions in the left- and right-handed representations can be constructed from spinors. 
For spin-$\frac{1}{2}$, the spin wave functions can be written as
\begin{equation}
    u_{\sigma}^{l0}(p;(\frac{1}{2},0),\frac{1}{2})\to U^l{}_{\alpha}\lambda^\alpha_{I},\qquad u_{\sigma}^{0r}(p;(0,\frac{1}{2}),\frac{1}{2})\to \tilde{\lambda}^{\dot{\alpha}}_{I},
\end{equation}
\begin{equation}
    \bar{u}^{\sigma}_{0r}(p;(\frac{1}{2},0),\frac{1}{2})\to\tilde{\lambda}^{I}_{\dot{\alpha}},\qquad
    \bar{u}^{\sigma}_{l0}(p;(0,\frac{1}{2}),\frac{1}{2})\to -(U^{-1})^\alpha_{\,\,l}\lambda^{I}_{\alpha},
\end{equation}
where $U^{l}_{\,\,\alpha}$ is a constant matrix that implements the change of notation between the indices $(l,r)$ and the spinor indices $(\alpha,\dot{\alpha})$. 
In the $(l,r)$ notation, the four-dimensional Pauli matrices are $(\sigma_{\mu})^{l}{}_{r}$, whereas in the spinor notation they are denoted by $(\sigma_{\mu})_{\alpha\dot{\alpha}}$.
With this convention, one chooses
\begin{equation}
    U^l{}_{\alpha}=\begin{pmatrix}
        0 & -1\\
        1 & 0
    \end{pmatrix},\qquad
    (U^{-1})^\alpha_{\,\,l}=\begin{pmatrix}
    0 & 1\\
    -1 & 0
\end{pmatrix}.
\end{equation}
More generally, the spin wave functions with arbitrary spin-$s$ can be constructed from basic spinors and written as
\begin{equation}\label{Eq:spinor wave function}
	\bar{u}^{\{I\}}_{\ell}(p; s) \equiv \left(|p\rangle^{2s_L} [p|^{2s_R}\right)^{\{I\}}_{\ell}, \qquad
	u_{\{I\}}^{\ell}(p;s) \equiv  \left(|p]^{2s_R}\langle p|^{2s_L}\right)_{\{I\}}^{\ell},
\end{equation}
where $[s_L,s_R]$ denotes the irreducible representation of the Lorentz group. 
For simplicity, the constant matrices $U$ that relate the $(l,r)$ indices to the spinor indices $(\alpha,\dot\alpha)$, as well as the normalization factors of the spin wave functions, are omitted here. 


Starting from the definition of
the spin wave functions, all the spin wave functions are written as functions of the parent momentum and the spin quantum numbers of the individual particles, in Eq.~\eqref{Eq:spinor wave function}. The rewritten amplitude is:
\begin{equation}\label{Eq:spinor covariant LS amplitude}
\begin{aligned}
\mathcal{A}_{\{I_{3}\}}^{\{I_{1}\}\{I_{2}\}}(p_{3},p_{1},p_{2};L,S)\equiv&\Gamma_{\ell_{3}}^{\ell_{1}\ell_{2}}(p_{3},p_{1},p_{2};L,S) (\tau^{2s_1}_1)^{\{I_{1}\}}_{\{J\}} (\tau^{2s_2}_2)^{\{I_{2}\}}_{\{K\}}\times\\
&u_{\{I_{3}\}}^{\ell_{3}}(p_{3};s_{3})\bar{u}_{\ell_{1}}^{\{J\}}(p_3;s_{1})\bar{u}_{\ell_{2}}^{\{K\}}(p_3;s_{2}),
\end{aligned}
\end{equation}
where the factor $\tau^{\{I\}}_{\{J\}}$ can be obtained from Eq.~\eqref{Eq:define tau 1}. The role of $\tau^{\{I\}}_{\{J\}}$ is analogous to the little group rotation part in Eq.~\eqref{Eq:any frame covariant LS amplitude}.
For the coupling structure $\Gamma_{\ell_{3}}^{\ell_{1}\ell_{2}}$, its definition is the same as in Eq.~\eqref{Eq:Coupling structure}, and the covariant projection tensor can be written as
\begin{equation}
	\begin{aligned}
		P_{\ell_3}^{\ell_1\ell_2}(p_3; s_3, s_1, s_2)
		&=\bar{u}^{\{I_3\}}_{\ell_3}(p_3, s_3)u_{\{I_1\}}^{\ell_1}(p_3, s_1)u_{\{I_2\}}^{\ell_2}(p_3, s_2)
		C_{s_3,\{I_3\}}^{s_1,\{I_1\}; s_2,\{I_2\}} \\
		&= (|3\rangle^{2s_{3L}} [ 3|^{2s_{3R}})^{\{I_3\}}_{\ell_3}(|3]^{2s_{1R}} \langle3|^{2s_{1L}})_{\{I_1\}}^{\ell_1} (|3]^{2s_{2R}} \langle3|^{2s_{2L}})_{\{I_2\}}^{\ell_2}
		C_{s_3,\{I_3\}}^{s_1,\{I_1\};s_2,\{I_2\}},
	\end{aligned}
\end{equation}
where the pair $[s_{iL},s_{iR}]\ (i=1,2,3)$ are the Lorentz group irreducible representation of the $[\ell_i]$.

For the orbital part $t_{\ell_{L}}^{L}$, one has
\begin{equation}
	t_{\ell_L}^{L}(p_3; p_1) = \mathcal{Y}^{L}_{\{I\}}(p_3, p_1) \times (|3\rangle^L [3|^L)^{\{I\}}_{\ell_L},
\end{equation}
where $\mathcal{Y}^{L}_{\{I\}}$ has already been defined in Eq.~\eqref{Eq:spinor Y}.

Having established the explicit structures, the covariant projection tensor amplitude in Eq.~\eqref{Eq:any frame covariant LS amplitude} and the canonical-spinor amplitude in Eq.~\eqref{Eq:spinor covariant LS amplitude} can now be compared. 
In any frame, the two amplitudes can be written as
\begin{equation}
\begin{aligned}
	A_{\sigma_{3}}^{\sigma_{1} \sigma_{2}} (p_3, p_1, p_2; L, S)=& \Gamma_{\ell_3}^{\ell_1 \ell_2}(k_3, p_1^*, p_2^*; L, S)
	u^{\ell_3}_{\sigma_{3}}(k_3;s_3)\\
    &\times
	\underbrace{D^{\ell^\prime_1}_{\,\,\ell_1}(R_{31}^{-1}) \bar{u}_{\ell^\prime_1}^{\sigma_{1}}(k_1;s_1)}_{\text{LG + wave function}}\
	\underbrace{D^{\ell^\prime_2}_{\,\,\ell_2}(R_{32}^{-1}) \bar{u}_{\ell^\prime_2}^{\sigma_{2}}(k_2;s_2)}_{\text{LG + wave function}} ,\\[1em]
	\mathcal{A}_{\{I_3\}}^{\{I_1\} \{I_2\}}(p_3, p_1, p_2; L, S) =& \Gamma_{\ell_3}^{\ell_1 \ell_2}(p_3, p_1, p_2; L, S)
	u^{\ell_3}_{\{I_3\}}(p_3; s_3)\\
    &\times
	\underbrace{\tau_{1\{J\}}^{\{I_1\}} \bar{u}^{\{J\}}_{\ell_1}(p_3; s_1)}_{\text{LG + wave function}}\,
	\underbrace{\tau_{2\{K\}}^{\{I_2\}} \bar{u}^{\{K\}}_{\ell_2}(p_3; s_2)}_{\text{LG + wave function}}.
\end{aligned}
\end{equation}
By comparison, the canonical-spinor based covariant projection tensor amplitude can keep covariance explicit, and it does not require boosting the event back to the two body COM frame. Moreover, the covariant projection tensor amplitude rewritten in terms of the canonical-spinor variables is equivalent to the canonical-spinor amplitude. This equivalence originates from fully contracting all Lorentz indices in the covariant projection tensor amplitude rewritten with canonical-spinor variables, and the result of this contraction is exactly the canonical-spinor amplitude. This point also guarantees that, when it is returned to the COM frame, the orbital and spin parts are completely separated, and it is consistent with the traditional-$LS$ amplitude.


\subsection{Examples of canonical-spinor amplitude calculation}
\label{subsec:Examples of spinor-canonical amplitude calculation}
In this subsection, several explicit examples of canonical-spinor amplitude calculations are presented. Using Eqs.\eqref{Eq:spinor canonical variables} and \eqref{Eq:spinor contraction basic}, the following basic
spinor structures are obtained:
\begin{equation}
p_{\alpha\dot{\alpha}}=|p_{\alpha I}\rangle[p^I_{\dot{\alpha}}|,\quad p_{\dot{\alpha}\alpha}=|p^{\dot{\alpha}I}]\langle p_I^{\alpha}|,
\end{equation}
\begin{equation}
    \langle p_Iq^J\rangle=-\langle q^Jp_I\rangle,\quad [ p_Jq^I]=-[ q^Ip_J],
\end{equation}
\begin{equation}
	\langle p^Ip_J\rangle=m\delta^I_J,\quad [p_Jp^I]=m\delta^I_J,\quad|p_I\rangle\langle p^I|=m\delta^{\beta}_{\alpha},\quad |p^I][p_I|=m\delta^{\dot{\beta}}_{\dot{\alpha}},
\end{equation}
\begin{equation}
	p|p^I]=-m|p^I\rangle,\quad p|p_I\rangle=-m|p_I].
\end{equation}
For on-shell three-point amplitudes, the following conditions must be satisfied:
\begin{equation}
    p_3=p_1+p_2,\quad (p_i)_{\alpha\dot{\alpha}}(p_j)^{\dot{\alpha}\beta} + (p_j)_{\alpha\dot{\alpha}}(p_i)^{\dot{\alpha}\beta}=2(p_i\cdot p_j)\delta^{\beta}_{\alpha},\quad p_i^2=m_i^2\quad(i,j=1,2,3).
\end{equation}
Here we choose an arbitrary reference frame and define the canonical-spinor variables for particle-1, particle-2, and particle-3 as
\begin{equation}
\begin{aligned}
    |1_{I}\rangle &= \sqrt{m_1}\begin{pmatrix}
        c^{2}_1e^{\frac{\eta_1}{2}}+|s_1|^{2}e^{-\frac{\eta_1}{2}} & c_1s^*_1(e^{\frac{\eta_1}{2}}-e^{-\frac{\eta_1}{2}})\\
        c_1s_1(e^{\frac{\eta_1}{2}}-e^{-\frac{\eta_1}{2}}) & |s_1|^{2}e^{\frac{\eta_1}{2}}+c^{2}_1e^{-\frac{\eta_1}{2}}
    \end{pmatrix}_{\alpha I}=\begin{pmatrix}
        a_1 & b_1\\
        b^*_1 & c_1
    \end{pmatrix}_{\alpha I},\\
    [1^{I}| &= \sqrt{m_1}\begin{pmatrix}
            c_1^{2}e^{\frac{\eta_1}{2}}+|s_1|^{2}e^{-\frac{\eta_1}{2}} & c_1s^*_1(e^{\frac{\eta_1}{2}}-e^{-\frac{\eta_1}{2}})\\
            c_1s_1(e^{\frac{\eta_1}{2}}-e^{-\frac{\eta_1}{2}}) & |s_1|^{2}e^{\frac{\eta_1}{2}}+c_1^{2}e^{-\frac{\eta_1}{2}}
        \end{pmatrix}_{\,\,\dot{\alpha}}^{I}=\begin{pmatrix}
        a_1 & b_1\\
        b^*_1 & c_1
    \end{pmatrix}_{\,\,\dot{\alpha}}^{I},\\
    |2_{I}\rangle &= \sqrt{m_2}\begin{pmatrix}
        c^{2}_2e^{\frac{\eta_2}{2}}+|s_2|^{2}e^{-\frac{\eta_2}{2}} & c_2s^*_2(e^{\frac{\eta_2}{2}}-e^{-\frac{\eta_2}{2}})\\
        c_2s_2(e^{\frac{\eta_2}{2}}-e^{-\frac{\eta_2}{2}}) & |s_2|^{2}e^{\frac{\eta_2}{2}}+c^{2}_2e^{-\frac{\eta_2}{2}}
    \end{pmatrix}_{\alpha I}=\begin{pmatrix}
        a_2 & b_2\\
        b^*_2 & c_2
    \end{pmatrix}_{\alpha I},\\
    [2^{I}| &= \sqrt{m_2}\begin{pmatrix}
            c_2^{2}e^{\frac{\eta_2}{2}}+|s_2|^{2}e^{-\frac{\eta_2}{2}} & c_2s^*_2(e^{\frac{\eta_2}{2}}-e^{-\frac{\eta_2}{2}})\\
            c_2s_2(e^{\frac{\eta_2}{2}}-e^{-\frac{\eta_2}{2}}) & |s_2|^{2}e^{\frac{\eta_2}{2}}+c_2^{2}e^{-\frac{\eta_2}{2}}
        \end{pmatrix}_{\,\,\dot{\alpha}}^{I}=\begin{pmatrix}
        a_2 & b_2\\
        b^*_2 & c_2
    \end{pmatrix}_{\,\,\dot{\alpha}}^{I},\\
    |3_{I}\rangle &= \sqrt{m_3}\begin{pmatrix}
        c^{2}_3e^{\frac{\eta_3}{2}}+|s_3|^{2}e^{-\frac{\eta_3}{2}} & c_3s^*_3(e^{\frac{\eta_3}{2}}-e^{-\frac{\eta_3}{2}})\\
        c_3s_3(e^{\frac{\eta_3}{2}}-e^{-\frac{\eta_3}{2}}) & |s_3|^{2}e^{\frac{\eta_3}{2}}+c^{2}_3e^{-\frac{\eta_3}{2}}
    \end{pmatrix}_{\alpha I}=\begin{pmatrix}
        a_3 & b_3\\
        b^*_3 & c_3
    \end{pmatrix}_{\alpha I},\\
    [3^{I}| &= \sqrt{m_3}\begin{pmatrix}
            c_3^{2}e^{\frac{\eta_3}{2}}+|s_3|^{2}e^{-\frac{\eta_3}{2}} & c_3s^*_3(e^{\frac{\eta_3}{2}}-e^{-\frac{\eta_3}{2}})\\
            c_3s_3(e^{\frac{\eta_3}{2}}-e^{-\frac{\eta_3}{2}}) & |s_3|^{2}e^{\frac{\eta_3}{2}}+c_3^{2}e^{-\frac{\eta_3}{2}}
        \end{pmatrix}_{\,\,\dot{\alpha}}^{I}=\begin{pmatrix}
        a_3 & b_3\\
        b^*_3 & c_3
    \end{pmatrix}_{\,\,\dot{\alpha}}^{I},
\end{aligned}
\end{equation}
where $c=\cos\frac{\theta_j}{2},s=e^{i\phi_j}\sin\frac{\theta_j}{2} ,\eta_j =\text{arctanh}{\frac{|\mathbf{p}_j|}{E_j}}\quad(j=1,2,3)$. Using the antisymmetric tensors $\varepsilon_{\alpha\beta}$ and $\varepsilon_{IJ}$ (and their inverses) to raise and lower Lorentz and little group indices, the remaining spinors follow straightforwardly from the above definitions.

As an illustration, the amplitude in Eq.~\eqref{Eq:amplitude example} can be written in explicit form as
\begin{equation}
\begin{aligned}
        \mathcal{A}_{\{I^{(3)}_{1}\}}^{\{I^{(1)}_{1}\}}(0,\frac{1}{2})\propto&\langle1^{I^{(1)}_{1}}3_{I^{(3)}_{1}}\rangle-[1^{I^{(1)}_{1}}3_{I^{(3)}_{1}}]\\
        =&\begin{pmatrix}
        c_1 & -b_1\\
        -b^*_1 & a_1
    \end{pmatrix}\begin{pmatrix}
        a_3 & b_3\\
        b^*_3 & c_3
    \end{pmatrix}-\begin{pmatrix}
        a_1 & b_1\\
        b^*_1 & c_1
    \end{pmatrix}\begin{pmatrix}
        -c_3 & b_3\\
        b^*_3 & -a_3
    \end{pmatrix}\\
    =&\begin{pmatrix}
        c_1a_3+a_1c_3-2b_1b_3^* & c_1b_3-b_1c_3-a_1b_3+b_1a_3\\
        a_1b^*_3-b_1^*a_3+b_1^*c_3-c_1b_3^* &a_1c_3+c_1a_3-2b_1^*b_3
    \end{pmatrix}^{I_1^{(1)}}_{\,\,I_1^{(3)}}.
\end{aligned}
\end{equation}
In what follows, canonical-spinor variables are denoted in \textbf{BOLD} for brevity, and the $\mathrm{SU}(2)$ little group indices are suppressed, since they are totally symmetric and would make the notation lengthy.

\begin{enumerate}
\item $s_3=\frac{1}{2},s_1=\frac{1}{2},s_2=0$. There are two $(L,S)$ combinations: $(0,\frac{1}{2})$ and $(1,\frac{1}{2})$.
\begin{enumerate}
\item $L=0,S=\frac{1}{2}:$
\begin{equation}
	\mathcal{A}_{\{I^{(3)}_{1}\}}^{\{I^{(1)}_{1}\}}(0,\frac{1}{2})=\mathcal{S}_{\{K_1\}}^{\{I^{(1)}_{1}\}}\mathcal{Y}^{0}C_{\frac{1}{2},\{I^{(3)}_{1}\}}^{\frac{1}{2},\{K_1\};0},
\end{equation}
where
\begin{equation}\label{Eq:S 1/2 to 1/2 0}
\begin{aligned}
    \mathcal{S}_{\{K_1\}}^{\{I^{(1)}_{1}\}} &=\frac{1}{\sqrt{m_3(E_{1\mathrm{com}}+m_1)}}(\tau_{1}^{1})_{\{J_{1}\}}^{\{I_{1}^{(1)}\}}C_{\frac{1}{2},\{K_{1}\}}^{\frac{1}{2},\{J_1\};0}\\
    &=\frac{1}{\sqrt{m_3(E_{1\mathrm{com}}+m_1)}}(\langle\mathbf{13}\rangle-[\mathbf{13}]),
\end{aligned}
\end{equation}
and
\begin{equation}\label{Eq:Y=0}
    \mathcal{Y}^{0}=\sqrt{\frac{m_{3}}{\pi|\mathbf{p}_{1\mathrm{com}}|}}.
\end{equation}
The amplitude is
\begin{equation}\label{Eq:amplitude example}
\begin{aligned}
    \mathcal{A}_{\{I^{(3)}_{1}\}}^{\{I^{(1)}_{1}\}}(0,\frac{1}{2})=\sqrt{\frac{m_{3}}{\pi|\mathbf{p}_{1\mathrm{com}}|}}\frac{1}{\sqrt{m_3(E_{1\mathrm{com}}+m_1)}}(\langle\mathbf{13}\rangle-[\mathbf{13}]).
\end{aligned}
\end{equation}
\item $L=1,S=\frac{1}{2}:$
\begin{equation}
	\mathcal{A}_{\{I^{(3)}_{1}\}}^{\{I^{(1)}_{1}\}}(1,\frac{1}{2})=\mathcal{S}_{\{K_1\}}^{\{I^{(1)}_{1}\}}\mathcal{Y}^{1}_{\{L_1L_2\}}C_{\frac{1}{2},\{I^{(3)}_{1}\}}^{\frac{1}{2},\{K_1\};1,\{L_1L_2\}},
\end{equation}
where $\mathcal{S}_{\{K_1\}}^{\{I^{(1)}_{1}\}}$ can be obtained from Eq.~\eqref{Eq:S 1/2 to 1/2 0} and
\begin{equation}\label{Eq:Y=1}
\begin{aligned}
        \mathcal{Y}^{1}_{\{L_{1}L_{2}\}}=\sqrt{\frac{3}{2\pi|\mathbf{p}_{1\mathrm{com}}|^{3}m_{3}}}\langle\mathbf{3}|p_1|\mathbf{3}].
    \end{aligned}
\end{equation}
The amplitude is
\begin{equation}
\begin{aligned}
    \mathcal{A}_{\{I^{(3)}_{1}\}}^{\{I^{(1)}_{1}\}}(1,\frac{1}{2})=&\sqrt{\frac{3}{2\pi|\mathbf{p}_{1\mathrm{com}}|^{3}m_{3}}}\frac{(m_1+m_2+m_3)(m_3+m_1-m_2)}{2\sqrt{m_3(E_{1\mathrm{com}}+m_1)}}\sqrt{\frac{2}{3}}\\
    &\times(\langle\mathbf{13}\rangle+[\mathbf{13}]).
\end{aligned}
\end{equation}
\end{enumerate}
\item $s_3=0,s_1=\frac{1}{2},s_2=\frac{1}{2}$. There are two $(L,S)$ combinations: $(0,0)$ and $(1,1)$.
\begin{enumerate}
\item $L=0,S=0:$
\begin{equation}
	\mathcal{A}^{\{I^{(1)}_{1}\},\{I^{(2)}_{1}\}}(0,0)=\mathcal{S}^{\{I^{(1)}_{1}\},\{I^{(2)}_{1}\}}\mathcal{Y}^{0}C_{0}^{0;0},
\end{equation}
where $\mathcal{Y}^{0}$ can be obtained from Eq.~\eqref{Eq:Y=0} and
\begin{equation}\label{Eq:S 0 to 1/2 1/2}
\begin{aligned}
    \mathcal{S}^{\{I^{(1)}_{1}\},\{I^{(2)}_{1}\}}&=\frac{1}{\sqrt{m_3(E_{1\mathrm{com}}+m_1)m_3(E_{2\mathrm{com}}+m_2)}}(\tau_{1}^{1})_{\{J_{1}\}}^{\{I_{1}^{(1)}\}}(\tau_{2}^{1})_{\{N_{1}\}}^{\{I_{1}^{(2)}\}}C_{0}^{\frac{1}{2},\{J_1\};\frac{1}{2},\{N_1\}}\\
    &=\sqrt{\frac{1}{2}}\frac{m_1+m_2+m_3}{\sqrt{m_3(E_{1\mathrm{com}}+m_1)m_3(E_{2\mathrm{com}}+m_2)}}(\langle\mathbf{12}\rangle-[\mathbf{12}]).
\end{aligned}
\end{equation}

The amplitude is
\begin{equation}
\begin{aligned}
    \mathcal{A}^{\{I^{(1)}_{1}\},\{I^{(2)}_{1}\}}(0,0)=&\sqrt{\frac{m_{3}}{\pi|\mathbf{p}_{1\mathrm{com}}|}}\sqrt{\frac{1}{2}}\frac{m_1+m_2+m_3}{\sqrt{m_3(E_{1\mathrm{com}}+m_1)m_3(E_{2\mathrm{com}}+m_2)}}\\
    &\times(\langle\mathbf{12}\rangle-[\mathbf{12}]).
\end{aligned}
\end{equation}
\item $L=1,S=1:$
\begin{equation}
	\mathcal{A}^{\{I^{(1)}_{1}\},\{I^{(2)}_{1}\}}(1,1)=\mathcal{S}^{\{I^{(1)}_{1}\},\{I^{(2)}_{1}\}}_{\{K_1K_2\}}\mathcal{Y}^{1}_{\{L_1L_2\}}C_{0}^{1,\{K_1K_2\};1,\{L_1L_2\}},
\end{equation}
where $\mathcal{Y}^{1}_{\{L_1L_2\}}$ can be obtained from Eq.~\eqref{Eq:Y=1} and 
\begin{equation}\label{Eq:S 1 to 1/2 1/2}
    \begin{aligned}
    \mathcal{S}^{\{I^{(1)}_{1}\},\{I^{(2)}_{1}\}}_{\{K_1K_2\}}&=\frac{1}{\sqrt{m_3(E_{1\mathrm{com}}+m_1)m_3(E_{2\mathrm{com}}+m_2)}}(\tau_{1}^{1})_{\{J_{1}\}}^{\{I_{1}^{(1)}\}}(\tau_{2}^{1})_{\{N_{1}\}}^{\{I_{1}^{(2)}\}}C_{1,\{K_1K_2\}}^{\frac{1}{2},\{J_1\};\frac{1}{2},\{N_1\}}\\
    &=\frac{1}{\sqrt{m_3(E_{1\mathrm{com}}+m_1)m_3(E_{2\mathrm{com}}+m_2)}}(\langle\mathbf{13}\rangle-[\mathbf{13}])(\langle\mathbf{23}\rangle-[\mathbf{23}]).
\end{aligned}
\end{equation}

The amplitude is
\begin{equation}
\begin{aligned}
    \mathcal{A}^{\{I^{(1)}_{1}\},\{I^{(2)}_{1}\}}(1,1)=&\sqrt{\frac{3}{2\pi|\mathbf{p}_{1\mathrm{com}}|^{3}m_{3}}}\frac{-(m_1+m_2+m_3)(m_3^2-(m_1-m_2)^2)}{\sqrt{m_3(E_{1\mathrm{com}}+m_1)m_3(E_{2\mathrm{com}}+m_2)}}\sqrt{\frac{1}{3}}\\
    &\times(\langle\mathbf{12}\rangle+[\mathbf{12}]).
\end{aligned}
\end{equation}
\end{enumerate}
\item $s_3=1,s_1=0,s_2=0$. There is one $(L,S)$ combinations: $(1,0)$.
\begin{enumerate}
\item $L=1,S=0:$
\begin{equation}
	\mathcal{A}_{\{I^{(3)}_{1}I^{(3)}_{2}\}}(1,0)=\mathcal{Y}^{1}_{\{L_1L_2\}}C_{1,\{I^{(3)}_{1}I^{(3)}_{2}\}}^{0;1,\{L_1L_2\}},
\end{equation}
where $\mathcal{Y}^{1}_{\{L_1L_2\}}$ can be obtained from Eq.~\eqref{Eq:Y=1}.

The amplitude is
\begin{equation}
\begin{aligned}
    \mathcal{A}_{\{I^{(3)}_{1}I^{(3)}_{2}\}}(1,0)=&\sqrt{\frac{3}{2\pi|\mathbf{p}_{1\mathrm{com}}|^{3}m_{3}}}\langle\mathbf{3}|p_1|\mathbf{3}].
\end{aligned}
\end{equation}
\end{enumerate}
\item $s_3=0,s_1=1,s_2=0$. There is one $(L,S)$ combinations: $(1,1)$.
\begin{enumerate}
\item $L=1,S=1:$
\begin{equation}
	\mathcal{A}^{\{I^{(1)}_{1}I^{(1)}_{2}\}}(1,1)=\mathcal{S}^{\{I^{(1)}_{1}I^{(1)}_{2}\}}_{\{K_1K_2\}}\mathcal{Y}^{1}_{\{L_1L_2\}}C_{0}^{1,\{K_1K_2\};1,\{L_1L_2\}},
\end{equation}
where $\mathcal{Y}^{1}_{\{L_1L_2\}}$ can be obtained from Eq.~\eqref{Eq:Y=1} and 
\begin{equation}\label{Eq:S 1 to 1 0}
\begin{aligned}
    \mathcal{S}^{\{I^{(1)}_{1}I^{(1)}_{2}\}}_{\{K_1K_2\}}&=\frac{1}{m_3(E_{1\mathrm{com}}+m_1)}(\tau_{1}^{2})_{\{J_{1}J_2\}}^{\{I_{1}^{(1)}I_2^{(1)}\}}C_{1,\{K_1K_2\}}^{1,\{J_1J_2\};0}\\
    &=\frac{1}{m_3(E_{1\mathrm{com}}+m_1)}(\langle\mathbf{13}\rangle-[\mathbf{13}])^2.
\end{aligned}
\end{equation}

The amplitude is
\begin{equation}
\begin{aligned}
    \mathcal{A}^{\{I^{(1)}_{1}I^{(1)}_{2}\}}(1,1)=&\sqrt{\frac{3}{2\pi|\mathbf{p}_{1\mathrm{com}}|^{3}m_{3}}}\frac{-(m_1+m_2+m_3)(m_3+m_1-m_2)}{m_3(E_{1\mathrm{com}}+m_1)}\sqrt{\frac{1}{3}}\langle\mathbf{1}|p_3|\mathbf{1}].
\end{aligned}
\end{equation}
\end{enumerate}
\item $s_3=1,s_1=1,s_2=0$. There are three $(L,S)$ combinations: $(0,1)$, $(1,1)$ and $(2,1)$.
\begin{enumerate}
\item $L=0,S=1:$
\begin{equation}
	\mathcal{A}^{\{I^{(1)}_{1}I_2^{(1)}\}}_{\{I^{(3)}_1I^{(3)}_2\}}(0,1)=\mathcal{S}^{\{I^{(1)}_{1}I^{(1)}_{2}\}}_{\{K_1K_2\}}\mathcal{Y}^{0}C_{1,\{I^{(3)}_1I^{(3)}_2\}}^{1,\{K_1K_2\};0},
\end{equation}
where $\mathcal{S}^{\{I^{(1)}_{1}I^{(1)}_{2}\}}_{\{K_1K_2\}}$ and $\mathcal{Y}^{0}$ can be obtained from Eqs.~\eqref{Eq:S 1 to 1 0} and \eqref{Eq:Y=0}, respectively.

The amplitude is
\begin{equation}
\begin{aligned}
    \mathcal{A}^{\{I^{(1)}_{1}I_2^{(1)}\}}_{\{I^{(3)}_1I^{(3)}_2\}}(0,1)=&\sqrt{\frac{m_{3}}{\pi|\mathbf{p}_{1\mathrm{com}}|}}\frac{1}{m_3(E_{1\mathrm{com}}+m_1)}(\langle\mathbf{13}\rangle-[\mathbf{13}])^2.
\end{aligned}
\end{equation}
\item $L=1,S=1:$
\begin{equation}
	\mathcal{A}^{\{I^{(1)}_{1}I_2^{(1)}\}}_{\{I^{(3)}_1I^{(3)}_2\}}(1,1)=\mathcal{S}^{\{I^{(1)}_{1}I^{(1)}_{2}\}}_{\{K_1K_2\}}\mathcal{Y}^{1}_{\{L_1L_2\}}C_{1,\{I^{(3)}_1I^{(3)}_2\}}^{1,\{K_1K_2\};1,\{L_1L_2\}},
\end{equation}
where $\mathcal{S}^{\{I^{(1)}_{1}I^{(1)}_{2}\}}_{\{K_1K_2\}}$ and $\mathcal{Y}^{1}_{\{L_1L_2\}}$ can be obtained from Eqs.~\eqref{Eq:S 1 to 1 0} and \eqref{Eq:Y=1}, respectively.

The amplitude is
\begin{equation}
\begin{aligned}
    \mathcal{A}^{\{I^{(1)}_{1}I_2^{(1)}\}}_{\{I^{(3)}_1I^{(3)}_2\}}(1,1)=&\sqrt{\frac{3}{2\pi|\mathbf{p}_{1\mathrm{com}}|^{3}m_{3}}}\frac{(m_1+m_2+m_3)(m_3+m_1-m_2)}{2m_3(E_{1\mathrm{com}}+m_1)}\\
    &\times(\langle\mathbf{13}\rangle^2-[\mathbf{13}]^2).
\end{aligned}
\end{equation}
\item $L=2,S=1:$
\begin{equation}
	\mathcal{A}^{\{I^{(1)}_{1}I_2^{(1)}\}}_{\{I^{(3)}_1I^{(3)}_2\}}(2,1)=\mathcal{S}^{\{I^{(1)}_{1}I^{(1)}_{2}\}}_{\{K_1K_2\}}\mathcal{Y}^{2}_{\{L_1L_2L_3L_4\}}C_{1,\{I^{(3)}_1I^{(3)}_2\}}^{1,\{K_1K_2\};2,\{L_1L_2L_3L_4\}},
\end{equation}
where $\mathcal{S}^{\{I^{(1)}_{1}I^{(1)}_{2}\}}_{\{K_1K_2\}}$ can be obtained from Eq.~\eqref{Eq:S 1 to 1 0} and
\begin{equation}\label{Eq:Y=2}
    \begin{aligned}
        \mathcal{Y}^2_{\{L_1L_2L_3L_4\}}=\sqrt{\frac{15}{8\pi|\mathbf{p}_{1\mathrm{com}}|^5m_3^3}}(\langle\mathbf{3}|p_1|\mathbf{3}])^2.
    \end{aligned}
\end{equation}

The amplitude is
\begin{equation}
\begin{aligned}
    \mathcal{A}^{\{I^{(1)}_{1}I_2^{(1)}\}}_{\{I^{(3)}_1I^{(3)}_2\}}(2,1)=&\sqrt{\frac{15}{8\pi|\mathbf{p}_{1\mathrm{com}}|^5m_3^3}}\frac{(m_1+m_2+m_3)(m_3+m_1-m_2)}{\sqrt{m_3(E_{1\mathrm{com}}+m_1)m_3(E_{2\mathrm{com}}+m_2)}}\sqrt{\frac{3}{5}}\\
    &\times\bigg(\frac{(m_1+m_2+m_3)(m_3+m_1-m_2)}{4}(\langle\mathbf{13}\rangle+[\mathbf{13}])^2\\
    &-\langle\mathbf{1}|p_3|\mathbf{1}]\langle\mathbf{3}|p_1|\mathbf{3}]\bigg).
\end{aligned}
\end{equation}
\end{enumerate}
\item $s_3=0,s_1=1,s_2=1$. There are three $(L,S)$ combinations: $(0,0)$, $(1,1)$ and $(2,2)$.
\begin{enumerate}
\item $L=0,S=0:$
\begin{equation}
	\mathcal{A}^{\{I^{(1)}_{1}I_2^{(1)}\},\{I^{(2)}_1I^{(2)}_2\}}(0,0)=\mathcal{S}^{\{I^{(1)}_{1}I_2^{(1)}\},\{I^{(2)}_1I^{(2)}_2\}}\mathcal{Y}^{0}C_{0}^{0;0},
\end{equation}
where $\mathcal{Y}^{0}$ can be obtained from Eq.~\eqref{Eq:Y=0}.
\begin{equation}\label{Eq:S 0 to 1 1}
    \begin{aligned}
        \mathcal{S}^{\{I^{(1)}_{1}I_2^{(1)}\},\{I^{(2)}_1I^{(2)}_2\}}=&\frac{1}{(m_3(E_{1\mathrm{com}}+m_1))(m_3(E_{1\mathrm{com}}+m_1))}(\tau^2_1)^{\{I^{(1)}_{1}I_2^{(1)}\}}_{\{J_1J_2\}}(\tau^2_2)^{\{I^{(2)}_{1}I_2^{(2)}\}}_{\{N_1N_2\}}\\
        &\times C^{1,\{J_1J_2\};1,\{N_1N_2\}}_0\\
        =&\sqrt{\frac{1}{3}}\frac{(m_1+m_2+m_3)^2}{(m_3(E_{1\mathrm{com}}+m_1))(m_3(E_{1\mathrm{com}}+m_1))}(\langle\mathbf{12}\rangle-[\mathbf{12}])^2.
    \end{aligned}
\end{equation}

The amplitude is
\begin{equation}
\begin{aligned}
    \mathcal{A}^{\{I^{(1)}_{1}I_2^{(1)}\},\{I^{(2)}_1I^{(2)}_2\}}(0,0)=&\sqrt{\frac{m_{3}}{\pi|\mathbf{p}_{1\mathrm{com}}|}}
    \sqrt{\frac{1}{3}}\frac{(m_1+m_2+m_3)^2}{(m_3(E_{1\mathrm{com}}+m_1))(m_3(E_{1\mathrm{com}}+m_1))}(\langle\mathbf{12}\rangle-[\mathbf{12}])^2.
\end{aligned}
\end{equation}
\item $L=1,S=1:$
\begin{equation}
	\mathcal{A}^{\{I^{(1)}_{1}I_2^{(1)}\},\{I^{(2)}_1I^{(2)}_2\}}(1,1)=\mathcal{S}^{\{I^{(1)}_{1}I_2^{(1)}\},\{I^{(2)}_1I^{(2)}_2\}}_{\{K_1K_2\}}\mathcal{Y}^{1}_{\{L_1L_2\}}C_{0}^{1,\{K_1K_2\};1,\{L_1L_2\}},
\end{equation}
where $\mathcal{Y}^{1}_{\{L_1L_2\}}$ can be obtained from Eq.~\eqref{Eq:Y=1} and 
\begin{equation}\label{Eq:S 1 to 1 1}
    \begin{aligned}
        \mathcal{S}^{\{I^{(1)}_{1}I_2^{(1)}\},\{I^{(2)}_1I^{(2)}_2\}}_{\{K_1K_2\}}=&\frac{1}{(m_3(E_{1\mathrm{com}}+m_1))(m_3(E_{1\mathrm{com}}+m_1))}(\tau^2_1)^{\{I^{(1)}_{1}I_2^{(1)}\}}_{\{J_1J_2\}}(\tau^2_2)^{\{I^{(2)}_{1}I_2^{(2)}\}}_{\{N_1N_2\}}\\
        &\times C^{1,\{J_1J_2\};1,\{N_1N_2\}}_{1,\{K_1K_2\}}\\
        =&\frac{(m_1+m_2+m_3)}{(m_3(E_{1\mathrm{com}}+m_1))(m_3(E_{1\mathrm{com}}+m_1))}\\
        &\times(\langle\mathbf{12}\rangle-[\mathbf{12}])(\langle\mathbf{13}\rangle-[\mathbf{13}])(\langle\mathbf{23}\rangle-[\mathbf{23}]).
    \end{aligned}
\end{equation}

The amplitude is
\begin{equation}
\begin{aligned}
    \mathcal{A}^{\{I^{(1)}_{1}I_2^{(1)}\},\{I^{(2)}_1I^{(2)}_2\}}(1,1)=&\sqrt{\frac{3}{2\pi|\mathbf{p}_{1\mathrm{com}}|^{3}m_{3}}}\frac{-(m_1+m_2+m_3)^2(m_3^2-(m_1-m_2)^2)}{(m_3(E_{1\mathrm{com}}+m_1))(m_3(E_{1\mathrm{com}}+m_1))}\sqrt{\frac{1}{3}}\\
    &\times(\langle\mathbf{12}\rangle^2-[\mathbf{12}]^2).
\end{aligned}
\end{equation}

\item $L=2,S=2:$
\begin{equation}
	\mathcal{A}^{\{I^{(1)}_{1}I_2^{(1)}\},\{I^{(2)}_1I^{(2)}_2\}}(2,2)=\mathcal{S}^{\{I^{(1)}_{1}I_2^{(1)}\},\{I^{(2)}_1I^{(2)}_2\}}_{\{K_1K_2K_3K_4\}}\mathcal{Y}^{2}_{\{L_1L_2L_3L_4\}}C_{0}^{2,\{K_1K_2K_3K_4\};2,\{L_1L_2L_3L_4\}},
\end{equation}
where $\mathcal{Y}^2_{\{L_1L_2L_3L_4\}}$ can be obtained from Eq.~\eqref{Eq:Y=2} and
\begin{equation}\label{Eq:S 2 to 1 1}
    \begin{aligned}
        \mathcal{S}^{\{I^{(1)}_{1}I_2^{(1)}\},\{I^{(2)}_1I^{(2)}_2\}}_{\{K_1K_2K_3K_4\}}=&\frac{1}{(m_3(E_{1\mathrm{com}}+m_1))(m_3(E_{1\mathrm{com}}+m_1))}(\tau^2_1)^{\{I^{(1)}_{1}I_2^{(1)}\}}_{\{J_1J_2\}}(\tau^2_2)^{\{I^{(2)}_{1}I_2^{(2)}\}}_{\{N_1N_2\}}\\
        &\times C^{1,\{J_1J_2\};1,\{N_1N_2\}}_{2,\{K_1K_2K_3K_4\}}\\
        =&\frac{1}{(m_3(E_{1\mathrm{com}}+m_1))(m_3(E_{1\mathrm{com}}+m_1))}\\
        &\times(\langle\mathbf{13}\rangle-[\mathbf{13}])^2(\langle\mathbf{23}\rangle-[\mathbf{23}])^2.
    \end{aligned}
\end{equation}

The amplitude is
\begin{equation}
\begin{aligned}
    \mathcal{A}^{\{I^{(1)}_{1}I_2^{(1)}\},\{I^{(2)}_1I^{(2)}_2\}}(2,2)=&\sqrt{\frac{15}{8\pi|\mathbf{p}_{1\mathrm{com}}|^5m_3^3}}\frac{(m_1+m_2+m_3)^2(m_3^2-(m_1-m_2)^2)}{(m_3(E_{1\mathrm{com}}+m_1))(m_3(E_{1\mathrm{com}}+m_1))}\sqrt{\frac{1}{5}}\\
    &\times\bigg(((m_1-m_2)^2-m_3^2)(\langle\mathbf{12}\rangle+[\mathbf{12}])^2+\langle\mathbf{1}|3|\mathbf{1}]\langle\mathbf{2}|3|\mathbf{2}]\bigg).\\
\end{aligned}
\end{equation}
\end{enumerate}

\item $s_{3}=\frac{1}{2}$, $s_{1}=\frac{3}{2}$, $s_{2}=0$. There are two $(L,S)$ combinations: $(1,\frac{3}{2})$ and $(2,\frac{3}{2})$.
\begin{enumerate}
    \item $L={1},S=\frac{3}{2}:$
\begin{equation}
    \mathcal{A}_{\{I^{(3)}_1\}}^{\{I^{(1)}_{1}I^{(1)}_{2}I^{(1)}_{3}\}}(1,\frac{3}{2})=\mathcal{S}_{\{K_1K_2K_3\}}^{\{I^{(1)}_{1}I^{(1)}_{2}I^{(1)}_{3}\}}\mathcal{Y}^{1}_{\{L_1L_2\}}C_{\frac{1}{2},\{I_{1}^{(3)}\}}^{\frac{3}{2},\{K_{1}K_{2}K_{3}\};1,\{L_{1}L_{2}\}}.
\end{equation}
where $\mathcal{Y}^{1}_{\{L_1L_2\}}$ can be obtained from Eq.~\eqref{Eq:Y=1}
\begin{equation}\label{Eq:S 3/2 to 3/2 0}
    \begin{aligned}
        \mathcal{S}_{\{K_1K_2K_3\}}^{\{I^{(1)}_{1}I^{(1)}_{2}I^{(1)}_{3}\}}=&\frac{1}{(m_3(E_{1\mathrm{com}}+m_1))^{\frac{3}{2}}}(\tau^{3}_1)_{\{J_1J_2J_3\}}^{\{I^{(1)}_{1}I^{(1)}_{2}I^{(1)}_{3}\}}C_{\frac{3}{2},\{K_1K_2K_3\}}^{\frac{3}{2},\{J_1J_2J_3\};0}\\
        =&\frac{1}{(m_3(E_{1\mathrm{com}}+m_1))^{\frac{3}{2}}}(\langle\mathbf{13}\rangle-[\mathbf{13}])^{3}.
    \end{aligned}
\end{equation}
The amplitude is
\begin{equation}
    \begin{aligned}
        \mathcal{A}_{\{I^{(3)}_1\}}^{\{I^{(1)}_{1}I^{(1)}_{2}I^{(1)}_{3}\}}(1,\frac{3}{2})=&
        \sqrt{\frac{3}{2\pi|\mathbf{p}_{1\mathrm{com}}|^{3}m_{3}}}\frac{-(m_1+m_2+m_3)(m_3+m_1-m_2)}{(m_3(E_{1\mathrm{com}}+m_1))^{\frac{3}{2}}}\sqrt{\frac{1}{2}}\\
        &\times(\langle\mathbf{13}\rangle-[\mathbf{13}])\langle\mathbf{1}|p_3|\mathbf{1}]
    \end{aligned}.
\end{equation}
\item $L={2},S=\frac{3}{2}:$
\begin{equation}
    \mathcal{A}_{\{I^{(3)}_1\}}^{\{I^{(1)}_{1}I^{(1)}_{2}I^{(1)}_{3}\}}(2,\frac{3}{2})=\mathcal{S}_{\{K_1K_2K_3\}}^{\{I^{(1)}_{1}I^{(1)}_{2}I^{(1)}_{3}\}}\mathcal{Y}^{2}_{\{L_1L_2L_3L_4\}}C_{\frac{1}{2},\{I_{1}^{(3)}\}}^{\frac{3}{2},\{K_{1}K_{2}K_{3}\};2,\{L_{1}L_{2}L_3L_4\}}.
\end{equation}
where $\mathcal{S}_{\{K_1K_2K_3\}}^{\{I^{(1)}_{1}I^{(1)}_{2}I^{(1)}_{3}\}}$ and $\mathcal{Y}^2_{\{L_1L_2L_3L_4\}}$ can be obtained from Eq.~\eqref{Eq:S 3/2 to 3/2 0} and \eqref{Eq:Y=2}, respectively.

The amplitude is
\begin{equation}
    \begin{aligned}
        \mathcal{A}_{\{I^{(3)}_1\}}^{\{I^{(1)}_{1}I^{(1)}_{2}I^{(1)}_{3}\}}(2,\frac{3}{2})=&
        \sqrt{\frac{15}{8\pi|\mathbf{p}_{1\mathrm{com}}|^{5}m^3_{3}}}\frac{-(m_1+m_2+m_3)^2(m_3+m_1-m_2)^2}{2(m_3(E_{1\mathrm{com}}+m_1))^{\frac{3}{2}}}\sqrt{\frac{2}{5}}\\
        &\times(\langle\mathbf{13}\rangle+[\mathbf{13}])\langle\mathbf{1}|p_3|\mathbf{1}]
    \end{aligned}.
\end{equation}
\end{enumerate}

\item $s_{3}=\frac{1}{2},s_{1}=\frac{1}{2},s_{2}=1$. There are four $(L,S)$ combinations: $(0,\frac{1}{2})$, $(1,\frac{1}{2})$, $(1,\frac{3}{2})$ and $(2,\frac{3}{2})$.
\begin{enumerate}
    \item $L={0},S=\frac{1}{2}$
\begin{equation}
    \mathcal{A}_{\{I^{(3)}_{1}\}}^{\{I^{(1)}_{1}\},\{I^{(2)}_{1}I^{(2)}_{2}\}}(0,\frac{1}{2})=\mathcal{S}_{\{K_{1}\}}^{\{I^{(1)}_{1}\},\{I^{(2)}_{1}I^{(2)}_{2}\}}\mathcal{Y}^{0}C_{\frac{1}{2},\{I^{(3)}_{1}\}}^{\frac{1}{2},\{K_{1}\};0},
\end{equation}
where $\mathcal{Y}^0$ can be obtained from Eq.~\eqref{Eq:Y=0} and
\begin{equation}\label{Eq:S 1/2 to 1/2 1}
\begin{aligned}
        \mathcal{S}_{\{K_{1}\}}^{\{I^{(1)}_{1}\},\{I^{(2)}_{1}I^{(2)}_{2}\}}=&\frac{1}{\sqrt{m_{3}(E_{1\mathrm{com}}+m_{1})(m_{3}(E_{2\mathrm{com}}+m_{2}))^{2}}}(\tau^{1}_{1})_{\{J_1\}}^{\{I^{(1)}_{1}\}}(\tau^{2}_{2})_{\{N_{1}N_{2}\}}^{\{I^{(2)}_{1}I^{(2)}_{2}\}}\\
        &\times C_{\frac{1}{2},\{K_1\}}^{\frac{1}{2},\{J_{1}\};1,\{N_{1}N_{2}\}}\\
        =&\sqrt{\frac{2}{3}}\frac{(m_1+m_2+m_3)}{\sqrt{m_{3}(E_{1\mathrm{com}}+m_{1})(m_{3}(E_{2\mathrm{com}}+m_{2}))^{2}}}\\
        &\times(\langle\mathbf{23}\rangle-[\mathbf{23}])(\langle\mathbf{12}\rangle-[\mathbf{12}]).
\end{aligned}
\end{equation}
The amplitude is
\begin{equation}
\begin{aligned}
        \mathcal{A}_{\{I^{(3)}_{1}\}}^{\{I^{(1)}_{1}\},\{I^{(2)}_{1}I^{(2)}_{2}\}}(0,\frac{1}{2})=&\sqrt{\frac{m_{3}}{\pi|\mathbf{p}_{1\mathrm{com}}|}}\sqrt{\frac{2}{3}}\frac{(m_1+m_2+m_3)}{\sqrt{m_{3}(E_{1\mathrm{com}}+m_{1})(m_{3}(E_{2\mathrm{com}}+m_{2}))^{2}}}\\
        &\times(\langle\mathbf{23}\rangle-[\mathbf{23}])(\langle\mathbf{12}\rangle-[\mathbf{12}]).
\end{aligned}
\end{equation}
\item $L={1},S=\frac{1}{2}:$
\begin{equation}
    \begin{aligned}
            \mathcal{A}_{\{I^{(3)}_{1}\}}^{\{I^{(1)}_{1}\},\{I^{(2)}_{1}I^{(2)}_{2}\}}(1,\frac{1}{2})=\mathcal{S}_{\{K_{1}\}}^{\{I^{(1)}_{1}\},\{I^{(2)}_{1}I^{(2)}_{2}\}}\mathcal{Y}^{1}_{\{L_1L_2\}}C_{\frac{1}{2},\{I^{(3)}_{1}\}}^{\frac{1}{2},\{K_{1}\};1,\{L_1L_2\}},
    \end{aligned}
\end{equation}
where $\mathcal{S}_{\{K_{1}\}}^{\{I^{(1)}_{1}\},\{I^{(2)}_{1}I^{(2)}_{2}\}}$ and $\mathcal{Y}^{1}_{\{L_1L_2\}}$ can be obtained from Eq.~\eqref{Eq:S 1/2 to 1/2 1} and Eq.~\eqref{Eq:Y=1}, respectively.

The amplitude is
\begin{equation}
    \begin{aligned}
            \mathcal{A}_{\{I^{(3)}_{1}\}}^{\{I^{(1)}_{1}\},\{I^{(2)}_{1}I^{(2)}_{2}\}}(1,\frac{1}{2})
            =&\sqrt{\frac{3}{2\pi|\mathbf{p}_{1\mathrm{com}}|^3m_3}}\frac{-(m_1+m_2+m_3)^2(m_3+m_2-m_1)}{\sqrt{m_{3}(E_{1\mathrm{com}}+m_{1})(m_{3}(E_{2\mathrm{com}}+m_{2}))^{2}}}\frac{2}{3}\\
            &\times(\langle\mathbf{23}\rangle+[\mathbf{23}])(\langle\mathbf{12}\rangle-[\mathbf{12}]).
    \end{aligned}
\end{equation}
\item $L={1},S=\frac{3}{2}:$
\begin{equation}
    \mathcal{A}_{\{I^{(3)}_{1}\}}^{\{I^{(1)}_{1}\},\{I^{(2)}_{1}I^{(2)}_{2}\}}(1,\frac{3}{2})=\mathcal{S}_{\{K_{1}K_2K_3\}}^{\{I^{(1)}_{1}\},\{I^{(2)}_{1}I^{(2)}_{2}\}}\mathcal{Y}^{1}_{\{L_1L_2\}}C_{\frac{1}{2},\{I^{(3)}_{1}\}}^{\frac{3}{2},\{K_{1}K_2K_3\};1,\{L_1L_2\}},
\end{equation}
where $\mathcal{Y}^{1}_{\{L_1L_2\}}$ can be obtained from Eq.~\eqref{Eq:Y=1} and 
\begin{equation}\label{Eq:S = 3/2 to 1/2 1}
\begin{aligned}
        \mathcal{S}_{\{K_{1}K_2K_3\}}^{\{I^{(1)}_{1}\},\{I^{(2)}_{1}I^{(2)}_{2}\}}=&\frac{1}{\sqrt{m_{3}(E_{1\mathrm{com}}+m_{1})(m_{3}(E_{2\mathrm{com}}+m_{2}))^{2}}}(\tau^{1}_{1})_{\{J_1\}}^{\{I^{(1)}_{1}\}}(\tau^{2}_{2})_{\{N_{1}N_{2}\}}^{\{I^{(2)}_{1}I^{(2)}_{2}\}}\\
        &\times C_{\frac{3}{2},\{K_1K_2K_3\}}^{\frac{1}{2},\{J_{1}\};1,\{N_{1}N_{2}\}}\\
        =&\frac{1}{\sqrt{m_{3}(E_{1\mathrm{com}}+m_{1})(m_{3}(E_{2\mathrm{com}}+m_{2}))^{2}}}\\
        &\times(\langle\mathbf{13}\rangle-[\mathbf{13}])(\langle\mathbf{23}\rangle-[\mathbf{23}])^{2}.
\end{aligned}
\end{equation}
The amplitude is
\begin{equation}
\begin{aligned}
        \mathcal{A}_{\{I^{(3)}_{1}\}}^{\{I^{(1)}_{1}\},\{I^{(2)}_{1}I^{(2)}_{2}\}}(1,\frac{3}{2})
        =&\sqrt{\frac{3}{2\pi|\mathbf{p}_{1\mathrm{com}}|^3m_3}}\frac{(m_1+m_2+m_3)(m_3+m_2-m_1)}{\sqrt{m_3(E_{1\mathrm{com}}+m_1)}(m_3(E_{2\mathrm{com}}+m_2))}\sqrt{\frac{1}{2}}\\
        &\times\bigg((m_2-m_{3}-m_1)([\mathbf{12}]+\langle\mathbf{12}\rangle)(\langle\mathbf{23}\rangle-[\mathbf{23}])\\
        &+\langle\mathbf{2}|1|\mathbf{2}](\langle\mathbf{13}\rangle-[\mathbf{13}])\bigg).
\end{aligned}
\end{equation}
\item $L={2},S=\frac{3}{2}:$
\begin{equation}
    \mathcal{A}_{\{I^{(3)}_{1}\}}^{\{I^{(1)}_{1}\},\{I^{(2)}_{1}I^{(2)}_{2}\}}(2,\frac{3}{2})=\mathcal{S}_{\{K_{1}K_2K_3\}}^{\{I^{(1)}_{1}\},\{I^{(2)}_{1}I^{(2)}_{2}\}}\mathcal{Y}^{2}_{\{L_1L_2L_3L_4\}}C_{\frac{1}{2},\{I^{(3)}_{1}\}}^{\frac{3}{2},\{K_{1}K_2K_3\};2,\{L_1L_2L_3L_4\}},
\end{equation}
where $\mathcal{S}_{\{K_{1}K_2K_3\}}^{\{I^{(1)}_{1}\},\{I^{(2)}_{1}I^{(2)}_{2}\}}$ and $\mathcal{Y}^{2}_{\{L_1L_2L_3L_4\}}$ can be obtained from Eq.~\eqref{Eq:S = 3/2 to 1/2 1} and Eq.~\eqref{Eq:Y=2}, respectively.
The amplitude is
\begin{equation}
\begin{aligned}
        \mathcal{A}_{\{I^{(3)}_{1}\}}^{\{I^{(1)}_{1}\},\{I^{(2)}_{1}I^{(2)}_{2}\}}(2,\frac{3}{2})
        =&\sqrt{\frac{15}{8\pi|\mathbf{p}_{1\mathrm{com}}|^5m_3^3}}\frac{(m_1+m_2+m_3)^2(m^2_3-(m_2-m_1)^2)}{\sqrt{(m_3(E_{1\mathrm{com}}+m_1))}m_3(E_{2\mathrm{com}}+m_2)}\sqrt{\frac{2}{5}}\\
        &\times\bigg((m_3+m_2-m_1)([\mathbf{12}]+\langle\mathbf{12}\rangle)([\mathbf{23}]+\langle\mathbf{23}\rangle)\\
        &+\langle\mathbf{2}|1|\mathbf{2}]([\mathbf{13}]+\langle\mathbf{13}\rangle)\bigg).
\end{aligned}
\end{equation}
\end{enumerate}

\item $s_3=1,s_1=\frac{1}{2},s_2=\frac{1}{2}$. There are four $(L,S)$ combinations: $(1,0)$, $(0,1)$, $(1,1)$ and $(2,1)$.
\begin{enumerate}
\item $L=1,S=0:$
\begin{equation}
	\mathcal{A}_{\{I^{(3)}_{1}I^{(3)}_{2}\}}^{\{I^{(1)}_{1}\},\{I^{(2)}_{1}\}}(1,0)=\mathcal{S}^{\{I^{(1)}_{1}\},\{I^{(2)}_{1}\}}\mathcal{Y}^{1}_{\{L_1L_2\}}C_{1,\{I^{(3)}_{1}I^{(3)}_{2}\}}^{0;1,\{L_1L_2\}},
\end{equation}
where $\mathcal{S}^{\{I^{(1)}_{1}\},\{I^{(2)}_{1}\}}$ and $\mathcal{Y}^{1}_{\{L_1L_2\}}$ can be obtained from Eqs.~\eqref{Eq:S 0 to 1/2 1/2} and \eqref{Eq:Y=1}, respectively.
The amplitude is
\begin{equation}
\begin{aligned}
    \mathcal{A}_{\{I^{(3)}_{1}I^{(3)}_{2}\}}^{\{I^{(1)}_{1}\},\{I^{(2)}_{1}\}}(1,0)=&\sqrt{\frac{3}{2\pi|\mathbf{p}_{1\mathrm{com}}|^{3}m_{3}}}\sqrt{\frac{1}{2}}\frac{m_1+m_2+m_3}{\sqrt{m_3(E_{1\mathrm{com}}+m_1)m_3(E_{2\mathrm{com}}+m_2)}}\\
    &\times(\langle\mathbf{12}\rangle-[\mathbf{12}])\langle\mathbf{3}|1|\mathbf{3}].
\end{aligned}
\end{equation}

\item $L=0,S=1:$
\begin{equation}
	\mathcal{A}_{\{I^{(3)}_{1}I^{(3)}_{2}\}}^{\{I^{(1)}_{1}\},\{I^{(2)}_{1}\}}(0,1)=\mathcal{S}_{\{K_1K_2\}}^{\{I^{(1)}_{1}\},\{I^{(2)}_{1}\}}\mathcal{Y}^{0}C_{1,\{I^{(3)}_{1}I^{(3)}_{2}\}}^{1,\{K_1K_2\};0},
\end{equation}
where $\mathcal{S}^{\{I^{(1)}_{1}\},\{I^{(2)}_{1}\}}_{\{K_1K_2\}}$ and $\mathcal{Y}^{0}$ can be obtained from Eqs.~\eqref{Eq:S 1 to 1/2 1/2} and \eqref{Eq:Y=0}, respectively.

The amplitude is
\begin{equation}
\begin{aligned}
    \mathcal{A}_{\{I^{(3)}_{1}I^{(3)}_{2}\}}^{\{I^{(1)}_{1}\},\{I^{(2)}_{1}\}}(0,1)=&\sqrt{\frac{m_{3}}{\pi|\mathbf{p}_{1\mathrm{com}}|}}\frac{1}{\sqrt{m_3(E_{1\mathrm{com}}+m_1)m_3(E_{2\mathrm{com}}+m_2)}}\\
    &\times(\langle\mathbf{13}\rangle-[\mathbf{13}])(\langle\mathbf{23}\rangle-[\mathbf{23}]).
\end{aligned}
\end{equation}

\item $L=1,S=1:$
\begin{equation}
	\mathcal{A}_{\{I^{(3)}_{1}I^{(3)}_{2}\}}^{\{I^{(1)}_{1}\},\{I^{(2)}_{1}\}}(1,1)=\mathcal{S}_{\{K_1K_2\}}^{\{I^{(1)}_{1}\},\{I^{(2)}_{1}\}}\mathcal{Y}^{1}_{\{L_1L_2\}}C_{1,\{I^{(3)}_{1}I^{(3)}_{2}\}}^{1,\{K_1K_2\};1,\{L_1L_2\}},
\end{equation}
where $\mathcal{S}^{\{I^{(1)}_{1}\},\{I^{(2)}_{1}\}}_{\{K_1K_2\}}$ and $\mathcal{Y}^{1}_{\{L_1L_2\}}$ can be obtained from Eqs.~\eqref{Eq:S 1 to 1/2 1/2} and \eqref{Eq:Y=1}, respectively.

The amplitude is
\begin{equation}
\begin{aligned}
    \mathcal{A}_{\{I^{(3)}_{1}I^{(3)}_{2}\}}^{\{I^{(1)}_{1}\},\{I^{(2)}_{1}\}}(1,1)=&\sqrt{\frac{3}{2\pi|\mathbf{p}_{1\mathrm{com}}|^{3}m_{3}}}\frac{(m_1+m_2+m_3)}{2\sqrt{m_3(E_{1\mathrm{com}}+m_1)m_3(E_{2\mathrm{com}}+m_2)}}\\
    &\times\bigg((m_3+m_1-m_2)(\langle\mathbf{13}\rangle+[\mathbf{13}])(\langle\mathbf{23}\rangle-[\mathbf{23}])\\
    &-(m_3+m_2-m_1)(\langle\mathbf{13}\rangle-[\mathbf{13}])(\langle\mathbf{23}\rangle+[\mathbf{23}])\bigg).
\end{aligned}
\end{equation}
\item $L=2,S=1:$
\begin{equation}
	\mathcal{A}_{\{I^{(3)}_{1}I^{(3)}_{2}\}}^{\{I^{(1)}_{1}\},\{I^{(2)}_{1}\}}(2,1)=\mathcal{S}_{\{K_1K_2\}}^{\{I^{(1)}_{1}\},\{I^{(2)}_{1}\}}\mathcal{Y}^{2}_{\{L_1L_2L_3L_4\}}C_{1,\{I^{(3)}_{1}I^{(3)}_{2}\}}^{1,\{K_1K_2\};2,\{L_1L_2L_3L_4\}},
\end{equation}
where $\mathcal{S}^{\{I^{(1)}_{1}\},\{I^{(2)}_{1}\}}_{\{K_1K_2\}}$ and $\mathcal{Y}^{2}_{\{L_1L_2L_3L_4\}}$ can be obtained from Eqs.~\eqref{Eq:S 1 to 1/2 1/2} and \eqref{Eq:Y=2}, respectively.

The amplitude is
\begin{equation}
\begin{aligned}
    \mathcal{A}_{\{I^{(3)}_{1}I^{(3)}_{2}\}}^{\{I^{(1)}_{1}\},\{I^{(2)}_{1}\}}(2,1)=&\sqrt{\frac{m_{3}}{\pi|\mathbf{p}_{1\mathrm{com}}|}}\frac{-(m_1+m_2+m_3)(m_3^2-(m_1^2-m_2^2))}{\sqrt{m_3(E_{1\mathrm{com}}+m_1)m_3(E_{2\mathrm{com}}+m_2)}}\\
    &\times\bigg((\langle\mathbf{12}\rangle+[\mathbf{12}])\langle\mathbf{3}|1|\mathbf{3}]\\
    &+\frac{(m_1+m_2+m_3)}{4}(\langle\mathbf{13}\rangle+[\mathbf{13}])(\langle\mathbf{23}\rangle+[\mathbf{23}]).
\end{aligned}
\end{equation}
\end{enumerate}
\item $s_3=1,s_1=1,s_2=1$. There are seven $(L,S)$ combinations: $(1,0)$, $(0,1)$, $(1,1)$, $(2,1)$, $(1,2)$, $(2,2)$ and $(3,2)$.
\begin{enumerate}
\item $L=1,S=0:$
\begin{equation}
	\mathcal{A}_{\{I^{(3)}_{1}I^{(3)}_{2}\}}^{\{I^{(1)}_{1}I_2^{(1)}\},\{I^{(2)}_{1}I_2^{(2)}\}}(1,0)=\mathcal{S}^{\{I^{(1)}_{1}I_2^{(1)}\},\{I^{(2)}_{1}I_2^{(2)}\}}\mathcal{Y}^{1}_{\{L_1L_2\}}C_{1,\{I^{(3)}_{1}I^{(3)}_{2}\}}^{0;1,\{L_1L_2\}},
\end{equation}
where $\mathcal{S}^{\{I^{(1)}_{1}I_2^{(1)}\},\{I^{(2)}_{1}I_2^{(2)}\}}$ and $\mathcal{Y}^{1}_{\{L_1L_2\}}$ can be obtained from Eqs.~\eqref{Eq:S 0 to 1 1} and \eqref{Eq:Y=1}, respectively.

The amplitude is
\begin{equation}
\begin{aligned}
    \mathcal{A}_{\{I^{(3)}_{1}I^{(3)}_{2}\}}^{\{I^{(1)}_{1}I_2^{(1)}\},\{I^{(2)}_{1}I_2^{(2)}\}}(1,0)=&\sqrt{\frac{3}{2\pi|\mathbf{p}_{1\mathrm{com}}|^{3}m_{3}}}\sqrt{\frac{1}{3}}\frac{(m_1+m_2+m_3)^2}{(m_3(E_{1\mathrm{com}}+m_1))(m_3(E_{1\mathrm{com}}+m_1))}\\
    &\times(\langle\mathbf{12}\rangle-[\mathbf{12}])^2\langle\mathbf{3}|1|\mathbf{3}].
\end{aligned}
\end{equation}

\item $L=0,S=1:$
\begin{equation}
	\mathcal{A}_{\{I^{(3)}_{1}I^{(3)}_{2}\}}^{\{I^{(1)}_{1}I_2^{(1)}\},\{I^{(2)}_{1}I_2^{(2)}\}}(0,1)=\mathcal{S}^{\{I^{(1)}_{1}I_2^{(1)}\},\{I^{(2)}_{1}I_2^{(2)}\}}_{\{K_1K_2\}}\mathcal{Y}^{0}C_{1,\{I^{(3)}_{1}I^{(3)}_{2}\}}^{1\{K_1K_2\};0},
\end{equation}
where $\mathcal{S}^{\{I^{(1)}_{1}I_2^{(1)}\},\{I^{(2)}_{1}I_2^{(2)}\}}_{\{K_1K_2\}}$ and $\mathcal{Y}^{0}$ can be obtained from Eqs.~\eqref{Eq:S 1 to 1 1} and \eqref{Eq:Y=0}, respectively.

The amplitude is
\begin{equation}
\begin{aligned}
    \mathcal{A}_{\{I^{(3)}_{1}I^{(3)}_{2}\}}^{\{I^{(1)}_{1}I_2^{(1)}\},\{I^{(2)}_{1}I_2^{(2)}\}}(0,1)=&\sqrt{\frac{m_{3}}{\pi|\mathbf{p}_{1\mathrm{com}}|}}\frac{(m_1+m_2+m_3)}{(m_3(E_{1\mathrm{com}}+m_1))(m_3(E_{1\mathrm{com}}+m_1))}\\
    &\times(\langle\mathbf{12}\rangle-[\mathbf{12}])(\langle\mathbf{13}\rangle-[\mathbf{13}])(\langle\mathbf{23}\rangle-[\mathbf{23}]).
\end{aligned}
\end{equation}

\item $L=1,S=1:$
\begin{equation}
	\mathcal{A}_{\{I^{(3)}_{1}I^{(3)}_{2}\}}^{\{I^{(1)}_{1}I_2^{(1)}\},\{I^{(2)}_{1}I_2^{(2)}\}}(1,1)=\mathcal{S}^{\{I^{(1)}_{1}I_2^{(1)}\},\{I^{(2)}_{1}I_2^{(2)}\}}_{\{K_1K_2\}}\mathcal{Y}^{1}_{\{L_1L_2\}}C_{1,\{I^{(3)}_{1}I^{(3)}_{2}\}}^{1,\{K_1K_2\};1,\{L_1L_2\}},
\end{equation}
where $\mathcal{S}^{\{I^{(1)}_{1}I_2^{(1)}\},\{I^{(2)}_{1}I_2^{(2)}\}}_{\{K_1K_2\}}$ and $\mathcal{Y}^{1}_{\{L_1L_2\}}$ can be obtained from Eqs.~\eqref{Eq:S 1 to 1 1} and \eqref{Eq:Y=1}, respectively.

The amplitude is
\begin{equation}
\begin{aligned}
    \mathcal{A}_{\{I^{(3)}_{1}I^{(3)}_{2}\}}^{\{I^{(1)}_{1}I_2^{(1)}\},\{I^{(2)}_{1}I_2^{(2)}\}}(1,1)=&\sqrt{\frac{3}{2\pi|\mathbf{p}_{1\mathrm{com}}|^{3}m_{3}}}\frac{(m_1+m_2+m_3)^2(\langle\mathbf{12}\rangle-[\mathbf{12}])}{2(m_3(E_{1\mathrm{com}}+m_1))(m_3(E_{1\mathrm{com}}+m_1))}\\
    &\times\bigg((m_3+m_1-m_2)(\langle\mathbf{13}\rangle+[\mathbf{13}])(\langle\mathbf{23}\rangle-[\mathbf{23}])\\
    &+(m_1-m_3-m_2)(\langle\mathbf{13}\rangle-[\mathbf{13}])(\langle\mathbf{23}\rangle+[\mathbf{23}]).
\end{aligned}
\end{equation}
\item $L=2,S=1:$
\begin{equation}
	\mathcal{A}_{\{I^{(3)}_{1}I^{(3)}_{2}\}}^{\{I^{(1)}_{1}I_2^{(1)}\},\{I^{(2)}_{1}I_2^{(2)}\}}(2,1)=\mathcal{S}^{\{I^{(1)}_{1}I_2^{(1)}\},\{I^{(2)}_{1}I_2^{(2)}\}}_{\{K_1K_2\}}\mathcal{Y}^{2}_{\{L_1L_2L_3L_4\}}C_{1,\{I^{(3)}_{1}I^{(3)}_{2}\}}^{1,\{K_1K_2\};2,\{L_1L_2L_3L_4\}},
\end{equation}
where $\mathcal{S}^{\{I^{(1)}_{1}I_2^{(1)}\},\{I^{(2)}_{1}I_2^{(2)}\}}_{\{K_1K_2\}}$ and $\mathcal{Y}^{2}_{\{L_1L_2L_3L_4\}}$ can be obtained from Eqs.~\eqref{Eq:S 1 to 1 1} and \eqref{Eq:Y=2}, respectively.

The amplitude is
\begin{equation}
\begin{aligned}
    \mathcal{A}_{\{I^{(3)}_{1}I^{(3)}_{2}\}}^{\{I^{(1)}_{1}I_2^{(1)}\},\{I^{(2)}_{1}I_2^{(2)}\}}(2,1)=&\sqrt{\frac{15}{8\pi|\mathbf{p}_{1\mathrm{com}}|^5m_3^3}}\frac{-(m_1+m_2+m_3)^2(m_3^2-(m_1-m_2)^2)}{(m_3(E_{1\mathrm{com}}+m_1))(m_3(E_{1\mathrm{com}}+m_1))}\sqrt{\frac{3}{5}}\\
    &\times\bigg(\frac{(m_1+m_2+m_3)}{4}(\langle\mathbf{12}\rangle-[\mathbf{12}])(\langle\mathbf{13}\rangle+[\mathbf{13}])(\langle\mathbf{23}\rangle+[\mathbf{23}])\\
    &+(\langle\mathbf{12}\rangle^2-[\mathbf{12}]^2)(\langle\mathbf{3}|1|\mathbf{3}]).
\end{aligned}
\end{equation}
\item $L=1,S=2:$
\begin{equation}
	\mathcal{A}_{\{I^{(3)}_{1}I^{(3)}_{2}\}}^{\{I^{(1)}_{1}I_2^{(1)}\},\{I^{(2)}_{1}I_2^{(2)}\}}(1,2)=\mathcal{S}^{\{I^{(1)}_{1}I_2^{(1)}\},\{I^{(2)}_{1}I_2^{(2)}\}}_{\{K_1K_2K_3K_4\}}\mathcal{Y}^{1}_{\{L_1L_2\}}C_{1,\{I^{(3)}_{1}I^{(3)}_{2}\}}^{2,\{K_1K_2K_3K_4\};1,\{L_1L_2\}},
\end{equation}
where $\mathcal{S}^{\{I^{(1)}_{1}I_2^{(1)}\},\{I^{(2)}_{1}I_2^{(2)}\}}_{\{K_1K_2K_3K_4\}}$ and $\mathcal{Y}^{1}_{\{L_1L_2\}}$ can be obtained from Eqs.~\eqref{Eq:S 2 to 1 1} and \eqref{Eq:Y=1}, respectively.

The amplitude is
\begin{equation}
\begin{aligned}
    \mathcal{A}_{\{I^{(3)}_{1}I^{(3)}_{2}\}}^{\{I^{(1)}_{1}I_2^{(1)}\},\{I^{(2)}_{1}I_2^{(2)}\}}(1,2)=&\sqrt{\frac{3}{2\pi|\mathbf{p}_{1\mathrm{com}}|^{3}m_{3}}}\frac{(m_1+m_2+m_3)}{(m_3(E_{1\mathrm{com}}+m_1))(m_3(E_{1\mathrm{com}}+m_1))}\sqrt{\frac{3}{5}}\\
    &\times\bigg(((m_1-m_2)^2-m_3^2)(\langle\mathbf{13}\rangle-[\mathbf{13}])(\langle\mathbf{23}\rangle-[\mathbf{23}])(\langle\mathbf{12}\rangle+[\mathbf{12}])\\
    &+(m_2-m_3-m_1)\langle\mathbf{1}|3|\mathbf{1}](\langle\mathbf{23}\rangle-[\mathbf{23}])^2\\
    &+(m_3+m_2-m_1)\langle\mathbf{2}|3|\mathbf{2}](\langle\mathbf{13}\rangle-[\mathbf{13}])^2\bigg)
.
\end{aligned}
\end{equation}
\item $L=2,S=2:$
\begin{equation}
	\mathcal{A}_{\{I^{(3)}_{1}I^{(3)}_{2}\}}^{\{I^{(1)}_{1}I_2^{(1)}\},\{I^{(2)}_{1}I_2^{(2)}\}}(2,2)=\mathcal{S}^{\{I^{(1)}_{1}I_2^{(1)}\},\{I^{(2)}_{1}I_2^{(2)}\}}_{\{K_1K_2K_3K_4\}}\mathcal{Y}^{2}_{\{L_1L_2L_3L_4\}}C_{1,\{I^{(3)}_{1}I^{(3)}_{2}\}}^{2,\{K_1K_2K_3K_4\};2,\{L_1L_2L_3L_4\}},
\end{equation}
where $\mathcal{S}^{\{I^{(1)}_{1}I_2^{(1)}\},\{I^{(2)}_{1}I_2^{(2)}\}}_{\{K_1K_2K_3K_4\}}$ and $\mathcal{Y}^{2}_{\{L_1L_2L_3L_4\}}$ can be obtained from Eqs.~\eqref{Eq:S 2 to 1 1} and \eqref{Eq:Y=2}, respectively.

The amplitude is
\begin{equation}
\begin{aligned}
    \mathcal{A}_{\{I^{(3)}_{1}I^{(3)}_{2}\}}^{\{I^{(1)}_{1}I_2^{(1)}\},\{I^{(2)}_{1}I_2^{(2)}\}}(2,2)=&\sqrt{\frac{15}{8\pi|\mathbf{p}_{1\mathrm{com}}|^5m_3^3}}\frac{(m_1+m_2+m_3)^2(m_3^2-(m_1-m_2)^2)}{2(m_3(E_{1\mathrm{com}}+m_1))(m_3(E_{1\mathrm{com}}+m_1))}\sqrt{\frac{4}{5}}\\
    &\times\bigg((m_2-m_3-m_1)(\langle\mathbf{13}\rangle+[\mathbf{13}])(\langle\mathbf{23}\rangle-[\mathbf{23}])(\langle\mathbf{12}\rangle+[\mathbf{12}])\\
    &+(m_3+m_2-m_1)(\langle\mathbf{13}\rangle-[\mathbf{13}])(\langle\mathbf{23}\rangle+[\mathbf{23}])(\langle\mathbf{12}\rangle+[\mathbf{12}])\\
    &+\langle\mathbf{1}|3|\mathbf{1}](\langle\mathbf{23}\rangle^2-[\mathbf{23}]^2)\\
    &+\langle\mathbf{2}|3|\mathbf{2}](\langle\mathbf{13}\rangle^2-[\mathbf{13}]^2)\bigg).
\end{aligned}
\end{equation}
\item $L=3,S=2:$
\begin{equation}
\begin{aligned}
    	\mathcal{A}_{\{I^{(3)}_{1}I^{(3)}_{2}\}}^{\{I^{(1)}_{1}I_2^{(1)}\},\{I^{(2)}_{1}I_2^{(2)}\}}(3,2)=&\mathcal{S}^{\{I^{(1)}_{1}I_2^{(1)}\},\{I^{(2)}_{1}I_2^{(2)}\}}_{\{K_1K_2K_3K_4\}}\mathcal{Y}^{3}_{\{L_1L_2L_3L_4L_5L_6\}}\\
    &\times C_{1,\{I^{(3)}_{1}I^{(3)}_{2}\}}^{2,\{K_1K_2K_3K_4\};3,\{L_1L_2L_3L_4L_5L_6\}},
\end{aligned}
\end{equation}
where $\mathcal{S}^{\{I^{(1)}_{1}I_2^{(1)}\},\{I^{(2)}_{1}I_2^{(2)}\}}_{\{K_1K_2K_3K_4\}}$ can be obtained from Eq.~\eqref{Eq:S 2 to 1 1} and 
\begin{equation}
    \begin{aligned}
        \mathcal{Y}^{3}_{\{L_1L_2L_3L_4L_5L_6\}}=\sqrt{\frac{35}{16\pi|\mathbf{p}_{1\mathrm{com}}|^7m^5_3}}(\langle\mathbf{3}|1|\mathbf{3}])^3.
    \end{aligned}
\end{equation}

The amplitude is
\begin{equation}
\begin{aligned}
    \mathcal{A}_{\{I^{(3)}_{1}I^{(3)}_{2}\}}^{\{I^{(1)}_{1}I_2^{(1)}\},\{I^{(2)}_{1}I_2^{(2)}\}}(3,2)=&\sqrt{\frac{35}{16\pi|\mathbf{p}_{1\mathrm{com}}|^7m^5_3}}\frac{(m_1+m_2+m_3)^2(m_3^2-(m_1-m_2)^2)}{(m_3(E_{1\mathrm{com}}+m_1))(m_3(E_{1\mathrm{com}}+m_1))}\sqrt{\frac{3}{7}}\\
    &\times\bigg(\big(((m_3^2-(m_1-m_2)^2))(\langle\mathbf{12}\rangle+[\mathbf{12}])^2-\langle\mathbf{1}|3|\mathbf{1}]\langle\mathbf{2}|3|\mathbf{3}]\big)\langle\mathbf{3}|1|\mathbf{3}]\\
    &+\frac{(m_1+m_2+m_3)}{4}\Big((m_1-m_3-m_2)(\langle\mathbf{23}\rangle+[\mathbf{23}])^2\langle\mathbf{1}|3|\mathbf{1}]\\
    &+(m_3+m_1-m_2)(\langle\mathbf{13}\rangle+[\mathbf{13}])^2\langle\mathbf{2}|3|\mathbf{2}]\\
    &+(m_3^2-(m_1-m_2)^2)(\langle\mathbf{12}\rangle+[\mathbf{12}])(\langle\mathbf{13}\rangle+[\mathbf{13}])(\langle\mathbf{23}\rangle+[\mathbf{23}])\Big)\bigg).\\
\end{aligned}
\end{equation}
\end{enumerate}

\end{enumerate}

\section{General analysis of cascade decay}
\label{sec:General analysis of cascade decay}

In the previous sections, both non-covariant and covariant methods have been introduced. 
In the non-covariant methods, two-body decay amplitudes are defined in the corresponding COM frames of the decays. Therefore, when the particle momenta are given in an arbitrary frame, such as the lab frame, one must boost the event to the relevant COM frames in order to evaluate the amplitudes. However, since such boosts are generally not along the momentum directions of the final-state particles, they usually induce nontrivial Wigner rotations on the spin states of the particles.

If one were to directly sum or contract amplitudes defined in different COM frames, incorrect results would be obtained, because the spin quantization axes in those frames are not aligned. It is therefore necessary to perform Wigner rotations to align the particle states, so that the same particle is described with a consistent choice of quantization axis. 
Only after this alignment can different two-body decay amplitudes be summed or contracted consistently.

Because the definitions of the spin quantization axes differ in the canonical and helicity amplitudes, the treatment in cascade decays is also different. In the canonical amplitude, the spin quantization axis is chosen along the $z$-axis of a selected frame, whereas in the helicity amplitude it is defined along the particle’s momentum direction in the chosen frame. Consequently, the canonical and helicity formalisms require different procedures when applied to cascade decays.

In contrast, the covariant canonical-spinor two-body decay amplitudes are defined in an arbitrary frame and are constructed directly from the four-momenta in that frame. 
The spin quantization axes are all chosen to point along the $z$-axis of the selected frame, so the spin indices can be contracted and summed directly. Moreover, the momenta entering each amplitude are taken directly from the chosen frame, and no boost back to the COM frame is required. 
This makes the formalism particularly convenient when treating cascade decays with multiple decay chains.

In the following, we discuss separately the treatment of cascade decays in the non-covariant approach and in the covariant approach.

\subsection{Non-covariant cascade decay}

In this subsection, we mainly discuss the application of non-covariant approaches to cascade decays. 
Here, “non-covariant” means that the amplitude is not written in a covariant form, namely it contains ingredients that must be evaluated by boosting back to the COM frame. In this sense, the previously introduced non-covariant amplitudes include the helicity amplitude, canonical amplitude, the covariant projection tensor amplitude in the PS-scheme , and the Zemach tensor amplitude. 
These constructions are all defined starting from the COM frame at the level of their amplitude expressions.
According to the definition of the spin quantization axes, these amplitudes can be further classified into amplitudes carrying canonical indices and amplitudes carrying helicity indices. 

Therefore, in the canonical cascade decay we discuss amplitudes with canonical indices, including the canonical amplitude, the covariant projection tensor amplitude in the PS-scheme, and the Zemach amplitude. 
In the helicity cascade decay, we discuss helicity amplitude.

\subsubsection{Canonical cascade decay}
\label{subsubsec:Canonical cascade decay}

In the previous sections, we presented several amplitude constructions that require boosting to the COM frame and ultimately carry canonical indices, including the \textbf{traditional-$LS$ amplitude}, \textbf{the covariant projection tensor amplitude in the PS-scheme}, and the \textbf{Zemach tensor amplitude}. In section~\ref{sec:Covariant tensor and Zemach tensor amplitudes}, we showed that these three amplitudes are completely equivalent in the COM frame. 
Under a boost, all three amplitude constructions induce the same Wigner rotation acting on the final-state $\sigma$ indices, and the rotation is identical for the three amplitudes. 
Therefore, in cascade decay analyses that require boosting back to the COM frame, they can be treated in the same way. 
In this subsubsection, we unify them under a single framework, which we refer to as the canonical cascade decay.

For a cascade decay, we first choose a global reference frame, such as the lab frame, and fix a set of axes $(x,y,z)$ in that frame. In the lab frame, all external momenta $p_i$ are expressed with respect to the same fixed coordinate system.

A key feature of the canonical basis is that the spin projection label $\sigma_i$ is always defined as the projection onto this fixed lab frame $z$-axis. In other words, the spin polarization direction associated with every particles is tied to the same quantization axis. When each two-body decay amplitude is computed by boosting to the corresponding parent rest frame, the accompanying Wigner $D$-matrices precisely transport the daughter spin polarization direction back to this common quantization axis. 
As a result, all spin polarization direction appearing in different decay steps refer to the same spin quantization axis, and the intermediate-state spin sum is a well-defined contraction in a common quantization axis.

This has an immediate consequence for cascade decays and for processes with multiple decay chains leading to the same final state. Since every decay chain amplitude is ultimately expressed with external spin polarization direction defined with respect to the same lab frame $z$-axis, different chains can be added coherently without introducing any additional alignment rotations between different decay chains. 

Next, we will give explicit examples for a single decay chain and for multiple decay chains, and provide a concrete method for calculating the cascade decay amplitude.

\paragraph{Canonical cascade decay amplitude for single decay chain}
As an explicit example, we consider the two-step cascade decay $X \to 3 + Y$, followed by $3 \to 1 + 2$.
We define the lab frame to be the global frame associated with the fixed coordinate system above.
In this lab frame, the momenta $\mathbf p_1$, $\mathbf p_2$, and $\mathbf p_Y$ of the final-state particles $1,2,Y$ are taken as the measured inputs.
The four-momentum of the intermediate particle-$3$ is then reconstructed from momentum conservation as
\begin{equation}
\mathbf p_3 = \mathbf p_1 + \mathbf p_2,
\qquad
\mathbf p_X = \mathbf p_3 + \mathbf p_Y .
\end{equation}

According to the general expression for the canonical two-body decay amplitude in any frame given in Eq.~\eqref{Eq:any frame canonical}, the two two-body amplitudes entering the cascade
$X\to 3+Y$ and $3\to 1+2$ can be written explicitly as follows

\begin{enumerate}
\item Two-body decay $X\to 3+Y$ in the lab frame.

For the first step $X\to 3+Y$, boosting from the lab frame to the $X$ rest frame by $L^{-1}_{X,c}(\mathbf p_X)$ gives the daughter momenta $\mathbf p_3^{*}$ and $\mathbf p_Y^{*}$. The corresponding amplitude in the lab frame is the two-body decay COM frame times Wigner $D$-matrices, which can be written as
\begin{equation}\label{Eq:any frame canonical Xto3Y}
A_{\sigma_X}^{\sigma_3\sigma_Y}(\mathbf p_X,\mathbf p_3,\mathbf p_Y)
=
\sum_{L_1S_1}\sum_{\sigma_3^\prime \sigma_Y^\prime}
A_{\sigma_X}^{\sigma_3^\prime\sigma_Y^\prime}(\mathbf k_X,\mathbf p_3^{*},\mathbf p_Y^{*};L_1,S_1)
D^{\sigma_3^\prime(s_3)*}_{\,\,\sigma_3}(R_{X3})
D^{\sigma_Y^\prime(s_Y)*}_{\,\,\sigma_Y}(R_{XY}),
\end{equation}
where $(L_1,S_1)$ denotes the allowed $LS$ combinations for the two-body decay. The Wigner rotations $R_{X3}$ and $R_{XY}$ are induced by the boost to the particle-$X$ rest frame, following Eq.~\eqref{Eq:canonical wiger rotation}, they can be written as
\begin{equation}
R_{X3}=R_{\hat{\mathbf{n}}_{X3}}R_z(\omega_{X3})R_{\hat{\mathbf{n}}_{X3}}^{-1},
\qquad
R_{XY}=R_{\hat{\mathbf{n}}_{XY}}R_z(\omega_{XY})R_{\hat{\mathbf{n}}_{XY}}^{-1},
\end{equation}
where the rotation axes are chosen as
\begin{equation}
\hat{\mathbf n}_{X3}\ \parallel\ -\mathbf p_X \times \mathbf p_3,
\qquad
\hat{\mathbf n}_{XY}\ \parallel\ -\mathbf p_X \times \mathbf p_Y.
\end{equation}
The corresponding rotation angles about the $z$ axis can be obtained from Eq.~\eqref{Eq:wigner angle}. Explicitly, for $i=3,Y$ one may write
\begin{equation}\label{Eq:WignerAngle_Xi}
\tan\frac{\omega_{Xi}}{2}
=
\frac{\sin\theta_{Xi}\sinh\left(\frac{\eta_X}{2}\right)\sinh\left(\frac{\eta_i}{2}\right)}
{\cosh\left(\frac{\eta_X}{2}\right)\cosh\left(\frac{\eta_i}{2}\right)
+\cos\theta_{Xi}\sinh\left(\frac{\eta_X}{2}\right)\sinh\left(\frac{\eta_i}{2}\right)},
\qquad i=3,Y,
\end{equation}
where $\eta_X$ is the rapidity of the canonical boost $L^{-1}_{X,c}(\mathbf p_X)$, $\eta_i$ is the rapidity associated with the standard canonical boost $L_{i,c}(\mathbf p_i)$ of particle-$i$, and $\theta_{Xi}$ is the angle between $\mathbf p_X$ and $\mathbf p_i$ in the lab frame.

After the Wigner rotation, the amplitude in the lab frame has its quantization axis aligned with the lab $z$-axis.

\item Two-body decay $3\to 1+2$ in the lab frame.

For the second step $3\to 1+2$, boosting from the lab frame to the $3$ rest frame by $L^{-1}_{3,c}(\mathbf p_3)$ gives the daughter momenta $\mathbf p_1^{*}$ and $\mathbf p_2^{*}$. The corresponding amplitude in the lab frame is the two-body decay COM frame times Wigner $D$-matrices, which can be written as
\begin{equation}\label{Eq:any frame canonical 3to12}
A_{\sigma_3}^{\sigma_1\sigma_2}(\mathbf p_3,\mathbf p_1,\mathbf p_2)
=
\sum_{L_2S_2}\sum_{\sigma_1^\prime \sigma_2^\prime}
A_{\sigma_3}^{\sigma_1^\prime\sigma_2^\prime}(\mathbf k_3,\mathbf p_1^{*},\mathbf p_2^{*};L_2,S_2)
D^{\sigma_1^\prime(s_1)*}_{\,\,\sigma_1}(R_{31})
D^{\sigma_2^\prime(s_2)*}_{\,\,\sigma_2}(R_{32}),
\end{equation}
where $(L_2,S_2)$ denotes the allowed $LS$ combinations for the two-body decay. The Wigner rotations $R_{31}$ and $R_{32}$ are induced by the boost to the particle-$3$ rest frame, following Eq.~\eqref{Eq:canonical wiger rotation}, they can be written as
\begin{equation}
R_{31}=R_{\hat{\mathbf{n}}_{31}}R_z(\omega_{31})R_{\hat{\mathbf{n}}_{31}}^{-1},
\qquad
R_{32}=R_{\hat{\mathbf{n}}_{32}}R_z(\omega_{32})R_{\hat{\mathbf{n}}_{32}}^{-1},
\end{equation}
where the rotation axes are chosen as
\begin{equation}
\hat{\mathbf n}_{31}\ \parallel\ -\mathbf p_3 \times \mathbf p_1,
\qquad
\hat{\mathbf n}_{32}\ \parallel\ -\mathbf p_3 \times \mathbf p_2.
\end{equation}
The corresponding rotation angles about the $z$ axis can be obtained from Eq.~\eqref{Eq:wigner angle}. Explicitly, for $i=1,2$ one may write
\begin{equation}\label{Eq:WignerAngle_3i}
\tan\frac{\omega_{3i}}{2}
=
\frac{\sin\theta_{3i}\sinh\left(\frac{\eta_3}{2}\right)\sinh\left(\frac{\eta_i}{2}\right)}
{\cosh\left(\frac{\eta_3}{2}\right)\cosh\left(\frac{\eta_i}{2}\right)
+\cos\theta_{3i}\sinh\left(\frac{\eta_3}{2}\right)\sinh\left(\frac{\eta_i}{2}\right)},
\qquad i=1,2,
\end{equation}
where $\eta_3$ is the rapidity of the canonical boost $L^{-1}_{3,c}(\mathbf p_3)$, $\eta_i$ is the rapidity associated with the standard canonical boost $L_{i,c}(\mathbf p_i)$ of particle-$i$, and $\theta_{3i}$ is the angle between $\mathbf p_3$ and $\mathbf p_i$ in the lab frame.

After the Wigner rotation, the amplitude in the lab frame has its quantization axis aligned with the lab $z$-axis.

\end{enumerate}

In each of the two-body decays above, we have aligned the spin quantization axes of the particles to the lab frame $z$-axis via Wigner rotations. Therefore, in both two-body decay amplitudes, the spin quantum number $\sigma_3$ of particle‑3 is always defined relative to the same direction, the lab frame $z$-axis.
With a sum over the canonical spin projection of the intermediate particle-3, the total cascade decay amplitude can be written as
\begin{equation}\label{Eq:canonical cascade decay}
	A_{\sigma_X}^{\sigma_{1}\sigma_{2}\sigma_{Y}}(\mathbf{p}_X,\mathbf{p}_1,\mathbf{p}_2,\mathbf{p}_Y)
	=
	\sum_{\sigma_3}
	A_{\sigma_X}^{\sigma_{3}\sigma_{Y}}(\mathbf{p}_X,\mathbf{p}_3,\mathbf{p}_Y)
	P_{3}
	A_{\sigma_{3}}^{\sigma_{1}\sigma_{2}}(\mathbf{p}_3,\mathbf{p}_1,\mathbf{p}_2),
\end{equation}
where $P_3$ is the resonant propagator of the intermediate particle-$3$. The amplitude for the cascade decay also has its quantization axis aligned with the same direction, namely the lab frame $z$-axis.

\begin{figure}[H]
\centering
\includegraphics[scale=1]{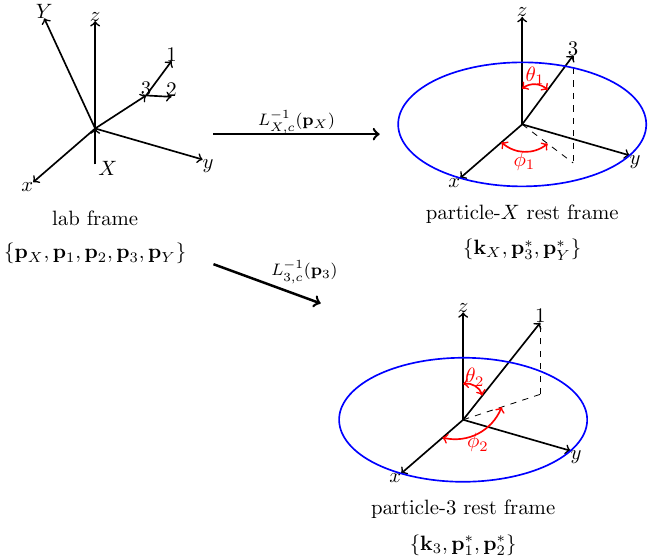}
\caption{$L^{-1}_{X,c}(\mathbf{p}_{X})$ is the canonical-standard boost of particle-$X$ from the lab frame to the rest frame of particle-$X$, the momenta are $\{\mathbf{k}_X,\mathbf{p}_3^*,\mathbf{p}_Y^*\}$. $L^{-1}_{3,c}(\mathbf{p}_{3})$ is the canonical-standard boost of particle-3 from the lab frame to the rest frame of particle-$3$, the momenta are $\{\mathbf{k}_3,\mathbf{p}^*_1,\mathbf{p}^*_2\}$. The angles shown correspond to those used in the amplitudes defined above.}
\label{fig:canonical cascade decay}
\end{figure}

The geometry of the cascade decay process is shown in Fig.~\ref{fig:canonical cascade decay}.

\paragraph{Canonical cascade decay amplitude for multiple decay chain}

For a cascade process with multiple decay chains, each chain is defined in the same lab frame. Since the canonical spin projections of all final-state particles are defined with respect to the same fixed lab-frame $z$-axis, the spin polarization directions are consistent among different chains. Therefore, one can first compute the cascade amplitude of each decay chain in the lab frame, and then obtain the total amplitude by directly summing the chain amplitudes.

As an explicit example, we consider two decay chains,
\begin{itemize} 
	\item Decay chain 1: $X \to 3 + Y$, followed by $3 \to 1 + 2$
	\item Decay chain 2: $X \to 4 + 2$, followed by $4 \to 1 + Y$
\end{itemize}
In the lab frame, following the single chain construction discussed above and using Eq.~\eqref{Eq:canonical cascade decay}, the corresponding cascade decay amplitudes are given by
\begin{equation}\label{Eq:chain_amplitudes_X1X2}
\begin{aligned}
A_{\sigma_X}^{\sigma_{1}\sigma_{2}\sigma_{Y}}(\mathbf{p}_X,\mathbf{p}_1,\mathbf{p}_2,\mathbf{p}_Y;3)
&=
\sum_{\sigma_{3}}
A_{\sigma_X}^{\sigma_{3}\sigma_Y}(\mathbf p_X,\mathbf p_{3},\mathbf p_Y)
P_{3}
A_{\sigma_{3}}^{\sigma_1\sigma_2}(\mathbf p_{3},\mathbf p_1,\mathbf p_2),\\
A_{\sigma_X}^{\sigma_{1}\sigma_{2}\sigma_{Y}}(\mathbf{p}_X,\mathbf{p}_1,\mathbf{p}_2,\mathbf{p}_Y;4)
&=
\sum_{\sigma_{4}}
A_{\sigma_X}^{\sigma_{4}\sigma_2}(\mathbf p_X,\mathbf p_{4},\mathbf p_2)
P_{4}
A_{\sigma_{4}}^{\sigma_1\sigma_Y}(\mathbf p_{4},\mathbf p_1,\mathbf p_Y).
\end{aligned}
\end{equation}
The total cascade amplitude for this process is obtained by directly summing the two chain amplitudes,
\begin{equation}\label{Eq:total_two_chains}
A_{\sigma_X}^{\sigma_{1}\sigma_{2}\sigma_{Y}}(\mathbf{p}_X,\mathbf{p}_1,\mathbf{p}_2,\mathbf{p}_Y)
=
A_{\sigma_X}^{\sigma_{1}\sigma_{2}\sigma_{Y}}(\mathbf{p}_X,\mathbf{p}_1,\mathbf{p}_2,\mathbf{p}_Y;3)
+
A_{\sigma_X}^{\sigma_{1}\sigma_{2}\sigma_{Y}}(\mathbf{p}_X,\mathbf{p}_1,\mathbf{p}_2,\mathbf{p}_Y;4).
\end{equation}

Finally, an advantage of the canonical cascade construction is that the spin polarization directions of the particles are consistent for each decay chain, so the total cascade amplitude can be obtained by directly summing the cascade amplitudes of all decay chains in the lab frame. 

However, in practical calculations, evaluating each two-body decay amplitude in any frame still requires computing the two-body amplitude in the corresponding COM frame together with the Wigner rotations. Once multiple decay chains are involved, the computational cost becomes too large, which is not very convenient in practice.

\subsubsection{Helicity cascade decay}
\label{subsubsec:Helicity cascade decay}
In this subsubsection, we discuss the treatment of helicity amplitudes~\eqref{Eq:helicity amplitude} in cascade decays.
Since helicity is defined with respect to the momentum direction in a given reference frame, different reference frames generally lead to different helicity definitions. 
In the following, we focus on the definitions of cascade decay amplitudes under two different helicity conventions.

\begin{itemize}
    \item \textbf{Two-body decay COM frame helicity (local helicity).}
    For each two-body decay $a\to b+c$ one defines the helicities $\lambda_b^{(a)}$ and $\lambda_c^{(a)}$ in the rest frame of particle-$a$.
    This choice yields the standard Jacob-Wick factorization of each two-body amplitude into the Wigner $D$-function of the decay angles and the helicity-coupling amplitude.
    
    In most of the literature, this definition is used to construct cascade decay amplitudes. When the cascade decay involves only a single decay chain, it is equivalent to the lab frame description and yields the correct results for physical observables. However, once multiple decay chains are present, one cannot directly add the amplitudes from different chains. This is because different decay chains correspond to different topologies, and in general the final-state helicities associated with different chains are not defined in the same way. A naive direct sum would therefore lead to incorrect results.
    
    The correct procedure is to choose one reference decay chain and then rotate the final-state helicities of the other chains so that they are aligned with those of the reference chain, thereby obtaining cascade amplitudes defined with the same final-state helicity convention~\cite{Chen:2017gtx,Marangotto:2019ucc,JPAC:2019ufm,Wang:2020giv,Gao:2023jtq,Habermann:2024sxs}. After this alignment rotation, the cascade amplitudes from different decay chains can be added directly, and the total cascade decay amplitude is obtained by summing over all chains.
    
    \item \textbf{Lab helicity (global helicity).}
    For the final-state particle-$i$ one may define the helicity $\lambda_i$ in a fixed global frame, for example the lab frame, by projecting the spin onto the momentum direction in the lab frame.

    Under this convention, each two-body decay amplitude is defined in the lab frame, and the helicities of all particles are defined along their respective momentum directions in the lab frame. For intermediate particles, this guarantees that the spin quantization axis is defined in the same way when they appear as daughter particles and as parent particles. Therefore, for a given cascade decay chain, one can directly sum over the helicity of the intermediate state and obtain the amplitude for that chain.

    Moreover, when multiple decay chains are present, the final-state helicities in every chain are defined in the same lab frame. The spin quantization axes for the same final-state particle in different chains are always taken along that particle’s lab frame momentum direction. As a consequence, there is no mismatch of final-state helicity conventions across different chains, and one can directly add the cascade decay amplitudes from different chains to obtain the total cascade decay amplitude.

\end{itemize}

Based on the two helicity conventions introduced above, we first use the case of a single decay chain to illustrate how the corresponding cascade decay amplitude is constructed under each convention. We then turn to the situation with multiple decay chains and explain how the cascade decay amplitude with several chains is constructed in each convention.

\paragraph{Helicity cascade decay amplitude for single decay chain}

Consider a cascade decay process $X \to 3 + Y$, followed by $3 \to 1 + 2$, where all particles are massive and carry spin.
According to the two helicity conventions introduced above, the corresponding cascade amplitude can be constructed in two different ways.

For the two-body decay COM frame helicity convention, we directly construct the cascade decay amplitude from the two-body amplitudes defined in the corresponding COM frames, and then obtain the total cascade amplitude as
\begin{equation}
A^{\lambda_1\lambda_2\lambda_Y}_{\lambda_X}(\mathbf{k}_X,\mathbf{q}^*_Y,\mathbf{q}^*_1,\mathbf{q}^*_2)
=
\sum_{\lambda_3}
A^{\lambda_3\lambda_Y}_{\lambda_X}(\mathbf{k}_X,\mathbf{q}^*_3,\mathbf{q}^*_Y)
P_3
A^{\lambda_1\lambda_2}_{\lambda_3}(\mathbf{k}_3,\mathbf{q}^*_1,\mathbf{q}^*_2),
\end{equation}
where each helicity $\lambda_i$ is defined along the momentum direction in the COM frame of the corresponding two-body decay.

In contrast, for the lab frame helicity convention, we construct the cascade decay amplitude using the two-body decay amplitudes written in the lab frame, and then obtain the total cascade amplitude as
\begin{equation}\label{eq:cascade_global_helicity}
A^{\lambda_1^\prime\lambda_2^\prime\lambda_Y^\prime}_{\lambda_X}(\mathbf{p}_X,\mathbf{p}_1,\mathbf{p}_2,\mathbf{p}_Y)
=
\sum_{\lambda_3^\prime}
 A^{\lambda_3^\prime\lambda_Y^\prime}_{\lambda_X}(\mathbf{p}_X,\mathbf{p}_3,\mathbf{p}_Y)
P_3
 A^{\lambda_1^\prime\lambda_2^\prime}_{\lambda_3^\prime}(\mathbf{p}_3,\mathbf{p}_1,\mathbf{p}_2),
\end{equation}
where each helicity $\lambda_i^\prime$ is defined along the momentum direction in the lab frame. In this cascade decay, the helicity of the parent particle-$X$ is the same in both conventions.

\begin{enumerate}
\item Two-body decay COM frame helicity convention

\begin{equation}\label{Eq:no little group one cascade decay}
A^{\lambda_1\lambda_2\lambda_Y}_{\lambda_X}(\mathbf{k}_X,\mathbf{q}^*_Y,\mathbf{q}^*_1,\mathbf{q}^*_2)
=
\sum_{\lambda_3}
A^{\lambda_3\lambda_Y}_{\lambda_X}(\mathbf{k}_X,\mathbf{q}^*_3,\mathbf{q}^*_Y)
P_3
A^{\lambda_1\lambda_2}_{\lambda_3}(\mathbf{k}_3,\mathbf{q}^*_1,\mathbf{q}^*_2),
\end{equation}
where each amplitude is defined in the helicity COM frame. The helicity COM frame is determined by starting from the initial COM frame and applying the helicity-standard boost to an intermediate particle produced from the two-body decay of the parent particle, bringing it to its rest frame. This frame is then defined as the COM frame for the subsequent two-body decay of that intermediate particle as the new parent. Proceeding step by step via successive helicity-standard boosts in this manner, we obtain the helicity COM frames for all two-body decays in the cascade.

Here, we can sum over the helicities of the intermediate state particle in the two amplitudes because, for the intermediate particle, the helicity quantization axes are consistent between the two helicity COM frames. Suppose we consider the first two-body decay, where the particle state is $|\mathbf{q}_3^*, \lambda_3\rangle$, and in the second two-body decay, the particle state is $|\mathbf{k}_3, \lambda_3^\prime\rangle$. The transformation between these two helicity COM frames is given by the helicity standard boost for particle-3 with momentum $\mathbf{q}^*_3$, which can be decomposed as 
\begin{equation}
    L^{-1}_{3,h}(\mathbf{q}^*_3)=R^{-1}(\phi_1,\theta_1,0)L^{-1}_{3,c}(\mathbf{q}^*_3),
\end{equation} 
where $L^{-1}_{3,c}(\mathbf{q}^*_3)$ is a pure boost along the momentum direction $\mathbf{q}^*_3$.
Due to the rotational invariance of helicity and its invariance under the pure boosts along the momentum direction, we ultimately have 
\begin{equation}
\begin{aligned}
     L^{-1}_{3,h}(\mathbf{q}^*_3)|\mathbf{q}_3^*, \lambda_3\rangle&=R^{-1}(\phi_1,\theta_1,0)L^{-1}_{3,c}(\mathbf{q}^*_3)|\mathbf{q}_3^*, \lambda_3\rangle\\
     &=|\mathbf{k}_3,\lambda_3\rangle,
\end{aligned}
\end{equation}
which indicates that $\lambda_3=\lambda_3^\prime$. Thus, the choice of helicity quantization axis for the intermediate particle state is consistent, and we can directly sum over them. This demonstrates that when defining the helicity COM frame, it must be obtained by boosting from the decay COM frame of its parent particle. This ensures that the helicity quantization axes for each intermediate particle are consistent, thus allowing summation over them.

Starting from the first step two-body decay $X \to 3 + Y$, in the particle-$X$ rest frame. From Eq.~\eqref{Eq:helicity amplitude}, the helicity amplitude in the COM frame can be written as
\begin{equation}
    A^{\lambda_3\lambda_Y}_{\lambda_X}(\mathbf{k}_X,\mathbf{q}^*_3,\mathbf{q}^*_Y)
    = \sqrt{\frac{2s_X+1}{4\pi}}
      D^{\lambda_{X}(s_X)*}_{\,\,\lambda_3-\lambda_Y}(\phi_{1},\theta_{1},0)
      H^{\lambda_{3}\lambda_{Y}},
\end{equation}
where $(\theta_{1},\phi_{1})$ are the corresponding helicity angles. 

For the second step two-body decay $3 \to 1 + 2$, in the particle-$3$ rest frame, the helicity amplitude can be written as 
\begin{equation}
    A^{\lambda_1\lambda_2}_{\lambda_3}(\mathbf{k}_3,\mathbf{q}^*_1,\mathbf{q}^*_2)
    = \sqrt{\frac{2s_3+1}{4\pi}}
      D^{\lambda_{3}(s_3)*}_{\,\,\lambda_1-\lambda_2}(\phi_{2},\theta_{2},0)
      H^{\lambda_{1}\lambda_{2}},
\end{equation}
where $(\theta_{2},\phi_{2})$ are the corresponding helicity angles. 
The geometric definitions of these two sets of angles can be found in Fig.~\ref{fig:decay-plane}.

\begin{figure}[H]
\centering

\includegraphics[scale=1.1]{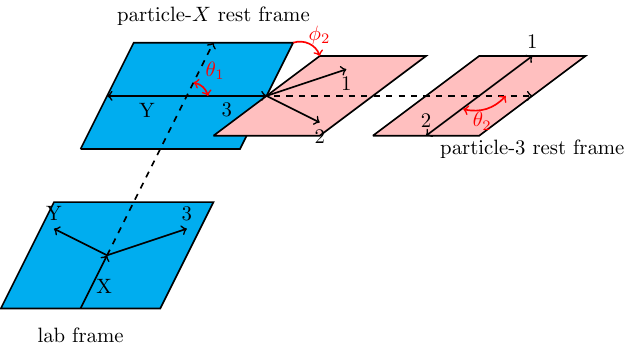}

\caption{Definitions of helicity angles for the decay chain $X \to 3 + Y$, $3 \to 1+2$.
For the first two-body decay $X \to 3 + Y$, we may choose the decay plane such that $\phi_1=0$. For a single decay chain this only introduces an overall phase and does not affect physical results. The corresponding helicity angles for this decay are therefore $(\theta_1,0)$. For the second two-body decay, the rest frame of particle-3 is obtained from the rest frame of particle-$X$ by applying the helicity-standard boost of particle-3, $L^{-1}_{3,h}(\mathbf q_3^{*})$. In this frame, the helicity angles of the decay $3 \to 1+2$ are $(\theta_2,\phi_2)$.}
\label{fig:decay-plane}
\end{figure}

\item Lab helicity convention

Under the lab frame convention, the helicities of all particles are defined with respect to the directions of their momenta in the lab frame. Therefore, one can compute each two-body subprocess in the cascade decay by using the lab frame helicity two-body decay formula, Eq.~\eqref{Eq:helicity in any frame}. 
In particular, for the intermediate resonance, its helicity is defined along its lab momentum direction both when it appears as a daughter particle and when it appears as a parent particle. 
Hence the helicity of the intermediate state is the same in the two roles, and the intermediate helicity can be summed over directly. As a result, the cascade decay amplitude can be written as 
\begin{equation}\label{Eq:helicity cascade decayy}
A^{\lambda_1^\prime\lambda_2^\prime\lambda_Y^\prime}_{\lambda_X}(\mathbf{p}_X,\mathbf{p}_1,\mathbf{p}_2,\mathbf{p}_Y)
=
\sum_{\lambda_3^\prime}
 A^{\lambda_3^\prime\lambda_Y^\prime}_{\lambda_X}(\mathbf{p}_X,\mathbf{p}_3,\mathbf{p}_Y)
P_3
 A^{\lambda_1^\prime\lambda_2^\prime}_{\lambda_3^\prime}(\mathbf{p}_3,\mathbf{p}_1,\mathbf{p}_2).
\end{equation}
We next write down the lab frame amplitudes for each two-body decay. 

For the two-body decay $X\to 3+Y$, the two-body decay amplitude in the lab frame can be written as 
\begin{equation}
A^{\lambda_3^\prime\lambda_Y^\prime}_{\lambda_X}(\mathbf{p}_X,\mathbf{p}_3,\mathbf{p}_Y)
=
\sum_{\tilde\lambda_3 \tilde\lambda_Y}
A^{\tilde\lambda_3\tilde\lambda_Y}_{\lambda_X}(\mathbf{k}_X,\mathbf{\tilde q}^*_3,\mathbf{\tilde q}^*_Y)
D^{\tilde\lambda_3(s_3)*}_{\,\,\lambda_3^\prime}(R_{X3}^\prime)
D^{\tilde\lambda_Y(s_Y)*}_{\,\,\lambda_Y^\prime}(R_{XY}^\prime),
\end{equation}
where the helicity COM frame associated with $A^{\tilde\lambda_3\tilde\lambda_Y}_{\lambda_X}(\mathbf{k}_X,\mathbf{\tilde q}^*_3,\mathbf{\tilde q}^*_Y)$ is obtained from the lab frame by applying the helicity-standard boost of particle $X$. 
This choice ensures that in the rest frame of particle-$X$, its spin quantization axis is the $z$-axis. 
When boosting the system $X\to 3+Y$, for the state of particle-$3$, helicity is invariant under rotations and also invariant under boosts along its momentum direction, so its helicity coincides with the helicity defined in the lab frame. 
For particle-$3$ and particle-$Y$, the helicity boost induces the Wigner rotation $R_{Xi}^\prime\ (i=3,Y)$. This is done to ensure that, in the final two-body decay amplitude written in the lab frame, the helicities of particle-3 and particle-$Y$ are defined along their momentum directions in the lab frame.
From Eq.~\eqref{Eq:helicity wiger rotation}, this Wigner rotation can be written as 
\begin{equation}
    R^{\prime}_{Xi}=R^{-1}(\mathbf{p}^*_i)R_{Xi}(\boldsymbol{\omega}_i)R(\mathbf{p}_X),
\end{equation}
where
\begin{equation}
\begin{aligned}
R_{Xi}(\boldsymbol{\omega}_i)=&R_{\hat{\mathbf{n}}}R_{z}(\omega_{Xi})R^{-1}_{\hat{\mathbf{n}}},\\
\tan\frac{\omega_{Xi}}{2}
=&
\frac{\sin\theta_{Xi}\sinh\left(\frac{\eta_X}{2}\right)\sinh\left(\frac{\eta_i}{2}\right)}
{\cosh\left(\frac{\eta_X}{2}\right)\cosh\left(\frac{\eta_i}{2}\right)
+\cos\theta_{Xi}\sinh\left(\frac{\eta_X}{2}\right)\sinh\left(\frac{\eta_i}{2}\right)},\\
\mathbf{p}_i^*=&R(\hat{\mathbf{p}}_X)\mathbf{\tilde q}^*_i,\\
\hat{\mathbf n}_{Xi}\parallel&-\mathbf p_X \times \mathbf p_i.
\end{aligned}
\end{equation}
The helicity amplitude in the lab frame can be expressed as 
\begin{equation}
A^{\lambda_3^\prime\lambda_Y^\prime}_{\lambda_X}(\mathbf{p}_X,\mathbf{p}_3,\mathbf{p}_Y)
=
\sum_{\tilde\lambda_3 \tilde\lambda_Y}
\sqrt{\frac{2s_X+1}{4\pi}}
      D^{\lambda_{X}(s_X)*}_{\,\,\tilde\lambda_3-\tilde\lambda_Y}(\phi_{1}^\prime,\theta_{1}^\prime,0)
      H^{\tilde\lambda_3\tilde\lambda_Y}
D^{\tilde\lambda_3(s_3)*}_{\,\,\lambda_3^\prime}(R_{X3}^\prime)
D^{\tilde\lambda_Y(s_Y)*}_{\,\,\lambda_Y^\prime}(R_{XY}^\prime).
\end{equation}

For the two-body decay $3\to 1+2$, the treatment here differs from the two-body decay COM frame helicity convention. 
In the two-body decay COM frame helicity convention, one needs to perform step-by-step helicity boosts to reach the corresponding helicity COM frame. 
Here, however, it is sufficient to start from the lab frame and apply a single helicity standard boost associated with the lab frame momentum of particle-3, which brings particle-3 to its rest frame. 
In this case, the two-body decay amplitude in lab frame can be written as 
\begin{equation}
A_{\lambda_3^\prime}^{\lambda_1^\prime\lambda_2^\prime}(\mathbf p_3,\mathbf p_1,\mathbf p_2)
=
\sum_{\tilde\lambda_1 \tilde\lambda_2}
A_{\lambda_3^\prime}^{\tilde\lambda_1\tilde\lambda_2}(\mathbf k_3,\mathbf{\tilde q}_1^{*},\mathbf{\tilde q}_2^{*})
D^{\tilde\lambda_1(s_1)*}_{\,\,\tilde\lambda_1^\prime}(R_{31}^\prime)
D^{\tilde\lambda_2(s_2)*}_{\,\,\lambda_2^\prime}(R_{32}^\prime).
\end{equation}
For particle-3, this helicity boost consists of a rotation followed by a pure boost along the momentum direction of particle-3. 
Therefore, the helicity of particle-3 in the two-body COM frame coincides with its helicity defined in the lab frame. For particle-1 and particle-2, one applies the Wigner rotation $R_{3i}\ (i=1,2)$ induced by the helicity boost, in order to ensure that in the final two-body decay amplitude written in the lab frame, the helicities of particle-1 and particle-2 are defined along their momentum directions in the lab frame.
From Eq.~\eqref{Eq:helicity wiger rotation}, this Wigner rotation can be written as 
\begin{equation}
    R^{\prime}_{3i}=R^{-1}(\mathbf{p}^*_i)R_{3i}(\boldsymbol{\omega}_i)R(\mathbf{p}_3),
\end{equation}
where
\begin{equation}
\begin{aligned}
R_{3i}(\boldsymbol{\omega}_i)=&R_{\hat{\mathbf{n}}}R_{z}(\omega_{3i})R^{-1}_{\hat{\mathbf{n}}},\\
\tan\frac{\omega_{3i}}{2}
=&
\frac{\sin\theta_{3i}\sinh\left(\frac{\eta_3}{2}\right)\sinh\left(\frac{\eta_i}{2}\right)}
{\cosh\left(\frac{\eta_3}{2}\right)\cosh\left(\frac{\eta_i}{2}\right)
+\cos\theta_{3i}\sinh\left(\frac{\eta_3}{2}\right)\sinh\left(\frac{\eta_i}{2}\right)},\\
\mathbf{p}_i^*=&R(\hat{\mathbf{p}}_3)\mathbf{\tilde q}^*_i,\\
\hat{\mathbf n}_{3i}\parallel&-\mathbf p_3 \times \mathbf p_i.
\end{aligned}
\end{equation}
The helicity amplitude in the lab frame can be expressed as 
\begin{equation}
A^{\lambda_1^\prime\lambda_2^\prime}_{\lambda_3^\prime}(\mathbf{p}_3,\mathbf{p}_1,\mathbf{p}_2)
=
\sum_{\tilde\lambda_1 \tilde\lambda_2}
\sqrt{\frac{2s_3+1}{4\pi}}
      D^{\lambda_{3}^\prime(s_3)*}_{\,\,\tilde\lambda_1-\tilde\lambda_2}(\phi_{2}^\prime,\theta_{2}^\prime,0)
      H^{\tilde\lambda_1\tilde\lambda_2}
D^{\tilde\lambda_1(s_1)*}_{\,\,\lambda_1^\prime}(R_{31}^\prime)
D^{\tilde\lambda_2(s_2)*}_{\,\,\lambda_2^\prime}(R_{32}^\prime).
\end{equation}

In each of the two-body decays discussed above, we use Wigner rotations to align the spin quantization axes with the momentum directions of the corresponding particles in the lab frame. 
Consequently, in the amplitudes of these two-body decays, the helicity of particle-3 is always defined along the momentum direction of particle-3 in the lab frame. 
By summing over the helicity of the intermediate particle-3, the total cascade decay amplitude can be expressed as
\begin{equation}
A^{\lambda_1^\prime\lambda_2^\prime\lambda_Y^\prime}_{\lambda_X}(\mathbf{p}_X,\mathbf{p}_1,\mathbf{p}_2,\mathbf{p}_Y)
=
\sum_{\lambda_3^\prime}
 A^{\lambda_3^\prime\lambda_Y^\prime}_{\lambda_X}(\mathbf{p}_X,\mathbf{p}_3,\mathbf{p}_Y)
P_3
 A^{\lambda_1^\prime\lambda_2^\prime}_{\lambda_3^\prime}(\mathbf{p}_3,\mathbf{p}_1,\mathbf{p}_2).
\end{equation}
The geometric of the cascade decay process shows in Fig.~\ref{fig:helicity lab cascade decay}.
\begin{figure}[H]
	\centering
	\includegraphics[scale=1]{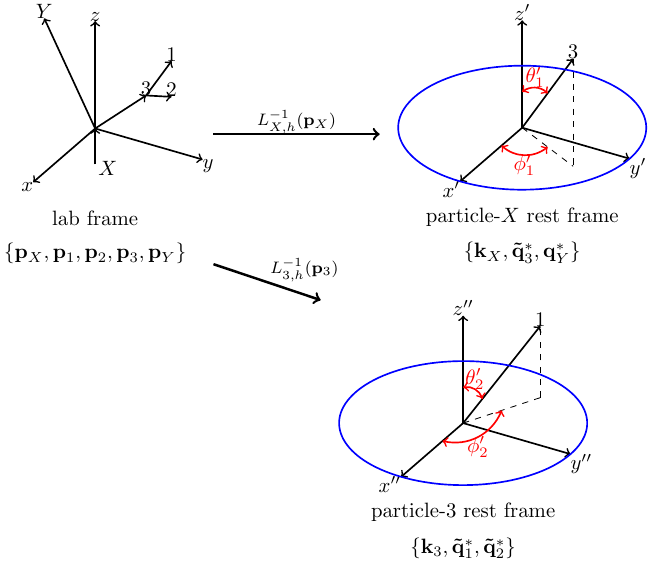}
	\caption{$L^{-1}_{X,h}(\mathbf p_X)$ is the helicity boost from the lab frame to the rest frame of particle-$X$. It ensures that after the boost the spin quantization axis of particle-$X$ points along the $z^\prime$ direction, with the associated coordinate system denoted by $\{x^\prime, y^\prime, z^\prime\}$. In the two-body decay $X\to 3+Y$ COM frame, the momenta of the particles in this frame are $\mathbf{k}_X,\mathbf{\tilde q}^*_3,\mathbf{\tilde q}^*_Y$, and the helicity angles are $(\phi^\prime_1,\theta^\prime_1,0)$. $L^{-1}_{3,h}(\mathbf p_3)$ is the helicity boost from the lab frame to the rest frame of particle-$3$. It ensures that after the boost the spin quantization axis of particle-$3$ points along the $z^{\prime\prime}$ direction, with the associated coordinate system denoted by $\{x^{\prime\prime}, y^{\prime\prime}, z^{\prime\prime}\}$. In the two-body decay $3\to 1+2$ COM frame, the momenta of the particles in this frame are $\mathbf{k}_3,\mathbf{\tilde q}^*_1,\mathbf{\tilde q}^*_2$, and the helicity angles are $(\phi^\prime_2,\theta^\prime_2,0)$.}
	\label{fig:helicity lab cascade decay}
\end{figure}

\end{enumerate}

The two constructions above differ only by a choice of helicity convention.
They are therefore physically equivalent, provided the corresponding Wigner $D$-matrices induced by non collinear boosts are treated consistently.
In particular, when we are concerned with some physical observables, we sum over the final-state helicities and take the modulus squared of the amplitude. One has
\begin{equation}
\sum_{\lambda_1^\prime\lambda_2^\prime\lambda_Y^\prime\lambda_X}
\big|A^{\lambda_1^\prime\lambda_2^\prime\lambda_Y^\prime}_{\lambda_X}\big|^2
=
\sum_{\lambda_1\lambda_2\lambda_Y\lambda_X}
\big|A^{\lambda_1\lambda_2\lambda_Y}_{\lambda_X}\big|^2,
\end{equation}
so both conventions lead to the same physical decay rate.

Despite this equivalence for a single decay chain, the situation becomes practically different when multiple decay chains contribute to the same final state. 
In the two-body COM frame helicity convention, the step-by-step helicity constructions are performed in different two-body rest frames, and different decay topologies generally lead to different helicity definitions (spin quantization axes) for the same external particle. 
As a result, there is a basis mismatch between chain amplitudes, and a direct coherent sum of the amplitudes from different chains is in general incorrect. 
To obtain a meaningful total amplitude, one must first bring all chain amplitudes into a common helicity basis by applying the appropriate alignment rotations.

In contrast, in the lab-frame helicity convention the helicities of all external particles are defined with respect to their momentum directions in the same lab frame. 
Therefore, the spin bases are already aligned across different decay chains, and the chain amplitudes can be coherently summed directly without any additional alignment.
In the following, we discuss these two treatments in detail.

\paragraph{Helicity cascade decay amplitude for multiple decay chain}

\begin{enumerate}
\item Two-body decay COM frame helicity convention for multiple decay chain

According to the convention of the two-body decay helicity COM frame, since the topological structure of each cascade decay chain differs, the step-by-step helicity boost procedure from the initial COM frame to the frame in which the final-state particles reside must differ. 
Therefore, when summing the amplitudes from different decay chains, the mismatch in the definition of the final-state particle helicities arises. 
To resolve this issue and correctly sum the contributions from each decay chain, we must introduce the Wigner rotation to align the helicities of the final-state particles. 
Upon alignment, the amplitudes from each decay chain can be coherently summed to obtain the total cascade decay amplitude. 
Next, we will discuss in detail the implementation of this alignment rotation.

As an explicit example, we consider two decay chains,
\begin{itemize} 
	\item Decay chain-1: $X \to 3 + Y$, followed by $3 \to 1 + 2$
	\item Decay chain-2: $X \to 4 + 2$, followed by $4 \to 1 + Y$
\end{itemize}

The helicity amplitude for the decay chain-1 is
\begin{equation}
A^{\lambda_1\lambda_2\lambda_Y}_{\lambda_X}(\mathbf{k}_X,\mathbf{q}^*_Y,\mathbf{q}^*_1,\mathbf{q}^*_2;3)
=
\sum_{\lambda_3}
A^{\lambda_3\lambda_Y}_{\lambda_X}(\mathbf{k}_X,\mathbf{q}^*_3,\mathbf{q}^*_Y)
P_3
A^{\lambda_1\lambda_2}_{\lambda_3}(\mathbf{k}_3,\mathbf{q}^*_1,\mathbf{q}^*_2).
\end{equation}
For the two-body decay $X\to 3+Y$, the helicity of particle-$X$ $(\lambda_X)$ is defined along the $z$-axis in the COM frame of this two-body decay, while the helicities of particle-3 and particle-$Y$ $(\lambda_3,\lambda_Y)$ are defined along their respective momentum directions in the same COM frame. 

For the two-body decay $3\to 1+2$, the helicity of particle-3 $(\lambda_3)$ is defined along the $z^\prime$-axis  in the COM frame of this two-body decay (This $z^\prime$-axis is fixed by applying the helicity boost from the COM frame of the decay $X\to 3+Y$.), while the helicities of particle-1 and particle-2 $(\lambda_1,\lambda_2)$ are defined along their respective momentum directions in that COM frame.

As for the helicity of particle-3 in these two frames, because helicity is invariant under rotations and also invariant under boosts along the direction of momentum, and because the transformation from the rest frame of $X$ to the rest frame of particle-3 is achieved by a single helicity boost, the helicity of particle-3 $\lambda_3$ remains unchanged between the two rest frames. Equivalently, the spin quantization axis associated with particle-3 is the same in both decays. 
Therefore, one can sum over the intermediate helicity of particle-3 to obtain the cascade decay amplitude. The amplitudes in these two COM frames can be written as 
\begin{equation}
    A^{\lambda_3\lambda_Y}_{\lambda_X}(\mathbf{k}_X,\mathbf{q}^*_3,\mathbf{q}^*_Y)
    = \sqrt{\frac{2s_X+1}{4\pi}}
      D^{\lambda_{X}(s_X)*}_{\,\,\lambda_3-\lambda_Y}(\phi_{1},\theta_{1},0)
      H^{\lambda_{3}\lambda_{Y}},
\end{equation}
 and
\begin{equation}
    A^{\lambda_1\lambda_2}_{\lambda_3}(\mathbf{k}_3,\mathbf{q}^*_1,\mathbf{q}^*_2)
    = \sqrt{\frac{2s_3+1}{4\pi}}
      D^{\lambda_{3}(s_3)*}_{\,\,\lambda_1-\lambda_2}(\phi_{2},\theta_{2},0)
      H^{\lambda_{1}\lambda_{2}},
\end{equation}
where $(\theta_{1},\phi_{1})$ and $(\theta_{2},\phi_{2})$ are the corresponding helicity angles show in Fig.~\ref{fig:decay-plane2}.

The helicity amplitude for the decay chain-2 is
\begin{equation}
A^{\bar\lambda_1\bar\lambda_2\bar\lambda_Y}_{\lambda_X}(\mathbf{k}_X,\mathbf{\bar q}^*_Y,\mathbf{\bar q}^*_1,\mathbf{\bar q}^*_2;4)
=
\sum_{\bar\lambda_4}
A^{\bar\lambda_4\bar\lambda_2}_{\lambda_X}(\mathbf{k}_X,\mathbf{\bar q}^{*}_4,\mathbf{\bar q}^{*}_2)
P_4
A^{\bar\lambda_1\bar\lambda_Y}_{\bar\lambda_4}(\mathbf{k}_4,\mathbf{\bar q}^{*}_1,\mathbf{\bar q}^{*}_Y).
\end{equation}
For the two-body decay $X\to 4+2$, the helicity of particle-$X$ $(\lambda_X)$ is defined along the $z$-axis in the COM frame of this two-body decay, while the helicities of particle-4 and particle-2 $(\bar\lambda_4,\bar\lambda_2)$ are defined along their respective momentum directions in the same COM frame. 

For the two-body decay $4\to 1+Y$, the helicity of particle-4 $(\bar\lambda_4)$ is defined along the $z^{\prime\prime}$-axis  in the COM frame of this two-body decay (This $z^{\prime\prime}$-axis is fixed by applying the helicity boost from the COM frame of the decay $X\to 4+2$.), while the helicities of particle-1 and particle-$Y$ $(\bar\lambda_1,\bar\lambda_Y)$ are defined along their respective momentum directions in that COM frame.

As for the helicity of particle-4 remains unchanged between the two COM frames. Equivalently, the spin quantization axis associated with particle-4 is the same in both decays. 
Therefore, one can sum over the intermediate helicity of particle-4 to obtain the cascade decay amplitude. The amplitudes in these two COM frames can be written as 
\begin{equation}
    A^{\bar\lambda_4\bar\lambda_2}_{\lambda_X}(\mathbf{k}_X,\mathbf{\bar q}^*_4,\mathbf{\bar q}^*_2)
    = \sqrt{\frac{2s_X+1}{4\pi}}
      D^{\lambda_{X}(s_X)*}_{\,\,\bar\lambda_4-\bar\lambda_2}(\bar\phi_{1},\bar\theta_{1},0)
      H^{\bar\lambda_{4}\bar\lambda_{2}},
\end{equation}
 and
\begin{equation}
    A^{\bar\lambda_1\bar\lambda_Y}_{\bar\lambda_4}(\mathbf{k}_4,\mathbf{\bar q}^*_1,\mathbf{\bar q}^*_Y)
    = \sqrt{\frac{2s_4+1}{4\pi}}
      D^{\bar\lambda_{4}(s_4)*}_{\,\,\bar\lambda_1-\bar\lambda_Y}(\bar\phi_{2},\bar\theta_{2},0)
      H^{\bar\lambda_{1}\bar\lambda_{Y}},
\end{equation}
where $(\bar\theta_{1},\bar\phi_{1})$ and $(\bar\theta_{2},\bar\phi_{2})$ are the corresponding helicity angles show in Fig.~\ref{fig:decay-plane1}.

\begin{figure}[H]
\centering

\includegraphics[scale=1.1]{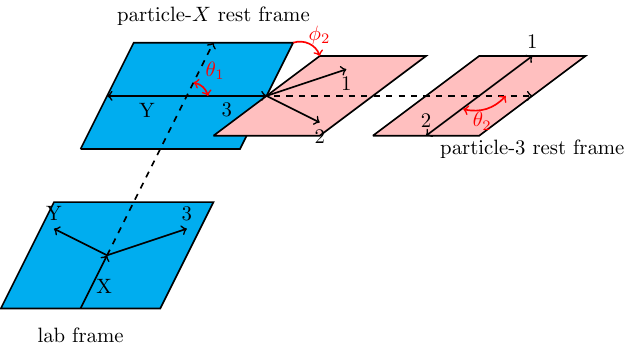}

\caption{Definitions of helicity angles for the first decay chain $X \to 3 + Y$, $3 \to 1+2$. This decay chain is taken as the reference chain. We define the rest frame of particle-$X$ as the coordinate system associated with the blue plane. For the first two-body decay $X \to 3 + Y$, the helicity angles are $(\theta_1,0)$. For the second two-body decay $3 \to 1+2$, defined in the rest frame of particle-3, the helicity angles are $(\theta_2,\phi_2)$.}
\label{fig:decay-plane2}
\end{figure}

\begin{figure}[H]
\centering

\includegraphics[scale=1.1]{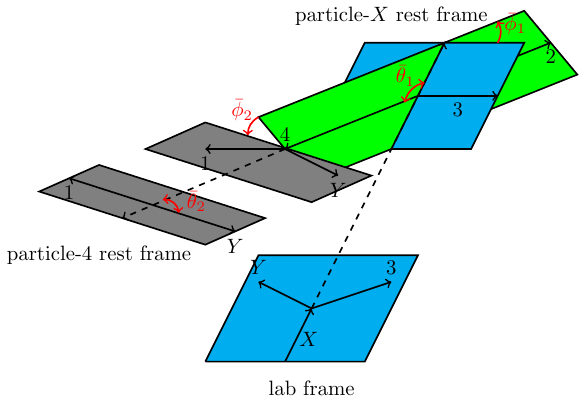}

\caption{Definitions of helicity angles for the second decay chain $X \to 2 + 4$, $4 \to 1+Y$. The rest frame of particle-$X$ is taken to be the coordinate system associated with the blue plane. For the first two-body decay $X \to 2 + 4$, the helicity angles are $(\bar\theta_1,\bar\phi_1)$. For the second two-body decay $4 \to 1+Y$, defined in the rest frame of particle-4, the helicity angles are $(\bar\theta_2,\bar\phi_2)$.}
\label{fig:decay-plane1}
\end{figure}

One can see that the two cascade decay amplitudes define the final-state helicities with different directions, namely they correspond to different spin quantization axes. To obtain a common set of quantization axes, we next align one decay chain to the reference chain. Here we choose decay chain-1 as the reference, and decay chain-2 is to be aligned to it. 

We first define the relations among the relevant reference frames in decay chain-1.
\begin{itemize} 
\item The decay $X\to 3+Y$ is defined in the rest frame of particle-$X$, which we denote by $\boldsymbol{S}_X$. 
\item In the rest frame of particle-$Y$, obtained by applying the helicity boost from the rest frame of particle-$X$, we define it as $\boldsymbol{S}_Y$.
\item The second two-body decay $3\to 1+2$ is defined in the rest frame of particle-3, which we denote by $\boldsymbol{S}_3$.
\item By boosting from the rest frame of particle-3 to the rest frames of the particle-1 and particle-2, we define the corresponding frames $\boldsymbol{S}_1$ and $\boldsymbol{S}_2$, respectively.
\end{itemize}
For decay chain-2, we denote the rest frames of the relevant particles by
\begin{itemize} 
\item The decay $X\to 2+4$ is defined in the rest frame of particle-$X$, which is the same as in decay chain-1, and we denote it by $\boldsymbol{S}_X$.
\item In the rest frame of particle-$2$, obtained by applying the helicity boost from the rest frame of particle-$X$, we define it as $\boldsymbol{\bar S}_2$.
\item The second two-body decay $4\to 1+Y$ is defined in the rest frame of particle-4, which we denote by $\boldsymbol{\bar S}_4$.
\item By boosting from the rest frame of particle-4 to the rest frames of the particle-1 and particle-$Y$, we define the corresponding frames $\boldsymbol{\bar S}_1$ and $\boldsymbol{\bar S}_Y$, respectively.
\end{itemize}
These different frames can be related via helicity boosts starting from $\boldsymbol{S}_X$. 
For decay chain-1, the relations among the frames are given by 
\begin{equation}\label{Eq:frame change1}
\begin{aligned}
    L^{-1}_{i,h}(\mathbf{q}^*_{i})\boldsymbol{S}_{X}=&\boldsymbol{S}_{i}\quad (i=3,Y),\\
    L^{-1}_{j,h}(\mathbf{q}^{*}_j)\boldsymbol{S}_{3}=&\boldsymbol{S}_j\quad  (j=1,2).
\end{aligned}
\end{equation}
For decay chain-2, the relations among the frames are given by 
\begin{equation}\label{Eq:frame change2}
\begin{aligned}
    L^{-1}_{i,h}(\mathbf{\bar q}^*_{i})\boldsymbol{S}_{X}=&\boldsymbol{\bar S}_{i}\quad (i=4,2),\\
    L^{-1}_{j,h}(\mathbf{\bar q}^{*}_j)\boldsymbol{\bar S}_{4}=&\boldsymbol{\bar S}_j\quad  (j=1,Y).
\end{aligned}
\end{equation}
Since the helicity of each final-state particle is defined along its momentum direction in the corresponding two-body decay COM frame, and since helicity is invariant under rotations and also invariant under boosts along that direction, the helicity of the particle remains unchanged after performing the helicity boost from the two-body decay COM frame to its rest frame. 
Equivalently, it becomes the $z$-axis direction in the particle’s own rest frame. 
Therefore, starting from the final-state particle rest frames in the reference decay chain, one can use the sequence of frame transformations introduced above to boost step by step to the final-state particle rest frames in the other decay chain. 
The net effect of this sequence of helicity boosts is a pure rotation. This rotation is precisely the alignment rotation that must be applied to the non-reference decay chain.

For example, using Eqs.~\eqref{Eq:frame change1} and \eqref{Eq:frame change2}, for particle-1, the relation between its rest frames in the two decay chains, $\boldsymbol{S}_1$ and $\boldsymbol{\bar S}_1$, is given by 
\begin{equation}
     L^{-1}_{1,h}(\mathbf{\bar q}^*_{1})L^{-1}_{4,h}(\mathbf{\bar q}^*_{4})L_{3,h}(\mathbf{q}^*_{3})L_{1,h}(\mathbf{q}^{*}_1)\boldsymbol{S}_1=\boldsymbol{\bar S}_1,
\end{equation}
From this, the relation between the two helicities $\lambda_1$ and $\bar{\lambda}_1$ can be determined by
\begin{equation}
   L^{-1}_{1,h}(\mathbf{\bar q}^*_{1})L^{-1}_{4,h}(\mathbf{\bar q}^*_{4})L_{3,h}(\mathbf{q}^*_{3})L_{1,h}(\mathbf{q}^{*}_1)|s_1\lambda_1\rangle=\sum_{\bar{\lambda}_1}D^{\bar{\lambda}_1}_{\,\,\lambda_1}(R_{\bar{1}1})|s_1\bar{\lambda}_1\rangle,
\end{equation}
where $R_{\bar{1}1}$ denotes the alignment rotation that aligns $\bar\lambda_1$ of the decay chain-2 with $\lambda_1$ of the decay chain-1, and can be written as
\begin{equation}
    R_{\bar{1}1}=L^{-1}_{1,h}(\mathbf{\bar q}^*_{1})L^{-1}_{4,h}(\mathbf{\bar q}^*_{4})L_{3,h}(\mathbf{q}^*_{3})L_{1,h}(\mathbf{q}^{*}_1).
\end{equation}
Similarly, the alignment rotations for particle-2 and particle-Y can be written as
\begin{align}
    R_{\bar{2}2}=&L^{-1}_{2,h}(\mathbf{\bar q}^*_2)L_{3,h}(\mathbf{q}^*_3)L_{2,h}(\mathbf{q}^*_2),\\
    R_{\bar{Y}Y}=&L^{-1}_{Y,h}(\mathbf{\bar q}^*_Y)L^{-1}_{4,h}(\mathbf{\bar q}^*_{4})L_{Y,h}(\mathbf{q}^*_Y).
\end{align}
After this alignment, the amplitude of decay chain-2 can be written as 
\begin{equation}
A^{\lambda_1\lambda_2\lambda_Y}_{\lambda_X}(\mathbf{k}_X,\mathbf{\bar q}^*_Y,\mathbf{\bar q}^*_1,\mathbf{\bar q}^*_2;4)=\sum_{\bar{\lambda}_1,\bar{\lambda}_2,\bar{\lambda}_Y}A^{\bar\lambda_1\bar\lambda_2\bar\lambda_Y}_{\lambda_X}(\mathbf{k}_X,\mathbf{\bar q}^*_Y,\mathbf{\bar q}^*_1,\mathbf{\bar q}^*_2;4)D^{\bar{\lambda}_1*}_{\,\,\lambda_1}(R_{\bar{1}1})D^{\bar{\lambda}_2*}_{\,\,\lambda_2}(R_{\bar{2}2})D^{\bar{\lambda}_Y*}_{\,\,\lambda_Y}(R_{\bar{Y}Y}).
\end{equation}
Finally, the total amplitude is obtained by directly adding the amplitudes of the two decay chains, and can be written as 
\begin{equation}
\begin{aligned}
A_{\lambda_{1}}^{\lambda_{2}\lambda_{3}\lambda_{4}}=&A^{\lambda_1\lambda_2\lambda_Y}_{\lambda_X}(\mathbf{k}_X,\mathbf{q}^*_Y,\mathbf{q}^*_1,\mathbf{q}^*_2;3)+A^{\lambda_1\lambda_2\lambda_Y}_{\lambda_X}(\mathbf{k}_X,\mathbf{\bar q}^*_Y,\mathbf{\bar q}^*_1,\mathbf{\bar q}^*_2;4)\\
=&A^{\lambda_1\lambda_2\lambda_Y}_{\lambda_X}(\mathbf{k}_X,\mathbf{q}^*_Y,\mathbf{q}^*_1,\mathbf{q}^*_2;3)\\
&+\sum_{\bar{\lambda}_1,\bar{\lambda}_2,\bar{\lambda}_Y}A^{\bar\lambda_1\bar\lambda_2\bar\lambda_Y}_{\lambda_X}(\mathbf{k}_X,\mathbf{\bar q}^*_Y,\mathbf{\bar q}^*_1,\mathbf{\bar q}^*_2;4)D^{\bar{\lambda}_1*}_{\,\,\lambda_1}(R_{\bar{1}1})D^{\bar{\lambda}_2*}_{\,\,\lambda_2}(R_{\bar{2}2})D^{\bar{\lambda}_Y*}_{\,\,\lambda_Y}(R_{\bar{Y}Y}).
\end{aligned}
\end{equation}

In a more general case with additional decay chains and more final particles, the procedure for handling the cascade decay amplitude is to choose one decay chain as the reference chain and align the helicities of the final-state particles in all other decay chains with the helicities in the reference chain. Concretely, starting from the rest frames of the particles in a given decay chain, successive helicity boosts are applied step by step to reach the reference frame of the initial parent particle; from this frame, helicity boosts are then applied step by step to reach the rest frames of the particles in the reference chain. The detailed procedure can be written as 
\begin{equation}
    \bar{S}_i\xrightarrow{L^{-1}_{i,h}(\mathbf{\bar q}^*_i)}\cdots\xrightarrow{L^{-1}_{j,h}(\mathbf{\bar q}_j^*)}S_X\xrightarrow{L_{k,h}(\mathbf{q}^*_{k})}\cdots\xrightarrow{L_{i,h}(\mathbf{q}^{*}_i)}S_i,
\end{equation}
and the corresponding alignment rotation is given by 
\begin{equation}
    R_{\bar{i}i}=L^{-1}_{i,h}(\mathbf{\bar q}^*_i)\cdots L^{-1}_{j,h}(\mathbf{\bar q}_j^*)L_{k,h}(\mathbf{q}^*_{k})\cdots L_{i,h}(\mathbf{q}^{*}_i).
\end{equation}
By applying the corresponding alignment rotations to the helicities of the final-state particles in the other decay chains, the helicities of all decay chains can be aligned. The total cascade decay amplitude is then obtained by summing the amplitudes with aligned helicities.

\item Lab helicity convention for multiple decay chain

For a cascade process with multiple decay chains, each chain is defined in the same lab frame. Since the lab helicity convention of all final-state particles are defined with respect to the momentum direction of particles, the spin polarization directions are consistent among different chains. Therefore, one can first compute the cascade amplitude of each decay chain in the lab frame, and then obtain the total amplitude by directly summing the chain amplitudes.

As an explicit example, we consider two decay chains,
\begin{itemize} 
	\item Decay chain-1: $X \to 3 + Y$, followed by $3 \to 1 + 2$
	\item Decay chain-2: $X \to 4 + 2$, followed by $4 \to 1 + Y$
\end{itemize}
In the lab frame, following the single chain construction discussed above and using Eq.~\eqref{Eq:helicity cascade decayy}, the corresponding cascade decay amplitudes are given by
\begin{equation}
\begin{aligned}
A_{\lambda_X}^{\lambda_{1}\lambda_{2}\lambda_{Y}}(\mathbf{p}_X,\mathbf{p}_1,\mathbf{p}_2,\mathbf{p}_Y;3)
&=
\sum_{\lambda_{3}}
A_{\lambda_X}^{\lambda_{3}\lambda_Y}(\mathbf p_X,\mathbf p_{3},\mathbf p_Y)
P_{3}
A_{\lambda_{3}}^{\lambda_1\lambda_2}(\mathbf p_{3},\mathbf p_1,\mathbf p_2),\\
A_{\lambda_X}^{\lambda_{1}\lambda_{2}\lambda_{Y}}(\mathbf{p}_X,\mathbf{p}_1,\mathbf{p}_2,\mathbf{p}_Y;4)
&=
\sum_{\lambda_{4}}
A_{\lambda_X}^{\lambda_{4}\lambda_2}(\mathbf p_X,\mathbf p_{4},\mathbf p_2)
P_{4}
A_{\lambda_{4}}^{\lambda_1\lambda_Y}(\mathbf p_{4},\mathbf p_1,\mathbf p_Y).
\end{aligned}
\end{equation}

Finally, the total cascade amplitude for this process is obtained by directly summing the two chain amplitudes,
\begin{equation}\label{Eq:total_two_chains2}
A_{\lambda_X}^{\lambda_{1}\lambda_{2}\lambda_{Y}}(\mathbf{p}_X,\mathbf{p}_1,\mathbf{p}_2,\mathbf{p}_Y)
=
A_{\lambda_X}^{\lambda_{1}\lambda_{2}\lambda_{Y}}(\mathbf{p}_X,\mathbf{p}_1,\mathbf{p}_2,\mathbf{p}_Y;3)
+
A_{\lambda_X}^{\lambda_{1}\lambda_{2}\lambda_{Y}}(\mathbf{p}_X,\mathbf{p}_1,\mathbf{p}_2,\mathbf{p}_Y;4).
\end{equation}

\end{enumerate}

Both conventions can be applied to cascade decays with multiple decay chains. 
In the two-body COM-frame helicity convention, the step-by-step construction within each individual chain is consistent by construction: adjacent stages are connected by a single helicity boost, so the intermediate particle helicity is unchanged between the two corresponding rest frames and no additional alignment is needed within the same chain. 
The mismatch only appears between different chains, and only on the external final-state helicity. 
Therefore, to coherently sum different chain amplitudes, it is sufficient to introduce, for each non-reference chain and for each final-state particle, a single alignment rotation $R_{\bar{i}i}$ and the associated Wigner $D$-matrices. 
Consequently, the number of alignment rotations to be determined scales roughly linearly with the number of non-reference chains times the number of final-state particles, and is typically modest.

In contrast, in the lab frame helicity convention all decay chains share the same helicity definition from the outset, since each final-state helicity is always defined with respect to its lab momentum direction. 
Hence different chains can be summed coherently without any extra chain-by-chain alignment rotations. 
The trade-off is that the building blocks written in the lab frame generally involve explicit Wigner rotations induced by non-collinear boosts, and such rotation factors may reappear at multiple steps in the cascade construction.

\subsection{Covariant cascade decay}
\label{subsec:Covariant cascade decay}

In this subsection, we mainly discuss the application of covariant approaches to cascade decays. Here, “covariant” means that the amplitudes are written directly in any frame and do not require additional input from a reference frame such as the COM frame. In our previous discussion, the covariant amplitude include the \textbf{canonical-spinor amplitude} and the \textbf{covariant tensor amplitude (GS-scheme)}.

Moreover, in these constructions the final-state particles are described using a common choice of spin quantization axis. Therefore, once a specific frame is selected in a cascade decay, for example the lab frame, the quantization axes for all two-body decays are taken to point in the same direction, such as the $z$-axis.
As a result, these amplitudes admit a unified treatment in cascade decays. In the following, we refer to this procedure as the covariant cascade decay.

In the covariant cascade decay, when the lab frame is chosen as the reference frame, the spin quantization axes of all two-body decay amplitudes are fixed to be the $z$-axis of this lab frame. Therefore, for a given decay chain, the two-body decay amplitude in which the intermediate particle appears as a daughter and the two-body amplitude in which it appears as a parent can be combined directly by summing over its spin index $\sigma$.
Moreover, for different decay chains, one can directly add the chain amplitudes computed in the same lab frame to obtain the total cascade amplitude. This shows that, in this construction, no complicated alignment procedure is required, in contrast to the non-covariant treatment.

As a concrete example, consider the cascade decay process $X \to 3 + Y$, followed by $3 \to 1 + 2$. In the lab frame, its cascade amplitude can be written as 
\begin{equation}\label{Eq:Covariant cascade decay}
\mathcal{A}_{\sigma_{X}}^{\sigma_{1}\sigma_{2}\sigma_{Y}}(p_X,p_1,p_2,p_Y) = \sum_{\sigma_3}\mathcal{A}_{\sigma_{X}}^{\sigma_{3}\sigma_{Y}}(p_X,p_3,p_Y)P_{3}\mathcal{A}_{\sigma_{3}}^{\sigma_{1}\sigma_{2}}(p_3,p_1,p_2),
\end{equation} 
where $P_3$ is the resonant propagator of the intermediate particle-$3$.
Each two-body decay amplitude can be written as
\begin{equation}
\begin{aligned}
    \mathcal{A}_{\sigma_{X}}^{\sigma_{3}\sigma_{Y}}(p_X,p_3,p_Y)=&\sum_{L_1S_1}\mathcal{A}_{\sigma_{X}}^{\sigma_{3}\sigma_{Y}}(p_X,p_3,p_Y;L_1,S_1),\\
    \mathcal{A}_{\sigma_{3}}^{\sigma_{1}\sigma_{2}}(p_3,p_1,p_2)=&\sum_{L_2S_2}\mathcal{A}_{\sigma_{3}}^{\sigma_{1}\sigma_{2}}(p_3,p_1,p_2;L_2,S_2),
\end{aligned}
\end{equation}
where $(L_1,S_1)$ and $(L_2,S_2)$ denote the possible $LS$ partial wave combinations for the two two-body decays.

\begin{figure}[H]
\centering
\begin{minipage}{0.45\textwidth}
\centering
\includegraphics[scale=1.1]{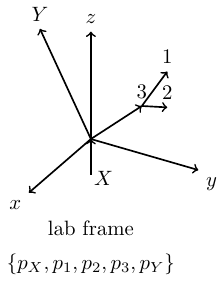}
\end{minipage}
\caption{Geometry of the cascade decay $X\to 3+Y$, $3\to 1+2$ in the lab frame. The five lab frame four-momenta ${p_X,p_1,p_2,p_3,p_Y}$ are used as the direct inputs of the two-body amplitudes.}
\label{fig:Covariant cascade decay}
\end{figure}

In the lab frame, the particle momenta are given by $p_X$, $p_1$, $p_2$, $p_Y$ and $p_3=p_1+p_2$. 
For each two-body decay, we directly substitute these momenta into the corresponding two-body amplitude. In this way, the cascade decay amplitude can be obtained directly.
The lab frame momenta $\{p_X,p_1,p_2,p_3,p_Y\}$ shown in Fig.~\ref{fig:Covariant cascade decay} can be used directly.

For processes with multiple decay chains, the amplitude for each chain can be obtained using the above procedure, without any additional boosts to intermediate COM frames and alignment rotation. 
As a result, the total amplitude is simply the coherent sum of the amplitudes from all decay chains.
As an explicit example, we consider two decay chains,
\begin{itemize} 
	\item Decay chain-1: $X \to 3 + Y$, followed by $3 \to 1 + 2$
	\item Decay chain-2: $X \to 4 + 2$, followed by $4 \to 1 + Y$
\end{itemize}
The total amplitude can be written as
\begin{equation}
\mathcal{A}_{\sigma_{X}}^{\sigma_{1}\sigma_{2}\sigma_{Y}}(p_X,p_1,p_2,p_Y) =  \mathcal{A}_{\sigma_{X}}^{\sigma_{1}\sigma_{2}\sigma_{Y}}(p_X,p_1,p_2,p_Y;3) +  \mathcal{A}_{\sigma_{X}}^{\sigma_{1}\sigma_{2}\sigma_{Y}}(p_X,p_1,p_2,p_Y;4),
\end{equation}
where the two terms denote the covariant cascade decay amplitudes associated with the two distinct resonances $3$ and $4$.

When more decay chains are present, one only needs to compute the amplitude for each chain in the lab frame and then add them directly to obtain the total amplitude. This shows that using covariant decay amplitudes as the basic building blocks for cascade decays leads to a more streamlined approach and is well suited for handling cascade processes with multiple decay chains.

\section{The cascade decay of \texorpdfstring{$\Lambda^{+}_{c}\to \Lambda \pi^{+}\pi^{0}$}{Lambda-c+ to Lambda pi+ pi0}}
\label{sec:The cascade decay of lambda}
To validate and compare different PWA schemes, the cascade decay $\Lambda_c^+ \to \Lambda \pi^+ \pi^0$ is used as a benchmark process, and fits are performed using different schemes. 
This cascade decay can proceed via several possible intermediate resonant states, with each resonance corresponding to a distinct decay chain.

In this work, the main objective is to test whether the coefficients in front of each partial wave, $g_{LS}$, and the resulting physical observables are consistent across different PWA schemes. 
For this reason, only the intermediate states listed in
Ref.~\cite{BESIII:2022udq} with statistical significance larger than $5\sigma$ are included. 
The selected components are the $\rho(770)^{+}$, $\Sigma(1385)^{+}$,
$\Sigma(1385)^{0}$, $\Sigma(1670)^{+}$, $\Sigma(1670)^{0}$,
$\Sigma(1750)^{+}$, and $\Sigma(1750)^{0}$ states, as well as the non-resonant $(NR_{1^{-}})$ component. 
These selected components can be organized into three distinct topological structures of decay chains, namely:
\begin{itemize}
	\item \textbf{Topology I} ($\rho^+$-type): $\Lambda_c^+ \to \Lambda R$, $R \to \pi^+ \pi^0$, $\Lambda \to p \pi^-$,\\
	where $R$ is a meson resonance decaying into $\pi^+ \pi^0$. The dominant contributions in this topology come from the \textbf{$\rho(770)^+$} and the non-resonant $J^P = 1^-$ component, denoted as $NR_{1^-}$.
	
	\item \textbf{Topology II} ($\Sigma^{*+}$-type): $\Lambda_c^+ \to \Sigma^{*+} \pi^0$, $\Sigma^{*+} \to \Lambda \pi^+$, $\Lambda \to p \pi^-$,\\
	where $\Sigma^{*+}$ is an excited baryon resonance. This topology includes the \textbf{$\Sigma(1385)^+$}, \textbf{$\Sigma(1670)^+$}, and \textbf{$\Sigma(1750)^+$} states.
	
	\item \textbf{Topology III} ($\Sigma^{*0}$-type): $\Lambda_c^+ \to \Sigma^{*0} \pi^+$, $\Sigma^{*0} \to \Lambda \pi^0$, $\Lambda \to p \pi^-$,\\
	where $\Sigma^{*0}$ is another excited baryon resonance. This topology includes the \textbf{$\Sigma(1385)^0$}, \textbf{$\Sigma(1670)^0$}, and \textbf{$\Sigma(1750)^0$} states.
\end{itemize}
To analyze these decay channels, the open-source package TF-PWA~\cite{tfpwa} is used to generate events with the $g_{LS}$ values given in Ref.~\cite{BESIII:2022udq}. 
The generated events are then fitted using the different partial wave amplitudes for comparison. 
Since only the consistency among the formalisms is of interest, the generated samples do not include background contributions or detector acceptance and efficiency effects.

\subsection{Calculate amplitudes in different ways}

In practice, we performed fits using the traditional-$LS$ amplitude, the canonical-spinor amplitude, and the helicity amplitude, and found that they give mutually consistent results. Therefore, for the components whose final-state quantum numbers are defined in the canonical basis, we take our canonical-spinor amplitude as an example and compare it with the helicity amplitude that is predominantly used in TF-PWA.

Next, we present several useful relations associated with the orbital angular momentum $L$. We consider the decay process $3\to 1+2$, where the particle masses are $m_3$, $m_1$, and $m_2$. The magnitude of the momentum in the COM frame can be written as 
\begin{equation}
	q=\frac{\sqrt{((m_{3}+m_{2})^{2}-m_{1}^{2})((m_{3}+m_{1})^{2}-m_{2}^{2})}}{2m_{3}}.
\end{equation}
The Blatt-Weisskopf barrier factoris $B_{L}^{\prime}(q,q_{0},d)$ is given by~\cite{VonHippel:1972fg} 
\begin{equation}\label{Eq:Blatt-Weisskopf barrier factor}
	\begin{aligned}
		B_{0}^{\prime}(q,q_{0},d)=&1,\\
		B_{1}^{\prime}(q,q_{0},d)=&\sqrt{\dfrac{1+(q_{0}d)^{2}}{1+(qd)^{2}}},\\
		B_{2}^{\prime}(q,q_{0},d)=&\sqrt{\dfrac{9+3(q_{0}d)^{2}+(q_{0}d)^{4}}{9+3(qd)^{2}+(qd)^{4}}},\\
		B_{3}^{\prime}(q,q_{0},d)=&\sqrt{\dfrac{225+45(q_{0}d)^{2}+6(q_{0}d)^{4}+(q_{0}d)^{6}}{225+45(qd)^{2}+6(qd)^{4}+(qd)^{6}}},\\
		B_{4}^{\prime}(q,q_{0},d)=&\sqrt{\dfrac{11025+1575(q_{0}d)^{2}+135(q_{0}d)^{4}+10(q_{0}d)^{6}+(q_{0}d)^{8}}{11025+1575(qd)^{2}+135(qd)^{4}+10(qd)^{6}+(qd)^{8}}}.
	\end{aligned}
\end{equation}
These functions effectively suppress the amplitude at high $L$, preventing unphysical growth of high-$L$ partial waves. 
The parameter $d$ is a hadronic scale that characterizes the size of the interaction region. 
In this work, the radius parameter is fixed to $d = 0.73~\text{fm}$, following Ref.~\cite{barrierSet}.

\subsubsection{Helicity scheme}
For the \textbf{helicity scheme}, the amplitude of each two-body decay in the COM frame is given by Eq.~\eqref{Eq:helicity amplitude}. 
In the cascade decay, each intermediate resonance is associated with a propagator, denoted generically by $P_{\rho}(m)$, $P_{\Sigma^{*+}}(m)$, $P_{\Sigma^{*0}}(m)$, $P_{NR_{1^-}}(m)$ and $P_{\Lambda}(m)$.

The propagator $P$ includes different models. 
For the $\Lambda$ and the non-resonant $NR_{1^-}$ contribution, the propagators are taken to be unity.
For the $\Sigma^{*+}$ and $\Sigma^{*0}$ resonances, the Breit-Wigner formalism is used,
\begin{equation}\label{Eq:Sigma resonance}
    P_{\Sigma^*}(m)=\frac{1}{m_{0}^{2}-m^2-i\,m_0\Gamma(m)},
\end{equation}
where $m$ is the invariant mass of the resonance and $m_0$ is the nominal resonance mass. The mass-dependent width is
\begin{equation}
    \Gamma(m)=\Gamma_0\left(\frac{q}{q_0}\right)^{2L+1}\frac{m_0}{m}B^{\prime2}_L(q,q_0,d).
\end{equation}

For the $\rho(770)^+$ resonance, the Gounaris-Sakurai model~\cite{GSmodel} is used:
\begin{equation}\label{Eq:rho resonance}
     P_{\rho}(m)=\frac{1+D\Gamma_0/m_0}{m_{0}^{2}-m^2+f(m)-im_0\Gamma(m)},
\end{equation}
where the $f(m)$ and $D$ are defined as
\begin{equation}
    f(m)=\Gamma_0\frac{m^2_0}{q^3_0}\left[q^2\left[h(m)-h(m_0)\right]+(m^2_0-m^2)q_0^2\frac{\mathrm{d}h}{\mathrm{d}m}|_{m_0}\right],
\end{equation}
\begin{equation}
    h(m) = \frac{2q}{\pi m}\ln\left({\frac{m+2q}{2m_\pi}}\right),
\end{equation}
\begin{equation}
    \frac{\mathrm{d}h}{\mathrm{d}m}|_{m_0}=h(m_0)\left[(8q_0^2)^{-1}-(2m_0^2)^{-1}\right]+(2\pi m_0^2)^{-1},
\end{equation}
\begin{equation}
    D = \frac{f(0)}{\Gamma_{0}m_0} = \frac{3}{\pi}\frac{m_{\pi}^{2}}{q^2_0}\ln{\left(\frac{m_0+2q_0}{2m_\pi}\right)}+\frac{m_0}{2\pi q_0} -\frac{m^2_\pi m _0}{\pi q^3_0}.
\end{equation}
Using the construction of helicity cascade decay amplitudes discussed in subsubsection~\ref{subsubsec:Helicity cascade decay}, the amplitude for each decay chain can be written explicitly as follows.

For the topology I, the corresponding helicity amplitude can be expressed as
\begin{equation}\label{Eq:helicity rho}
A^{\lambda_{p}}_{\lambda_{\Lambda^+_c}}(\rho) = \sum_{\lambda_{\Lambda},\lambda_{\rho}} A_{\lambda_{\Lambda^+_c}}^{\lambda_{\Lambda}\lambda_{\rho}}P_{\rho}(M_{\pi^{+}\pi^{0}})A_{\lambda_{\rho}}A_{\lambda_{\Lambda}}^{\lambda_{p}},
\end{equation}
\begin{equation}\label{Eq:helicity NR1-}
A_{\lambda_{\Lambda^+_c}}^{\lambda_{p}}(NR_{1^-}) = \sum_{\lambda_{\Lambda},\lambda_{NR_{1^-}}}A_{\lambda_{\Lambda^+_c}}^{\lambda_{\Lambda}\lambda_{NR_{1^-}}}A_{\lambda_{NR_{1^-}}}A_{\lambda_{\Lambda}}^{\lambda_{p}}.
\end{equation}
For the Topology II, the corresponding helicity amplitude can be expressed as
\begin{equation}\label{Eq:helicity Sigma+}
    A_{\lambda_{\Lambda^+_c}}^{\lambda_{p}^{(1)}}(\Sigma^{*+}) = \sum _{\lambda_{\Sigma^{*+}},\lambda_{\Lambda}}A_{\lambda_{\Lambda^+_c}}^{\lambda_{\Sigma^{*+}}}P_{\Sigma^{*+}}(M_{\Lambda\pi^+})A_{\lambda_{\Sigma^{*+}}}^{\lambda_{\Lambda}}A_{\lambda_{\Lambda}}^{\lambda_{p}^{(1)}}.
\end{equation}
For the Topology III, the corresponding amplitude can be expressed as
\begin{equation}\label{Eq:helicity Sigma0}
    A_{\lambda_{\Lambda^+_c}}^{\lambda_{p}^{(2)}}(\Sigma^{*0}) = \sum _{\lambda_{\Sigma^{*0}},\lambda_{\Lambda}}A_{\lambda_{\Lambda^+_c}}^{\lambda_{\Sigma^{*0}}}P_{\Sigma^{*0}}(M_{\Lambda\pi^0})A_{\lambda_{\Sigma^{*0}}}^{\lambda_{\Lambda}}A_{\lambda_{\Lambda}}^{\lambda_{p}^{(2)}}.
\end{equation}

Based on the discussion in subsubsection~\ref{subsubsec:Helicity  cascade decay}, 
when constructing helicity amplitudes for processes with multiple decay chains, it is necessary to choose one decay chain as a reference and align all the other 
decay chains accordingly. In the present analysis, we take the topology I decay chain as the reference chain and align the remaining decay chains with respect to it. Since among the final stable particles only the proton carries nonzero spin, for the topology II decay chain the helicity boost associated with proton is given by 
\begin{equation}
    S^{(1)}_p\xrightarrow{L^{-1}_h(\mathbf{p}_p^{\Lambda (1)})}S_{\Lambda}^{(1)}\xrightarrow{L_h^{-1}(\mathbf{p}^{\Sigma^{*+}}_{\Lambda (1)})}S^{(1)}_{\Sigma^{*+}}\xrightarrow{L^{-1}_h(\mathbf{p}_{\Sigma^{*+}}^{\Lambda^+_c})}S_{\Lambda^+_c}\xrightarrow{L_h(\mathbf{p}_{\Lambda}^{\Lambda^+_c})}S_{\Lambda}\xrightarrow{L_h(\mathbf{p}^{\Lambda}_{p})}S_p,
\end{equation}
and the corresponding alignment rotation is
\begin{equation}
    R_{\Sigma^{*+}\rho}=L^{-1}_h(\mathbf{p}_p^{\Lambda (1)})L^{-1}_h(\mathbf{p}^{\Sigma^{*+}}_{\Lambda (1)})L^{-1}_h(\mathbf{p}_{\Sigma^{*+}}^{\Lambda^+_c})L_h(\mathbf{p}_{\Lambda}^{\Lambda^+_c})L_h(\mathbf{p}^{\Lambda}_{p}).
\end{equation}
For the topology III decay chain, the helicity boost of proton is given by 
\begin{equation}
    S^{(2)}_p\xrightarrow{L^{-1}_h(\mathbf{p}_p^{\Lambda (2)})}S_{\Lambda}^{(2)}\xrightarrow{L^{-1}_h(\mathbf{p}^{\Sigma^{*0}}_{\Lambda (2)})}S^{(2)}_{\Sigma^{*0}}\xrightarrow{L^{-1}_h(\mathbf{p}_{\Sigma^{*0}}^{\Lambda^+_c})}S_{\Lambda^+_c}\xrightarrow{L_h(\mathbf{p}_{\Lambda}^{\Lambda^+_c})}S_{\Lambda}\xrightarrow{L_h(\mathbf{p}^{\Lambda}_{p})}S_p,
\end{equation}
and the corresponding alignment rotation is 
\begin{equation}
    R_{\Sigma^{*0}\rho}=L^{-1}_h(\mathbf{p}_p^{\Lambda (2)})L^{-1}_h(\mathbf{p}^{\Sigma^{*0}}_{\Lambda (2)})L^{-1}_h(\mathbf{p}_{\Sigma^{*0}}^{\Lambda^+_c})L_h(\mathbf{p}_{\Lambda}^{\Lambda^+_c})L_h(\mathbf{p}^{\Lambda}_{p}).
\end{equation}
Therefore, after applying these alignment rotations to 
Eqs.~\eqref{Eq:helicity Sigma+} and \eqref{Eq:helicity Sigma0}, 
and then summing them with the contributions of 
Eqs.~\eqref{Eq:helicity rho} and \eqref{Eq:helicity NR1-}, 
the total helicity amplitude for the cascade decay can be written as
\begin{equation}
\begin{aligned}
    A_{\lambda_{\Lambda^+_c}}^{\lambda_p} =& \left(A_{\lambda_{\Lambda^+_c}}^{\lambda_p}(\rho)+A_{\lambda_{\Lambda^+_c}}^{\lambda_p}(NR_{1^-})\right)\\
    &+\sum_{\lambda_p^{(1)}}\left(\sum_{\Sigma^{*+}} A_{\lambda_{\Lambda^+_c}}^{\lambda_p^{(1)}}(\Sigma^{*+})\right)D^{\lambda_p^{(1)}(\frac{1}{2})*}_{\,\,\lambda_p}( R_{\Sigma^{*+}\rho})\\
    &+\sum_{\lambda_p^{(2)}}\left(\sum_{\Sigma^{*0}} A_{\lambda_{\Lambda^+_c}}^{\lambda_p^{(2)}}(\Sigma^{*0})\right)D^{\lambda_p^{(2)}(\frac{1}{2})*}_{\,\,\lambda_p}(R_{\Sigma^{*0}\rho}).
\end{aligned}
\end{equation}
Then this cascade decay amplitude can be fitted using the method described in subsection~\ref{subsec:Likelihood Function and Fit Fraction} in order to determine the corresponding $LS$ partial wave coefficients $g_{LS}$.

\subsubsection{Canonical-spinor scheme}
The construction of the cascade decay amplitude in the \textbf{canonical-spinor scheme} is now discussed. 
The canonical-spinor two-body decay amplitudes used here are those of Eq.~\eqref{Eq:amplitude index conversion} with the normalization factor $N$ set to 1. The corresponding spin and orbital parts $\mathcal{S}$ and $\mathcal{Y}$ are
given in Eqs.~\eqref{Eq:spinor Sprime} and~\eqref{Eq:Yprime}, respectively.
The explicit analytic form of each two-body decay amplitude is given in subsection~\ref{subsec:Examples of spinor-canonical amplitude calculation} and can be identified according to the spin assignments of the particles involved.

In this scheme, the four-momenta of all particles in any frame are inserted directly into the corresponding canonical-spinor two-body decay amplitudes to obtain the contribution of each decay chain. 
The canonical-spinor cascade decay amplitudes for each decay chain can be written as
\begin{equation}\label{Eq:rho spinor canonical}
\mathcal{A}_{\sigma_{\Lambda^+_c}}^{\sigma_{p}}(\rho) = \sum_{\sigma_{\Lambda},\sigma_{\rho}} \mathcal{A}_{\sigma_{\Lambda^+_c}}^{\sigma_{\Lambda}\sigma_{\rho}}P_{\rho}(M_{\pi^{+}\pi^{0}})\mathcal{A}_{\sigma_{\rho}}\mathcal{A}_{\sigma_{\Lambda}}^{\sigma_{p}},
\end{equation}
\begin{equation}\label{Eq:NR spinor canonical}
    \mathcal{A}_{\sigma_{\Lambda^+_c}}^{\sigma_{p}}(NR_{1^-}) = \sum_{\sigma_{\Lambda},\sigma_{NR_{1^-}}}\mathcal{A}_{\sigma_{\Lambda^+_c}}^{\sigma_{\Lambda}\sigma_{NR_{1^-}}}\mathcal{A}_{\sigma_{NR_{1^-}}}\mathcal{A}_{\sigma_{\Lambda}}^{\sigma_{p}},
\end{equation}
\begin{equation}\label{Eq:sigmaplus spinor canonical}
    \mathcal{A}_{\sigma_{\Lambda^+_c}}^{\sigma_{p}}(\Sigma^{*+}) = \sum _{\sigma_{\Sigma^{*+}},\sigma_{\Lambda}}\mathcal{A}_{\sigma_{\Lambda^+_c}}^{\sigma_{\Sigma^{*+}}}P_{\Sigma^{*+}}(M_{\Lambda\pi^+})\mathcal{A}_{\sigma_{\Sigma^{*+}}}^{\sigma_{\Lambda}}\mathcal{A}_{\sigma_{\Lambda}}^{\sigma_{p}},
\end{equation}
\begin{equation}
    \mathcal{A}_{\sigma_{\Lambda^+_c}}^{\sigma_{p}}(\Sigma^{*0}) = \sum _{\sigma_{\Sigma^{*0}},\sigma_{\Lambda}}\mathcal{A}_{\sigma_{\Lambda^+_c}}^{\sigma_{\Sigma^{*0}}}P_{\Sigma^{*0}}(M_{\Lambda\pi^0})\mathcal{A}_{\sigma_{\Sigma^{*0}}}^{\sigma_{\Lambda}}\mathcal{A}_{\sigma_{\Lambda}}^{\sigma_{p}}.
\end{equation}
The propagators of the intermediate resonances have already been given in Eq.~\eqref{Eq:Sigma resonance} and Eq.~\eqref{Eq:rho resonance}.

Following the discussion in subsection~\ref{subsec:Covariant cascade decay}, the total amplitude can be written as 
\begin{equation}
    \mathcal{A}_{\sigma_{\Lambda^+_c}}^{\sigma_p} = \mathcal{A}_{\sigma_{\Lambda^+_c}}^{\sigma_{p}}(\rho)+\mathcal{A}_{\sigma_{\Lambda^+_c}}^{\sigma_{p}}(NR_{1^-})
    +\sum_{\Sigma^{*+}} \mathcal{A}_{\sigma_{\Lambda^+_c}}^{\sigma_{p}}(\Sigma^{*+})
    +\sum_{\Sigma^{*0}} \mathcal{A}_{\sigma_{\Lambda^+_c}}^{\sigma_{p}}(\Sigma^{*0}).
\end{equation}
Using this cascade decay amplitude, a fit is performed following the procedure outlined in subsection~\ref{subsec:Likelihood Function and Fit Fraction}, from which the corresponding $LS$ partial wave coefficients and physical observables are extracted.

\subsection{Numerical implementation of the canonical-spinor amplitude}
In practical PWA, canonical-spinor amplitudes are calculated numerically. Some explicit examples are given in subsection~\ref{subsec:Examples of spinor-canonical amplitude calculation}; more generally, amplitudes for arbitrary spins are obtained by a direct numerical implementation of Eq.~\eqref{Eq:sphe1}.

For the CGCs part, the three spins $(s_1,s_2,s_3)$ determine the integers $a,b,c$ appearing in Eq.~\eqref{Eq:SU(2) CGC}. In the numerical implementation, the expression in Eq.~\eqref{Eq:SU(2) CGC} is expanded into a sum of products containing $(a+c)$ identity tensors $\delta^{a+c}$ and $b$ antisymmetric tensors $\varepsilon^b$. The indices of each term are first reordered to match the ordering used in the expression, and each group of indices enclosed in $(\cdots)$ is then symmetrized. Multiplying by the overall prefactor in Eq.~\eqref{Eq:SU(2) CGC} finally yields the CGCs tensor with the desired set of $\mathrm{SU}(2)$ indices.

For the orbital part $\mathcal{Y}$, the starting point is the basic $\langle3_{I_1}|1|3_{I_2}]$ tensor. Its explicit form can be obtained by substituting the canonical-spinor variables, and can be written as 
\begin{equation}
    \langle 3_{I_1}|1|3_{I_2}]=\lambda^{~\alpha}_{I_1}(p_3)p_{1\alpha\dot\alpha}\tilde{\lambda}^{\dot{\alpha}}_{~I_2}(p_3),
\end{equation}
where the momenta are taken from Eq.~\eqref{Eq:bispinor momentun} and the spinors are defined in Eq.~\eqref{Eq:spinor canonical variables} with $\tilde{\lambda}^{\dot{\alpha}}_{~I}=\epsilon_{IJ}\tilde{\lambda}^{\dot{\alpha}J}$ between them.
For higher $L$, a new tensor $\mathcal{Y}_{\{L_1\cdots L_{2L}\}}$ is constructed by taking the tensor product of $L$ such $\langle 3_{I_1}|1|3_{I_2}]$ tensors, which yields an object with $2L$ indices. Symmetrizing over all $2L$ indices and multiplying by the overall prefactor in Eq.~\eqref{Eq:Yprime} then gives the orbital tensor.

For the spin part with total spin-$S$, a tensor $\tau$ is introduced. The basic tensor $\tau$ is defined in Eq.~\eqref{Eq:define tau}, and its explicit form in terms of canonical-spinor variables can be written as 
\begin{equation}
    (\tau^\prime_i)^{ I}_{J}(p_3,p_i)=\frac{1}{\sqrt{2m_3(E_{i\mathrm{com}}+m_i)}}\Big(\lambda^{I\alpha}(p_i)\lambda_{\alpha J}(p_3)-\tilde{\lambda}^{I}_{~\dot{\alpha}}(p_i)\tilde{\lambda}_{~J}^{\dot{\alpha}}(p_3)\Big)\quad(i=1,2).
\end{equation} 
The corresponding tensor $\tau^{2s_1}$ is constructed as the tensor product of $2s_1$ basic tensors $\tau_1$, followed by complete symmetrization over its upper and lower indices. Particle-2 is treated analogously, yielding $\tau^{2s_2}$. The tensors $\tau^{2s_1}$ and $\tau^{2s_2}$ are then contracted with the CGCs tensor over the appropriate indices to form the spin tensor $\mathcal{S}$.

Finally, $\mathcal{S}$ and $\mathcal{Y}$ are contracted with the external CGCs tensor to produce the amplitude with explicit $\mathrm{SU}(2)$ indices. Using Eq.~\eqref{Eq:amplitude index conversion}, this amplitude is then converted into the familiar form with $\sigma$ indices.

\subsection{Likelihood function and fit fraction}
\label{subsec:Likelihood Function and Fit Fraction}
To perform the amplitude analysis, the TF-PWA package is used to generate phase-space (PHSP) samples for the multi-body final state, following the results given in Ref.~\cite{BESIII:2022udq}. 
These samples do not include background contributions or detector efficiency effects. 
The various amplitude formalisms discussed in this work are then fitted to these PHSP samples. 
The fitting procedure is as follows.

For each event, the probability density function is written as
\begin{equation}
    P(x) = \frac{|A(x)|^2}{\int |A(x)|^2 \, d\Phi},
\end{equation}
where $x$ denotes a point in the final-state phase space. The squared amplitude is averaged over the spin states of the initial baryon $\Lambda_c^+$ as
\begin{equation}
    |A(x)|^2 = \frac{1}{2} \sum_{\sigma_{{\Lambda_c^+}}, \sigma_p} \left| A_{\sigma_{\Lambda^+_c}}^{\sigma_p}(x) \right|^2,
\end{equation}
with the factor $1/2$ accounting for the assumption of unpolarized $\Lambda_c^+$.

The integration is calculated with a Monte Carlo method using the generated PHSP samples:
\begin{equation}
    \int |A(x)|^2 d\Phi \approx \frac{1}{N_{\text{MC}}} \sum_{i=1}^{N_{\text{MC}}} |A(x_i)|^2,
\end{equation}
where $x_i$ represents the $i$-th phase space point, and $N_{\text{MC}}$ is the total number of MC samples.

The negative log-likelihood (NLL) is then constructed from the data as
\begin{equation}
    -\ln L = - \sum_{i=1}^{N_{\text{data}}} \ln P(x_i),
\end{equation}
where the summation runs over all simulated data events. Minimizing this function yields the best-fit parameters of the model.

The parameter error matrix is calculated by the inverse of the Hessian matrix
\begin{equation}\label{Eq:error matrix}
    V^{-1}_{ij} = -\frac{\partial^2 \ln L}{\partial X_i \partial X_j},
\end{equation}
where $X_i$ is the $i$-th floating parameter in the fit.

The fit fraction (FF) for each resonant component can be calculated as
\begin{equation}
    \text{FF}_i = \frac{\int |A_i(x)|^2 d\Phi}{\int \left| \sum_k A_k(x) \right|^2 d\Phi},
\end{equation}
where $A_i(x)$ is the amplitude of the $i$-th component, and the integration is performed using the same PHSP samples.

In cases where interference between two components $i$ and $j$ is relevant, the interference fraction is defined as
\begin{equation}
    \text{FF}_{ij} = \frac{\int |A_i(x) + A_j(x)|^2 d\Phi}{\int \left| \sum_k A_k(x) \right|^2 d\Phi} - \text{FF}_i - \text{FF}_j.
\end{equation}

The statistical uncertainties for the FFs are obtained using the standard form of error propagation:
\begin{equation}
    \sigma_Y^2 = \sum_{i,j} \left(\frac{\partial Y}{\partial X_i}\right)_{X=\mu}\cdot V_{ij} \cdot\left(\frac{\partial Y}{\partial X_j}\right)_{X=\mu},
\end{equation}
where $Y$ is the variable whose uncertainty is to be estimated, $X = \{X_i\}$  denotes the set of floating parameters, $V_{ij}$ is the corresponding error matrix defined in Eq.~\eqref{Eq:error matrix}, and $\mu$ is the results of floating variables $X$.

\subsection{Fitting result}
To simplify the PWA fit, the global factor of the total amplitude for the full decay chain involving the $\rho(770)^+$ is fixed to have a magnitude of 1 and a phase of 0 in all schemes. 
Within each decay chain, one of the partial wave coefficients $g_{LS}$ is likewise fixed to have a magnitude of 1 and a phase of 0. For the states $\Lambda_c^+$, $\rho(770)^+$, $\Sigma(1385)^+$, and $\Sigma(1385)^0$, the nominal mass and width are taken from the corresponding world average values~\cite{ParticleDataGroup:2020ssz}. 
In contrast, the resonance parameters of the other $\Sigma^*$ states are  adopted from the most recent measurements~\cite{Sarantsev:2019xxm}.

After these conventions, each scheme involves a total of 38 floating parameters. 
These include 14 parameters associated with the magnitudes and phases of the global factor of the total amplitudes for each decay chain, and 24 parameters corresponding to partial wave coefficients.

The final fit results obtained in all schemes are mutually consistent. 
The fitted parameters and fit fractions are summarized in Table~\ref{tab:fit_comparison} and Table~\ref{tab:helicity2}. 
The interference fractions and invariant mass spectra are shown in Table~\ref{tab:interference} and Fig.~\ref{fig:nominal1}.

\begin{table}[H]
	\setlength{\abovecaptionskip}{0.2cm}
	\setlength{\belowcaptionskip}{0.2cm}
	\centering
	\caption{The numerical results of the total amplitude of different components and FFs. The total FF is 152.0\%. Only list the statistical uncertainties. }
	\label{tab:fit_comparison}
	\begin{tabular}{c c c c}
		\hline\hline
		Process & Magnitude & Phase $\phi$ (rad) & FF ($\%$) \\
		\hline
		$\Lambda\rho(770)^+$    & $1.0\ (\text{fixed})$ & $0.0\ (\text{fixed})$ & $58.3 \pm 2.4$ \\
		$\Sigma(1385)^+\pi^0$   & $0.49\pm 0.05$ & -0.11 $\pm$ 0.11 & $8.6 \pm 0.5$ \\
		$\Sigma(1385)^0\pi^+$   & $0.42\pm 0.05$ & 3.06 $\pm$ 0.13 & $9.5 \pm 0.5$ \\
		$\Sigma(1670)^+\pi^0$   & $0.34\pm 0.06$ & -0.81 $\pm$ 0.16 & $4.0 \pm 0.5$ \\
		$\Sigma(1670)^0\pi^+$   & $0.35\pm 0.05$ & 2.89 $\pm$ 0.17 & $2.6 \pm 0.4$ \\
		$\Sigma(1750)^+\pi^0$   & $1.89\pm 0.15$ & -1.71 $\pm$ 0.08 & $17.7 \pm 1.8$ \\
		$\Sigma(1750)^0\pi^+$   & $1.96\pm 0.17$ & 1.39 $\pm$ 0.09 & $19.4 \pm 1.9$ \\
		$\Lambda + NR_{1^-}$    & $3.99\pm 0.29$ & 2.17 $\pm$ 0.08 & $31.9 \pm 2.9$ \\
		\hline\hline
	\end{tabular}
\end{table}

\vspace{-0.3cm} 

\begin{table}[H]
\caption{Numerical results of the partial wave coefficients $g_{LS}$ for
  different resonances in the nominal fit. Only statistical
  uncertainties are listed.}  \setlength{\abovecaptionskip}{1.2cm}
\setlength{\belowcaptionskip}{0.2cm}
\label{tab:helicity2}
\begin{center}
\vspace{-0.0cm}
\begin{tabular}{c c c | c c c}
		\hline \hline
		\multicolumn{3}{c|}{$\frac{1}{2}^{+}(\Lambda^{+}_{c})\to\frac{3}{2}^{+}(\Sigma(1385)^+)+0^-(\pi^0)$} 
		& \multicolumn{3}{c}{$\frac{1}{2}^{+}(\Lambda^{+}_{c})\to\frac{3}{2}^{+}(\Sigma(1385)^0)+0^-(\pi^+)$}\\
		Amplitude & Magnitude & Phase $\phi$ (rad)
		& Amplitude & Magnitude & Phase $\phi$ (rad)\\
		\hline
		$g^{\Sigma(1385)^+}_{1,\frac{3}{2}}$   & $1.0$ (fixed) & $0.0$ (fixed)
		& $g^{\Sigma(1385)^0}_{1,\frac{3}{2}}$ & $1.0$ (fixed) & $0.0$ (fixed) \\
		$g^{\Sigma(1385)^+}_{2,\frac{3}{2}}$   & $ 1.22 \pm 0.17 $ & $ 2.67 \pm 0.12 $
		& $g^{\Sigma(1385)^0}_{2,\frac{3}{2}}$ & $ 1.61 \pm 0.26 $ & $ 2.47 \pm 0.13 $\\
		\hline\hline
		
		\multicolumn{3}{c|}{$\frac{1}{2}^{+}(\Lambda^{+}_{c})\to\frac{3}{2}^{-}(\Sigma(1670)^+)+0^-(\pi^0)$} 
		& \multicolumn{3}{c}{$\frac{1}{2}^{+}(\Lambda^{+}_{c})\to\frac{3}{2}^{-}(\Sigma(1670)^0)+0^-(\pi^+)$}\\
		Amplitude & Magnitude & Phase $\phi$ (rad)
		& Amplitude & Magnitude & Phase $\phi$ (rad)\\
		\hline
		$g^{\Sigma(1670)^+}_{1,\frac{3}{2}}$   & $1.0$ (fixed)  & $0.0$ (fixed)
		& $g^{\Sigma(1670)^0}_{1,\frac{3}{2}}$ & $1.0$ (fixed)  & $0.0$ (fixed) \\
		$g^{\Sigma(1670)^+}_{2,\frac{3}{2}}$   & $ 1.54 \pm 0.29 $ & $ 0.99 \pm 0.15 $
		& $g^{\Sigma(1670)^0}_{2,\frac{3}{2}}$ & $ 1.01 \pm 0.20 $ & $ 0.25 \pm 0.19 $\\
		\hline\hline
		
		\multicolumn{3}{c|}{$\frac{1}{2}^{+}(\Lambda^{+}_{c})\to\frac{1}{2}^{-}(\Sigma(1750)^+)+0^-(\pi^0)$} 
		& \multicolumn{3}{c}{$\frac{1}{2}^{+}(\Lambda^{+}_{c})\to\frac{1}{2}^{-}(\Sigma(1750)^0)+0^-(\pi^+)$}\\
		Amplitude & Magnitude & Phase $\phi$ (rad)
		& Amplitude & Magnitude & Phase $\phi$ (rad)\\
		\hline
		$g^{\Sigma(1750)^+}_{0,\frac{1}{2}}$   & $1.0$ (fixed)  & $0.0$ (fixed)
		& $g^{\Sigma(1750)^0}_{0,\frac{1}{2}}$ & $1.0$ (fixed)  & $0.0$ (fixed) \\
		$g^{\Sigma(1750)^+}_{1,\frac{1}{2}}$   & $ 0.35 \pm 0.07 $ & $ -2.26 \pm 0.15 $
		& $g^{\Sigma(1750)^0}_{1,\frac{1}{2}}$ & $ 0.37 \pm 0.07 $ & $ -1.84 \pm 0.14 $\\
		\hline\hline
		
		\multicolumn{3}{c|}{$\frac{1}{2}^{+}(\Lambda^{+}_{c})\to\frac{1}{2}^{+}(\Lambda)+1^-(\rho(770)^+)$} 
		& \multicolumn{3}{c}{$\frac{1}{2}^{+}(\Lambda^{+}_{c})\to\frac{1}{2}^{+}(\Lambda)+1^-(N\!R_{1^{-}})$}\\
		Amplitude & Magnitude & Phase $\phi$ (rad) 
		& Amplitude & Magnitude & Phase $\phi$ (rad)\\
		\hline
		$g^{\rho}_{0,\frac{1}{2}}$ & $1.0$ (fixed) & $0.0$ (fixed)
		& $g^{N\!R}_{0,\frac{1}{2}}$ & $1.0$ (fixed) & $0.0$ (fixed) \\
		$g^{\rho}_{1,\frac{1}{2}}$ & $ 0.46 \pm 0.08 $ & $ -1.69 \pm 0.09 $
		& $g^{N\!R}_{1,\frac{1}{2}}$ & $ 1.00 \pm 0.08 $ & $ -0.45 \pm 0.09 $\\
		$g^{\rho}_{1,\frac{3}{2}}$ & $ 0.89 \pm 0.07 $ & $ 0.54 \pm 0.08 $
		& $g^{N\!R}_{1,\frac{3}{2}}$ & $ 0.18 \pm 0.06 $ & $ -3.0 \pm 0.4 $ \\
		$g^{\rho}_{2,\frac{3}{2}}$ & $ 0.55 \pm 0.08 $ & $ -0.14 \pm 0.12 $
		& $g^{N\!R}_{2,\frac{3}{2}}$ & $ 0.29 \pm 0.08 $ & $ -2.22 \pm 0.23 $\\
		\hline\hline
		
		\multicolumn{3}{c|}{$\frac{1}{2}^{+}(\Lambda)\to\frac{1}{2}^{+}(p)+0^-(\pi^-)$}\\
		Amplitude & Magnitude & Phase $\phi$ (rad)\\
		\cline{1-3}
		$g^{\Lambda}_{0,\frac{1}{2}}$ & $1.0$ (fixed) & $0.0$ (fixed)\\
		$g^{\Lambda}_{1,\frac{1}{2}}$ & $0.435376$ (fixed) & $0.0$ (fixed)\\
		\hline\hline
\end{tabular}
\end{center}
\end{table}

\begin{table*}[!htbp]
\caption{Interference fractions (I.F.) between $\Lambda^+_c$ amplitudes in units of percentage. The uncertainties are statistical only.}
\setlength{\abovecaptionskip}{1.2cm}
\setlength{\belowcaptionskip}{0.2cm}
\label{tab:interference}
\begin{center}
\vspace{-0.0cm}
\resizebox{\textwidth}{16mm}{
\begin{tabular}{c | c c c c c c c}
	\hline \hline
	I.F. & $\Lambda +N\!R_{1^-}$ & $\Sigma(1385)^0\pi^+$ & $\Sigma(1385)^+\pi^0$ & $\Sigma(1670)^0\pi^+$ & $\Sigma(1670)^+\pi^0$ & $\Sigma(1750)^0\pi^+$ & $\Sigma(1750)^+\pi^0$\\
	\hline
    $\Sigma(1385)^0\pi^+$ & $ -0.17 \pm 0.28 $ \\
    $\Sigma(1385)^+\pi^0$ & $ -0.57 \pm 0.28 $ & $ -0.03 \pm 0.02 $ \\
    $\Sigma(1670)^0\pi^+$ & $ -0.42 \pm 0.12 $ & $ -0.00 \pm 0.00 $ & $ -0.43 \pm 0.05 $  \\
    $\Sigma(1670)^+\pi^0$ & $ -0.40 \pm 0.12 $ & $ -0.42 \pm 0.06 $ & $ 0.00 \pm 0.00 $ & $ 0.03 \pm 0.01 $ \\
    $\Sigma(1750)^0\pi^+$ & $ -11.0 \pm 2.2 $ & $ 0.03 \pm 0.01 $ & $ 0.36 \pm 0.04 $ & $ -0.0 \pm 0.0 $ & $ -0.03 \pm 0.03 $ \\
   $\Sigma(1750)^+\pi^0$ & $ -9.2 \pm 2.1 $ & $ 0.26 \pm 0.04 $ & $ -0.06 \pm 0.01 $ & $ 0.11 \pm 0.04 $ & $ 0.07 \pm 0.01 $ & $ -7.1 \pm 0.7 $ \\
    $\Lambda\rho(770)^+$ & $ -7.4 \pm 2.5 $ & $ -6.7 \pm 0.4 $ & $ -6.73 \pm 0.33 $ & $ 0.56 \pm 0.24 $ & $ 1.55 \pm 0.25 $ & $ -2.0 \pm 0.9 $ & $ -2.3 \pm 0.9 $ \\
	\hline\hline
\end{tabular}
}
\end{center}
\end{table*}
\vspace{-0.0cm}

\vspace{-0.0cm}
\begin{figure}[H]
\setlength{\abovecaptionskip}{-1pt}
\setlength{\belowcaptionskip}{10pt}
\centering
\includegraphics[trim = 9mm 0mm 0mm 0mm, width=0.45\textwidth]{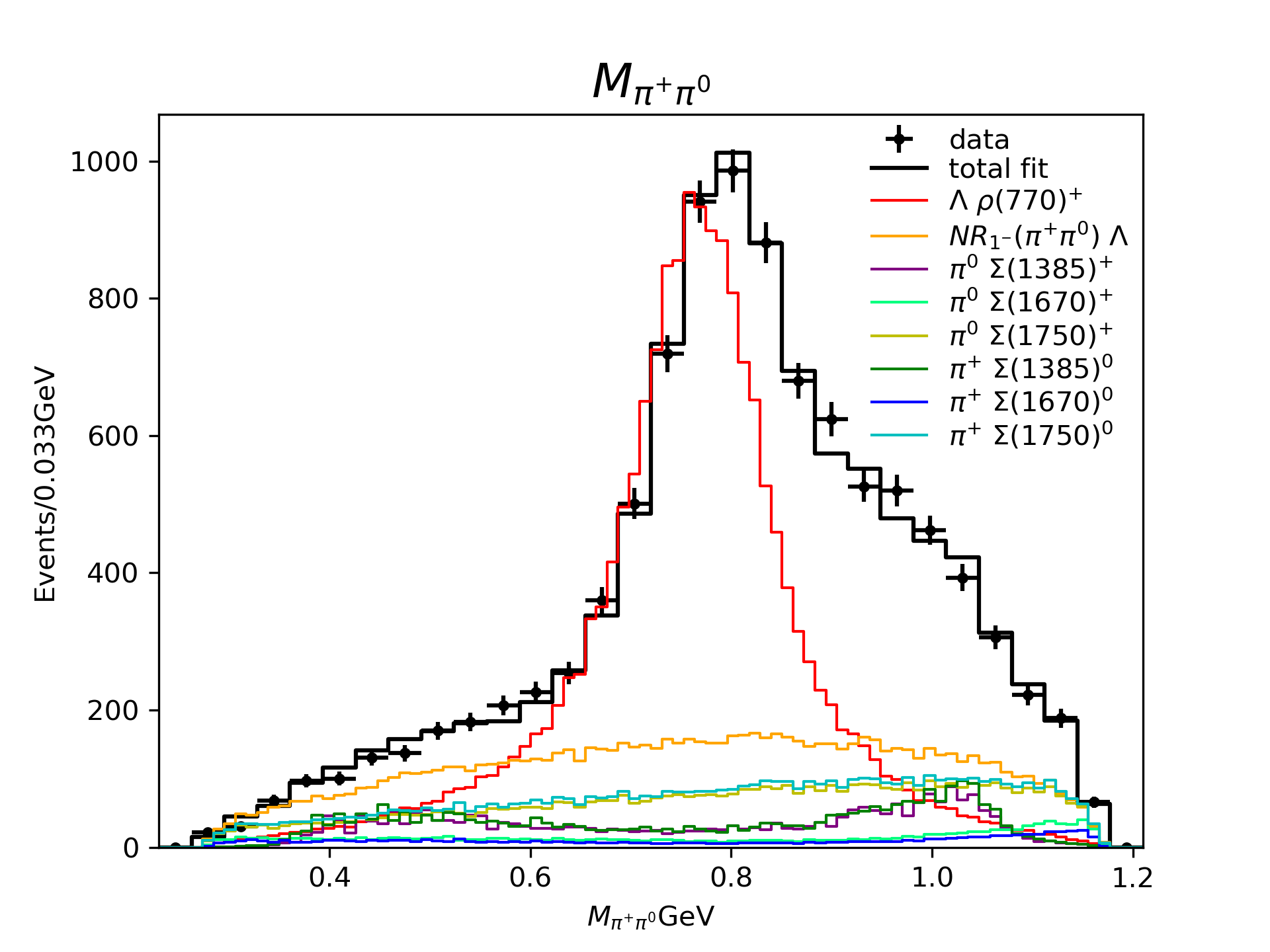}
\includegraphics[trim = 9mm 0mm 0mm 0mm, width=0.45\textwidth]{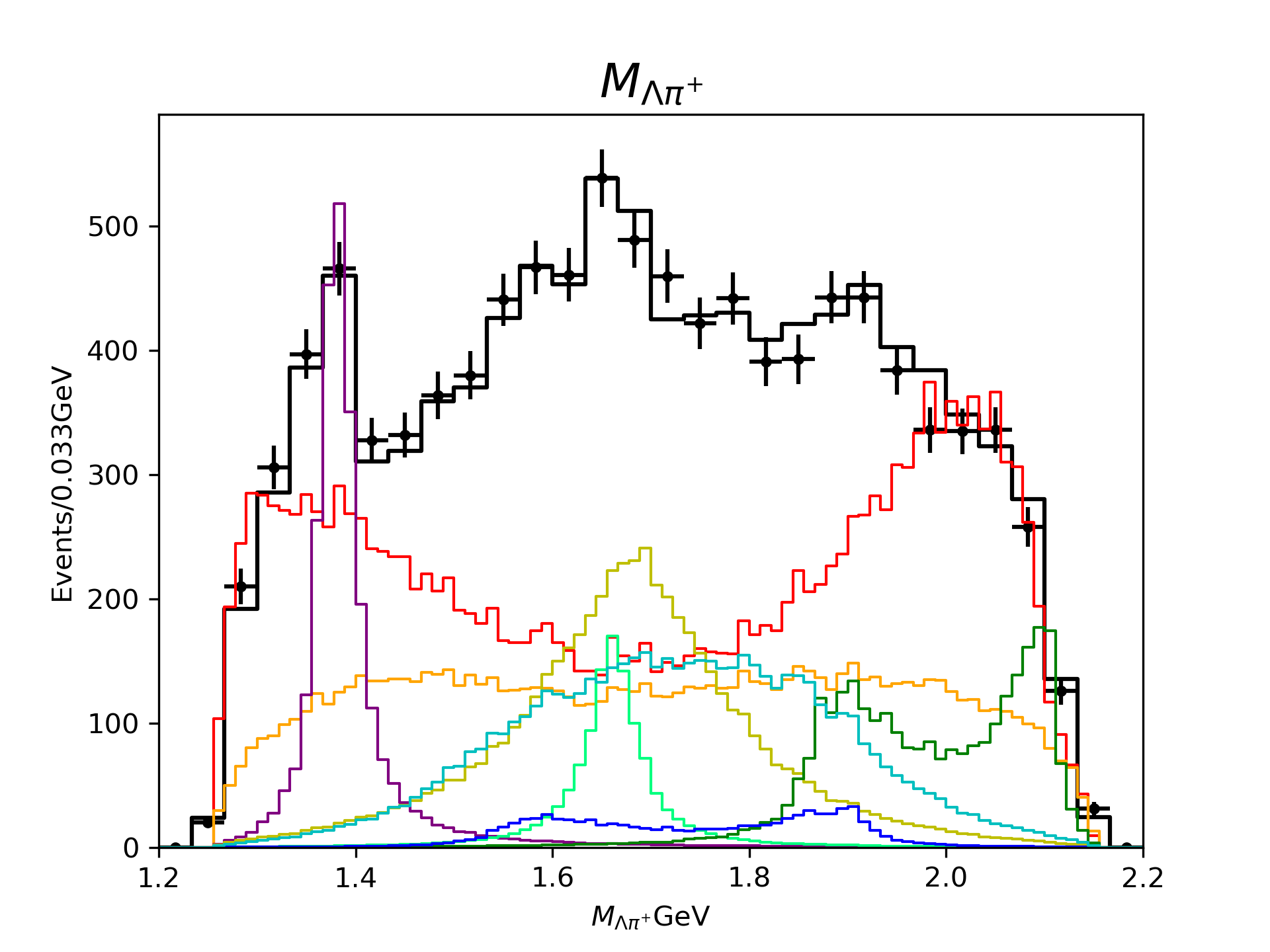}\\
\includegraphics[trim = 9mm 0mm 0mm 0mm, width=0.45\textwidth]{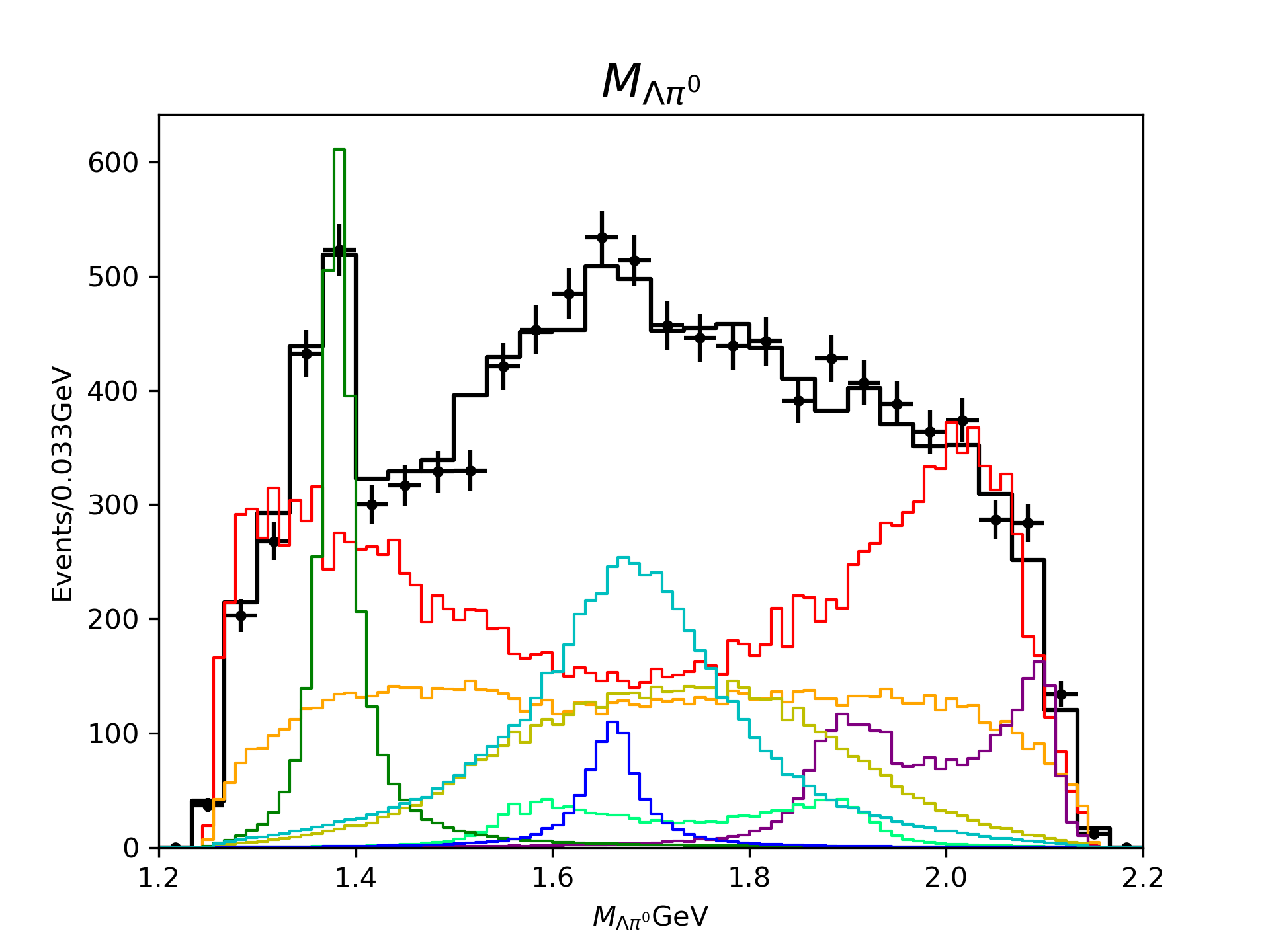}
\caption{Projections of the fit results in the invariant mass spectra
  $M_{\pi^+\pi^0}$, $M_{\Lambda\pi^+}$ and $M_{\Lambda\pi^0}$. Points with
  error bars denote data. Different styles of the curves denote
  different components.}
\label{fig:nominal1}
\end{figure}

\section{Conclusion}
In this paper, we study covariant amplitude constructions for PWA, aiming for decay amplitudes that are directly evaluable in any frame while maintaining a clear $LS$ partial wave expansion.

We first review the helicity and the canonical single-particle states and the corresponding two-particle states. In the two-body COM frame, when the two-particle state is coupled into a state with fixed total angular momentum $J$, the helicity and the canonical constructions correspond to two different coupling formalisms. In the helicity  formalism, the coupling is performed directly in the helicity basis, whereas in the canonical formalism the coupling is realized through the $LS$ coupling.
Consequently, for the two-body decay $3\to1+2$, there are two commonly used partial wave expansions of the decay amplitude, namely the helicity partial wave expansion and the $LS$ partial wave expansion. Although these two expansions are written in different forms, they are physically equivalent. This equivalence follows from the completeness of the helicity two-particle states and the canonical two-particle states, which allows one basis to be expanded in the other.

We then derive the decay amplitudes in any frame for both the canonical formalism and the helicity formalism. Since the partial wave definitions are made in the COM frame, any frame evaluation requires boosting the momenta back to the COM frame. For non-collinear boosts, the boost induces nontrivial Wigner rotations, so that the amplitude in any frame is given by the COM frame amplitude accompanied by the Wigner rotations generated by the boost. In particular, in the $LS$ partial wave framework that is central to this work, the $LS$ coupling is defined in the COM frame. Therefore, the construction of the corresponding partial waves requires returning to the COM frame, and the definition is not manifestly Lorentz covariant at the level of construction.

To restore Lorentz covariance, several methods, including Zemach tensor, covariant tensor, and covariant projection tensor methods, have been developed to construct Lorentz covariant $LS$ tensors, in which the angular momentum coupling is implemented in Lorentz representations of $\mathrm{SO}(3,1)$. In practice, tensor constructions of $LS$ amplitudes can be organized into two classes, referred to as the PS-scheme and the GS-scheme.
\begin{itemize}
	\item \textbf{PS-scheme:} In the COM frame of the PS-scheme, the partial wave amplitude can be decomposed into the covariant $LS$ coupling structure and the pure-spin part,
    \begin{equation}
  A^{\sigma_1\sigma_2}_{\sigma_3}({k}_3,{p}_1^*,{p}_2^*;L,S) =\underbrace{\Gamma^{\ell_1\ell_2}_{\ell_{3}}({k}_3,{p}_1^*,{p}_2^*;L,S)}_{\text{covariant $LS$ coupling structure}}\times \underbrace{u^{\ell_3}_{\sigma_3}({k}_3;s_3) \bar{u}^{\sigma_1}_{\ell_1}({k}_1;s_1) \bar{u}^{\sigma_2}_{\ell_2}({k}_2;s_2)}_{\text{pure-spin part}},
\end{equation}
    where the pure-spin part does not depend on the final-state relative momentum. As a result, the spin coupling is equivalent to coupling the intrinsic spins defined at the rest frames of the individual particles, and it matches the traditional-$LS$ method. For the orbital part, a Lorentz tensor structure in $\mathrm{SO}(3,1)$ is constructed, which is equivalent, in the COM frame, to the spherical harmonic structure in the $\mathrm{SO}(3)$ little group representation. Therefore, in the PS-scheme the orbital-spin separation is consistent with the traditional-$LS$ method, realizing a clean separation between orbital and spin. The drawback is that the definition still relies on the COM frame, and hence it is not manifestly covariant at the level of construction.
    
	\item \textbf{GS-scheme:} In the GS-scheme, the partial wave amplitude can be decomposed into the covariant $LS$ coupling structure and the general-spin part,
\begin{equation}
  \mathcal{C}^{\sigma_1\sigma_2}_{\sigma_3}({p}_3,{p}_1,{p}_2;L,S) = \underbrace{\Gamma_{\ell_{3}}^{\ell_{1}\ell_{2}}({p}_3,{p}_1,{p}_2;L,S)}_{\text{covariant $LS$ coupling structure}}\times \underbrace{u_{\sigma_{3}}^{\ell_{3}}({p}_3;s_{3})\bar{u}_{\ell_{1}}^{\sigma_{1}}({p}_1;s_{1})\bar{u}_{\ell_{2}}^{\sigma_{2}}({p}_2;s_{2})}_{\text{general-spin part}},
\end{equation}
    which is manifestly covariant and can be evaluated directly in any frame, without boosting to the COM frame. However, when one returns to the COM frame, the general-spin part typically depends on the final-state momenta, so that orbital information enters the spin coupling and the orbital-spin separation is not complete. Consequently, this construction does not realize the full orbital-spin separation of the traditional-$LS$ method, and the physical meaning of the orbital-spin decomposition is modified. As a result, a fixed $(L,S)$ component in the two definitions need not correspond to the same partial wave, and this may lead to different partial wave coefficients. Only in the non-relativistic limit, the spin part is reduced to rest frame spin wave functions, which is the purely intrinsic-spin objects, and the tensor constructions of $LS$ amplitudes in the GS-scheme would have clear $LS$ separation.  
    \end{itemize}
Therefore, current tensor constructions of $LS$ amplitudes face a trade-off. It can be seen that neither scheme simultaneously provides a manifestly covariant construction that avoids boosts to the two-body COM frame and an $LS$ separation that is identical to the traditional $LS$ definition.
In addition, for both schemes, when massless particles are involved, one must impose gauge invariance constraints explicitly and select a linearly independent set of vertex structures, which further increases the complexity of practical implementations.


Motivated by this trade-off, the goal is to construct an amplitude that satisfies the following requirements simultaneously. First, it is manifestly Lorentz covariant at the level of construction and can be evaluated directly in any frame. Second, it preserves, in physical meaning, the same orbital-spin separation as in the traditional-$LS$ method, so that a fixed $(L,S)$ component corresponds term by term to the traditional-$LS$ decomposition. Third, the construction is given in a general formula applicable to particles of arbitrary spin, including the case with massless particles.

To achieve this, we present an on-shell construction of covariant orbital-spin coupling amplitudes refer to as the canonical-spinor amplitudes, which can be written as
\begin{equation}
    \begin{aligned}
        \mathcal{A}_{\{I^{(3)}\}}^{\{I^{(1)}\},\{I^{(2)}\}}(p_{3},p_{1},p_{2};L,S)=&\underbrace{\mathcal{Y}^{L}_{\{I\}}}_{\text{orbital part}}\underset{\text{spin part}}{\underbrace{\mathcal{S}_{\{K\}}^{\{I^{(1)}\},\{I^{(2)}\}}C_{s_{3},\{I^{(3)}\}}^{S,\{K\};L,\{I\}}}}.\\
    \end{aligned}
\end{equation}
In this scheme, the canonical-spinor variables are used as the basic building blocks to assemble the orbital part and the spin part. The angular momentum coupling is carried out in the little group $\mathrm{SU}(2)$ representation, while the canonical-spinor variables carry $\mathrm{SL}(2,\mathbb C)$ Lorentz indices, which ensures the covariance of the construction. The resulting canonical-spinor amplitudes can be evaluated in any frame by directly substituting the four-momenta in that frame, without boosting each event back to the two-body COM frame. More importantly, when returning to the two-body COM frame, the $LS$ definition in this construction coincides with that of the traditional-$LS$ amplitude. This guarantees that the fitted $g_{LS}$ coefficients agree with the traditional-$LS$ method term by term and can therefore be compared directly.

In section~\ref{sec:General analysis of cascade decay}, the treatment of cascade decays in non-covariant and covariant approaches is discussed. 
In non-covariant approaches, each two-body amplitude is defined in its own rest frame, so an explicit alignment of spin quantization axis is required when assembling a multistep decay.
In the canonical case, the spin quantization axis is fixed to the $z$-axis. The two-body amplitudes at different steps can therefore be aligned to a common reference $z$-axis, and different decay chains can be summed coherently. 
In the non-covariant helicity case, two standard treatments are used. One is to assemble the COM helicity amplitudes step by step and contract the intermediate-state indices. When multiple decay chains are present, the final-state helicity conventions can differ between chains, so an additional alignment is still needed before a coherent sum is formed. The other is to define all helicities with respect to a single reference frame, such as the lab, so that different chains can be added directly, in the same way as in the canonical treatment.
In covariant approaches, the two-body building blocks are already expressed in a common covariant form and can be evaluated directly from the four-momenta. In practice, each chain amplitude is computed directly from the four-momenta, and the chain amplitudes are summed coherently without additional alignment steps. 
This leads to a simpler implementation, especially for processes with multiple decay chains.

As a benchmark, we study the cascade decay $\Lambda_c^+\to\Lambda\pi^+\pi^0$~\cite{BESIII:2022udq} and implement the amplitudes in TF-PWA. We include the intermediate states with statistical significance larger than $5\sigma$ and perform fits using the likelihood function, fit fraction, and interference fraction definitions. The fitted magnitudes, phases, fit fractions, and interference fractions are consistent between the helicity amplitude, the traditional-$LS$ amplitude and the canonical-spinor amplitude. This numerical study supports that the canonical-spinor amplitude is practical for PWA.

\section*{Acknowledgments}
We would like to thank Hao-Jie Jing, Jia-Jun Wu for valuable comments on the draft, Gang Li, Wen-Bin Qian and Bing-Song Zou for valuable discussions. This work is supported by the National Science Foundation of China under Grants No. 12347105, No. 12375099 and No. 12047503, and the National Key Research and Development Program of China Grant No. 2020YFC2201501, No. 2021YFA0718304.

\appendix
\appendix
\section{\texorpdfstring{$D$}{D}-functions}
\label{app:D-functions}
This appendix summarizes the definition and basic properties of $D$-functions used throughout. 
The $D$-functions with spin $j$ is defined as
\begin{equation}
	D^{\sigma^\prime(j)}_{\,\,\sigma}(\alpha,\beta,\gamma) = \langle \mathbf{k},j\sigma^{\prime} | U[R(\alpha,\beta,\gamma)] |\mathbf{k},j\sigma \rangle = e^{-i\sigma^{\prime}\alpha }d^{\sigma^\prime(j)}_{\,\,\sigma}(\beta)e^{-i\sigma\gamma }.
\end{equation}
The matrices $D^{\sigma^\prime(j)}_{\,\,\sigma}$ are unitary and satisfy the group property:
\begin{equation}
    \sum_{\sigma}D^{\sigma^\prime(j)}_{\,\,\sigma}(R)D^{\sigma^{\prime\prime}(j)*}_{\,\,\sigma}(R)=\delta_{\sigma^{\prime\prime}}^{\sigma^\prime},
\end{equation}
\begin{equation}
    D^{\sigma^\prime(j)}_{\,\,\sigma^{\prime\prime}}(R_2R_1)=\sum_{\sigma} D^{\sigma^\prime(j)}_{\,\,\sigma}(R_2) D^{\sigma(j)}_{\,\,\sigma^{\prime\prime}}(R_1).
\end{equation}
The functions $d^{\sigma^\prime(j)}_{\,\,\sigma}(\beta)$ can be expressed as~\cite{Rose1995AngularMomentum}
\begin{equation}
\begin{aligned}
    d^{\sigma^\prime(j)}_{\,\,\sigma}(\beta)=&\sum_t (-1)^t\frac{[(j+\sigma^\prime)!(j-\sigma^\prime)!(j+\sigma)!(j-\sigma)!]^{\frac{1}{2}}}{t!(j+\sigma^\prime-t)!(j-\sigma-t)!(t+\sigma-\sigma^\prime)!}\\
    &\times \left(\cos{\frac{\beta}{2}}\right)^{2(j-t)+\sigma^\prime-\sigma}\left(\sin{\frac{\beta}{2}}\right)^{2t+\sigma-\sigma^\prime}.
\end{aligned}
\end{equation}
The sum is taken over all values of t which lead to non negative
factorials. 

The functions $d^{\sigma^\prime(j)}_{\,\,\sigma}(\beta)$ have the following symmetry properties:
\begin{align}
    d^{\sigma^\prime(j)}_{\,\,\sigma}(\beta)=&(-1)^{\sigma^\prime-\sigma}d^{\sigma(j)}_{\,\,\sigma^\prime}(\beta)=(-1)^{\sigma^\prime-\sigma}d^{\sigma^\prime(j)}_{\,\,\sigma}(-\beta),\\
    d^{\sigma^\prime(j)}_{\,\,\sigma}(\beta)=&(-1)^{\sigma^\prime-\sigma}d^{-\sigma^\prime(j)}_{\,\,-\sigma}(\beta),\\ \label{Eq:d}
    d^{\sigma^\prime(j)}_{\,\,\sigma}(\pi-\beta)=&(-1)^{j+\sigma^\prime}d^{\sigma^\prime(j)}_{\,\,-\sigma}(\beta),\\
    d^{\sigma^\prime(j)}_{\,\,\sigma}(\pi)=&(-1)^{j-\sigma}\delta^{\sigma^\prime}_{-\sigma}.
\end{align}
Using Eq.~\eqref{Eq:d}, the $D$-functions have
\begin{equation}
    D^{\sigma^\prime(j)*}_{\,\,\sigma}(R)=D^{\sigma(j)}_{\,\,\sigma^\prime}(R^{-1})=(-1)^{\sigma^\prime-\sigma}D^{-\sigma^\prime(j)}_{\,\,-\sigma}(R).
\end{equation}
Using the identity
\begin{equation}
    R(\pi+\alpha,\pi-\beta,\pi-\gamma)=R(\alpha,\beta,\gamma)R(0,\pi,0),
\end{equation}
one obtains the corresponding relation for $D$-functions:
\begin{equation}
    D^{\sigma^\prime(j)}_{\,\,\sigma}(\pi+\alpha,\pi-\beta,\pi-\gamma)=(-1)^{j-\sigma}D^{\sigma^\prime(j)}_{\,\,-\sigma}(\alpha,\beta,\gamma),
\end{equation}
and
\begin{equation}
    D^{\sigma^\prime(j)}_{\,\,\sigma}(\pi+\phi,\pi-\theta,0)=(-1)^{j}D^{\sigma^\prime(j)}_{\,\,-\sigma}(\phi,\theta,0).
\end{equation}
The spherical harmonics $Y^L_{\sigma_L}(\phi,\theta)$ are related to the $D$-functions via
\begin{equation}\label{Eq:define spherical harmonics}
    D^{\sigma_L(j)*}_{\,\,0}(\phi,\theta,0)=\sqrt{\frac{4\pi}{2L+1}}Y^L_{\sigma_L}(\theta,\phi).
\end{equation}
The $D$-functions satisfy the following coupling rule:
\begin{equation}\label{Eq:D coupling 1}
D^{\sigma_1^\prime(j_1)}_{\,\,\sigma_1}D^{\sigma^\prime_2(j_2)}_{\,\,\sigma_2}=\sum_{j_3}C^{j_1,\sigma_1^\prime;j_2,\sigma_2^\prime}_{j_3,\sigma_3^\prime}C^{j_3,\sigma_3}_{j_1,\sigma_1;j_2\sigma_2}D^{\sigma_3^\prime(j_3)}_{\,\,\sigma_3},
\end{equation}
where
\begin{equation}
|j_1-j_2|\leq j_3\leq j_1+j_2,\quad\sigma_3=\sigma_1+\sigma_2,\quad\sigma_3^\prime=\sigma_1^\prime+\sigma_2^\prime.
\end{equation}
Or, equivalently,
\begin{equation}\label{Eq:D coupling 2}
D^{\sigma_1^\prime(j_1)}_{\,\,\sigma_1}D^{\sigma_3^\prime(j_3)*}_{\,\,\sigma_3}=\sum_{j_2}\frac{2j_2+1}{2j_3+1}C^{j_1,\sigma_1^\prime;j_2,\sigma_2^\prime}_{j_3,\sigma_3^\prime}C^{j_3,\sigma_3}_{j_1,\sigma_1;j_2\sigma_2}D^{\sigma^\prime_2(j_2)*}_{\,\,\sigma_2},
\end{equation}
where
\begin{equation}
|j_1-j_3|\leq j_2\leq j_1+j_3,\quad\sigma_2=\sigma_3-\sigma_1,\quad\sigma_2^\prime=\sigma_3^\prime-\sigma_1^\prime.
\end{equation}

The action of a rotation on a particle state is given by:
\begin{equation}
    U(R)|j\sigma\rangle=\sum_{\sigma^\prime}D^{\sigma^\prime(j)}_{\,\,\sigma}(R)|j\sigma^\prime\rangle,
\end{equation}
or
\begin{equation}
    \langle j\sigma |U^\dagger(R)=\sum_{\sigma^\prime}D^{\sigma^\prime(j)*}_{\,\,\sigma}(R)\langle j\sigma^\prime|=\sum_{\sigma^\prime}D^{\sigma(j)}_{\,\,\sigma^\prime}(R^{-1})\langle j\sigma^\prime|.
\end{equation}

\section{Normalization of one- and two-particle states}
\label{app:Normalization of One- and Two-Particle States}
One-particle momentum eigenstates $| \mathbf{p}, j \sigma \rangle$ and $| \mathbf{p}, j \lambda \rangle$ are normalized as
\begin{equation}
  \langle \mathbf{p}^\prime, j^\prime \sigma^\prime | \mathbf{p}, j \sigma \rangle
  = \tilde{\delta}(\mathbf{p}^\prime - \mathbf{p})\,\delta_{j j^\prime} \delta_{\sigma \sigma^\prime},
  \qquad
  \langle \mathbf{p}^\prime, j^\prime \lambda^\prime | \mathbf{p}, j \lambda \rangle
  = \tilde{\delta}(\mathbf{p}^\prime - \mathbf{p})\,\delta_{j j^\prime} \delta_{\lambda \lambda^\prime},
\end{equation}
where the Lorentz invariant $\delta$-function is given by
\begin{equation}
  \tilde{\delta}(\mathbf{p}^\prime - \mathbf{p})
  = (2\pi)^3 (2E)\, \delta^{(3)}(\mathbf{p}^\prime - \mathbf{p}),
\end{equation}
and the invariant volume element is defined by
\begin{equation}
  \tilde{d}p
  = \frac{d^3\mathbf{p}}{(2\pi)^3\, 2E}.
\end{equation}
With these conventions, the completeness relations take the form
\begin{equation}
  \sum_{j,\sigma} \int \tilde{d}p \, | \mathbf{p}, j \sigma \rangle \langle \mathbf{p}, j \sigma \rvert = \mathbf{I},
  \qquad
  \sum_{j,\lambda} \int \tilde{d}p \, | \mathbf{p}, j \lambda \rangle \langle \mathbf{p}, j \lambda \rvert = \mathbf{I},
\end{equation}
where $\mathbf{I}$ denotes the identity operator.

For the discussion of two-particle kinematics in this appendix, only spinless particles are needed.  
The spin label is therefore suppressed and the notation $| \mathbf{p} \rangle$ is used for one-particle states.  
Two-particle states are defined by
\begin{equation}
  | \mathbf{p}_1 \mathbf{p}_2 \rangle
  \equiv | \mathbf{p}_1 \rangle | \mathbf{p}_2 \rangle ,
\end{equation}
and satisfy the normalization
\begin{equation}\label{Eq:normalization p1p2p1p2}
  \langle \mathbf{p}_1^\prime \mathbf{p}_2^\prime | \mathbf{p}_1 \mathbf{p}_2 \rangle
  = \tilde{\delta}(\mathbf{p}_1^\prime - \mathbf{p}_1)\tilde{\delta}(\mathbf{p}_2^\prime - \mathbf{p}_2).
\end{equation}

A two-particle system with four-momenta $p_1$ and $p_2$ can be described equivalently by the total four-momentum
\begin{equation}
  p = p_1 + p_2 ,
\end{equation}
and by the orientation $\Omega$ of the relative three-momentum in the COM frame.  
The invariant mass of the two-particle system is denoted by
\begin{equation}
  m = E_{1\mathrm{com}} + E_{2\mathrm{com}} ,
\end{equation}
and the magnitude of the three-momentum in the COM frame is denoted by $q$, such that
\begin{equation}
  \mathbf{p}_1^{*} = -\mathbf{p}_2^{*},
  \qquad
  q=|\mathbf{p}_1^{*}| = |\mathbf{p}_2^{*}|  .
\end{equation}

Two-particle states labelled by the total four-momentum $p$ and the solid angle $\Omega$ of the relative momentum are introduced as
\begin{equation}\label{Eq:normalization omega to pp}
  | p,\Omega \rangle
  = N|\mathbf{p}_1\mathbf{p}_2\rangle ,
\end{equation}
where $N$ is the normalization constant.  
The desired normalization of these states is
\begin{equation}\label{Eq:normalization pomega pomega}
  \langle p^\prime,\Omega^\prime | p,\Omega \rangle
  = (2\pi)^4 \delta^{(4)}(p^\prime - p)\,\delta^{(2)}(\Omega^\prime - \Omega),
\end{equation}
where $\delta^{(2)}(\Omega^\prime - \Omega)$ denotes the delta function on the unit sphere, normalized according to
\begin{equation}
  \int d\Omega  \delta^{(2)}(\Omega^\prime - \Omega) = 1.
\end{equation}
The relation between $|\mathbf{p}_1 \mathbf{p}_2 \rangle$ and $| p,\Omega \rangle$ is fixed by the two-body phase-space measure.  
In the COM frame, the two-body phase space is defined by
\begin{equation}\label{Eq:normalization Phi}
\begin{aligned}
d\Phi=& (2\pi)^4 \delta^{(4)}(p_1 + p_2 - p)
\tilde{d}p_1\,\tilde{d}p_2\\
=&\frac{1}{(4\pi)^2} \frac{q}{m}\, d\Omega.
\end{aligned}
\end{equation}
Combining this with $\delta^{(2)}(\Omega^\prime - \Omega)$ and integrating over these variables gives
\begin{equation}\label{Eq:normalization omega Phi}
\begin{aligned}
\int\delta^{(2)}(\Omega^\prime - \Omega)\, d\Phi
=& \int\delta^{(2)}(\Omega^\prime - \Omega)\,
\frac{1}{(4\pi)^2} \frac{q}{m}\, d\Omega\\
=& \frac{1}{(4\pi)^2} \frac{q}{m}.
\end{aligned}
\end{equation}
Multiplying Eqs.~\eqref{Eq:normalization p1p2p1p2} and \eqref{Eq:normalization pomega pomega} by the invariant volume element $d\tilde{p}_1\,d\tilde{p}_2$, Eq.~\eqref{Eq:normalization p1p2p1p2} integrates to 1. 
For Eq.~\eqref{Eq:normalization pomega pomega}, using Eqs.~\eqref{Eq:normalization omega to pp}, \eqref{Eq:normalization Phi} and \eqref{Eq:normalization omega Phi} one obtains the normalization constant
\begin{equation}
    N=\frac{1}{4\pi}\sqrt{\frac{q}{m}}.
\end{equation}

\section{Clebsch-Gordon Coefficients}

\subsection{CGCs calculation}
Consider two canonical states $\left|s_1 \sigma_1\right\rangle$ and $\left|s_2 \sigma_2\right\rangle$. Our goal is to couple these two states into a total spin-$S$, such that,
\begin{equation}
  S = s_1 + s_2.
\end{equation}
The states are related by a linear relation given by
\begin{equation}
  \left|S \sigma\right\rangle
  = \sum_{\sigma_1 \sigma_2} C^{s_1,\sigma_1;s_2,\sigma_2}_{S,\sigma}\left|s_1 \sigma_1\right\rangle \left|s_2 \sigma_2\right\rangle,
\end{equation}
where the coefficients $C^{s_1,\sigma_1;s_2,\sigma_2}_{S,\sigma}$ are called CGCs.
We have that $\sigma = \sigma_1 + \sigma_2$.
An expression for the CG coefficients was also derived by Wigner~\cite{Rose1995AngularMomentum}
\begin{equation}
\begin{aligned}
      C^{s_1,\sigma_1;s_2,\sigma_2}_{S,\sigma}
  =& \delta_{\sigma}^{\sigma_1+\sigma_2}\\
  &\times
  \left[
    (2S+1)
    \frac{
      (S+s_1-s_2)!(S-s_1+s_2)!(s_1+s_2-S)!(S+\sigma)!(S-\sigma)!
    }{
      (S+s_1+s_2+1)!(s_1-\sigma_1)!(s_1+\sigma_1)!(s_2-\sigma_2)!(s_2+\sigma_2)!
    }
  \right]^{\frac12}\\
  &\times
  \sum_{t}
  \frac{(-1)^{t+S+\sigma}}{t!}
  \frac{
    (S+\sigma_1+s_2-t)!(s_1-\sigma_1+t)!
  }{
    S-(s_1+s_2-t)!(S+\sigma-t)!(t+s_1-s_2-\sigma)!
  }.
\end{aligned}
\end{equation}
The sum is taken over all values of $t$ which lead to non negative factorials.
Using the orthogonality relations, we can get the inverse relations
\begin{equation}
  \left|s_1 \sigma_1\right\rangle \left|s_2 \sigma_2\right\rangle
  = \sum_{S} C_{s_1,\sigma_1;s_2,\sigma_2}^{S,\sigma} \left|S \sigma\right\rangle.
\end{equation}
The CGCs have the following symmetry properties:
\begin{align}
  C^{s_1,\sigma_1;s_2,\sigma_2}_{S,\sigma}
  &= (-1)^{s_1+s_2-S}C^{s_1,-\sigma_1;s_2,-\sigma_2}_{S,-\sigma},\\
  C^{s_1,\sigma_1;s_2,\sigma_2}_{S,\sigma}
  &= (-1)^{s_1+s_2-S}C^{s_2,\sigma_2;s_1,\sigma_1}_{S,\sigma},\\
  C^{s_1,\sigma_1;s_2,\sigma_2}_{S,\sigma}
  &= (-1)^{s_1-\sigma_1}
  \left[\frac{2S+1}{2s_2+1}\right]^{\frac12}C^{s_1,\sigma_1;S,-\sigma}_{s_2,-\sigma_2}.
\end{align}

\subsection{Examples of SU(2) CGCs calculation}
\label{subsec:Examples of CGC calculation}
We first present some examples of SU(2) CGC calculations. The SU(2) CGCs are
\begin{equation}
	C_{s,\{J\}}^{s_{1},\{I\};s_{2},\{K\}}\equiv\sqrt{\dfrac{(a+b)!(b+c)!(a+c+1)!}{b!(a+b+c+1)!a!c!}}\delta_{(J_{1}\cdots J_{a}}^{(I_{1}\cdots I_{a}}\varepsilon^{I_{a+1}\cdots I_{a+b}),(K_{1}\cdots K_{b}}\delta_{J_{a+1}\cdots J_{a+c})}^{K_{b+1}\cdots K_{b+c})},
\end{equation}
with
\begin{equation}
	\begin{aligned}
		a=&s_{1}+s-s_{2},\\
		b=&s_{1}+s_{2}-s,\\
		c=&s_{2}+s-s_{1},\\
		\varepsilon^{I_{1}\cdots I_{b},K_{1}\cdots K_{b}}=&\varepsilon^{I_{1}K_{1}}\cdots \varepsilon^{I_{b}K_{b}}.
	\end{aligned}
\end{equation}

\begin{enumerate}
\item $s=\frac{1}{2}, s_1=\frac{1}{2}, s_2=0$ with $a=1,b=0,c=0$. The CGC is
\begin{equation}
	C_{\frac{1}{2},\{J_{1}\}}^{\frac{1}{2},\{I_{1}\};0}=\delta_{J_{1}}^{I_{1}}.
\end{equation}
\item $s=1,s_1=1,s_2=0$ with $a=2,b=0,c=0$. The CGC is
\begin{equation}
\begin{aligned}
    C_{1,\{J_{1}J_2\}}^{1,\{I_{1}I_2\};0}=&\delta_{(J_{1}J_2)}^{(I_{1}I_2)}\\
    =&\frac{1}{2}(\delta_{J_{1}}^{I_{1}}\delta_{J_2}^{I_2}+\delta_{J_{1}}^{I_{2}}\delta_{J_{2}}^{I_{1}}).
\end{aligned}
\end{equation}
\item $s=\frac{3}{2},s_1=\frac{3}{2},s_2=0$ with $a=3,b=0,c=0$. The CGC is
\begin{equation}
    \begin{aligned}
        C_{\frac{3}{2},\{J_{1}J_2J_3\}}^{\frac{3}{2},\{I_1I_2I_3\};0}=&\delta_{(J_1J_2J_3)}^{(I_1I_2I_3)}\\
        =&\frac{1}{6}(\delta_{J_1}^{I_1}\delta_{J_2}^{I_2}\delta_{J_3}^{I_3}+\delta_{J_1}^{I_1}\delta_{J_2}^{I_3}\delta_{J_3}^{I_2}+\delta_{J_1}^{I_2}\delta_{J_2}^{I_1}\delta_{J_3}^{I_3}\\
        &+\delta_{J_1}^{I_2}\delta_{J_2}^{I_3}\delta_{J_3}^{I_1}+\delta_{J_1}^{I_3}\delta_{J_2}^{I_1}\delta_{J_3}^{I_2}+\delta_{J_1}^{I_3}\delta_{J_2}^{I_2}\delta_{J_3}^{I_1}).
    \end{aligned}
\end{equation}
\item $s=\frac{1}{2},s_1=\frac{1}{2},s_2=1$ with $a=0,b=1,c=1$. The CGC is
\begin{equation}
\begin{aligned}
    	C_{\frac{1}{2},\{J_{1}\}}^{\frac{1}{2},\{I_{1}\};1,\{K_{1}K_{2}\}}=&\sqrt{\frac{2}{3}}\varepsilon^{I_{1}(K_{1}}\delta_{J_{1}}^{K_{2})}\\
        =&\frac{1}{2}\sqrt{\frac{2}{3}}\left(\varepsilon^{I_{1}K_{1}}\delta_{J_{1}}^{K_{2}}+\varepsilon^{I_{1}K_{2}}\delta_{J_{1}}^{K_{1}}\right).
\end{aligned}
\end{equation}
\item $s=\frac{3}{2},s_1=\frac{1}{2},s_2=1$ with $a=1,b=0,c=2$. The CGC is
\begin{equation}
\begin{aligned}
    	C_{\frac{3}{2},\{J_{1}J_2J_3\}}^{\frac{1}{2},\{I_{1}\};1,\{K_{1}K_{2}\}}=&\delta_{(J_1J_2J_3)}^{I_1(K_1K_2)}\\
        =&\frac{1}{6}(\delta_{J_1}^{I_1}\delta_{J_2}^{K_1}\delta_{J_3}^{K_2}+\delta_{J_1}^{I_1}\delta_{J_3}^{K_1}\delta_{J_2}^{K_2}+\delta_{J_2}^{I_1}\delta_{J_1}^{K_1}\delta_{J_3}^{K_2}\\
        &+\delta_{J_2}^{I_1}\delta_{J_3}^{K_1}\delta_{J_1}^{K_2}+\delta_{J_3}^{I_1}\delta_{J_1}^{K_1}\delta_{J_2}^{K_2}+\delta_{J_3}^{I_1}\delta_{J_2}^{K_1}\delta_{J_1}^{K_2}).
\end{aligned}
\end{equation}
\item $s=\frac{1}{2},s_1=\frac{3}{2},s_2=1$ with $a=1,b=2,c=0$. The CGC is
\begin{equation}
\begin{aligned}
    	C_{\frac{1}{2},\{J_{1}\}}^{\frac{3}{2},\{I_{1}I_2I_3\};1,\{K_{1}K_{2}\}}=&\sqrt{\frac{1}{2}}\delta_{J_1}^{(I_{1}}\varepsilon^{I_{2}I_{3}),(K_{1}K_{2})}\\
        =&\frac{1}{6}\sqrt{\frac{1}{2}}(\delta_{J_1}^{I_{1}}\varepsilon^{I_{2}K_{1}}\varepsilon^{I_3K_2} +\delta_{J_1}^{I_{1}}\varepsilon^{I_{3}K_{1}}\varepsilon^{I_2K_2} +\delta_{J_1}^{I_{2}}\varepsilon^{I_{1}K_{1}}\varepsilon^{I_{3}K_{2}}\\
        &+\delta_{J_1}^{I_{2}}\varepsilon^{I_{3}K_{1}}\varepsilon^{I_{1}K_{2}} +\delta_{J_1}^{I_{3}}\varepsilon^{I_{1}K_{1}}\varepsilon^{I_{2}K_{2}} +\delta_{J_1}^{I_{3}}\varepsilon^{I_{2}K_{1}}\varepsilon^{I_{1}K_{2}}.\\
\end{aligned}
\end{equation}
\item $s=1,s_1=1,s_2=1$ with $a=1,b=1,c=1$. The CGC is
\begin{equation}
\begin{aligned}
    	C_{1,\{J_{1}J_2\}}^{1,\{I_{1}I_2\};1,\{K_{1}K_{2}\}}=&\delta_{(J_1}^{(I_{1}}\varepsilon^{I_{2})(K_{1}}\delta_{J_{2})}^{K_{2})}\\
        =&\frac{1}{8}(\delta_{J_1}^{I_{1}}\varepsilon^{I_{2}K_{1}}\delta_{J_{2}}^{K_{2}} + \delta_{J_1}^{I_{1}}\varepsilon^{I_{2}K_{2}}\delta_{J_{2}}^{K_{1}} +\delta_{J_1}^{I_{2}}\varepsilon^{I_{1}K_{1}}\delta_{J_{2}}^{K_{2}} +\delta_{J_1}^{I_{2}}\varepsilon^{I_{1}K_{2}}\delta_{J_{2}}^{K_{1}}\\
        &+\delta_{J_2}^{I_{1}}\varepsilon^{I_{2}K_{1}}\delta_{J_{1}}^{K_{2}} +\delta_{J_2}^{I_{1}}\varepsilon^{I_{2}K_{2}}\delta_{J_{1}}^{K_{1}} +\delta_{J_2}^{I_{2}}\varepsilon^{I_{1}K_{1}}\delta_{J_{1}}^{K_{2}} + \delta_{J_2}^{I_{2}}\varepsilon^{I_{1}K_{2}}\delta_{J_{1}}^{K_{1}}).
\end{aligned}
\end{equation}
\end{enumerate}

\section{Numerical bases-transformation}
\label{app:Numerically relate different bases}

In this appendix, numerical procedures for relating different partial wave bases are summarized. Although amplitudes may take different explicit forms depending on the chosen basis, the underlying physical content remains the same. In particular, the same scattering or decay process can be described in different schemes, such as the helicity basis or the traditional-$LS$ basis.

Consider two coupling bases with partial wave amplitudes denoted by $C_{LS}$ and $H_{L^\prime S^\prime}$, respectively. They are connected by a linear
transformation with basis-change coefficients $t_{LS;L^\prime S^\prime}$,
\begin{equation}
  C_{LS} = \sum_{L^\prime S^\prime} t_{LS;L^\prime S^\prime }\, H_{L^\prime S^\prime },
  \label{Eq:basis-transform-CLS-HLS}
\end{equation}
where $t_{LS;L^\prime S^\prime }$ is the transformation matrix element from the $H_{L^\prime S^\prime }$ basis to the $C_{LS}$ basis. In practical calculations, one possible method to determine the transformation matrix $t_{LS;L^{\prime}S^{\prime}}$ is through sampling in phase space. If there are $N$ independent $LS$ combinations, one can select $N$ sets of kinematically allowed momenta from the physical phase space, denoted as:
\begin{equation}
	\{p_{3}^{1}p_{1}^{1}p_{2}^{1}\} , \quad\{p_{3}^{2}p_{1}^{2}p_{2}^{2}\},\quad  \cdots ,\quad \{p_{3}^{N}p_{1}^{N}p_{2}^{N}\}.
\end{equation}
By calculating each amplitudes at these sampled momentum points, one can construct and solve a system of linear equations to extract the numerical values of $t_{LS;L^{\prime}S^{\prime}}$. 
Since the physical amplitude $A$ is basis-independent, it can be equivalently written in either representation: 
\begin{equation}
	A = \sum_{LS}g_{LS}C_{LS} =\sum_{L^{\prime}S^{\prime}}g_{L^{\prime}S^{\prime}}H_{L^{\prime}S^{\prime}},
\end{equation}
here, $g_{L,S}$ and $g_{L^{\prime}S^{\prime}}$ are the energy depend coefficients in the respective basis. Assuming that $g_{LS}$ are known in the $C_{LS}$ basis, one can straightforwardly compute the transformed coefficients $g_{L^{\prime}S^{\prime}}$ in the $H_{L^{\prime}S^{\prime}}$ basis via: 
\begin{equation}
	g_{L^{\prime}S^{\prime}}^{\prime} = \sum_{LS} g_{LS}t_{LS;L^{\prime}S^{\prime}}.
\end{equation}
This transformation ensures the physical amplitude remains invariant under the change of basis, thereby guaranteeing consistency of physical predictions such as decay widths, angular distributions, and differential cross sections.

\bibliographystyle{JHEP}
\bibliography{ref}

@article{Li:2022qff,
    author = "Li, Xiao-yu and Dong, Xiang-Kun and Jing, Hao-Jie",
    title = "{Spin-orbit amplitudes for decays with arbitrary spin}",
    eprint = "2212.06417",
    archivePrefix = "arXiv",
    primaryClass = "hep-ph",
    doi = "10.1016/j.nuclphysa.2023.122761",
    journal = "Nucl. Phys. A",
    volume = "1040",
    pages = "122761",
    year = "2023"
}

@article{Jing:2023rwz,
    author = "Jing, Hao-Jie and Ben, Di and Wu, Shu-Ming and Wu, Jia-Jun and Zou, Bing-Song",
    title = "{Covariant orbital-spin scheme for any spin based on irreducible tensor}",
    eprint = "2301.01575",
    archivePrefix = "arXiv",
    primaryClass = "hep-ph",
    doi = "10.1007/JHEP06(2023)039",
    journal = "JHEP",
    volume = "06",
    pages = "039",
    year = "2023"
}

@article{Jing:2024mag,
    author = "Jing, Hao-Jie and Wu, Shu-Ming and Wu, Jia-Jun",
    title = "{Algorithms for partial wave amplitudes in the covariant L-S scheme}",
    eprint = "2405.06576",
    archivePrefix = "arXiv",
    primaryClass = "hep-ph",
    doi = "10.1103/PhysRevD.110.016014",
    journal = "Phys. Rev. D",
    volume = "110",
    number = "1",
    pages = "016014",
    year = "2024"
}

@article{Nakayama:2002mu,
    author = "Nakayama, K. and Speth, J. and Lee, T. S. H.",
    title = "{eta meson production in NN collisions}",
    eprint = "nucl-th/0202012",
    archivePrefix = "arXiv",
    doi = "10.1103/PhysRevC.65.045210",
    journal = "Phys. Rev. C",
    volume = "65",
    pages = "045210",
    year = "2002"
}

@article{Benmerrouche:1994uc,
    author = "Benmerrouche, M. and Mukhopadhyay, Nimai C. and Zhang, J. F.",
    title = "{Effective Lagrangian approach to the theory of eta photoproduction in the N* (1535) region}",
    eprint = "hep-ph/9412248",
    archivePrefix = "arXiv",
    reportNumber = "RPI-N97-94, SAL-TH-94-04",
    doi = "10.1103/PhysRevD.51.3237",
    journal = "Phys. Rev. D",
    volume = "51",
    pages = "3237--3266",
    year = "1995"
}

@article{Benmerrouche:1996ij,
    author = "Benmerrouche, M. and Mukhopadhyay, Nimai C. and Zhang, J. F.",
    title = "{Model independent extraction of the N* (1535) electrostrong form-factor from eta electroproduction}",
    eprint = "nucl-th/9611032",
    archivePrefix = "arXiv",
    reportNumber = "RPI-96-N110",
    doi = "10.1103/PhysRevLett.77.4716",
    journal = "Phys. Rev. Lett.",
    volume = "77",
    pages = "4716--4719",
    year = "1996"
}

@article{Marangotto:2019ucc,
    author = "Marangotto, Daniele",
    title = "{Helicity Amplitudes for Generic Multibody Particle Decays Featuring Multiple Decay Chains}",
    eprint = "1911.10025",
    archivePrefix = "arXiv",
    primaryClass = "hep-ph",
    doi = "10.1155/2020/6674595",
    journal = "Adv. High Energy Phys.",
    volume = "2020",
    pages = "6674595",
    year = "2020"
}

@article{VonHippel:1972fg,
    author = "Von Hippel, F. and Quigg, C.",
    title = "{Centrifugal-barrier effects in resonance partial decay widths, shapes, and production amplitudes}",
    doi = "10.1103/PhysRevD.5.624",
    journal = "Phys. Rev. D",
    volume = "5",
    pages = "624--638",
    year = "1972"
}

@article{tfpwa,
title = {Open-source framework {TF-PWA} package},
journal = {GitHub link: https://github.com/jiangyi15/tf-pwa},
year = {2020},
url = {https://github.com/jiangyi15/tf-pwa},
author = {Y. Jiang and others}
}

@article{Wang:2020giv,
    author = "Wang, Mengzhen and Jiang, Yi and Liu, Yinrui and Qian, Wenbin and Lyu, Xiaorui and Zhang, Liming",
    title = "{A novel method to test particle ordering and final state alignment in helicity formalism}",
    eprint = "2012.03699",
    archivePrefix = "arXiv",
    primaryClass = "hep-ex",
    doi = "10.1088/1674-1137/abf139",
    journal = "Chin. Phys. C",
    volume = "45",
    number = "6",
    pages = "063103",
    year = "2021"
}

@article{JPAC:2019ufm,
    author = "Mikhasenko, M. and others",
    collaboration = "JPAC",
    title = "{Dalitz-plot decomposition for three-body decays}",
    eprint = "1910.04566",
    archivePrefix = "arXiv",
    primaryClass = "hep-ph",
    reportNumber = "JLAB-THY-19-3070",
    doi = "10.1103/PhysRevD.101.034033",
    journal = "Phys. Rev. D",
    volume = "101",
    number = "3",
    pages = "034033",
    year = "2020"
}

@article{BESIII:2022udq,
    author = "Ablikim, Medina and others",
    collaboration = "BESIII",
    title = "{Partial wave analysis of the charmed baryon hadronic decay $ {\Lambda}_c^{+} ${\textrightarrow} {\ensuremath{\Lambda}}{\ensuremath{\pi}}$^{+}${\ensuremath{\pi}}$^{0}$}",
    eprint = "2209.08464",
    archivePrefix = "arXiv",
    primaryClass = "hep-ex",
    doi = "10.1007/JHEP12(2022)033",
    journal = "JHEP",
    volume = "12",
    pages = "033",
    year = "2022"
}

@article{BESIII:2023wfi,
    author = "Ablikim, Medina and others",
    collaboration = "BESIII",
    title = "{Determination of Spin-Parity Quantum Numbers of X(2370) as 0-+ from J/{\ensuremath{\psi}}{\textrightarrow}{\ensuremath{\gamma}}KS0KS0{\ensuremath{\eta}}'}",
    eprint = "2312.05324",
    archivePrefix = "arXiv",
    primaryClass = "hep-ex",
    doi = "10.1103/PhysRevLett.132.181901",
    journal = "Phys. Rev. Lett.",
    volume = "132",
    number = "18",
    pages = "181901",
    year = "2024"
}

@article{BESIII:2023sbq,
    author = "Ablikim, Medina and others",
    collaboration = "BESIII",
    title = "{Measurement of the cross sections for e+e-{\textrightarrow}{\ensuremath{\eta}}{\ensuremath{\pi}}+{\ensuremath{\pi}}- at center-of-mass energies between 2.00 and 3.08~GeV}",
    eprint = "2310.10452",
    archivePrefix = "arXiv",
    primaryClass = "hep-ex",
    doi = "10.1103/PhysRevD.108.L111101",
    journal = "Phys. Rev. D",
    volume = "108",
    number = "11",
    pages = "L111101",
    year = "2023"
}

@article{BESIII:2023xac,
    author = "Ablikim, Medina and others",
    collaboration = "BESIII",
    title = "{Measurement of the e$^{+}$e$^{-}$ {\textrightarrow} $ {K}_S^0{K}_L^0 ${\ensuremath{\pi}}$^{0}$ cross sections from $ \sqrt{s} $ = 2.000 to 3.080 GeV}",
    eprint = "2309.13883",
    archivePrefix = "arXiv",
    primaryClass = "hep-ex",
    doi = "10.1007/JHEP01(2024)180",
    journal = "JHEP",
    volume = "01",
    pages = "180",
    year = "2024"
}

@article{BESIII:2023htx,
    author = "Ablikim, M. and others",
    collaboration = "BESIII",
    title = "{Observation of D+{\textrightarrow}KS0a0(980)+ in the Amplitude Analysis of D+{\textrightarrow}KS0{\ensuremath{\pi}}+{\ensuremath{\eta}}}",
    eprint = "2309.05760",
    archivePrefix = "arXiv",
    primaryClass = "hep-ex",
    doi = "10.1103/PhysRevLett.132.131903",
    journal = "Phys. Rev. Lett.",
    volume = "132",
    number = "13",
    pages = "131903",
    year = "2024"
}

@article{barrierSet,
  title = {Partial wave analysis of $\ensuremath{\psi}(3686)\ensuremath{\rightarrow}{K}^{+}{K}^{\ensuremath{-}}\ensuremath{\eta}$},
  author = {Ablikim, M. and others},
  collaboration = {BESIII Collaboration},
  journal = {Phys. Rev. D},
  volume = {101},
  issue = {3},
  pages = {032008},
  numpages = {12},
  year = {2020},
  month = {Feb},
  publisher = {American Physical Society},
  doi = {10.1103/PhysRevD.101.032008},
  url = {https://link.aps.org/doi/10.1103/PhysRevD.101.032008}
}

@article{GSmodel,
  title = {Finite-{W}idth {C}orrections to the {V}ector-{M}eson-{D}ominance {P}rediction for $\ensuremath{\rho}\ensuremath{\rightarrow}{e}^{+}{e}^{\ensuremath{-}}$},
  author = {Gounaris, G. J. and Sakurai, J. J.},
  journal = {Phys. Rev. Lett.},
  volume = {21},
  issue = {4},
  pages = {244--247},
  numpages = {0},
  year = {1968},
  month = {Jul},
  publisher = {American Physical Society},
  doi = {10.1103/PhysRevLett.21.244},
  url = {https://link.aps.org/doi/10.1103/PhysRevLett.21.244}
}

@article{Chung:1998,
  title = {General formulation of covariant helicity-coupling amplitudes},
  author = {Chung, S. U.},
  journal = {Phys. Rev. D},
  volume = {57},
  issue = {1},
  pages = {431--442},
  numpages = {0},
  year = {1998},
  month = {Jan},
  publisher = {American Physical Society},
  doi = {10.1103/PhysRevD.57.431},
  url = {https://link.aps.org/doi/10.1103/PhysRevD.57.431}
}

@article{Filippini:1995yc,
    author = "Filippini, V. and Fontana, A. and Rotondi, A.",
    title = "{Covariant spin tensors in meson spectroscopy}",
    doi = "10.1103/PhysRevD.51.2247",
    journal = "Phys. Rev. D",
    volume = "51",
    pages = "2247--2261",
    year = "1995"
}

@article{Arkani-Hamed:2017jhn,
    author = "Arkani-Hamed, Nima and Huang, Tzu-Chen and Huang, Yu-tin",
    title = "{Scattering amplitudes for all masses and spins}",
    eprint = "1709.04891",
    archivePrefix = "arXiv",
    primaryClass = "hep-th",
    reportNumber = "NCTS-TH/1714, NCTS-TH-1714",
    doi = "10.1007/JHEP11(2021)070",
    journal = "JHEP",
    volume = "11",
    pages = "070",
    year = "2021"
}

@article{Dulat:2011rn,
    author = "Dulat, Sayipjamal and Wu, Jia-Jun and Zou, B. S.",
    title = "{Proposal and theoretical formalism for studying baryon radiative decays from $J/\psi \to B^*\bar{B} + B\bar{B^*} \to \gamma B \bar{B}$}",
    eprint = "1103.5810",
    archivePrefix = "arXiv",
    primaryClass = "hep-ph",
    doi = "10.1103/PhysRevD.83.094032",
    journal = "Phys. Rev. D",
    volume = "83",
    pages = "094032",
    year = "2011"
}

@article{Dulat:2005in,
    author = "Dulat, Sayipjamal and Zou, Bing-Song",
    title = "{Covariant tensor formalism for partial wave analyses of $\psi$ decays into $\gamma B\bar B$, $\gamma\gamma V$ and $\psi(2S)\to\gamma\chi_{c0,1,2}$ with $\chi_{c0,1,2}\to K\bar K \pi^+\pi^- $ and $2\pi^+2\pi^-$}",
    eprint = "hep-ph/0508087",
    archivePrefix = "arXiv",
    doi = "10.1140/epja/i2005-10140-1",
    journal = "Eur. Phys. J. A",
    volume = "26",
    pages = "125--134",
    year = "2005",
    note = "[Erratum: Eur.Phys.J.A 56, 275 (2020)]"
}

@article{Zou:2002ar,
    author = "Zou, B. S. and Bugg, D. V.",
    title = "{Covariant tensor formalism for partial wave analyses of psi decay to mesons}",
    eprint = "hep-ph/0211457",
    archivePrefix = "arXiv",
    doi = "10.1140/epja/i2002-10135-4",
    journal = "Eur. Phys. J. A",
    volume = "16",
    pages = "537--547",
    year = "2003"
}

@article{Zou:2002yy,
    author = "Zou, B. S. and Hussain, F.",
    title = "{Covariant L-S scheme for the effective N*NM couplings}",
    eprint = "hep-ph/0210164",
    archivePrefix = "arXiv",
    reportNumber = "BIHEP-TH-2002-29",
    doi = "10.1103/PhysRevC.67.015204",
    journal = "Phys. Rev. C",
    volume = "67",
    pages = "015204",
    year = "2003"
}

@article{Liang:2002tk,
    author = "Liang, W. H. and Shen, P. N. and Wang, J. X. and Zou, B. S.",
    title = "{Theoretical formalism and Monte Carlo study of partial wave analysis for J / psi --{\ensuremath{>}} p anti-p omega}",
    doi = "10.1088/0954-3899/28/2/311",
    journal = "J. Phys. G",
    volume = "28",
    pages = "333--343",
    year = "2002"
}

@article{Chung:1993da,
    author = "Chung, S. U.",
    title = "{Helicity coupling amplitudes in tensor formalism}",
    doi = "10.1103/PhysRevD.56.4419",
    journal = "Phys. Rev. D",
    volume = "48",
    pages = "1225--1239",
    year = "1993",
    note = "[Erratum: Phys.Rev.D 56, 4419 (1997)]"
}

@book{Weinberg_1995, place={Cambridge}, title={The Quantum Theory of Fields}, publisher={Cambridge University Press}, author={Weinberg, Steven}, year={1995}}

@book{Gourgoulhon2013SRGF,
  author    = {Gourgoulhon, \'Eric},
  title     = {{Special Relativity in General Frames}},
  doi       = {10.1007/978-3-642-37276-6},
  publisher = {Springer},
  address   = {Berlin, Heidelberg},
  series    = {Graduate Texts in Physics},
  year      = {2013}
}

@book{Rose1995AngularMomentum,
  author    = {Rose, M. E.},
  title     = {Elementary Theory of Angular Momentum},
  publisher = {Dover Publications},
  year      = {1995}
}

@article{Zemach:1965ycj,
    author = "Zemach, Charles",
    title = "{Use of angular momentum tensors}",
    doi = "10.1103/PhysRev.140.B97",
    journal = "Phys. Rev.",
    volume = "140",
    pages = "B97--B108",
    year = "1965"
}

@article{Jacob:1959at,
    author = "Jacob, M. and Wick, G. C.",
    title = "{On the General Theory of Collisions for Particles with Spin}",
    doi = "10.1006/aphy.2000.6022",
    journal = "Annals Phys.",
    volume = "7",
    pages = "404--428",
    year = "1959"
}

@article{Rarita:1941mf,
    author = "Rarita, William and Schwinger, Julian",
    title = "{On a theory of particles with half integral spin}",
    doi = "10.1103/PhysRev.60.61",
    journal = "Phys. Rev.",
    volume = "60",
    pages = "61",
    year = "1941"
}

@article{Chung:2007nn,
    author = "Chung, Suh-Urk and Friedrich, Jan",
    title = "{Covariant helicity-coupling amplitudes: A New formulation}",
    eprint = "0711.3143",
    archivePrefix = "arXiv",
    primaryClass = "hep-ph",
    reportNumber = "TU-E18-06-0801",
    doi = "10.1103/PhysRevD.78.074027",
    journal = "Phys. Rev. D",
    volume = "78",
    pages = "074027",
    year = "2008"
}

@article{ParticleDataGroup:2020ssz,
    author = "Zyla, P. A. and others",
    collaboration = "Particle Data Group",
    title = "{Review of Particle Physics}",
    doi = "10.1093/ptep/ptaa104",
    journal = "PTEP",
    volume = "2020",
    number = "8",
    pages = "083C01",
    year = "2020"
}

@article{Sarantsev:2019xxm,
    author = "Sarantsev, A. V. and Matveev, M. and Nikonov, V. A. and Anisovich, A. V. and Thoma, U. and Klempt, E.",
    title = "{Hyperon II: Properties of excited hyperons}",
    eprint = "1907.13387",
    archivePrefix = "arXiv",
    primaryClass = "nucl-ex",
    doi = "10.1140/epja/i2019-12880-5",
    journal = "Eur. Phys. J. A",
    volume = "55",
    number = "10",
    pages = "180",
    year = "2019"
}

@article{Parke:1986gb,
    author = "Parke, Stephen J. and Taylor, T. R.",
    title = "{An Amplitude for $n$ Gluon Scattering}",
    reportNumber = "FERMILAB-PUB-86-042-T",
    doi = "10.1103/PhysRevLett.56.2459",
    journal = "Phys. Rev. Lett.",
    volume = "56",
    pages = "2459",
    year = "1986"
}

@article{Bern:1996je,
    author = "Bern, Zvi and Dixon, Lance J. and Kosower, David A.",
    title = "{Progress in one loop QCD computations}",
    eprint = "hep-ph/9602280",
    archivePrefix = "arXiv",
    reportNumber = "SLAC-PUB-7111, UCLA-96-TEP-5, SACLAY-SPH-T-96-10",
    doi = "10.1146/annurev.nucl.46.1.109",
    journal = "Ann. Rev. Nucl. Part. Sci.",
    volume = "46",
    pages = "109--148",
    year = "1996"
}

@inproceedings{Dixon:1996wi,
    author = "Dixon, Lance J.",
    title = "{Calculating scattering amplitudes efficiently}",
    booktitle = "{Theoretical Advanced Study Institute in Elementary Particle Physics (TASI 95): QCD and Beyond}",
    eprint = "hep-ph/9601359",
    archivePrefix = "arXiv",
    reportNumber = "SLAC-PUB-7106",
    pages = "539--584",
    month = "1",
    year = "1996"
}

@article{Elvang:2013cua,
    author = "Elvang, Henriette and Huang, Yu-tin",
    title = "{Scattering Amplitudes}",
    eprint = "1308.1697",
    archivePrefix = "arXiv",
    primaryClass = "hep-th",
    month = "8",
    year = "2013"
}

@inproceedings{Cheung:2017pzi,
    author = "Cheung, Clifford",
    title = "{TASI lectures on scattering amplitudes.}",
    booktitle = "{Theoretical Advanced Study Institute in Elementary Particle Physics}: {Anticipating the Next Discoveries in Particle Physics}",
    eprint = "1708.03872",
    archivePrefix = "arXiv",
    primaryClass = "hep-ph",
    reportNumber = "CALT-TH-2017-041",
    doi = "10.1142/9789813233348_0008",
    pages = "571--623",
    year = "2018"
}

@article{Travaglini:2022uwo,
    author = "Travaglini, Gabriele and others",
    title = "{The SAGEX review on scattering amplitudes}",
    eprint = "2203.13011",
    archivePrefix = "arXiv",
    primaryClass = "hep-th",
    reportNumber = "SAGEX-22-01",
    doi = "10.1088/1751-8121/ac8380",
    journal = "J. Phys. A",
    volume = "55",
    number = "44",
    pages = "443001",
    year = "2022"
}

@article{Badger:2023eqz,
    author = "Badger, Simon and Henn, Johannes and Plefka, Jan Christoph and Zoia, Simone",
    title = "{Scattering Amplitudes in Quantum Field Theory}",
    eprint = "2306.05976",
    archivePrefix = "arXiv",
    primaryClass = "hep-th",
    doi = "10.1007/978-3-031-46987-9",
    journal = "Lect. Notes Phys.",
    volume = "1021",
    pages = "pp.",
    year = "2024"
}

@article{Benincasa:2007xk,
    author = "Benincasa, Paolo and Cachazo, Freddy",
    title = "{Consistency Conditions on the S-Matrix of Massless Particles}",
    eprint = "0705.4305",
    archivePrefix = "arXiv",
    primaryClass = "hep-th",
    reportNumber = "UWO-TH-07-09",
    month = "5",
    year = "2007"
}

@article{Witten:2003nn,
    author = "Witten, Edward",
    title = "{Perturbative gauge theory as a string theory in twistor space}",
    eprint = "hep-th/0312171",
    archivePrefix = "arXiv",
    doi = "10.1007/s00220-004-1187-3",
    journal = "Commun. Math. Phys.",
    volume = "252",
    pages = "189--258",
    year = "2004"
}

@article{Britto:2004ap,
    author = "Britto, Ruth and Cachazo, Freddy and Feng, Bo",
    title = "{New recursion relations for tree amplitudes of gluons}",
    eprint = "hep-th/0412308",
    archivePrefix = "arXiv",
    doi = "10.1016/j.nuclphysb.2005.02.030",
    journal = "Nucl. Phys. B",
    volume = "715",
    pages = "499--522",
    year = "2005"
}

@article{Britto:2005fq,
    author = "Britto, Ruth and Cachazo, Freddy and Feng, Bo and Witten, Edward",
    title = "{Direct proof of tree-level recursion relation in Yang-Mills theory}",
    eprint = "hep-th/0501052",
    archivePrefix = "arXiv",
    doi = "10.1103/PhysRevLett.94.181602",
    journal = "Phys. Rev. Lett.",
    volume = "94",
    pages = "181602",
    year = "2005"
}

@article{Kleiss:1985yh,
    author = "Kleiss, R. and Stirling, W. James",
    title = "{Spinor Techniques for Calculating p anti-p ---{\ensuremath{>}} W+- / Z0 + Jets}",
    reportNumber = "CERN-TH-4186-85",
    doi = "10.1016/0550-3213(85)90285-8",
    journal = "Nucl. Phys. B",
    volume = "262",
    pages = "235--262",
    year = "1985"
}

@article{Hagiwara:1985yu,
    author = "Hagiwara, Kaoru and Zeppenfeld, D.",
    title = "{Helicity Amplitudes for Heavy Lepton Production in $e^+ e^-$ Annihilation}",
    reportNumber = "DESY-85-133, DTP/85/24",
    doi = "10.1016/0550-3213(86)90615-2",
    journal = "Nucl. Phys. B",
    volume = "274",
    pages = "1--32",
    year = "1986"
}

@article{Kleiss:1988xr,
    author = "Kleiss, R. and Stirling, W. James",
    title = "{TOP QUARK PRODUCTION AT HADRON COLLIDERS: SOME USEFUL FORMULAE}",
    doi = "10.1007/BF01548856",
    journal = "Z. Phys. C",
    volume = "40",
    pages = "419--423",
    year = "1988"
}

@article{Dittmaier:1998nn,
    author = "Dittmaier, Stefan",
    title = "{Weyl-van der Waerden formalism for helicity amplitudes of massive particles}",
    eprint = "hep-ph/9805445",
    archivePrefix = "arXiv",
    reportNumber = "CERN-TH-98-143",
    doi = "10.1103/PhysRevD.59.016007",
    journal = "Phys. Rev. D",
    volume = "59",
    pages = "016007",
    year = "1998"
}

@article{Schwinn:2005pi,
    author = "Schwinn, Christian and Weinzierl, Stefan",
    title = "{Scalar diagrammatic rules for Born amplitudes in QCD}",
    eprint = "hep-th/0503015",
    archivePrefix = "arXiv",
    reportNumber = "MZ-TH-05-05",
    doi = "10.1088/1126-6708/2005/05/006",
    journal = "JHEP",
    volume = "05",
    pages = "006",
    year = "2005"
}

@article{Schwinn:2006ca,
    author = "Schwinn, Christian and Weinzierl, Stefan",
    title = "{SUSY ward identities for multi-gluon helicity amplitudes with massive quarks}",
    eprint = "hep-th/0602012",
    archivePrefix = "arXiv",
    reportNumber = "MZ-TH-06-01, PITHA-06-01",
    doi = "10.1088/1126-6708/2006/03/030",
    journal = "JHEP",
    volume = "03",
    pages = "030",
    year = "2006"
}

@article{Badger:2005zh,
    author = "Badger, S. D. and Glover, E. W. Nigel and Khoze, V. V. and Svrcek, P.",
    title = "{Recursion relations for gauge theory amplitudes with massive particles}",
    eprint = "hep-th/0504159",
    archivePrefix = "arXiv",
    reportNumber = "IPPP-05-13, DCPT-05-26, PUPT-2158",
    doi = "10.1088/1126-6708/2005/07/025",
    journal = "JHEP",
    volume = "07",
    pages = "025",
    year = "2005"
}

@article{Badger:2005jv,
    author = "Badger, S. D. and Glover, E. W. Nigel and Khoze, Valentin V.",
    title = "{Recursion relations for gauge theory amplitudes with massive vector bosons and fermions}",
    eprint = "hep-th/0507161",
    archivePrefix = "arXiv",
    reportNumber = "IPPP-05-39, DCPT-05-78",
    doi = "10.1088/1126-6708/2006/01/066",
    journal = "JHEP",
    volume = "01",
    pages = "066",
    year = "2006"
}

@article{Conde:2016vxs,
    author = "Conde, Eduardo and Marzolla, Andrea",
    title = "{Lorentz Constraints on Massive Three-Point Amplitudes}",
    eprint = "1601.08113",
    archivePrefix = "arXiv",
    primaryClass = "hep-th",
    doi = "10.1007/JHEP09(2016)041",
    journal = "JHEP",
    volume = "09",
    pages = "041",
    year = "2016"
}

@article{Conde:2016izb,
    author = "Conde, Eduardo and Joung, Euihun and Mkrtchyan, Karapet",
    title = "{Spinor-Helicity Three-Point Amplitudes from Local Cubic Interactions}",
    eprint = "1605.07402",
    archivePrefix = "arXiv",
    primaryClass = "hep-th",
    doi = "10.1007/JHEP08(2016)040",
    journal = "JHEP",
    volume = "08",
    pages = "040",
    year = "2016"
}

@article{Basile:2024ydc,
    author = "Basile, Thomas and Joung, Euihun and Mkrtchyan, Karapet and Mojaza, Matin",
    title = "{Spinor-helicity representations of particles of any mass in dS4 and AdS4 spacetimes}",
    eprint = "2401.02007",
    archivePrefix = "arXiv",
    primaryClass = "hep-th",
    reportNumber = "Imperial-TP-KM-2024-01",
    doi = "10.1103/PhysRevD.109.125003",
    journal = "Phys. Rev. D",
    volume = "109",
    number = "12",
    pages = "125003",
    year = "2024"
}

@article{Chen:2017gtx,
    author = "Chen, Hong and Ping, Rong-Gang",
    title = "{Coherent helicity amplitude for sequential decays}",
    eprint = "1704.05184",
    archivePrefix = "arXiv",
    primaryClass = "hep-ph",
    doi = "10.1103/PhysRevD.95.076010",
    journal = "Phys. Rev. D",
    volume = "95",
    number = "7",
    pages = "076010",
    year = "2017"
}

@article{Habermann:2024sxs,
    author = "Habermann, Kai and Mikhasenko, Mikhail",
    title = "{Wigner rotations for cascade reactions}",
    eprint = "2409.06913",
    archivePrefix = "arXiv",
    primaryClass = "hep-ph",
    doi = "10.1103/PhysRevD.111.056015",
    journal = "Phys. Rev. D",
    volume = "111",
    number = "5",
    pages = "056015",
    year = "2025"
}

@article{Dong:2020mbk,
    author = "Dong, Xiang and Su, Kexin and Cai, Hao and Zhu, Kai and Gao, Yonggui",
    title = "{Tensor Amplitudes for Partial Wave Analysis of within Helicity Frame}",
    eprint = "2012.10913",
    archivePrefix = "arXiv",
    primaryClass = "hep-th",
    doi = "10.3390/universe10090376",
    journal = "Universe",
    volume = "10",
    number = "9",
    pages = "376",
    year = "2024"
}

@article{Zhu:1999pu,
    author = "Zhu, Jie-Jie and Yan, Mu-Lin",
    title = "{Covariant amplitudes for mesons}",
    eprint = "hep-ph/9903349",
    archivePrefix = "arXiv",
    month = "3",
    year = "1999"
}

@article{Richman:1984gh,
    author = "Richman, Jeffrey D.",
    title = "{An Experimenter's Guide to the Helicity Formalism}",
    reportNumber = "CALT-68-1148",
    month = "6",
    year = "1984"
}

@article{Anisovich:2024cmb,
    author = "Anisovich, A. V. and Nikonov, K. V. and Sarantsev, A. V.",
    title = "{The Covariant Approach for the Radiative Decay of the ${J/}\Psi$ Meson and Relation to Amplitudes Extracted in the Helicity Basis}",
    eprint = "2406.01172",
    archivePrefix = "arXiv",
    primaryClass = "hep-ph",
    doi = "10.1134/S1063778824700467",
    journal = "Phys. Atom. Nucl.",
    volume = "87",
    number = "4",
    pages = "498--504",
    year = "2024"
}

@article{Salgado:2013dja,
    author = "Salgado, Carlos W. and Weygand, Dennis P.",
    title = "{On the Partial-Wave Analysis of Mesonic Resonances Decaying to Multiparticle Final States Produced by Polarized Photons}",
    eprint = "1310.7498",
    archivePrefix = "arXiv",
    primaryClass = "nucl-ex",
    reportNumber = "JLAB-PHY-13-1821",
    doi = "10.1016/j.physrep.2013.11.005",
    journal = "Phys. Rept.",
    volume = "537",
    pages = "1--58",
    year = "2014"
}

@article{Pascalutsa:1998kcd,
    author = "Pascalutsa, V.",
    title = "{Covariant description of pion nucleon dynamics}",
    journal = "Hadronic J. Suppl.",
    volume = "16",
    pages = "1--123",
    year = "2001"
}

@article{Lutz:2003fm,
    author = "Lutz, M. F. M. and Kolomeitsev, E. E.",
    title = "{On meson resonances and chiral symmetry}",
    eprint = "nucl-th/0307039",
    archivePrefix = "arXiv",
    reportNumber = "GSI-PREPRINT-2003-19",
    doi = "10.1016/j.nuclphysa.2003.11.009",
    journal = "Nucl. Phys. A",
    volume = "730",
    pages = "392--416",
    year = "2004"
}

@article{JPAC:2017vtd,
    author = "Mikhasenko, M. and Pilloni, A. and Nys, J. and Albaladejo, M. and Fernandez-Ramirez, C. and Jackura, A. and Mathieu, V. and Sherrill, N. and Skwarnicki, T. and Szczepaniak, A. P.",
    collaboration = "JPAC",
    title = "{What is the right formalism to search for resonances?}",
    eprint = "1712.02815",
    archivePrefix = "arXiv",
    primaryClass = "hep-ph",
    reportNumber = "JLAB-THY-17-2606",
    doi = "10.1140/epjc/s10052-018-5670-y",
    journal = "Eur. Phys. J. C",
    volume = "78",
    number = "3",
    pages = "229",
    year = "2018"
}

@article{JPAC:2018dfc,
    author = "Pilloni, A. and Nys, J. and Mikhasenko, M. and Albaladejo, M. and Fern{\'a}ndez-Ram{\'\i}rez, C. and Jackura, A. and Mathieu, V. and Sherrill, N. and Skwarnicki, T. and Szczepaniak, A. P.",
    collaboration = "JPAC",
    title = "{What is the right formalism to search for resonances? II. The pentaquark chain}",
    eprint = "1805.02113",
    archivePrefix = "arXiv",
    primaryClass = "hep-ph",
    reportNumber = "JLAB-THY-18-2700",
    doi = "10.1140/epjc/s10052-018-6177-2",
    journal = "Eur. Phys. J. C",
    volume = "78",
    number = "9",
    pages = "727",
    year = "2018"
}

@article{Peters:2004qw,
    author = "Peters, Klaus J.",
    editor = "Melchiorri, F. and Rephaeli, Y.",
    title = "{A Primer on partial wave analysis}",
    eprint = "hep-ph/0412069",
    archivePrefix = "arXiv",
    doi = "10.1142/S0217751X06034811",
    journal = "Int. J. Mod. Phys. A",
    volume = "21",
    pages = "5618--5624",
    year = "2006"
}

@article{Chung:1971ri,
    author = "Chung, Suh Urk",
    title = "{SPIN FORMALISMS}",
    reportNumber = "CERN-71-08",
    doi = "10.5170/CERN-1971-008",
    month = "3",
    year = "1971"
}

@inproceedings{Haber:1994pe,
    author = "Haber, Howard E.",
    title = "{Spin formalism and applications to new physics searches}",
    booktitle = "{21st Annual SLAC Summer Institute on Particle Physics: Spin Structure in High-energy Processes (School: 26 Jul - 3 Aug, Topical Conference: 4-6 Aug) (SSI 93)}",
    eprint = "hep-ph/9405376",
    archivePrefix = "arXiv",
    reportNumber = "SCIPP-93-49, NSF-ITP-94-30",
    pages = "231--272",
    month = "4",
    year = "1994"
}

@article{Choi:2019aig,
    author = "Choi, Seong Youl and Jeong, Jae Hoon and Song, Ji Ho",
    title = "{General Spin Analysis from Angular Correlations in Two-Body Decays}",
    eprint = "1903.00166",
    archivePrefix = "arXiv",
    primaryClass = "hep-ph",
    doi = "10.1140/epjp/s13360-020-00132-1",
    journal = "Eur. Phys. J. Plus",
    volume = "135",
    number = "2",
    pages = "210",
    year = "2020"
}

@article{Gratrex:2015hna,
    author = "Gratrex, James and Hopfer, Markus and Zwicky, Roman",
    title = "{Generalised helicity formalism, higher moments and the $B \to K_{J_K}(\to K \pi) \bar{\ell}_1 \ell_2$ angular distributions}",
    eprint = "1506.03970",
    archivePrefix = "arXiv",
    primaryClass = "hep-ph",
    reportNumber = "CP3-ORIGINS-2015-017, DIAS-2015-17, CP3-Origins-2015-017 DNRF90, DIAS-2015-17",
    doi = "10.1103/PhysRevD.93.054008",
    journal = "Phys. Rev. D",
    volume = "93",
    number = "5",
    pages = "054008",
    year = "2016"
}

@article{Wigner:1939cj,
    author = "Wigner, Eugene P.",
    editor = "Kim, Y. S. and Zachary, W. W.",
    title = "{On Unitary Representations of the Inhomogeneous Lorentz Group}",
    doi = "10.2307/1968551",
    journal = "Annals Math.",
    volume = "40",
    pages = "149--204",
    year = "1939"
}

@article{Zemach:1963bc,
    author = "Zemach, Charles",
    title = "{Three pion decays of unstable particles}",
    doi = "10.1103/PhysRev.133.B1201",
    journal = "Phys. Rev.",
    volume = "133",
    pages = "B1201",
    year = "1964"
}

@article{Anisovich:2006bc,
    author = "Anisovich, A. V. and Sarantsev, A. V.",
    title = "{Partial decay widths of baryons in the spin-momentum operator expansion method}",
    eprint = "hep-ph/0605135",
    archivePrefix = "arXiv",
    doi = "10.1140/epja/i2006-10102-1",
    journal = "Eur. Phys. J. A",
    volume = "30",
    pages = "427--441",
    year = "2006"
}

@article{Wick:1962zz,
    author = "Wick, G. C.",
    title = "{Angular momentum states for three relativistic particles}",
    doi = "10.1016/0003-4916(62)90059-3",
    journal = "Annals Phys.",
    volume = "18",
    pages = "65--80",
    year = "1962"
}

@article{Berman:1965gi,
    author = "Berman, S. M. and Jacob, Maurice",
    title = "{SYSTEMATICS OF ANGULAR POLARIZATION DISTRIBUTIONS IN THREE-BODY DECAYS}",
    reportNumber = "SLAC-PUB-0073",
    doi = "10.1103/PhysRev.139.B1023",
    journal = "Phys. Rev.",
    volume = "139",
    pages = "B1023--B1038",
    year = "1965"
}

@article{Anisovich:2004zz,
    author = "Anisovich, A. and Klempt, E. and Sarantsev, A. and Thoma, U.",
    title = "{Partial wave decomposition of pion and photoproduction amplitudes}",
    eprint = "hep-ph/0407211",
    archivePrefix = "arXiv",
    doi = "10.1140/epja/i2004-10125-6",
    journal = "Eur. Phys. J. A",
    volume = "24",
    pages = "111--128",
    year = "2005"
}

@article{Anisovich:2001ra,
    author = "Anisovich, A. V. and Anisovich, V. V. and Markov, V. N. and Matveev, M. A. and Sarantsev, A. V.",
    title = "{Moment operator expansion for the two meson, two photon and fermion anti-fermion states}",
    eprint = "hep-ph/0105330",
    archivePrefix = "arXiv",
    doi = "10.1088/0954-3899/28/1/302",
    journal = "J. Phys. G",
    volume = "28",
    pages = "15--32",
    year = "2002"
}

@article{Ajaltouni:2019log,
    author = "Ajaltouni, Z. J. and Di Salvo, E.",
    title = "{Polarization in (quasi-)two-body decays and new physics}",
    eprint = "1904.02442",
    archivePrefix = "arXiv",
    primaryClass = "hep-ph",
    doi = "10.1140/epjc/s10052-021-09284-5",
    journal = "Eur. Phys. J. C",
    volume = "79",
    number = "12",
    pages = "989",
    year = "2019",
    note = "[Erratum: Eur.Phys.J.C 81, 487 (2021)]"
}

@article{Arunprasath:2016tfq,
    author = "Arunprasath, V. and Godbole, Rohini M. and Singh, Ritesh K.",
    title = "{Polarization of a top quark produced in the decay of a gluino or a stop in an arbitrary frame}",
    eprint = "1612.03803",
    archivePrefix = "arXiv",
    primaryClass = "hep-ph",
    doi = "10.1103/PhysRevD.95.076012",
    journal = "Phys. Rev. D",
    volume = "95",
    number = "7",
    pages = "076012",
    year = "2017"
}

@article{Boudjema:2009fz,
    author = "Boudjema, Fawzi and Singh, Ritesh K.",
    title = "{A Model independent spin analysis of fundamental particles using azimuthal asymmetries}",
    eprint = "0903.4705",
    archivePrefix = "arXiv",
    primaryClass = "hep-ph",
    reportNumber = "LAPTH-1322",
    doi = "10.1088/1126-6708/2009/07/028",
    journal = "JHEP",
    volume = "07",
    pages = "028",
    year = "2009"
}

@article{Gao:2010qx,
    author = "Gao, Yanyan and Gritsan, Andrei V. and Guo, Zijin and Melnikov, Kirill and Schulze, Markus and Tran, Nhan V.",
    title = "{Spin Determination of Single-Produced Resonances at Hadron Colliders}",
    eprint = "1001.3396",
    archivePrefix = "arXiv",
    primaryClass = "hep-ph",
    reportNumber = "FERMILAB-PUB-10-011-E",
    doi = "10.1103/PhysRevD.81.075022",
    journal = "Phys. Rev. D",
    volume = "81",
    pages = "075022",
    year = "2010"
}

@article{Aguilar-Saavedra:2012bvs,
    author = "Aguilar-Saavedra, J. A. and Herrero-Hahn, R. V.",
    title = "{Model-independent measurement of the top quark polarisation}",
    eprint = "1208.6006",
    archivePrefix = "arXiv",
    primaryClass = "hep-ph",
    doi = "10.1016/j.physletb.2012.11.031",
    journal = "Phys. Lett. B",
    volume = "718",
    pages = "983--987",
    year = "2013"
}

@article{Gao:2023jtq,
    author = "Gao, Yuanning and Rong, Tianze and Yang, Zhenwei and Zhang, Chenjia and Zhang, Yanxi",
    title = "{A scheme to fix multiple solutions in amplitude analyses*}",
    eprint = "2302.13862",
    archivePrefix = "arXiv",
    primaryClass = "hep-ph",
    doi = "10.1088/1674-1137/ad2674",
    journal = "Chin. Phys. C",
    volume = "48",
    number = "5",
    pages = "053001",
    year = "2024"
}

@article{Harshman:2004bw,
    author = "Harshman, N. L. and Licata, N.",
    title = "{Clebsch-Gordan coefficients for the extended quantum-mechanical Poincare group and angular correlations of decay products}",
    eprint = "hep-ph/0407299",
    archivePrefix = "arXiv",
    doi = "10.1016/j.aop.2004.11.013",
    journal = "Annals Phys.",
    volume = "317",
    pages = "182--202",
    year = "2005"
}

@article{Matveev:2018gry,
    author = "Matveev, M. A. and Sarantsev, A. V. and Semenov-Tian-Shansky, K. M. and Semenova, A. N.",
    title = "{Differential technique for the covariant orbital angular momentum operators}",
    eprint = "1802.07999",
    archivePrefix = "arXiv",
    primaryClass = "hep-ph",
    doi = "10.1140/epja/i2018-12539-9",
    journal = "Eur. Phys. J. A",
    volume = "54",
    number = "6",
    pages = "108",
    year = "2018"
}
\end{document}